# A New Single Equation of State to describe the Dynamic Viscosity and Self-Diffusion Coefficient for all Fluid Phases of Water from 200 K to 1800 K based on a New Original Microscopic Model

F. Aitken, F. Volino
Univ. Grenoble Alpes, CNRS, Grenoble INP, G2ELab, F-38000 Grenoble, France.

**Abstract**. A microscopic model able to describe simultaneously the dynamic viscosity and the self-diffusion coefficient of fluids is presented. This model is shown to emerge from the introduction of fractional calculus in a usual model of condensed matter physics based on an elastic energy functional. The main feature of the model is that all measurable quantities are predicted to depend in a non-trivial way on external parameters (e.g. the experimental set-up geometry, in particular the sample size). On the basis of an unprecedented comparative analysis of a collection of published experimental data, the modeling is applied to the case of water in all its fluid phases, in particular in the supercooled phase. It is shown that the discrepancies in the literature data are only apparent and can be quantitavely explained by the different experimental configurations (e.g. geometry, calibration). This approach makes it possible to reproduce the water viscosity with a better accuracy than the 2008 IAPWS formulation and also with a more physically satisfying modeling of the isochors. Moreover, it also allows the modeling within experimental accuracy of the translational self-diffusion data available in the literature in all water fluid phases. Finally, the formalism of the model makes it possible to understand the "anomalies" observed on the dynamic viscosity and self-diffusion coefficient and their possible links.

## 1 Introduction

The relationship between transport properties such as viscosity and self-diffusion coefficient is fundamental, for example, to understand the mobility measurements of microscopic objects (e.g. electrons, ions, etc., Refs. 1-2) in fluids. In addition, the dynamic viscosity is a property used for characterizing fluid flow regimes, i.e. systems out of the thermodynamic equilibrium while self-diffusion characterizes systems very close to thermodynamic equilibrium. Their knowledge is fundamental for understanding the physics of liquids and for the development of a large number of technical and industrial applications. However, the determination of these quantities requires many complex experimental sets-up in order to cover wide fluid domains. Therefore, the experimental data are often parceled or very scarce on a scale of two to three times the critical parameters especially for self-diffusion coefficients. Developing theoretical models is both a way of linking data to one another but it is also a means for understanding the microscopic structure, or at least to propose some image of this structure. Various approaches have been developed for constructing an equation of state for the viscosity of fluids from some highly theoretical basis to simple empirical correlations; extensive critical reviews of such models are readily available (e.g. Refs. 3-5). More often the best representation of a large set of experimental data (i.e. covering a very large domain of the phase diagram) is obtained by developing semi-empirical (or semi-theoretical) equations of state. The semi-theoretical descriptions based on thermostatistical concepts require the knowledge of the intermolecular forces, which can be very complex for dense fluids and often partly unknown, have difficulties or simply are not able to account for some experimental results such as:

1. The irreducible differences between different experiments (i.e. the error bars between authors do not overlap), and also the variations of these transport coefficients versus different parameters that differ significantly following the authors. As long as the differences can be attributed in some cases to calibration methods, the variations differences are no longer related to the apparatus and reveal different physical behaviors.
2. The existence of a measurable static shear elasticity for sub-millimeter liquid volumes sample sizes, or the increase of the shear viscosity at sufficiently high shear rates in rotating viscometers.

These difficulties of molecular theories are due to the fact that they ignore the interactions at large distance which play an important role in flow phenomena and in the notion of elasticity.

To circumvent these difficulties, we will extend to all the fluid states the approach described in Ref. 6 which was developed to describe usual liquid states. The basic idea of the model developed in Ref. 6 is to replace the microscopic description in the real space (where the relevant variables are the positions and the molecular pair potentials) by a description in the reciprocal space where the positions are replaced by wave-vectors and the intermolecular energies by energy densities. The variables in the reciprocal space are global variables that affect the whole sample contrary to the local variables of the thermostatistical approach. This kind of description is the one made in solids with normal modes of vibration and elastic constants but it cannot be applied in its usual form to liquids where the position fluctuations are very large compared to the intermolecular distances. Here the model developed in Ref. 6 allows describing the properties of liquids and more generally of fluids (i.e. phase behavior and transitions) by introducing fractional derivatives, in particular the displacement gradients is replaced by fractional gradients as shown in Appendix A.

In this paper we will focus on point 1 raised previously while the second point concerning the shear elasticity will be the subject of a forthcoming publication.

We have chosen here to present the application of this model to the case of water since a very large number of rather accurate experimental data exists from the supercooled liquid to supercritical states at high temperatures (~1200 K) and pressures from very low values until 6 GPa. Moreover, for some fluid states, many experiments carried out by different authors with different set-up exists which allow showing the difficulties mentioned above concerning the usual approaches. Water is also known as a complex fluid due to “abnormal” behavior that it exhibits. For example, it is well-known that, compared to most liquids, liquid water behaves anomalously below about 30 °C, such that increasing pressure along an isotherm reduces viscosity instead of increasing it (i.e. the viscosity passes through a pressure minimum on isotherms and then increases). Equally, this “abnormal” behaviors is reflected on the self-diffusion coefficient which increases with density (within the range of about 0.9 g/cm$^3$ up to about 1.1 g/cm$^3$, at temperatures below 42 °C) in contrast to usual liquids where increasing density decreases self-diffusion. These behaviors have been explained in the frame of the two-state model with bending and breakage of the hydrogen bonds and can be reproduced by the current reference equation of state for viscosity given in Ref. 7. But the form of this current reference equation of state (i.e. Eq. (2) in Ref. 7) for viscosity has not the "usual" additive decomposition with a dilute-gas limit part added to a residual part because of a particular behavior of the isochors with this kind of approach: in the gaseous phase, some states have a lower viscosity value than the viscosity of the dilute-gas limit part so a crossing exists between isochors and the dilute-gas representation. This particularity is very little discussed and no clear explanation has been proposed (see for example Ref. 8). As the present approach is entirely based on the description of the viscosity isochors, we will show that the crossing of isochors is an "artifact" due to the particular choice of the decomposition. So, the "abnormal" behaviors of water discussed above can be interpreted as the consequence of the set of behaviors of viscosity isochors.
By using this model, we will also show that it is possible to propose:

- a good representation of the viscosity current reference equation of state from Ref. 7 with smaller uncertainties that its own uncertainty in its domain of definition (see fig. 24 of Ref. 7),
- a better representation of most of the experimental data (*e.g.* Abramson, Cappi, Bonilla *et al.*) and to explain the differences between the authors by taking into account the specificities of the experimental set-up,
- a simultaneous description of the self-diffusion coefficient and viscosity for all fluid states which allows in particular to understand why the self-diffusion coefficient has no divergent variation close to the critical point contrary to the viscosity behavior.

An extensive review of the viscosity data of water has been done in Ref. 7 and will not be reproduced here. On the other hand, the data analyzed here will be processed in the light of the authors' various writings in order not only to interpret as little as possible the reading of these authors but also to discuss some critical analysis. This paper therefore includes an important historical part necessary to put all the analyzed data in perspective. This leads to a rather voluminous paper but it is necessary to detail the analysis that is done on each set of data since here we do not simply assign weights to each of these sets and then treat them as a weighted whole. In other words, each data set requires its own analysis because the analysis depends entirely on how the data were obtained as we will show.

In a first part we will detail the general modeling which is common to all liquids then in a second part we will particularize the equations of state in the case of water: we will begin by analyzing the viscosity data and then we will finish with the self-diffusion coefficient data. In addition, some Appendices are included that strongly support the theoretical developments presented here.

## 2 The General Basis of the Model

The general concepts of this microscopic approach have already been described in Ref. 6, hence we present an improved summary of the main results.

The model can be identified with a lattice model where uniaxial objects $j$ ($j = 1, 2, \ldots, \mathcal{N}$) are located at each site $\vec{r}_{f,j}$ of a regular lattice and point along one preferred direction $z$ (called the director). At mechanical equilibrium (no flow), the position and orientation of these objects fluctuate about their equilibrium positions $\vec{r}_{f,j}$ and equilibrium orientation $\theta_{f,j}$ due to thermal motions. The deviations from the equilibrium positions are described by random variables $\vec{u}_j$ of components $(u_{x,j}, u_{y,j}, u_{z,j})$ and $\theta_j$.

In this model, the uniaxial objects must not be identified with the molecules in the general case, but to a set of $n_B$ molecules called a "basic unit", where, in the *normal*[1] liquid phase, $n_B$ is the number of molecules in the unit cells of the crystal which exists at lower temperature, under the same pressure (i.e. under these conditions we can identify the basic unit to the unit cell). Indeed, the unit cell is the minimum entity which has the symmetry of the phase, generally apolar and not optically active (whereas individual and sufficiently complex molecules are very often polar and chiral). The unit cell allows reproducing the crystal by translation. It turns out that in many usual liquids, $n_B$ is equal to 4.

As a consequence, $\vec{u}_j$ describes the displacement of the center of mass of the basic units with respect to their equilibrium position $\vec{r}_{f,j}$ (external motions) but not the displacement of the center of mass of the molecules inside the basic unit (internal motions). The description of the internal motions is not useful for the properties we are considering in this paper. An example of such modeling of internal motions, usually studied by spectroscopic methods, can be found in Ref. 6 (Part VI, section 7).

### 2.1. Thermodynamic Equilibrium Properties

The starting point of the model is the assumption that the instantaneous displacements $\vec{u}_j(\vec{r})$ and $\theta_j(\vec{r})$ can be developed into Fourier series (whose coefficients we shall later refer to as *elastic modes*) on the lattice where $\vec{r}$ is a vector whose origin is at the equilibrium position $\vec{r}_{f,j}$. For the sake of shortness, we will limit now the presentation to translational aspects, because only these modes will be useful for the description of shear viscosity and self-diffusion coefficients. The treatment of rotational aspects is very similar. For component $u_{x,j}$ of $\vec{u}_j$, we have:

$$u_{x,j}(\vec{r}) = \sum_{\vec{q}} u_{x,j}(\vec{q}) e^{i\vec{q}\cdot\vec{r}} = \sum_{\vec{q}} u_{xq,j} e^{i\vec{q}\cdot\vec{r}} \tag{1}$$

where the amplitudes $u_{x,j}(\vec{q})$ are new statistically independent random variables. Each mode is characterized by its wave-vector $\vec{q}$ and its polarization. Since $u_{x,j}$ is the sum of a large number of independent random variables, the statistics of $u_{x,j}$ is Gaussian according to the central limit theorem, and the variance of its distribution function is $< u_{x,j}^2 >$. Throughout the rest of this paper, it will be assumed that the distribution function is the same for all the basic units and therefore the index $j$ will now be omitted.

Assuming isotropy of the reciprocal space, the wave-vector moduli $q$ are limited at short length scales by a cut-off wave-vector $q_c$, and towards long length scales by a wave-vector $q_c / N$. The quantity $N/q_c$ will be called the "*coherence length*" or the "*fluctuative distance*" since it is the size of the sample volume where such modes, which describe long range interactions between objects, exist. There are two transverse modes associated with shear and one longitudinal mode associated with compression. The density in reciprocal space of such modes for a volume $V$ is $3V/(2\pi)^3$. For shear viscosity only tranverse modes are pertinent so their density in reciprocal space is $2V/(2\pi)^3$.

[1] The liquid states where the density is close to the density of the liquid at the triple point throughout this article will systematically be referred to as the "normal liquid phase".

If one assumes that all lattice sites are occupied by one basic unit (as is the case in normal liquid phase), it is easy to show that we have:

$$q_c^3 \approx q_{c0}^3 = \frac{6\pi^2 n_V}{n_B} = \frac{6\pi^2 \rho\, \mathfrak{N}_a}{M\, n_B} \tag{2}$$

where $n_V$ is the number of molecules per unit volume ($n_V/n_B$ is thus the number of basic units per unit volume), $\rho$ the mass per unit volume, $M$ the molar mass and $\mathfrak{N}_a$ the Avogadro number. Thus, in the normal liquid phase the cut-off wave-vector modulus $q_c$ depends on thermodynamic variables as $\rho^{1/3}$.

From the energetic point of view, the effect of the displacements of the objects from equilibrium position is to increase the energy of the sample compared to the perfectly ordered state at zero Kelvin. In Ref. 6 a functional form has been postulated for this excess elastic energy: it was assumed that to each mode $\vec{q}$ one can associate an individual elastic constant $K_q$ which depends on $q$ according to the following power law $K_q = K\left(\frac{q}{q_c}\right)^{\nu}$ where the coefficient $K$ is a global elastic constant and the exponent $\nu$ is a real number, both being functions of thermodynamic variables. It is shown in Appendix A that this assumption is mathematically equivalent of using fractional gradients in the definition of the excess energy functional form. By combining Eq. (A.5) with the assumption of the equipartition of thermal energy, namely that the average energy per mode is $k_B T/2$, where $k_B$ is the Boltzmann constant and $T$ the absolute temperature, and integrating over all $q$ modes, the following results has been obtained for the expression of the full mean square displacement $<\left|u_x^2\right|>$ due to transverse modes only:

$$<\left|u_x^2\right|> = \frac{k_B T q_c}{\pi^2 K} H_N(\nu) \tag{3}$$

Due to isotropy of the model and for the sake of simplification, we shall later note $<\left|u_x^2\right|>$ simply $\left\langle u^2 \right\rangle$.

The quantity $H_N(\nu)$ is a function which has the following definition:

$$H_N(\nu) = \frac{N^{\nu-1} - 1}{\nu - 1} \tag{4}$$

Note that the function $H_N(\nu)$ is nothing but the real Riemann zeta function series limited to a finite value $N$, where the sum has been replaced by an integral. It plays a fundamental role in the model and admits very distinct behaviors with $N$ depending on the values taken by $\nu$. For example: $H_N(-\infty)=0$, $H_N(0)=1-\frac{1}{N}$, $H_N(1)=\ln(N)$, $H_N(2)=N-1$, $H_N(\infty)=\infty$. It follows that a singularity appears for $v=1$. Indeed for $N$ going to infinity, the function $H_N(v)$ and consequently the mean square displacement $\langle u^2\rangle$ tends to infinity for $v\geq 1$, while it remains finite and tends to $1/(1-v)$ for $v<1$. Assuming that $v$ is an increasing function of temperature, the temperature for which $v=1$, should naturally be identified with a translational order-disorder phase transition temperature, probably the glass transition.

It has been empirically demonstrated for nematic liquid crystals (which correspond to the rotational version of the model, Ref. 6, parts I and II) that the dependence of exponent $v$ versus thermodynamic parameters is:

$$1-v=\left(\frac{T_t}{T}-1\right)^{\frac{1}{2}} \qquad \text{for } T<T_t \text{ (i.e. ordered phase)} \tag{5a}$$

and

$$v-1=\left(1-\frac{T_t}{T}\right)^{\frac{1}{4}} \qquad \text{for } T>T_t \text{ (i.e. disordered phase)} \tag{5b}$$

where $k_BT_t$ corresponds to a microscopic characteristic elastic energy which depends, *a priori*, on temperature $T$, on density $\rho$, and possibly on other intensive variables such as static magnetic field and/or electric field. $T_t$ is associated with the nematic-isotropic transition temperature which occurs when $T_t = T$. According to Eq. (5), it is seen that despite the different functional forms on each side of the transition, the function $v$ versus $T/T_t$ is continuous and varies from $-\infty$ to $2$ when $T/T_t$ varies from 0 to $\infty$. Moreover, it is interesting to note that all derivatives of $v$ with respect to the variable $T/T_t$ are infinite at the transition, and gives the $v$ function an auto-similarity property around the transition.

Again, by analogy with the rotational case, the transition temperature $T_t$ for the present translational case can be defined by the following expression:

$$T_t(\rho,T,\ldots)=\frac{K(\rho,T,\ldots)}{2k_Bq_{c0}^3} \tag{6}$$

In order to characterize the equilibrium properties, we need to know the parameter $n_B$ and the following thermodynamic functions: $K(\rho,T,...)$, $q_c(\rho,T,...)$ and $N(\rho,T,...)$.

The only dimensionless parameters are $N$ and $n_B$. Therefore it would be interesting to obtain a scaling for the other two parameters. By taking as a scaling for $T_t$ the critical temperature $T_c$, and $q_{c0,\text{crit}}=\sqrt[3]{\frac{6\pi^2\rho_c\mathcal{N}_a}{Mn_B}}$ for $q_c$, Eq. (6) leads us to the following natural scale for the quantity $K(\rho,T,...)$:

$$K_0=\frac{12\pi^2}{n_B}\frac{R_gT_c\rho_c}{M} \tag{7}$$

where $R_g$ represents the perfect gas constant, and ($T_c$, $\rho_c$) are the critical parameters of the fluid (*i.e.* absolute temperature and density respectively). This scaling is interesting because it turns out that for many liquids $K_0$ is numerically close to the extrapolated latent heat of vaporization per unit volume of the corresponding liquid at $T = 0$ K (i.e. the cohesion energy density, see Appendix B).

Finally, the four following non-dimensional parameters ($K^* = K/K_0$, $q_c^* = q_c/q_{c0,\mathrm{crit}}$, $N$, $n_B$) are necessary to represent the equilibrium properties of the medium. The above scaling is such that $K^*$ and $q_c^*$ are close to unity in the normal liquid phase. With these scaling, Eq. (6) can be rewritten on the following non-dimensional form: $T_t^* = \dfrac{T_t}{T_c} = \dfrac{K^*}{\left(q_{c0}^*\right)^3}$. By using Eq. (2) for the definition of $q_{c0}$, $T_t^*$ can be simply written: $T_t^* = K^* \dfrac{\rho_c}{\rho}$.

$K^*$ is globally an increasing function of the density $\rho$. It turns out that the shear elastic parameter $K^*$ can be written as:

$$K^*(\rho,T) = K_0^*(\rho)\, f_K(\rho,T) \tag{8}$$

$K_0^*(\rho)$ represents the elastic parameter density dependence for all the fluid phases whereas the function $f_K(\rho,T)$ plays only a role in the supercooled or superfluid liquid phases and can be set to unity everywhere else. In order to have a finite value of the transition temperature $T_t^*$ when the density tends to zero, it is necessary that the function $K_0^*(\rho)$ tends to zero at this limit. We also deduce that the transition temperature $T_t^*$ for all the usual fluid phases is only a function of the density given by $K_0^*(\rho)\dfrac{\rho_c}{\rho}$.

By definition the parameter $N$ corresponds to a reduced characteristic macroscopic distance given in term of $q_c^{-1}$. In order to have a better physical understanding of this parameter, we introduce the new characteristic distance $d_N$ related to $N$ by:

$$N - 1 = d_N \frac{q_{c0,\mathrm{crit}}}{2\pi} \tag{9}$$

The length $q_{c0,\mathrm{crit}}/2\pi$ is an arbitrary choice which allows obtaining values for the coherence parameter $d_N$ of the order of the sample size, at least in the normal liquid phase.

At this point, it is interesting to introduce the quantity $\xi = N_{\mathrm{corr}}/q_c$ which represents the distance over which the fluctuations of the unit cells are correlated. This notion is important for analyzing data associated with collective motions such as light scattering or coherent neutron scattering, but also useful for our present purpose as we will see in section 3. $N_{\mathrm{corr}}$ is defined as (Ref. 6, part I):

$$N_{\text{corr}} = \left(\frac{N-1}{H_N(v)}\right)^{\frac{1}{2}} \qquad \text{for } T < T_t \text{ (ordered phase)} \tag{10a}$$

and

$$N_{\text{corr}} = \left(\frac{N-1}{H_N(v)}\right)^{\frac{1}{4}} \qquad \text{for } T > T_t \text{ (disordered phase)} \tag{10b}$$

For $T >> T_t$, $N_{\text{corr}}$ is very close to 1 which means that the fluctuations between neighboring objects located on the lattice are no more correlated. On the contrary for $T << T_t$, $N_{\text{corr}}$ is very large, in practice it is limited by the sample size and the fact that no real experiment can be made at $T = 0$ K. It is important to note that Eq. (10) defines the correlation length in the bulk. Close to a wall, it is possible in some experiments to find that this length is locally different from the value in the bulk. For example, this is the case for some viscosity experiments in the supercooled phase (see Fig. 32).
To illustrate this correlation length, we present in Appendix C an example for the particular case of water, which is the fluid used to test the model in this paper.

## 2.2. Transport Properties

In the previous section we have considered that the system is at rest in thermodynamic equilibrium conditions. To analyze transport properties, we must consider that the system has been set out of equilibrium by an external perturbation which transfers energy to the system. For example, this external perturbation could be related to the stress applied to the system to produce flow for studying the viscosity of a fluid or the magnetic field gradient for studying the translational self-diffusion of molecules in a fluid by NMR. It will be admitted here without any justification that the influence of external disturbances can be ignored for the analysis of the usual viscosity and diffusion coefficient measurement experiments. This will be justified in another forthcoming paper which will detail how these external perturbations must be taken into account in the model by introducing a new term in the elastic energy functionnal (see Ref. 6, part III, for the introduction of this concept).

Transport coefficients require the introduction of a time scale $\tau$, or equivalently of a new *macroscopic* distance $d$ defined by $\tau = d/c_0$, where $c_0$ is the characteristic shear elastic celerity given by $c_0 = \sqrt{K/\rho}$. The distance $d$ will be called the "*dissipative distance*" because it represents the characteristic region of the sample volume where the velocity gradient is important and thus where the dissipation mainly occurs.

The introduction of the distance $d$ allows us to express the coherence parameter $d_N$ as a function that depends on the geometry of the experiment such as:

$$d_N = f_N(\rho,T)\,d \tag{11}$$

where $f_N(\rho,T)$ is a functional form which depends only on the fluid properties. It should be stressed that $N$ being a property of the medium at thermodynamical equilibrium, it cannot depend on quantities that reflect the out-of-equilibrium state of the system associated with the measuring process. Thus, the distance $d$ in Eq. (11), that was introduced above to define a characteristic time, should be considered in the expression of $d_N$ as a typical distance of the system. For example, in a Poiseuille flow in a cylindrical tube, $d$ is both the distance over which velocity gradient exists and the radius of the tube.

When the system is out of equilibrium, it evolves in time, and one can define the conditional density probability function $p(u,t \mid u_0,t_0)$, which represents the probability to find the basic unit at position $u$ at time $t$, if it was at position $u_0$ at time $t = t_0$. It is well known that $p(u,t \mid u_0,t_0)$, averaged over all possible initial positions $u_0$, is the solution of the standard diffusion equation, characterized by the translational self-diffusion coefficient $D_t$. This equation should be solved by taking into account the experimental conditions, in particular the sample geometry. In the present modeling, it is assumed that in all cases, $D_t$ is given by the following relation:

$$D_t = \frac{3}{2}\frac{\langle u^2 \rangle}{\tau} = \frac{3k_B\,T\,q_c\,H_N(v)}{2\pi^2 K\tau} = \frac{3k_B\,T\,q_c\,H_N(v)}{2\pi^2 d\sqrt{K\rho}} \tag{12}$$

where the coefficient 3/2 comes from the fact that there are three kinds of modes for the diffusion instead of 2 for shear.

The analogy with the Stokes-Einstein (SE) law suggests that a dynamic shear viscosity can be defined by the expression:

$$\eta_l = \frac{K}{H_N(v)}\tau = \frac{d}{H_N(v)}\sqrt{\rho K} \tag{13}$$

where $K_N = K/H_N(v)$ represents a *macroscopic* static shear elastic modulus which characterizes both the deformation of the lattice and the displacement of the objects on this lattice when the sample is submitted to a shear strain. It is interesting to note that $K_N\xi/d$ can be identified with the threshold shear stress $\sigma_T$ measured in a rheology experiment performed at very low frequency (typically lower than 1 Hz). Such shear elasticity has already been measured in sub-millimeter sample sizes (Ref. 9), and these data on water will be analyzed in a forthcoming paper. When the applied shear stress is much greater than $\sigma_T$, as is the case in the vast majority of viscosity experiments, one recovers the usual situation where the fluid behaves as a newtonian fluid and in this case the viscosity experimental data can be analyzed by using solutions of Navier-Stokes equations.
By combining relations (12) and (13) one can rewrite Eq. (12) in the following SE form:

$$D_t = \frac{k_B T}{\frac{2\pi^2}{3}\eta_l q_c^{-1}} \tag{14}$$

So, Eq. (13) can be understandable as the "liquid term" of the viscosity and this explains the subscript $l$ in this relation.

In the case of usual approaches, a "gaseous term" is added to take into account experimental data in the gas phase. But in the present modeling, the gaseous part represents the gas (*i.e.* isolated molecules or atoms) released by the action of the shear stresses, i.e. the description of the medium consists in admitting a two-fluid model where this gas exists in a state of equilibrium and cannot be dissociated from the liquid part in this state but any setting out of equilibrium makes it possible to allow different properties of this gas to emerge from the liquid part (i.e. at equilibrium the Gibbs phase rule always applies so that the variance for a pure fluid is equal to 2). Since this gas diffuses in the velocity gradient, it contributes to the medium transport, thus to dissipation through an additional viscosity term. It can be visualized as a gas which diffuses with a diffusion coefficient $(D_t)_{Knu}$ between the layers that slip on each other in a conventional image of laminar flow. This gas spreads in the whole sample and has its own density $\rho_{Knu}$. As long as the density of the released gas is very low compared to the fluid density $\rho$, the flow corresponding to this gas is in a Knudsen regime such that its full mean square displacement $\langle u^2 \rangle_{Knu}$ is limited by a given distance $\delta$ defined by the experimental set-up. In case of a Poiseuille flow, $\delta$ is the radius of the tube, so that in this experimental configuration we expect $\delta = d$.
The contribution of this gaz to the total viscosity is related to the self-diffusion coefficient $(D_t)_{Knu}$ of this gas along the velocity gradient by the well-known formula:

$$\eta_{Knu} = \rho_{Knu}(D_t)_{Knu} = \rho_{Knu}\left(\frac{\langle u^2 \rangle}{\tau}\right)_{Knu} = \frac{\rho_{Knu}\delta^2}{\tau_{Knu}} \tag{15}$$

where $\rho_{Knu}$ is the density of this gas. Considering $\tau_{Knu} = \delta / c_{0,Knu}$ with $c_{0,Knu} = \sqrt{R_g T / M}$, Eq. (15) becomes:

$$\eta_{Knu} = \rho_{Knu}\delta\sqrt{\frac{R_g T}{M}} \tag{16}$$

In order to have only one unknown variable, Eq. (16) is rewritten on the following way by using the characteristic length scale $q_{c0,\mathrm{crit}}$:

$$\eta_{Knu} = \tilde{\rho}_{Knu}\sqrt{\frac{R_g T}{M}}\frac{2\pi}{q_{c0,\mathrm{crit}}} \tag{17}$$

where $\tilde{\rho}_{Knu} = \rho_{Knu}\ \delta\, q_{c0,\mathrm{crit}} / (2\pi)$.

The expression for determining the experimental value of the viscosity of a fluid considered as a Newtonian fluid is thus obtained by adding Eq. (13) and Eq. (17):

$$\eta(\rho,T) = \eta_l(\rho,T) + \eta_{Knu}(\rho,T) \tag{18}$$

As we will see in section 3.1, Eq. (18) tends towards a constant at the zero-density limit. However, within this limit it is necessary that $\eta$ tends towards zero when density tends towards zero: indeed, no dissipation occurs if there is no molecule/atom in the system. Therefore, Eq. (18) can only be an approximation of a more general model. It is then important to note that Eq. (18) is only valid as long as the following two conditions are true: $\rho >> \rho_{Knu}$ and $\langle u^2 \rangle << d^2$. Otherwise, a more general expression of the following form should be considered:

$$\eta(\rho,T) = \eta_l(\rho,T)\frac{\rho}{\rho+\rho_{l0}} + \eta_{Knu}(\rho,T)\frac{\rho}{\rho+\rho_{g0}} \tag{19}$$

where $\rho_{l0}$ and $\rho_{g0}$ represents the densities for which the following equalities $\langle u^2 \rangle = d^2$ and $\rho_{Knu}(\rho,T) = \rho$ are fulfilled, respectively. The relevance of Eq. (19) can only be demonstrated in media where experiments with very low densities can be carried out. An example of the application of Eq. (19) can be found in section 4.1.3 with Fig. 68.

For usual self-diffusion experiments, the fluid is in a state closer to equilibrium states than with viscosity experiments such that there is no released gas (*i.e.* because there is no linear velocity gradient or no flow). As a consequence, the translational self-diffusion coefficient for the fluid is simply described by Eq. (12). In Eq. (12) the parameter $q_c$ can be approximated by Eq. (2) for the normal liquid states but everywhere else $q_c$ is a more complex function of $\rho$ and $T$ such that:

$$q_c(\rho,T) = f_{q_c}(\rho,T)\; q_{c0}(\rho) \tag{20}$$

where $f_{q_c}(\rho,T)$ is a functional form which depends only on the fluid properties.

This viscosity modeling with a decomposition into two terms while only one of these term accounts for the self-diffusion coefficient allows in particular to understand and describe why the viscosity “seems to diverge” in the critical region while the self-diffusion coefficient shows no particular behavior.

In summary, the description of the transport properties introduces two new parameters in addition to those necessary for the description of thermodynamic equilibrium properties (i.e. the transport properties are determined on both equilibrium properties and some properties specific to the experiment conducted for their measurements). These two parameters are $d$ and $\tilde{\rho}_{Knu}$. The distance $d$ represents a distance scale characteristic of the dissipative processes in the sample associated with the method of measurement (as a consequence $d$ depends on the flow geometry) and cannot be determined *a priori* in the frame of the theory. But it turns out that $d$ has a numerical value close to a simple distance in the experimental set-up used to measure the viscosity (typically the radius of the tube in a Poiseuille flow). Eq. (11) implies that $d_N$ cannot be determined either in the frame of the present theory but it turns out that $d_N$ represents a characteristic length of the order of magnitude of the sample size. The parameter $\tilde{\rho}_{Knu}$ is a functional form of $\rho$ and $T$ which depends only on the fluid properties.

The remaining of the paper will be entirely devoted to the application of this general modeling to the analysis of viscosity and self-diffusion coefficients datasets in all fluid phases of water.

## 3 Application to the Dynamic Viscosity of Water

Since Ice I is hexagonal non compact with 4 molecules per unit cell, for water we thus have $n_B = 4$.

For all the existing water viscosity measurements in the literature, the approximation $\rho_{Knu} << \rho$ is always verified and thus only Eq. (18) can be considered (see section 4.1.3). We must also consider that the flow regime is always Newtonian for the present modeling to be valid (i.e. the shear stress is always much greater than the threshold shear stress $\sigma_T$ ).

The viscosity isochors given by the equation of state from Ref. 7 has been used to determine the different parameters (and their evolutions with density and temperature) corresponding to the authors' model below the triple point liquid density (the current reference equation of state will be named IAPWS08). Above this density value, the data from Cappi (Ref. 10) and Abramson (Ref. 11) have been used instead of the IAPWS08 formulation which does not provide a correct representation of these data (see §3.2). The data from Cappi are the only ones which have an enough large extension with pressure and temperature to cover the gap between the IAPWS08 formulation and the data from Abramson (Ref. 11). The data from Cappi (Ref. 10) are given in the form of relative values of viscosity. In Ref. 12 the relative data of Cappi where normalized at atmospheric pressure with those of Bingham *et al.* (Ref. 13). But here, in order to have a consistent representation of these data we have renormalized them by using the IAPWS08 formulation at atmospheric pressure. The data from Abramson (Ref. 11) are given in the form of absolute values which are numerically consistent with those of Cappi in their overlapping area. So, to keep consistency with the data from Cappi, these data have been also renormalized. It can be noticed that the differences generated are globally very small.

Also close to the critical point, we have used the re-evaluated data of Rivkin *et al.* (Ref. 14) given in Table 4 of Ref. 7 instead of the IAPWS08 formulation to determine the function $\tilde{\rho}_{Knu,0,\mathrm{crit}}(\rho,T)$ which allows to describe the local viscosity increase due to an increase of the released gas in this critical region compared to the "normal" behavior of the fluid in all other regions.

The equations describing the evolution of the different parameters for the particular case of water are grouped in the Tables 1-3 (the calculation programs with the documentation can be freely downloaded by following Ref. 15). Some parameters are represented by functions defined by parts with a coupling to the maximum density on the atmospheric isobar because it is extremely difficult to find a single function that covers seven orders of magnitude and which has limits imposed at both ends. This makes it possible to obtain simpler mathematical functions from both sides of the connection line. Some functions require the knowledge of the density along the saturated vapor pressure curve: in this case we have used the equations given in Ref. 7.

| *Name* (unit) | *Value or formula* |
|---|---|
| $n_B$ | 4 |
| $K_0$ (GPa) | 2.8474 |
| $f_K$ | $f_K(\rho,T)=1-\dfrac{(1-f_{K1}(T))\exp\left(-(T/T_{\mathrm{tr}})^{500}\right)}{1+\left(\dfrac{\lvert\rho-\rho_{1\mathrm{atm}}(T)\rvert}{0.303+0.185/\lvert T-201\rvert}\right)^{\gamma_K(T)}}$ with $\gamma_K(T)=\dfrac{31.36}{\left(1+\dfrac{\lvert T-203.44\rvert}{18}\right)\left(1+\left(\dfrac{427.28}{T+220}\right)^{160}\right)}$, $f_{K1}(T)=0.84313+(1-0.84313)\left(1-\dfrac{1}{1+(T/229.46)^{52.16}}\right)\exp\left(-\left(\dfrac{220}{T}\right)^{200}\right)$ |
| $K_0^*$ | See Appendix D for the expressions of $\rho_{1\mathrm{atm}}(T)$<br>$K_0^*(\rho)=\begin{cases} K_{0,\mathrm{lim}}^*(\rho)+K_{0,\mathrm{res}}^*(\rho) & \text{if } \rho \le \rho_{\mathrm{max,1atm}} \\ K_{0,\mathrm{high}}^*(\rho) & \text{if } \rho > \rho_{\mathrm{max,1atm}} \end{cases}$<br>with<br>$K_{0,\mathrm{lim}}^*(\rho)=0.001232\left(\frac{\rho}{\rho_c}\right)^3,$<br>$K_{0,\mathrm{res}}^*(\rho)=\frac{K_1^*}{1+\left(\frac{0.795202}{\rho}\right)^{3.87805}} + 0.00041231\exp\left(-3.681\left\|\ln\left(\frac{\rho}{0.1984}\right)\right\|^{1.44508}\right)\exp\left(-\left(\frac{\rho}{0.261567}\right)^{50}-\left(\frac{0.041}{\rho}\right)^{7}\right),$<br>$K_1^*=0.399491\exp\left(-\left\|\frac{\rho-0.9826}{0.14825}\right\|^{1.73526}\right)+0.99941\exp\left(-\left\|\frac{\rho-\rho_{\mathrm{max},1atm}}{0.78739}\right\|^{8.43}\right)$<br>$+0.024272\exp\left(-\left\|\frac{\rho-\rho_{\mathrm{max},1atm}}{0.031521}\right\|^{3.23}\right)$<br>$K_{0,\mathrm{high}}^*(\rho)=1.03858+4.0997(\rho-\rho_{\mathrm{max,1atm}})^2+74.7(\rho-\rho_{\mathrm{max,1atm}})^{0.613}\|\rho-1.1104\|^{1.564}$<br>$-74.2361(\rho-\rho_{\mathrm{max,1atm}})^{0.61419}\|\rho-1.11056\|^{1.56311}$ |

Table 1. Parameters for Eq. (8): $T$ in K and $\rho$ in g/cm$^3$. $T_{\mathrm{tr}}$ = 273.16 K represents the triple point temperature and $\rho_{\mathrm{max,1atm}}$ = 0.999972 g/cm$^3$ represents the maximum density value along the atmospheric isobar.

| *Name* | *formula* |
|---|---|
| $f_N$ | $f_N(\rho)=\begin{cases} f_{N,\mathrm{lim}}(\rho)\times(1+f_{N,\mathrm{res}}(\rho)) & \text{if } \rho > \rho_{\mathrm{max,1atm}} \\ f_{N,\mathrm{high}}(\rho) & \text{if } \rho > \rho_{\mathrm{max,1atm}} \end{cases}$<br>with<br>$f_{N,\mathrm{lim}}(\rho)=6.563\left(\frac{\rho}{\rho_c}\right)^2,$<br>$f_{N,\mathrm{res}}(\rho)=\left\{1.30504\left(\frac{\rho}{1.0003}\right)^{2.64}\exp\left(-\left\|\frac{\rho-1.0003}{0.11696}\right\|^{1.5886}\right)\right.$<br>$\left.+0.27124\left(\frac{0.99174}{\rho}\right)^{2.66}\exp\left(-\left\|\frac{\rho-0.99174}{0.02065}\right\|^{2.2084}\right)\right\}\exp\left(-\left(\frac{0.74699}{\rho}\right)^{23.9}\right),$<br>$f_{N,\mathrm{high}}(\rho)=f_{N,\mathrm{Cappi}}(\rho)+f_{N,\mathrm{Abramson}}(\rho)$<br>and<br>$f_{N,\mathrm{Cappi}}(\rho)=80.481+(160.46-80.481)\exp\left(-\left\|(\rho-\rho_{\mathrm{max,1atm}})/0.0646\right\|^{1.1416}\right),$ |
| | $f_{N,\text{Abramson}}(\rho)=\dfrac{55.14\left(1-\exp\left(-\left|(\rho-1.233)/0.22466\right|^{5.9}\right)\right)-39.54\left|\rho-1.233\right|\exp\left(-\left|(\rho-1.2924)/0.062\right|^{3}\right)}{1+\left(\dfrac{1.233}{\rho}\right)^{1000}}$ |

Table 2. Parameters for Eq. (11): $T$ in K and $\rho$ in g/cm$^3$. $\rho_{\text{max,1atm}}=0.999972\,\text{g/cm}^3$ represents the maximum density value along the atmospheric isobar.

| *Name* (unit) | *Value or formula* |
|---|---|
| $d$ (μm) | 100 |
| $\tilde{\rho}_{Knu}$ (g/cm$^3$) | $\tilde{\rho}_{Knu}(\rho,T)=\tilde{\rho}_{Knu,0}\exp\left(-\left(\dfrac{T_{Knu}}{T}\right)^{\gamma_{Knu}}\right)\times\left\{1+f_{Knu}\ln\left(1+\dfrac{T}{T_{Knu}}\right)\left[1-\left(1-\exp\left(-\left|\dfrac{T-2T_c}{975}\right|^{1.5}\right)\right)\exp\left(-\left(\dfrac{2T_c}{T}\right)^{100}\right)\right]\right\}$ |
| $\gamma_{Knu}$ | $\gamma_{Knu}(\rho)=1+\left\{3.3638\left(1-\exp\left(-\dfrac{\rho}{0.28849}\right)\right)+0.71646\exp\left(-\left|\dfrac{\rho-1.04}{0.26}\right|^{4.24}\right)\right\}\times\exp\left(-\left(\dfrac{\rho}{1.02139}\right)^{43}-\left(\dfrac{\rho}{1.00282}\right)^{300}\right)$ |
| $T_{Knu}$ (K) | $T_{Knu}(\rho)=-35\exp\left(-1.4\left|\ln\left(\dfrac{\rho}{0.04}\right)\right|^{1.75}-\left(\dfrac{\rho}{1.9}\right)^{10}\right)+\dfrac{52.721}{1+\dfrac{0.0022134}{\rho}}\exp\left(-\left(\dfrac{\rho}{0.23008}\right)^{1.3089}\right)+\left\{589.93+174.03\left(\dfrac{\rho}{1.0512}\right)^{3.411}\right\}\exp\left(-\left(\dfrac{\rho}{1.0512}\right)^{7.01}-\left(\dfrac{\rho}{1.01}\right)^{82}\right)$ |
| $f_{Knu}$ | $f_{Knu}(\rho)=\dfrac{0.27973}{1+(\rho/0.005579)^{0.8631}}$ |
| $\tilde{\rho}_{Knu,0}$ (g/cm$^3$) | $\tilde{\rho}_{Knu,0}(\rho,T)=\begin{cases}\tilde{\rho}_{Knu,0,\text{crit}}(\rho,T)+\tilde{\rho}_{Knu,0,\text{low}}(\rho) & \text{if } \rho<\rho_{\text{max,1atm}}\\ \tilde{\rho}_{Knu,0,\text{high}}(\rho) & \text{if } \rho\geq\rho_{\text{max,1atm}}\end{cases}$ with $\tilde{\rho}_{Knu,0,\text{crit}}(\rho,T)=\alpha_{\text{crit}}(T)\exp\left(-\left(\dfrac{\lvert\rho-\rho_c\rvert}{\varepsilon_{\text{crit}}(T)}\right)^{\frac{3}{2}}\right)$, $\alpha_{\text{crit}}(T)=\dfrac{0.011344}{1+\left(\dfrac{T-T_c}{1.111}\right)^2}+\dfrac{0.001795}{1+\left|\dfrac{T-T_c}{10}\right|}$, $\varepsilon_{\text{crit}}(T)=0.0603\left(1+\left|\dfrac{T-T_c}{3.413}\right|^{\frac{4}{3}}\right)$, |

$$\tilde{\rho}_{Knu,0,\mathrm{high}}(\rho) = 0.132303 + 0.0537583\exp\left(-\left|\frac{\rho-1.00718}{0.00996}\right|^{1.8906}\right)$$

$$+0.132887\left(1-\exp\left(-\left|\frac{\rho-\rho_{\mathrm{max,1atm}}}{0.17004}\right|^{3.517}\right)\right)+0.004008\exp\left(-\left|\frac{\rho-\rho_{\mathrm{max,1atm}}}{0.001}\right|\right),$$

$$+37.76\frac{\left|-1+\rho/1.233\right|^{2.443}}{1+(1.233/\rho)^{1000}}$$

$$\tilde{\rho}_{Knu,0,\mathrm{low}}(\rho) = \tilde{\rho}_{\mathrm{Knu1}}(\rho) + \tilde{\rho}_{\mathrm{Knu2}}(\rho) - \tilde{\rho}_{\mathrm{Knu3}}(\rho),$$

$$\tilde{\rho}_{\mathrm{Knu1}}(\rho) = \frac{\exp\left(-(\rho/1.9857)^{0.6588}-(\rho/0.6568)^{3.364}\right)}{\left(1+(\rho/0.058037)^{4.774}\right)^{0.0183188}}$$

$$\times\left\{0.0567491+0.00641937\left(\frac{\rho}{0.063819}\right)^{0.4407}+0.0098665\exp\left(-0.237369\left|\ln\left(\frac{\rho}{0.0143185}\right)\right|^{1.69637}\right)\right\},$$

$$\tilde{\rho}_{\mathrm{Knu2}}(\rho) = \frac{650\dfrac{\exp(-7.94263/\rho)}{1+(0.7749/\rho)^{17.84}}+0.201243\exp\left(-8.92\left|\ln(\rho/0.58775)\right|^{3.2924}\right)}{1+(0.879984/\rho)^{4.4569}},$$

$$\tilde{\rho}_{\mathrm{Knu3}}(\rho) = \frac{0.00173736\dfrac{\exp\left(-\left|\dfrac{\rho-0.968188}{0.013654}\right|^{1.43548}\right)}{\left(\dfrac{\rho+5\times10^{-5}}{0.968188}\right)^{67}}+0.0222429\dfrac{\exp\left(-\left|\dfrac{\rho-0.999277}{0.00918515}\right|^{1.73324}\right)}{\left(\dfrac{\rho+5\times10^{-5}}{0.999277}\right)^{54}}}{(1+0.918472/\rho)^{1000}}$$

Table 3. Parameters for the properties of the released gas: $T$ in K and $\rho$ in g/cm$^3$. $T_{\mathrm{tr}} = 273.16$ K represents the triple point temperature and $\rho_{\mathrm{max,1atm}} = 0.999972$ g/cm$^3$ represents the maximum density value along the atmospheric isobar.

From Ref. 7 the isochors of viscosity are supposed to have a correct representation from the saturated vapor pressure curve until a temperature of 1173.15 K. Now it is interesting to notice that the isochors data provided by the website of NIST (Ref. 16) are possibly given until 1275 K but the two representations differ completely at a given temperature on isochors as can be seen for example on Fig. 1. The divergence occurs where no more experimental data exist. So, to determine the different parameters of the model developed here we have considered only the region below this divergence temperature point. As can be seen on Fig. 1 the extrapolation of the present modeling gives a curve which is close to the IAPWS08 formulation until the latter reaches a maximum and then continues in the supercritical phase with a gas-like behavior until it crosses the dilute-gas limit curve of the IAPWS08 formulation. The fact that the IAPWS08 formulation reaches a maximum and then a minimum at a very high temperature in the supercritical phase seems physically difficult to understand and no explanation have been found in the literature on the justification of this behavior.

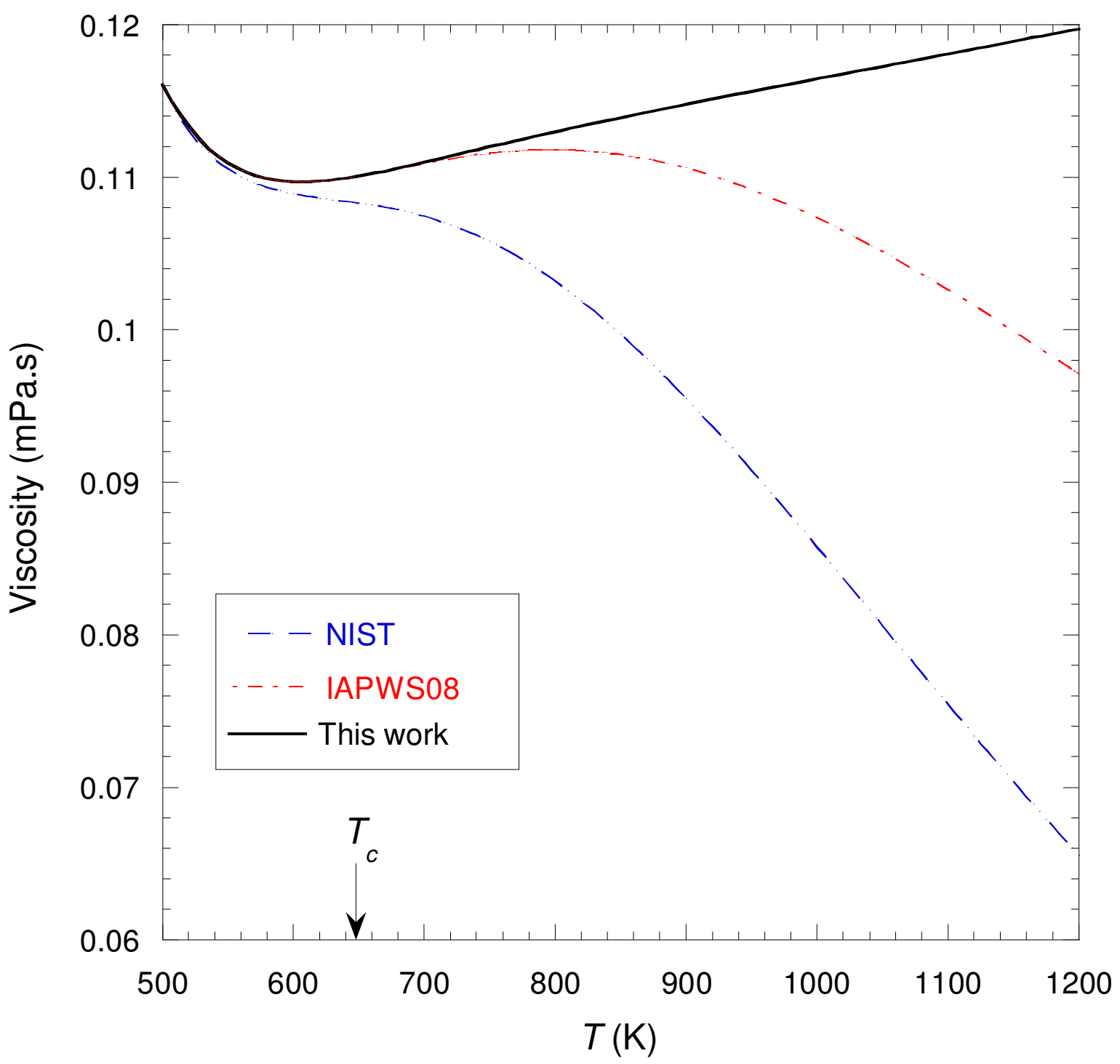

Fig. 1. Representation of the water viscosity isochor for $\rho = 0.825$ g/cm$^3$ as function of temperature for 3 different modelings.

It is interesting to have a look to the variation of $K_0^*$. The blue dash-dotted line on Fig. 2 shows that for $\rho$ smaller than 0.1 g/cm$^3$ the variation of $K_0^*$ is very close to a power law proportional to $\rho^3$. In this region of density, it means that $c_0$ is almost proportional to the density (see the black dashed curve on Fig. 2) and the transition temperature is such that $T_t^* \propto \rho^2$ showing that both of these parameters tend to zero in the zero-density limit. It appears also on Fig. 2 that $K_0^*$ is close to 1 in the normal liquid phase. Overall, the variation law of $K_0^*$ shows that the cohesion of the medium decreases when the density decreases, which corresponds well to the expected idea of a liquid that transits traditionally to a gas. It can also be seen from Fig. 2 that the celerity $c_0$ has typical values of the sound velocity in water.

Now, it can be noticed that in the supercooled phase at very low temperatures, the introduction of the function $f_K$ shows that $K^*$ must decrease faster than the law given by $K_0^*$ regardless of the density. In other words, the supercooled phase at low temperature is less cohesive than the normal liquid phase at the same density which is a coherent result with the isothermal compressibility data (see Appendix D).

Taking into account the evolution of the various functions with temperature, we deduce that at $T = 0$ K, the viscosity is such that:

$$\eta(\rho, T = 0\mathrm{K}) = \eta_l(\rho, T = 0\mathrm{K}) \to \infty .$$

On the contrary, when $T$ tends towards infinity, we have:

$$\eta(\rho,T \to \infty) = \frac{d\sqrt{\rho K}}{N-1} + \eta_{Knu}(\rho,T \to \infty) \to \infty\,,$$

and thus within these two limits the value of the viscosity is quasi-independent of the experimental device. We observe here a kind of "symmetry" on the behavior of the visocsity.

Fig. 3 shows the variation law of $f_N$ and therefore of the parameter $N$-1. We notice that this function is almost identical to the limit function $f_{N,\lim}$ except in the liquid region where a peak appears around the triple point liquid density $\rho_{\text{tr,Liq}}$. Taking a value of $d$ = 100 µm as an example, we can see that the maximum value of $d_N$ is around 1 cm, taking into account the arbitrary decomposition choice of Eq. (11). More fundamentally, the variation law of $f_N$ shows that the coherence of the medium decreases when the density decreases, which again corresponds well to the expected idea that one can have when switching from liquid to gas.

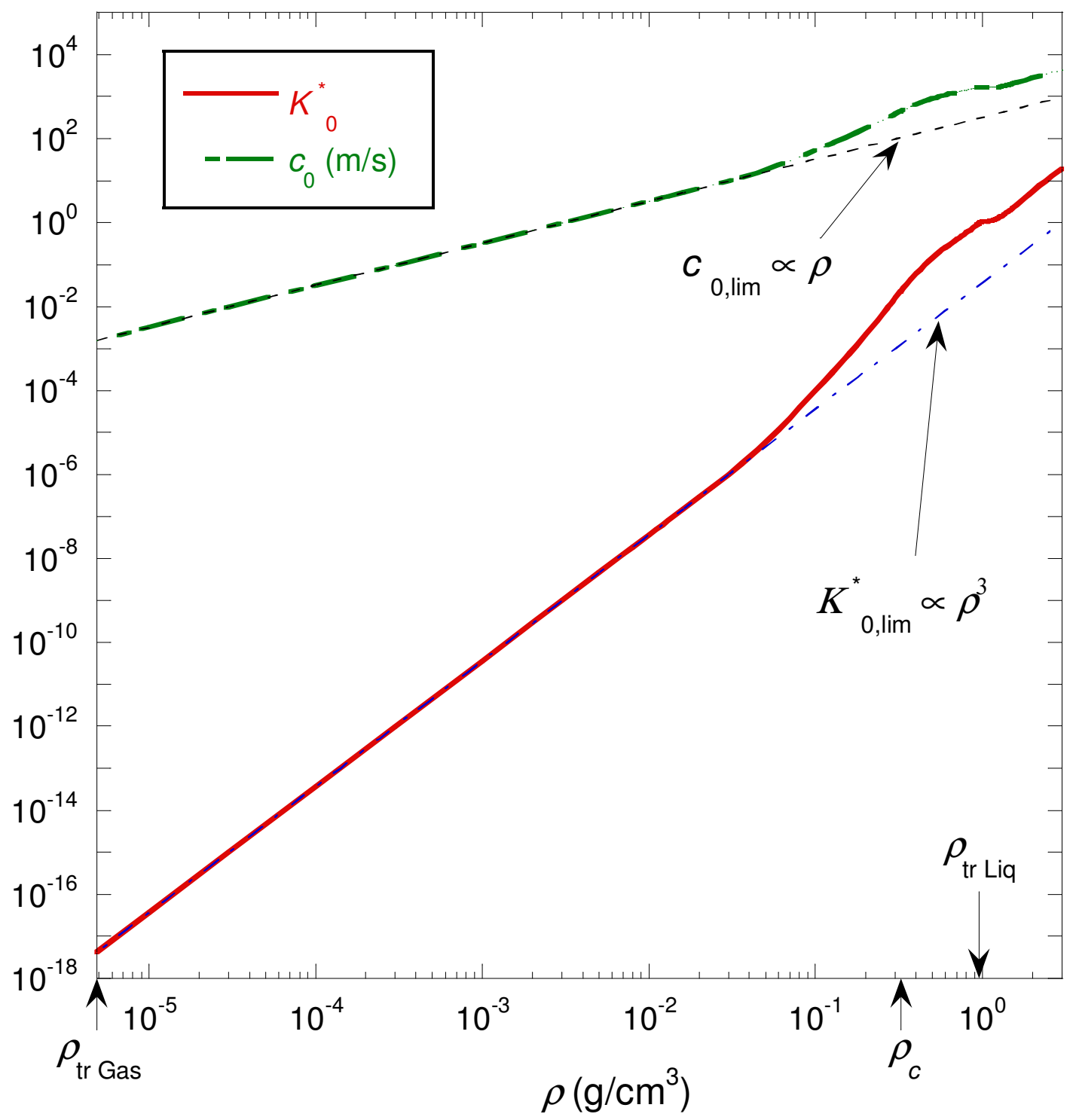

Fig. 2. Logarithmic plot of $K_0^*$ (red curve) and $c_0$ (green dashed curve) variations with density for water from the triple point gas density $\rho_{\text{tr,Gas}}$ up to 3 times the triple point liquid density $\rho_{\text{tr,Liq}}$.

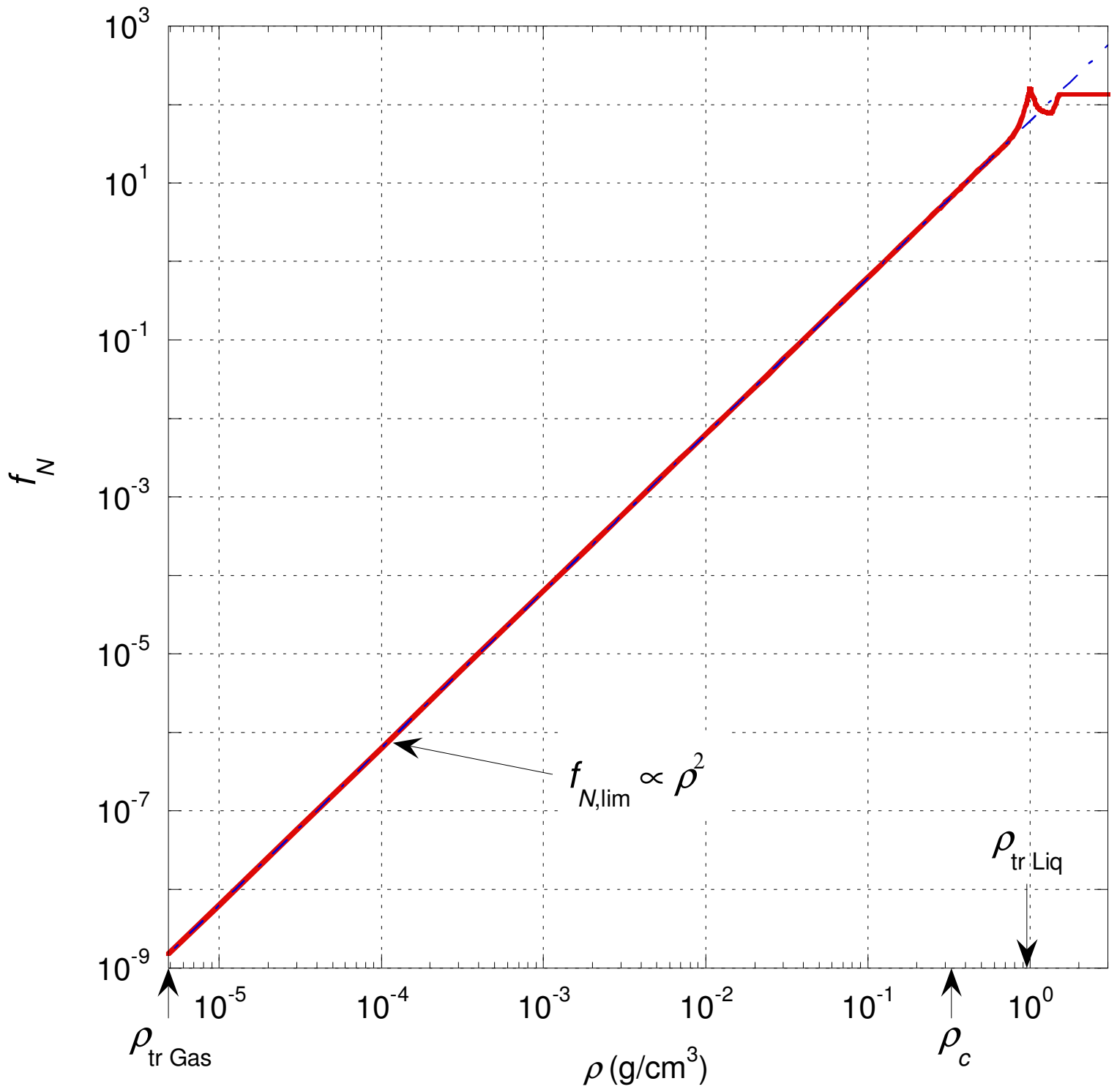

Fig. 3. Logarithmic plot of the variations with density of $f_N$ (red curve) for water from the triple point gas density $\rho_{\mathrm{tr,Gas}}$ up to 3 times the triple point liquid density $\rho_{\mathrm{tr,Liq}}$. The blue dot-dashed curve represents a power law $\propto \rho^2$.

Now we notice that the term corresponding to the amount of gas released by the shear flow is such that $\lim_{T\to\infty} \tilde{\rho}_{Knu}(\rho,T) = \tilde{\rho}_{Knu,0}$ and then that the parameter $\tilde{\rho}_{Knu,0}$ is multiplied by two temperature-dependent terms. Therefore, it is useful to discuss here the importance of these two terms. Fig. 4 shows that the logarithmic term is only significant for densities below the critical density (i.e. in the gaseous phase) and mainly at very low densities; otherwise in the liquid phase the term containing the logarithm function is close to a constant equal to 1. Fig. 4 shows that the term $T_{Knu}$ is always smaller than $T_c$ and decreases quickly to almost zero value beyond $\rho \sim 1$ g/cm$^3$, so the exponential term also becomes close to a constant equal to 1. Finally, whatever the temperature, it appears that beyond $\rho \sim 1$ g/cm$^3$ we have $\tilde{\rho}_{Knu}(\rho,T) \cong \tilde{\rho}_{Knu,0,\mathrm{high}}(\rho)$.

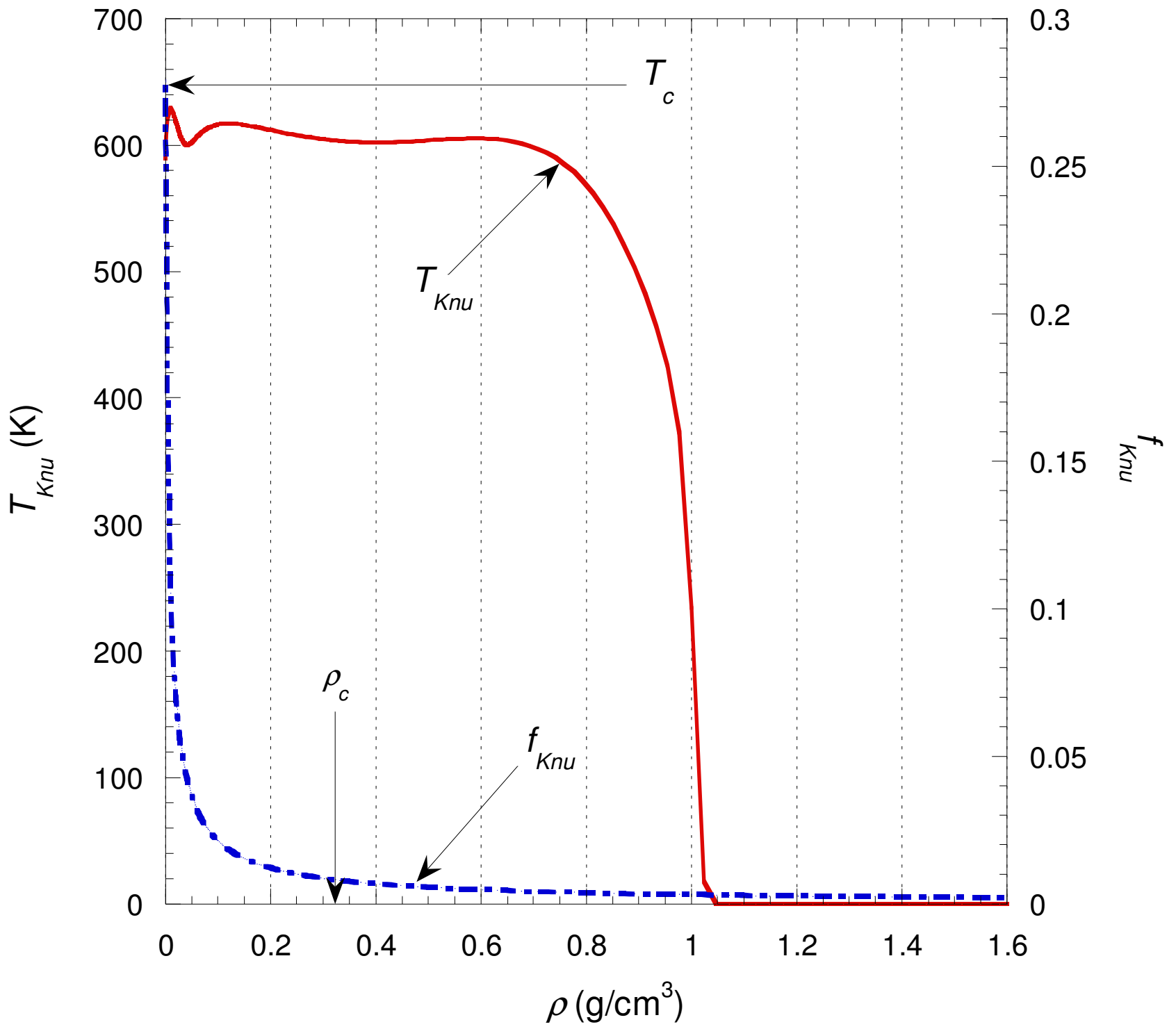

Fig. 4. Variations with density of $T_{Knu}$ (red curve) and $f_{Knu}$ (blue dot-dashed curve) for water.

It is also useful to look at the evolution of the different parameters within the zero-density limit: these evolutions and their consequences are analyzed in the following section.

### 3.1. Analysis of the Viscosity Isochors in the dilute-gas limit

As it has been stated in the introduction, for some viscosity isochors in the low-density fluid phase a crossing of viscosity isochors with the dilute-gas limit part is observed (see Fig. 5). In the IAPWS08 formulation the dilute-gas limit is defined as the behavior of a "perfect gas" whose viscosity depends only on the temperature and not on density. This idea comes from the fact that many experimental results seem to show that, for low enough densities, the viscosity becomes density independent. Fine analysis of the water viscosity data corresponding to the lowest experimental densities (see section 4.1.3) always show a dependence on temperature and density and these data are not correctly reproduced by the IAPWS08 formulation. This could explain why the extrapolation to the perfect gas behavior is not rigorously correct.

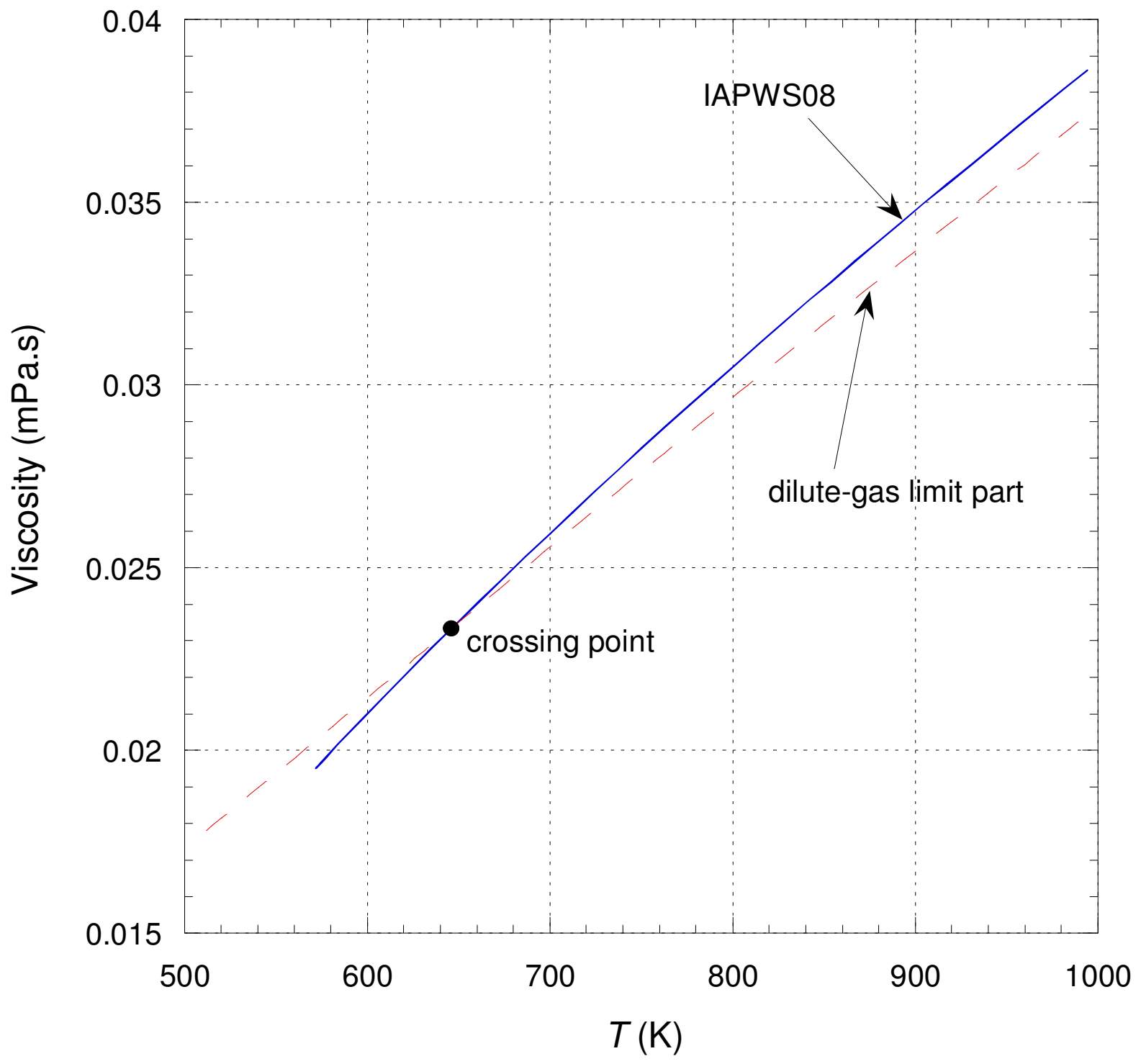

Fig. 5. Crossing of the water viscosity isochor at $\rho = 0.045$ g/cm$^3$, calculated from the IAPWS08 formulation (blue curve), with the IAPWS08 dilute-gas limit part (red dashed curve).

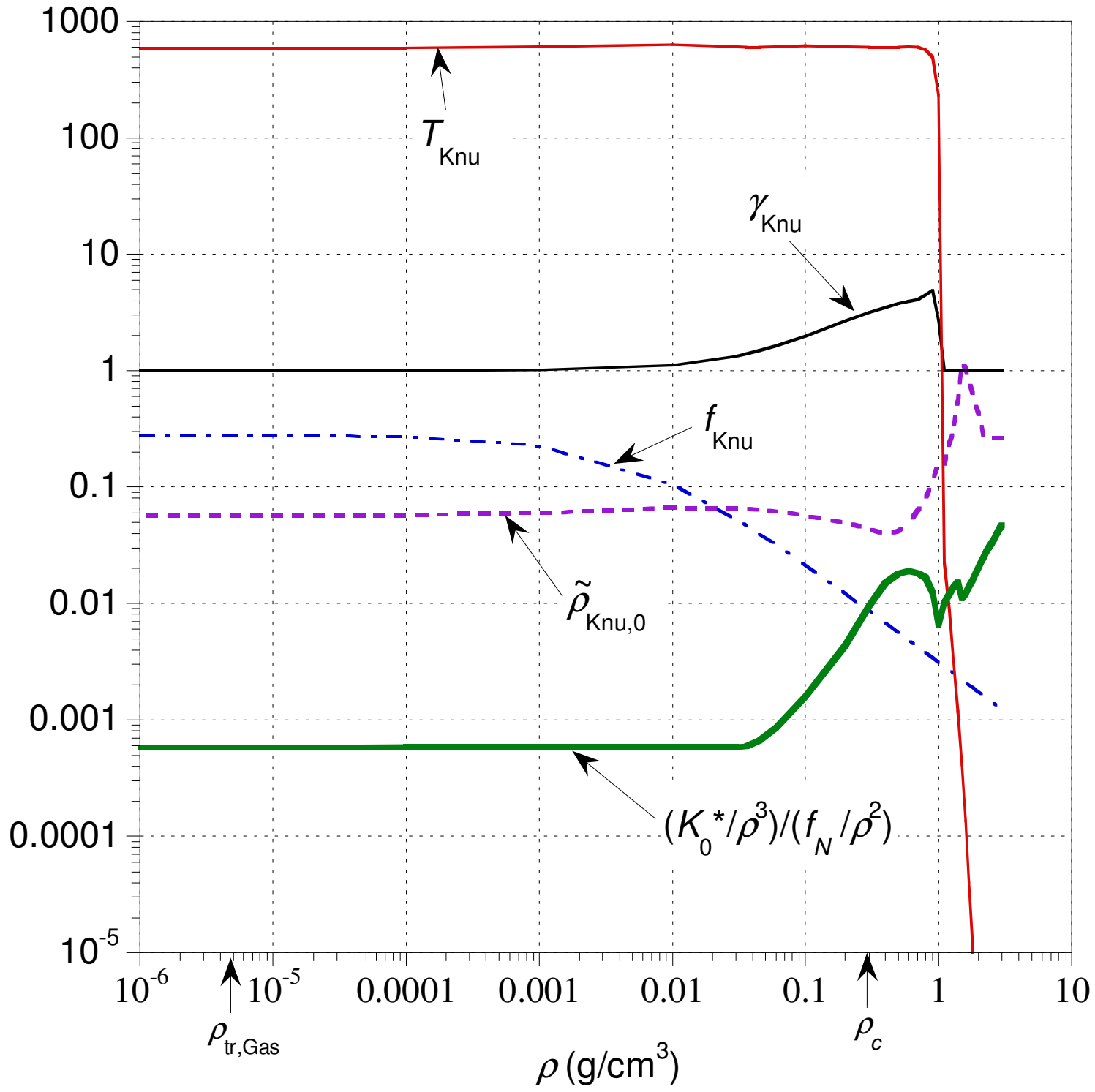

Fig. 6. Logarithmic plot of variations with density of all the parameters in Table 3.

In the present modeling there is no such "perfect gas" limit term hence there is no such crossing phenomenon. But it is interesting to determine what this model leads within the density limit $\rho \to 0$. Fig. 6 shows that all parameters describing the gas released become constant for a density lower than $10^{-4}$ g/cm$^3$. For this zero-density limit it appears that the term $\eta_{Knu}$ depends only on the temperature and of the set-up parameter *δ*.
We will now determine the limit of the liquid term. We have shown previously that the asymptotic form of $K_0^*$ (for temperatures greater than 250 K, *i.e.* $f_K(T > 250\,\mathrm{K}) \approx 1$ whatever the values of *ρ*) is such that $K_0^* \cong K_{0,\lim}^*(\rho) = c_{K0}(\rho/\rho_c)^3$ and the asymptotic form of $f_N$ is such that $f_N \cong f_{N,\lim}(\rho) = c_{N0}(\rho/\rho_c)^2$. The green curve on Fig. 6 shows that the ratio $(K_0^*/K_{0,\lim}^*)/(f_N/f_{N,\lim})$ tends well towards a constant at the zero-density limit. For the zero-density limit, the thermodynamic function $v(\rho,T) \to 2$ then $\lim_{\rho\to 0} H_N(v) = H_N(2) = N-1 \propto \left(\frac{\rho}{\rho_c}\right)^2 d\,\frac{q_{c0,\mathrm{crit}}}{2\pi}$. With these limits, we can deduce the asymptotic limit of the viscosity liquid term:

$$\eta_{\lim} = \lim_{\rho\to 0}\eta_l = 2\pi\frac{\sqrt{c_{K0}}}{c_{N0}}\frac{\sqrt{K_0\rho_c}}{q_{c0,\mathrm{crit}}} \tag{21a}$$

It finally appears that the zero-density viscosity limit of the liquid term is a constant that no longer depends on the experimental set-up but only on the intrinsic properties of the fluid. The numerical application for water leads to $\eta_{\lim} = 5.9349\times 10^{-3}$ mPa.s.
It is important to note that the value of $\eta_{\lim}$ can be approximated to less than 0.5% by the following expression:

$$\eta_{\lim} = 2\pi\frac{\hbar}{2\mathcal{V}_{\mathrm{mol}}} \tag{21b}$$

where $\hbar$ is the reduced Planck constant and $\mathcal{V}_{\mathrm{mol}}$ represents the geometric mean of the two characteristic molecular volumes of the medium which are the critical volume $\mathcal{V}_{\mathrm{c}}$ and the volume at zero temperature $\mathcal{V}_0$, i.e. $\mathcal{V}_{\mathrm{mol}} = \sqrt{\mathcal{V}_{\mathrm{c}}\mathcal{V}_0}$. The value of $\mathcal{V}_0$ is determined from Eq. (D.2) of Appendix D. The small difference between Eq. (21a) and Eq. (21b) can be attributed to the uncertainties in the proportionality constants of $K_{0,\lim}^*(\rho)$ and $f_{N,\lim}(\rho)$, as well as to the extrapolation of Eq. (D.2) to determine $\mathcal{V}_0$. It can therefore be assumed that the two expressions are equal. This result is a validation in itself of the choice of Eq. (9), Eq. (11) and the extrapolation of Eq. (D.2).
The fact that $\eta_{\lim}$ can be written in two different forms implies a relationship between the parameters $c_{K0}$ and $c_{N0}$ such that: $\frac{\sqrt{c_{K0}}}{c_{N0}} = \frac{\hbar}{2m}\sqrt{\frac{\rho_{00}}{K_0}}q_{c0,\mathrm{crit}}$ with $\rho_{00} = m/\mathcal{V}_0$ and $m = M/\mathcal{N}_a$. In other words, the values of these two parameters are not independent. Note that $\frac{\hbar}{2m}\sqrt{\frac{\rho_{00}}{K_0}}$ represents a characteristic distance whose physical interpretation will be given in another paper.

Finally, considering Eq. (19), we deduce that in real experiments (where $d$ and $\delta$ have finite values) the zero-density viscosity limit is zero whatever the temperature.

### 3.2. Comparison with the Current Reference Equation of State

The evaluation of the performances of our microscopic model makes sense only in comparison with all the experimental data. But since the IAPWS08 formulation reproduces a number of these experimental data with an uncertainty described in Fig. 24 of Ref. 7, it is useful to present the proposed model in relation to this representation. Fig. 7 to Fig. 9 show percentage deviations of the present modeling with the IAPWS08 formulation. In these figures the subregions are defined and labeled according to those defined in Fig. 24 of Ref. 7. This shows that the present modeling deviation is globally much smaller than the uncertainties of the IAPWS08 formulation for a large set of experimental data.

One could believe on Fig. 8 that the isobar at 100 MPa is very poorly represented but this isobar is the boundary between subregion 3 and subregion 2. This is highlighted in Fig. 9 where we can see that the latter is perfectly consistent with the estimated uncertainty.

The isobar at 350 MPa is on the boundary of subregion 2 and appears poorly represented but this latter corresponds to a region determined by the data of Cappi and Abramson. In this region, however, the IAPWS08 formulation represents the data of Cappi and Abramson rather poorly, whereas as we show a little further, we can represent these data within their experimental uncertainties. It is therefore relevant to observe such deviations here. It is also interesting to notice that the present modeling provides results closer to NIST data (Ref. 16) in subregion 3 than to the IAPWS08 formulation as can be seen on Fig. 10 for example for the isobar equal to 100 MPa whereas it is the contrary for subregion 5. However, in subregion 5 the deviation with NIST remains in the uncertainty; but this is a region where there are only few experimental data, so it is not *a priori* possible to say that the data from NIST are incompatible with the experimental data.

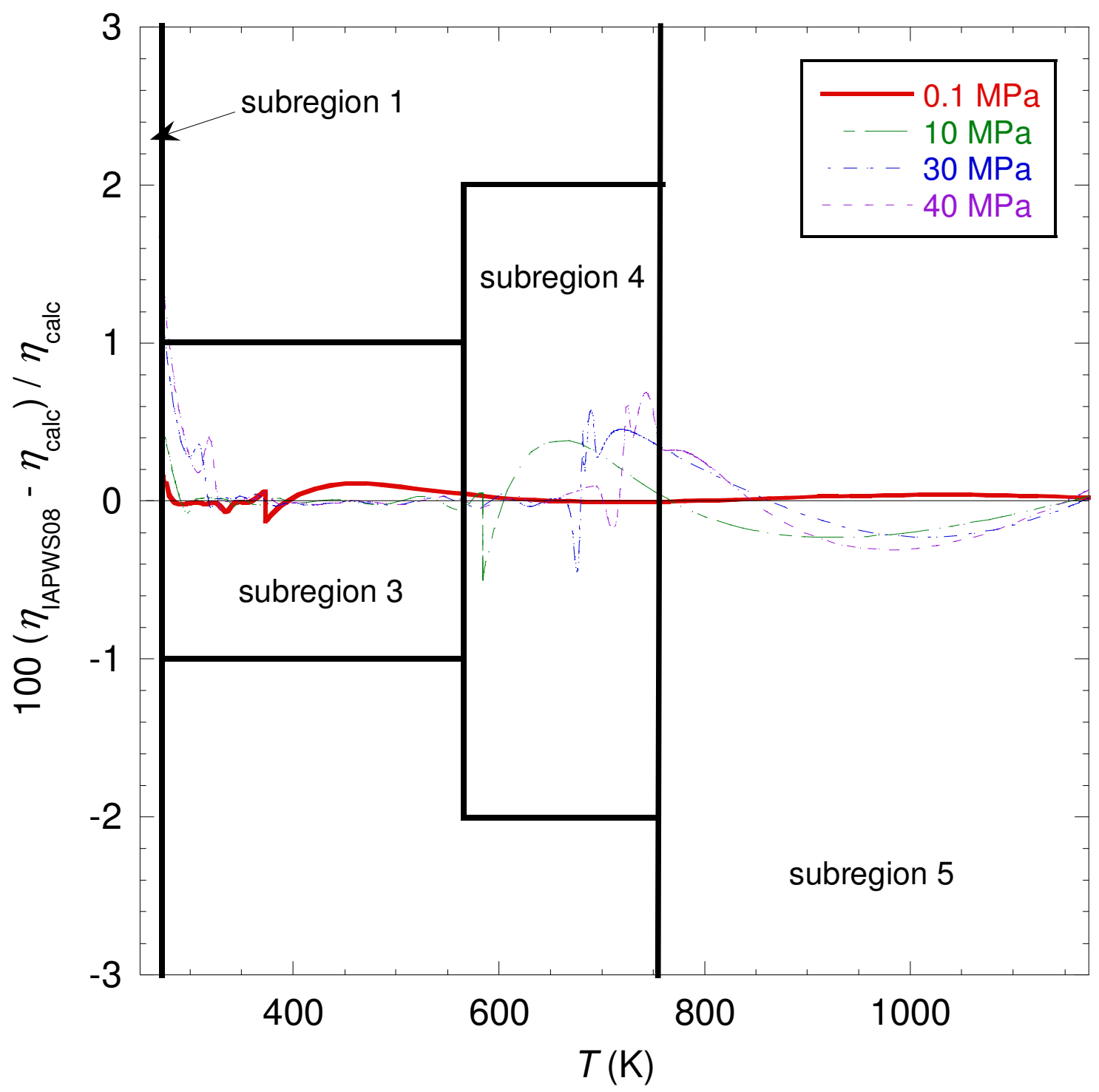

Fig. 7. Percentage deviations of the present modeling from the IAPWS08 formulation (Ref. 7) of water as a function of temperature for low pressure isobars.

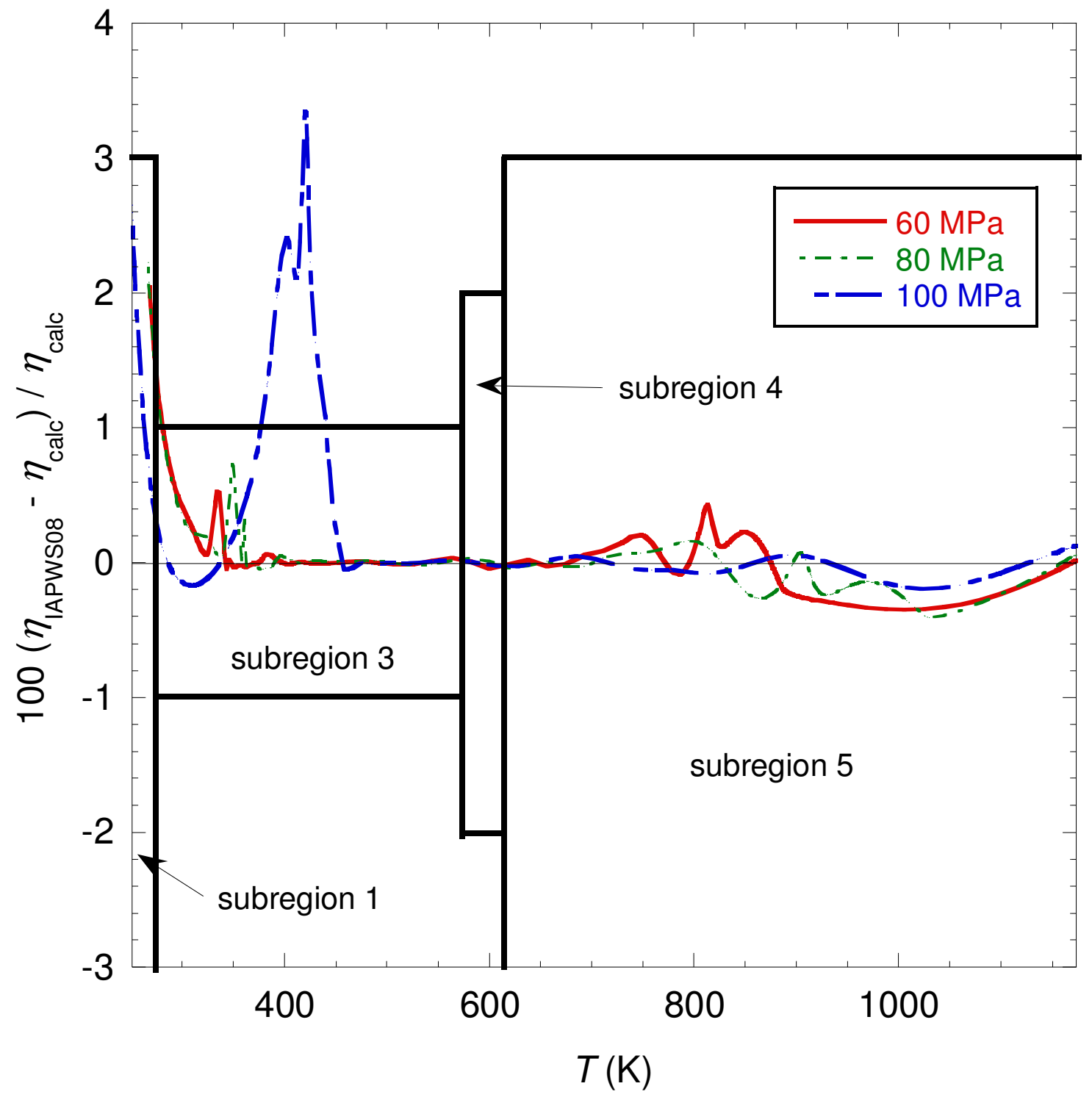

Fig. 8. Percentage deviations of the present modeling from the IAPWS08 formulation (Ref. 7) of water as a function of the temperature for middle pressure isobars.

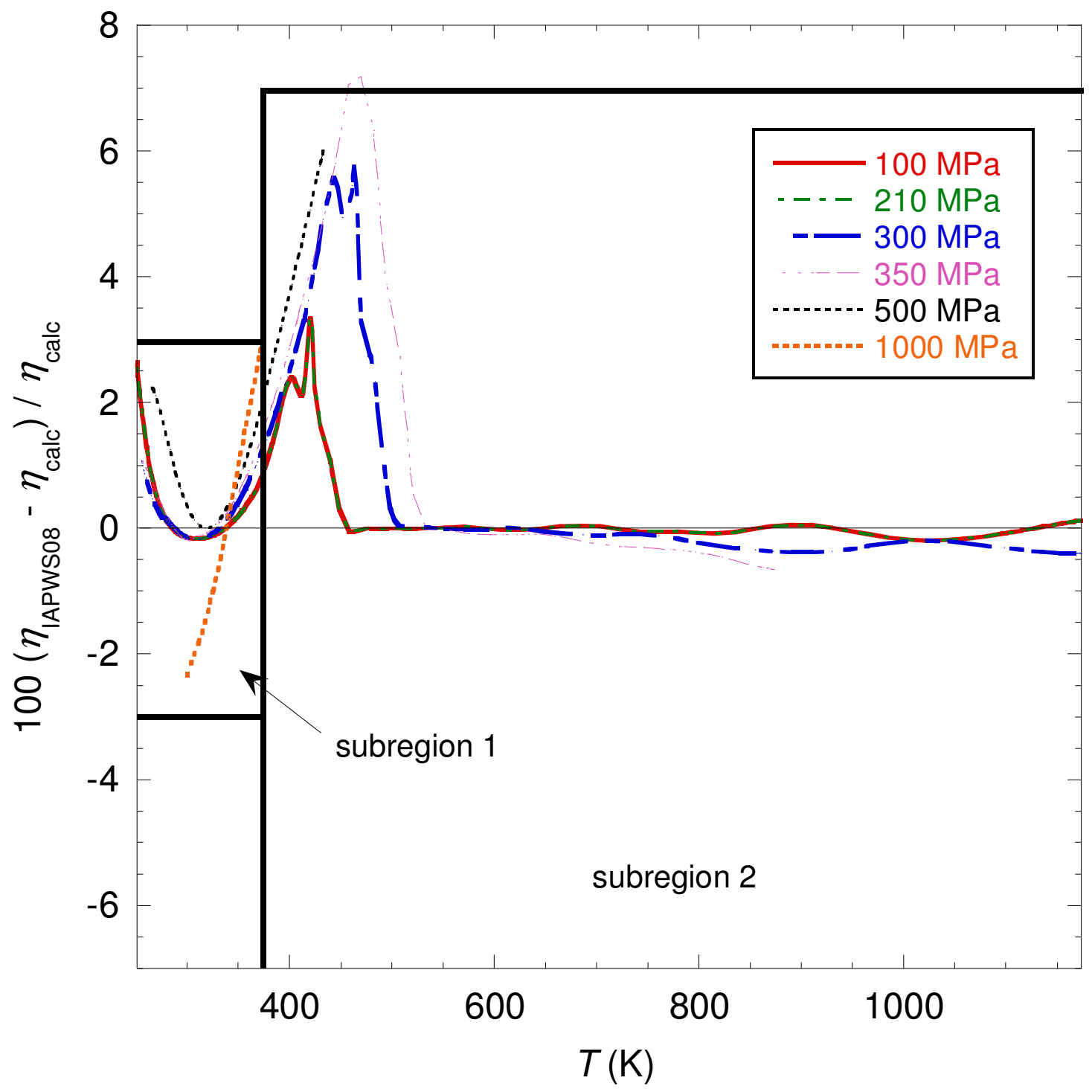

Fig. 9. Percentage deviations of the present modeling from the IAPWS08 formulation (Ref. 7) of water as a function of the temperature for high pressure isobars.

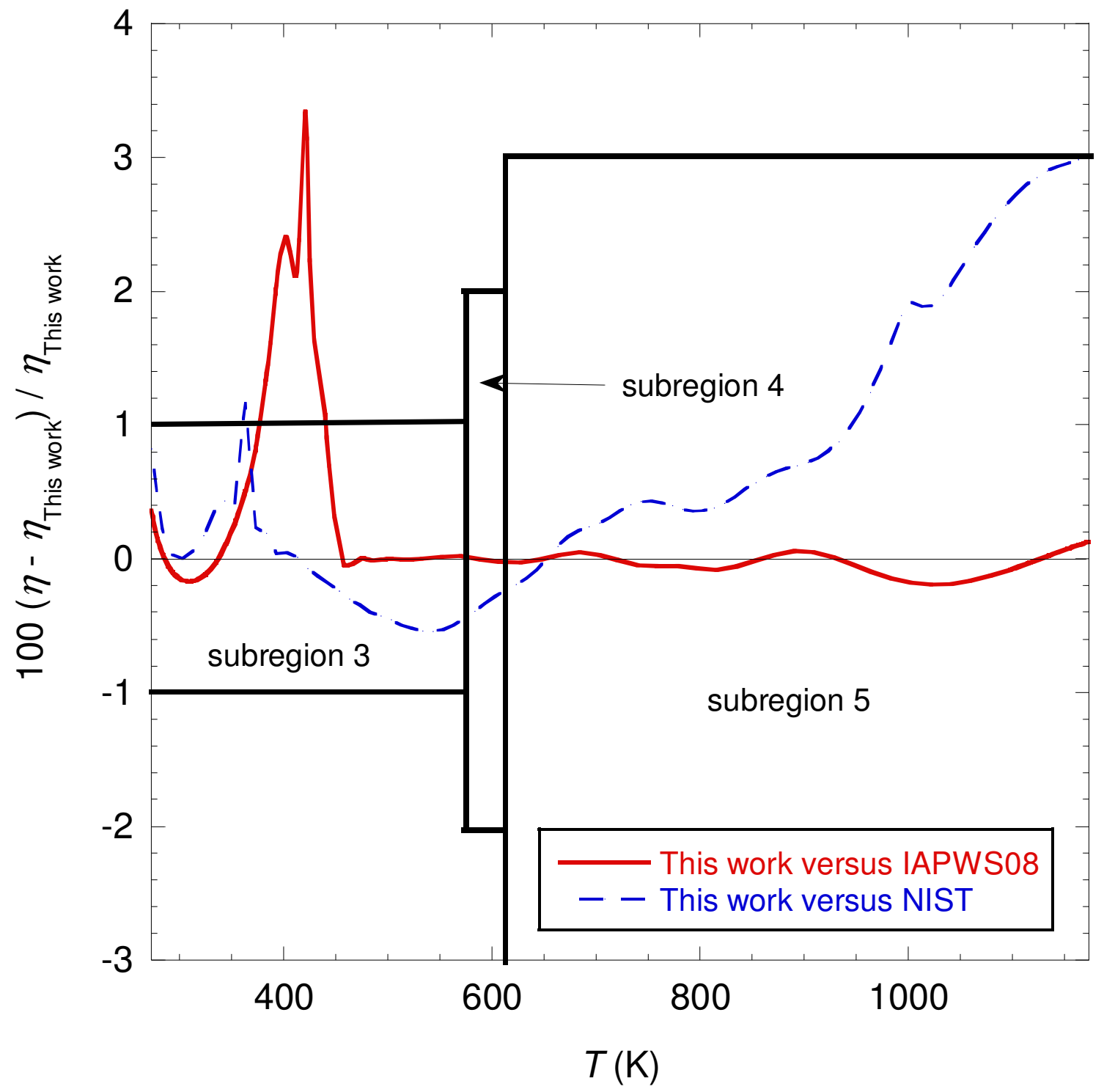

Fig. 10: Percentage deviations of the present water modeling from the IAPWS08 formulation (Ref. 7) and NIST data (Ref. 16) as a function of the temperature for the isobar equal to 100 MPa.

As we have used Cappi's data (Ref. 10) to determine some parameters variations, we present on Fig. 11 the deviation between the present modeling and the IAPWS08 formulation with the smoothed data of Cappi. As can be seen, the present modeling exhibits a better agreement with the smoothed data than that of the IAPWS08 formulation, especially the deviations are well distributed around the unit value unlike the IAPWS08 formulation. All points are included in the ±1% experimental uncertainty except one point on the boundary; we will see in section 4.1.2 that this point on the isotherm corresponding to 293.15 K is a smoothing artifact and that the deviation with the raw data is actually lower.

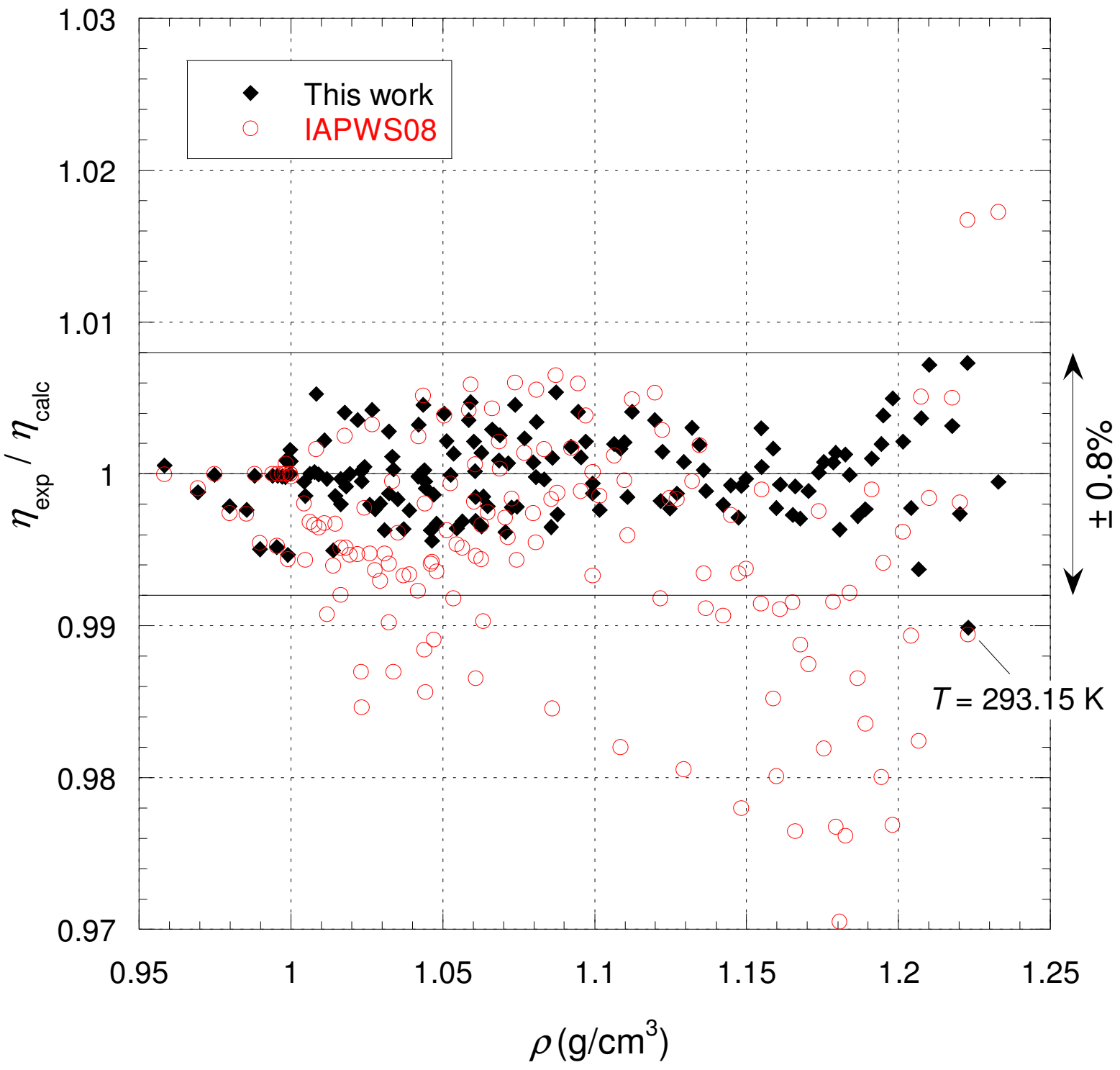

Fig. 11. Ratio of Cappi's smoothed data (Ref. 10) with the IAPWS formulation (empty circles) and the present modeling (black diamonds) as a function of density.

Experimental data from Abramson (Ref. 11) were obtained using rolling spheres having diameters between 30 and 60 μm in a high-pressure diamond-anvil cell. So in this experiment the size of the fluid velocity gradient regions is different from the equivalent size of the IAPWS08 formulation (i.e. equal to $d$ = 100 μm). Therefore, to analyze the Abramson's data (Ref. 11), it is necessary to take into account the geometric parameters of his diamond-anvil cell, which implies to take $d_{\text{Abramson}} = 55.4\,\mu\text{m}$ and to multiply $d_N$ by a constant $C_N = 0.825$. The fact that this coefficient $C_N$ is smaller than 1 indicates that the size of the coherent volume is smaller in Abramson's experiment than in Cappi's experiment, which is consistent with the geometric parameters of these two experiments.

Many experimental Abramson's data correspond to fluid states that are outside the computation domain of the 1995 IAPWS state equation formulation (Ref. 17) for density calculations. For these fluid states we have considered mainly two possibilities of density calculations: for pressures greater than 1000 MPa we used both the data of Bridgman (Ref. 18) and/or Wiryana *et al.* (Ref. 20) to interpolate and/or extrapolate density values. In high pressures regions, the resulting uncertainties will depend strongly on the calculated density values as it is shown on Fig. 12: it appears that for low temperature isotherms, the extrapolated density values from Bridgman's data lead to a good agreement with the experimental data whereas for the highest isotherms the extrapolated density values from Wiryana *et al.*'s data lead to a better agreement; in all cases the data can be reported with an uncertainty in accordance with the experimental uncertainty (*i.e.* the deviation between different experimental datasets reported on Fig. 3 of Ref. 11 is within 9% on the quasi-isotherm ~294.15 K). However, whatever the temperature of the isotherm, the IAPWS08 formulation is very far from the experimental data at high pressures and this large discrepancy cannot be attributed to uncertainties in the calculation of density.

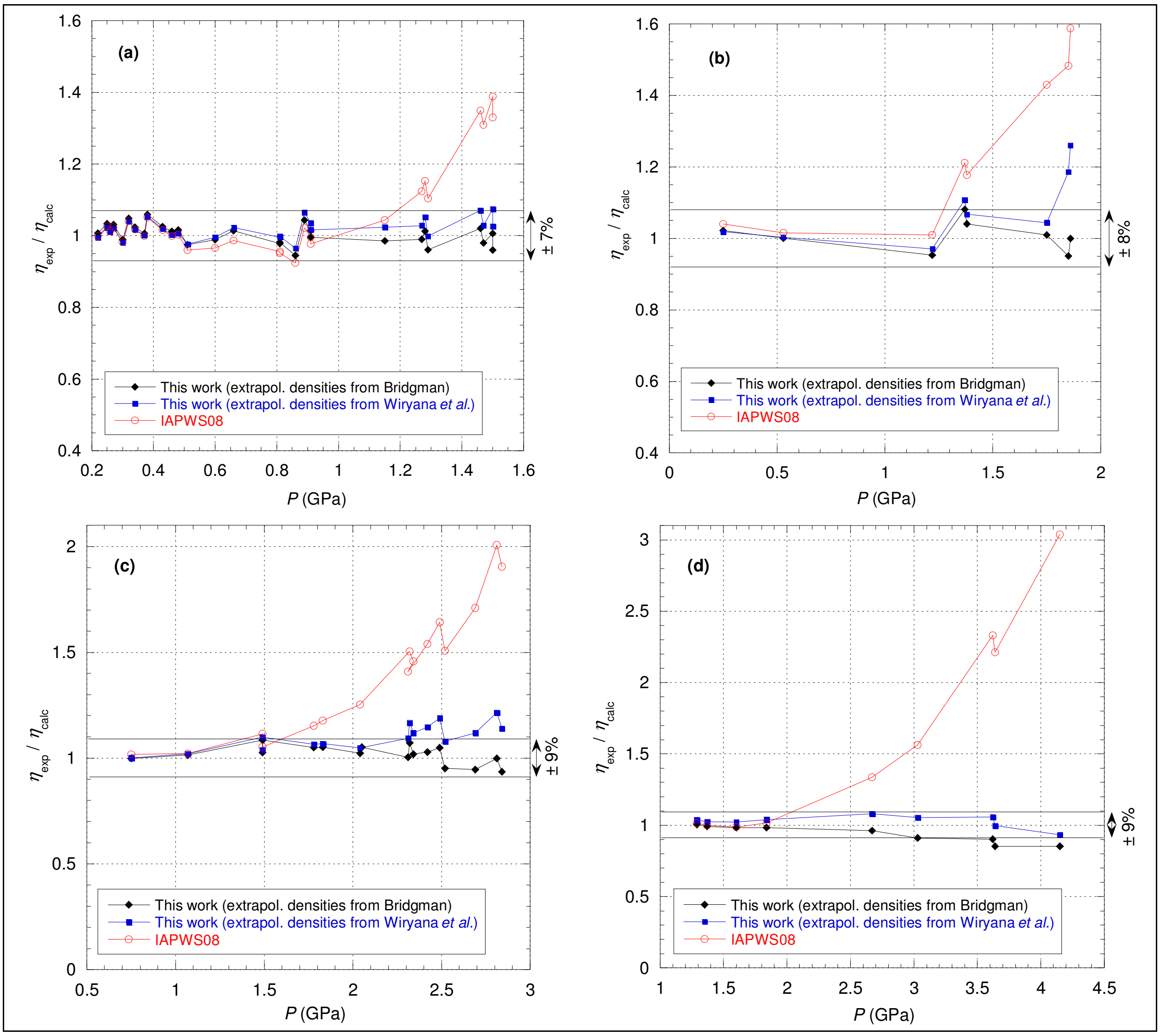

Fig. 12. Ratio of Abramson's data (Ref. 11) with the IAPWS08 formulation (empty circles) and the present modeling with $d = 55.4$ µm (black diamonds and blue squares) as a function of pressure along different quasi-isotherms: (a) ~ 294.15 K, (b) ~ 323.15 K, (c) ~ 373.15 K, (d) ~ 473.15 K. The lines are eye guides.

Abramson's last quasi-isotherm around 673.15 K contains only few data that are widely dispersed. Presenting the deviation curve in this case makes little sense. Also, the choice of the equation of state to calculate densities is not very important because the effect is very small compared to the experimental uncertainty. Fig. 13 shows that the present model is compatible with the data within 20% while the IAPWS08 formulation has opposite variations.

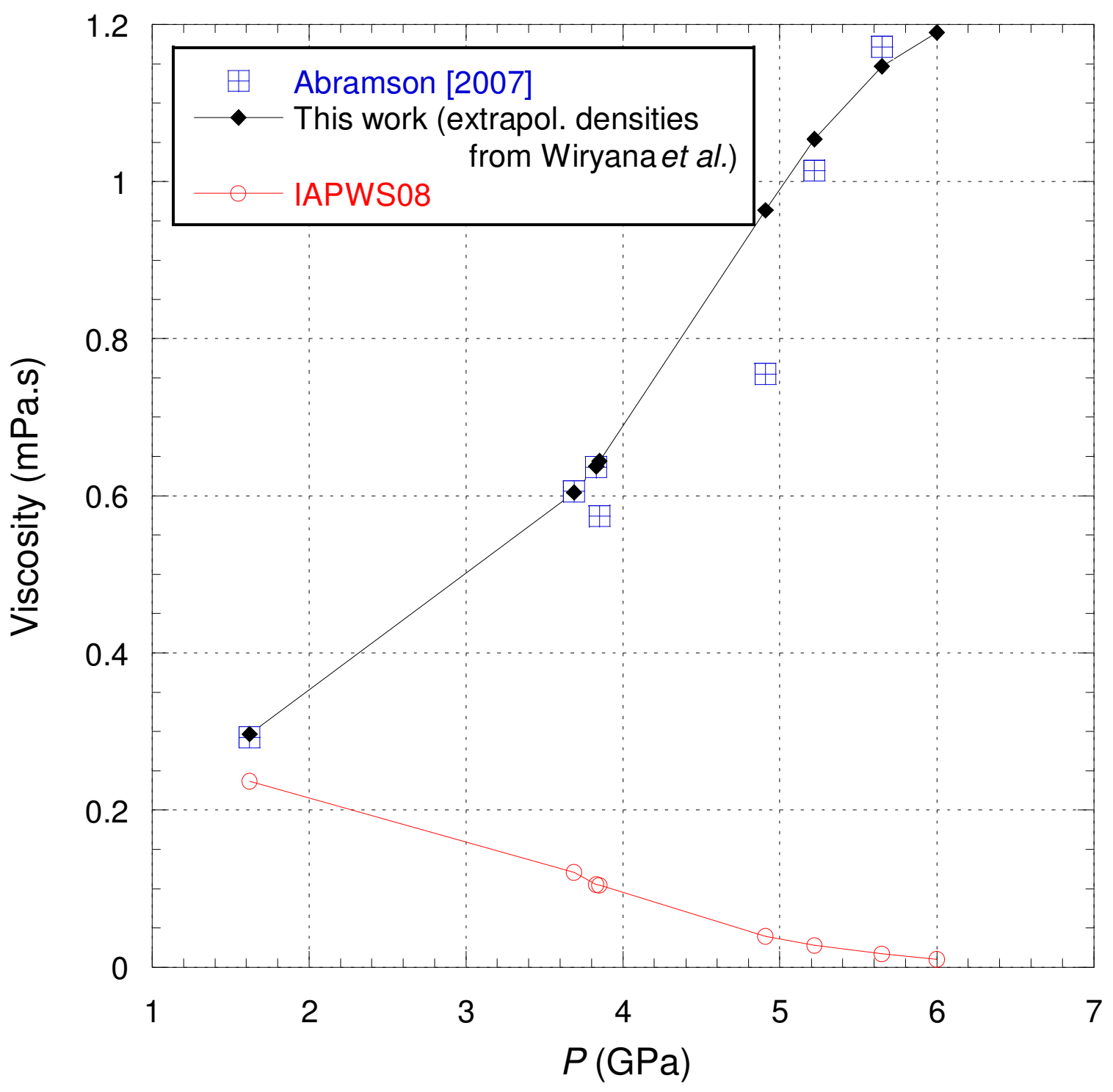

Fig. 13. Abramson's viscosity data of water (Ref. 11) with the IAPWS08 formulation (empty circles) and the present modeling with $d$ = 55.4 µm (black diamonds) as a function of the pressure along the quasi- isotherm ~673.15 K. The lines are eye guides.

The specific behavior of isochors near the critical point is described by the data of Rivkin *et al.* (Ref. 14) reanalyzed by Huber *et al.* (Ref. 7). Rivkin *et al.*'s experimental data were obtained using a 150 µm inner radius tube while the IAPWS08 formulation would correspond to an equivalent tube with an inner radius of 100 µm. Therefore, to analyze Rivkin *et al.*'s data, it is necessary to take into account the geometric parameters of his tube, which implies to take $d_{\text{Rivkin } et\,al.} = 150.435\,\mu\text{m}$ and to multiply $d_N$ by a constant $C_N = 1.01725$. By using these geometrical characteristics, Fig. 14 shows that both descriptions are almost equivalent: the data are included in the ±1% experimental uncertainty, except for two isolated points.

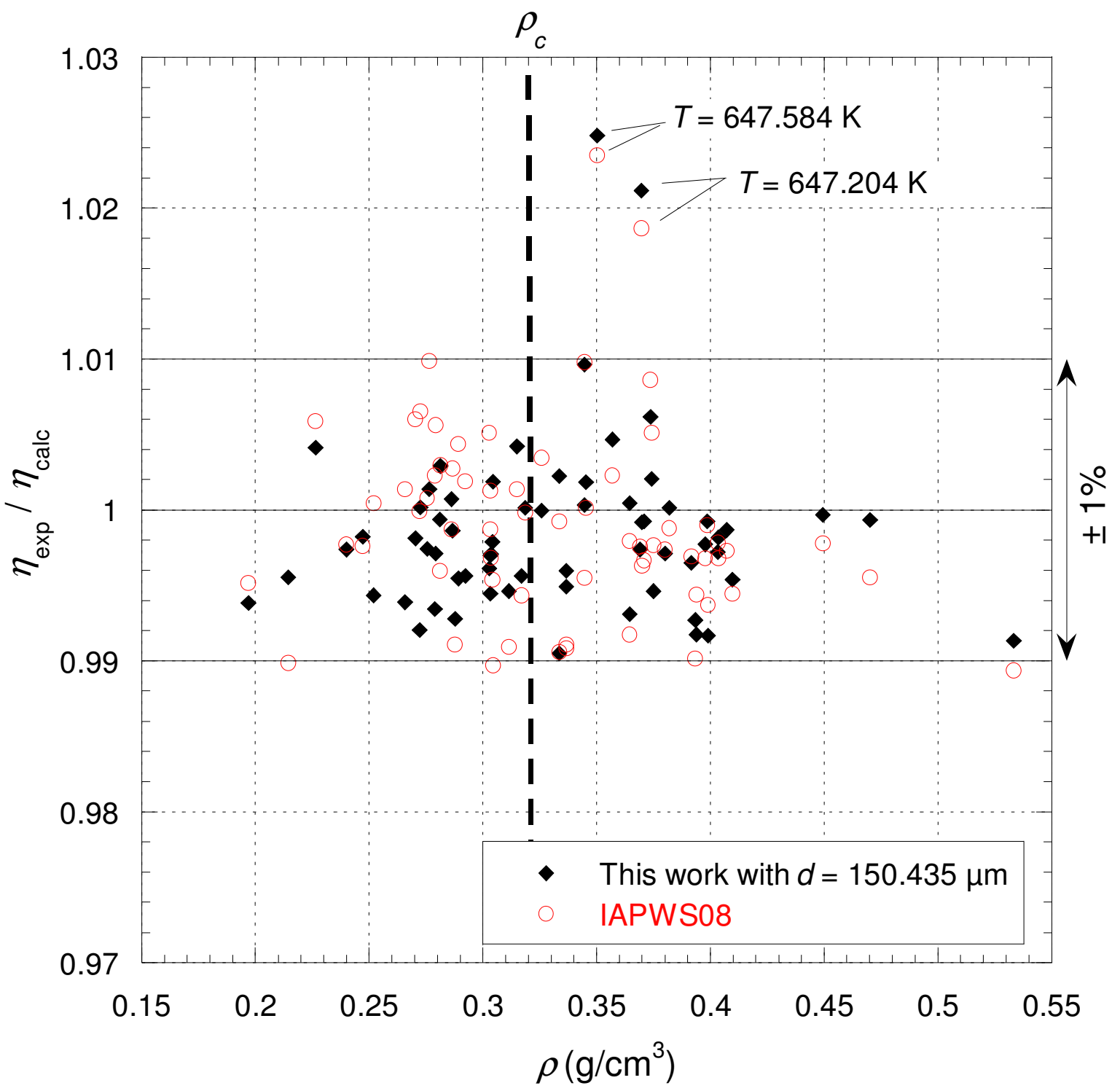

Fig. 14. Ratio of Rivkin *et al.*'s data of water (from Ref. 7, Table 4) with the IAPWS08 formulation (empty circles) and the present modeling (black diamonds) as a function of density.

Fig. 14 tends to show that the two models are almost equivalent. Rivkin *et al.*'s data are established along practically quasi-isotherms but variations near the critical point are better observed in an isochoric plot. We have therefore chosen to represent as an example a particular isochor near the critical density. Since there is no Rivkin *et al.*'s data along an isochor, we show on Fig. 15 an isochor such that we find as many points as possible fairly close to the latter. It is then interesting to observe on Fig. 15(a) that the two models are not equivalent and in particular they do not extrapolate in the same way on the saturated vapor pressure curve (SVP): Indeed, the IAPWS08 formulation diverges by construction while the present modeling does not diverge at all on SVP. However, Rivkin *et al.*'s data do not allow us to separate the two models here. We also observe that the curvature of the present modeling around 650 K seems more realistic than that of the IAPWS08 formulation.

Fig. 15(b) shows the evolution of the two terms constituting Eq. (18) along this isochor: it can be seen that they are of about the same order of magnitude but that the variation is essentially due to the gas-like term (i.e. Knudsen term), in particular the rise near SVP; therefore this variation effect does not occur on the self-diffusion coefficient close to SVP.

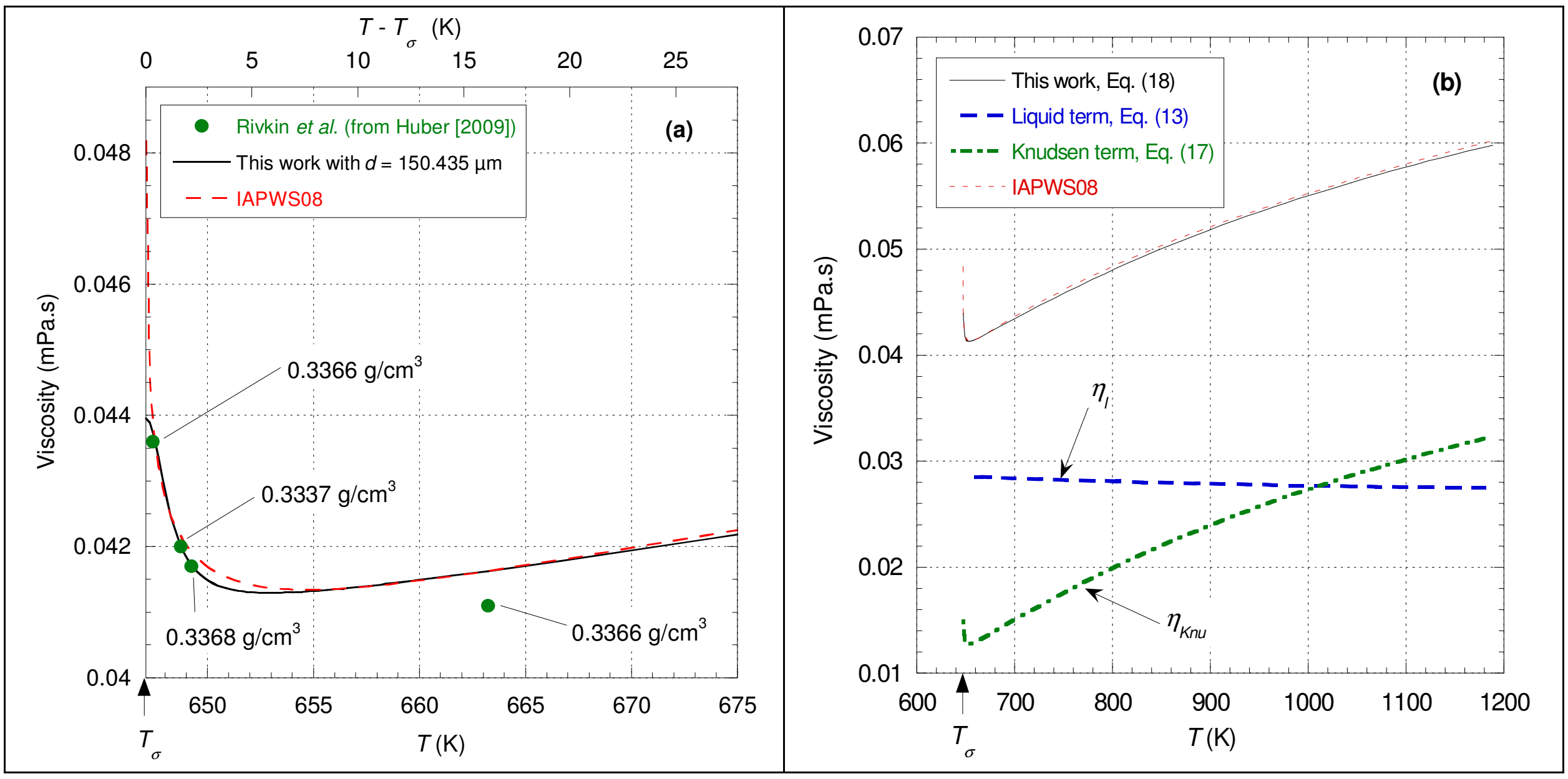

Fig. 15. Representation of the viscosity isochor of water for $\rho = 0.3351$ g/cm$^3$ as function of temperature: (a) partial representation of the isochor over a temperature range of 28 K from the temperature $T_\sigma$; (b) the corresponding viscous terms $\eta_l$ and $\eta_{Knu}$ in the present modeling. $T_\sigma$ represents the temperature on the saturated vapor pressure curve for this particular isochor.

Finally, the current reference equation of state can be viewed has an “experimental dataset” corresponding to a virtual set-up (i.e. a virtual viscometer). So, in the context of the model presented here, some of the parameters which describe this reference equation of state must be changed in a certain way to understand the real datasets coming from very different experimental set-up configurations and geometries. This will be the object of the next section.

## 4 Analysis of some viscosity datasets with emphasis on experimental set-up characteristics

It is very conventional to ignore experimental data that are sufficiently different from a set of supposedly consistent data. These data are generally discarded because of experimental errors that supposedly “escaped” to the corresponding authors. Indeed, most measurement methods only obtain viscosity relative values. To convert them into absolute values this requires the use of models in which device or calibration constants have to be determined. Depending on the methods chosen, this generates systematic pure discrepancies between different datasets. On the other hand, if the offsets are of little physical interest, the variations of these datasets according to various parameters (e.g. density, temperature) are representative of the underlying physics of the medium being studied. In a usual approach, different variations between authors for the same states in the phase diagram are not tolerable if these variations are greater than the respective uncertainties. It does indeed appear that for all the datasets analyzed here, the variations are not *a priori* compatible from one author to another if we take into account each author’s error bars/uncertainties.

We now give results showing the generality of the formalism emphasizing both its ability to describe phase behavior far and close to a transition and its relevance to account for observed sample size dependencies and pressure memory effects. In a general way we will see that to reproduce all the data it is enough to rescale the values of $d$ and $d_N$ with a constant: the two scaling factors will be named $C_d$ and $C_N$, respectively. Many experimental data are obtained in a relative form and very often they are calibrated using data from other experiments that do not correspond to the same experimental configurations. To take these experimental effects into account, it is necessary to start from relative data and renormalize them by taking into account their own experimental configuration. A renormalization constant $C$ (*i.e.* a proportionality constant in front of the right-hand side of Eq. (18) and/or Eq. (12)) must be considered to reproduce the different problems associated with calibration. Hence $C$ is a scaling factor (whose value is close to 1) which does not change the variation laws.

In this section we do not claim displaying all the experimental data exhaustively but only a significant set to support the present approach.

To determine the density values, we used the 1995 IAPWS formulation (i.e. named in the following IAPWS-95 formulation) in general (Ref. 17). For the data in the supercooled phase we used both the data of Bridgman (Ref. 18) and Mishima (Ref. 19) (see Appendix D) while for pressures greater than 1000 MPa we used both the data of Bridgman (Ref. 18) and Wiryana *et al.* (Ref. 20) to interpolate and/or extrapolate density values. Calculations with two determinations allow us to control whether or not there is a significant impact of these determinations on the result. When no clarification is provided on this point, it is because the impact of the different determinations is negligible on the final result.

Throughout the remaining of this article, the comparisons of the experimental data with the present modeling will all be illustrated by plots that we will call "deviation plots" in which the y-axis called "deviation" simply represents the ratio $\eta_{\exp}/\eta_{\text{calc}}$ for the viscosity and $D_{t_{\exp}}/D_{t_{\text{calc}}}$ for the self-diffusion coefficient.

### 4.1. Dynamic Viscosity

To support our purpose, we will scan the different regions of the phase diagram.

#### *4.1.1. Viscosity of Water at Atmospheric Pressure*

Due to importance of the liquid water, a very large number of data is available for temperatures ranging from -35 °C to 100 °C (i.e. 238.15 K to 373.15 K). On this isobar most of the authors give viscosity tables with 4 digits which means that the experimental data must be reproduced with a great precision. From Ref. 7 all the data between 0 °C and 100 °C (i.e. from 273.15 K to 373.15 K) can be reproduced within an uncertainty of ±1%. This uncertainty is globally much greater than the own uncertainty of each dataset. Now if we observe the data from different authors by taking into account their own uncertainty, all these data are *a priori* not consistent with one another (i.e. their error bars does not overlap). For this entire paragraph the temperature scales is displayed both in Celsius degree with the symbol $t$ and in Kelvin with the symbol $T$.

Before analyzing in detail the different data sets, it is interesting to consider the variations of the shear elastic constant versus temperature along the atmospheric isobar. Fig. 16a shows that $K^*$ and $K_0^*$ have very close values above 250 K and these values are close to 1. Further, it is seen that the variation $K_0^*(T)$ is similar to that of the density. Now Fig. 16b shows that the released gas density $\tilde{\rho}_{Knu}$ also follows the same variation as that of the density, with a maximum at a temperature close to the density maximum (i.e. at 3.98 °C). It is also observed that below 250 K in the supercooled phase, no more gaz is released by the shear associated with the viscosity measurement. This is consistent with the fact that the size of the basic units becomes larger and larger on cooling as we will see in section 4.2.1. We stress that the actual density $\rho_{Knu}$ of the relesead gaz is much smaller than $\tilde{\rho}_{Knu}$ by a factor $2\pi/(\delta\, q_{c0,\mathrm{crit}})$, typically by a factor $10^{-5}$ dępending on the experimental set up characterised by the distance $\delta$ (which corresponds typically to the radius of the tube in a Poiseuille flow).

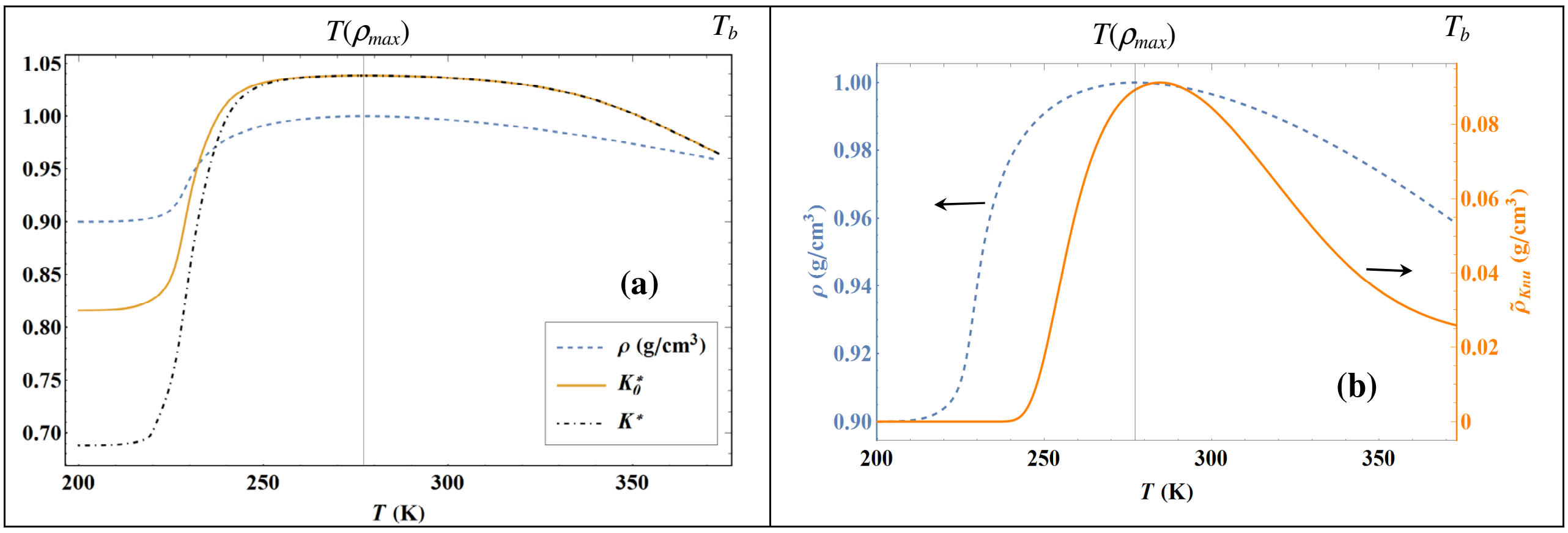

Fig. 16. Temperature variations of different parameters from 200 K to the boiling point along the atmospheric isobar: (a) density $\rho$ (blue dashed curve), $K_0^*$ (orange curve) and $K^*$ (black dot-dashed curve); (b) densities $\rho$ (blue dashed curve) and $\tilde{\rho}_{Knu}$ (orange curve). $T_b$ represents the temperature of the boiling point and $T(\rho_{\max})$ represents the temperature corresponding to the maximum density.

We will first show that all datasets between 0 °C and 100 °C (i.e. from 273.15 K to 373.15 K) can be reproduced correctly with the present modeling by rescaling very slightly the values of $d$ and $d_N$. Fig. 17 to Fig. 19 show that the IAPWS08 formulation is not able to reproduce some experimental data in the limit of ±1% in particular the variations with the temperature are particularly incorrect which is an important inconsistency.

Let's start the analysis with Leroux's data (Ref. 22). It is interesting to note that Dorsay in its monography (Ref. 21) disregarded Leroux' data because of possible errors on the temperatures values as he wrote:

> "P. Leroux ascribes an uncertainty of not over 1 in 200 to his elaborate determinations in the range 1.5 °C to 44.5 °C; but their variation with the temperature is quite different from that of the values obtained by others. It is believed that this discrepancy is due to errors in the temperatures, as the method by which he inferred the temperature of the water is not satisfactory, and the discrepancy is such as would exist if the recorded temperature were, in each case, intermediate between the actual temperature of the water and that of the room, […]. His values are omitted from this compilation".

Fig. 17 shows that Leroux' data are perfectly consistent with the temperature variations and that the uncertainties of these data are within ±0.6% in agreement with the experimental uncertainty.

Fig. 18 and Fig. 19 show that Slotte's corrected data (corrected data from Ref. 23) and Sprung's corrected data (corrected data from Ref. 23) are more accurate than those of Leroux except for the first point at 0 °C from Slotte's dataset which seems to present a too large deviation: this might be interpreted as an artifact due to the correction made by Thorpe *et al.* (Ref. 23) on the original measurements. Since the coefficient $C_N$ is always smaller than 1 for these experimental datasets it means that for these experiments the coherence volume is always smaller than the one used for representing the IAPWS08 formulation.

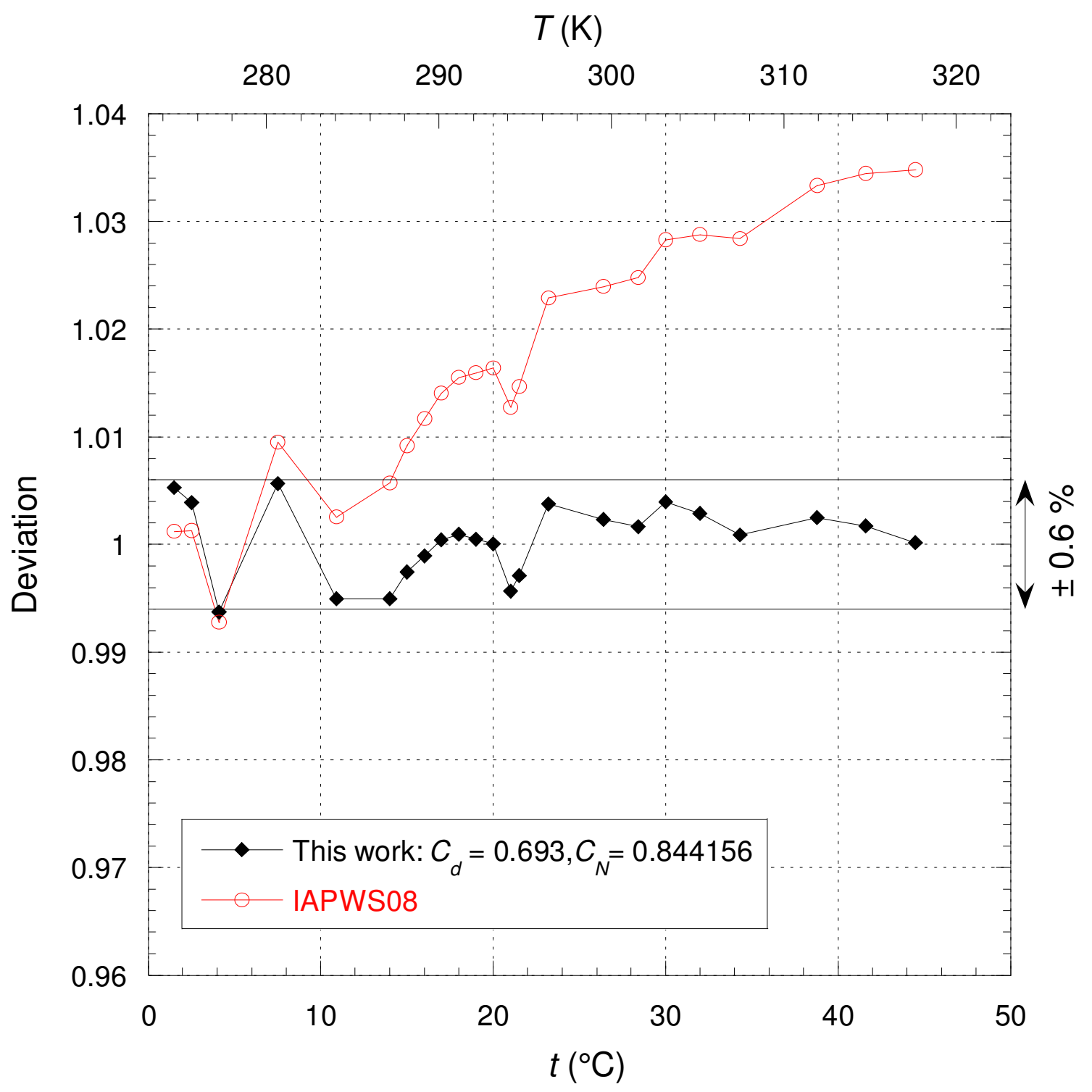

Fig. 17. Ratio of Leroux's data (Ref. 22) with the IAPWS08 formulation (empty circles) and the present modeling (black diamonds) as a function of the temperature at atmospheric pressure. The lines are eye guides.

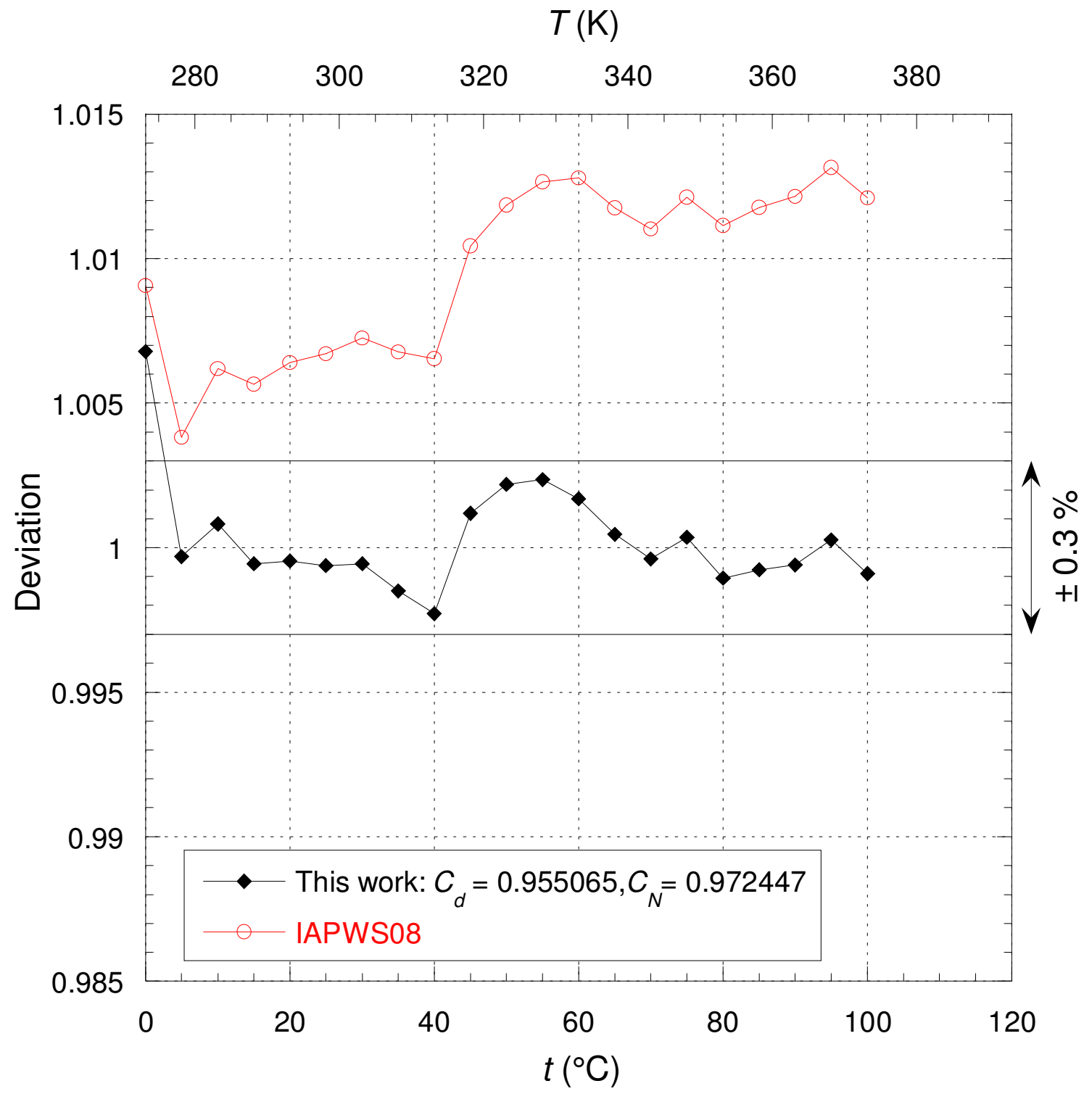

Fig. 18. Ratio of Slotte's data (from Ref. 23) with the IAPWS08 formulation (empty circles) and the present modeling (black diamonds) as a function of temperature at atmospheric pressure. The lines are eye guides.

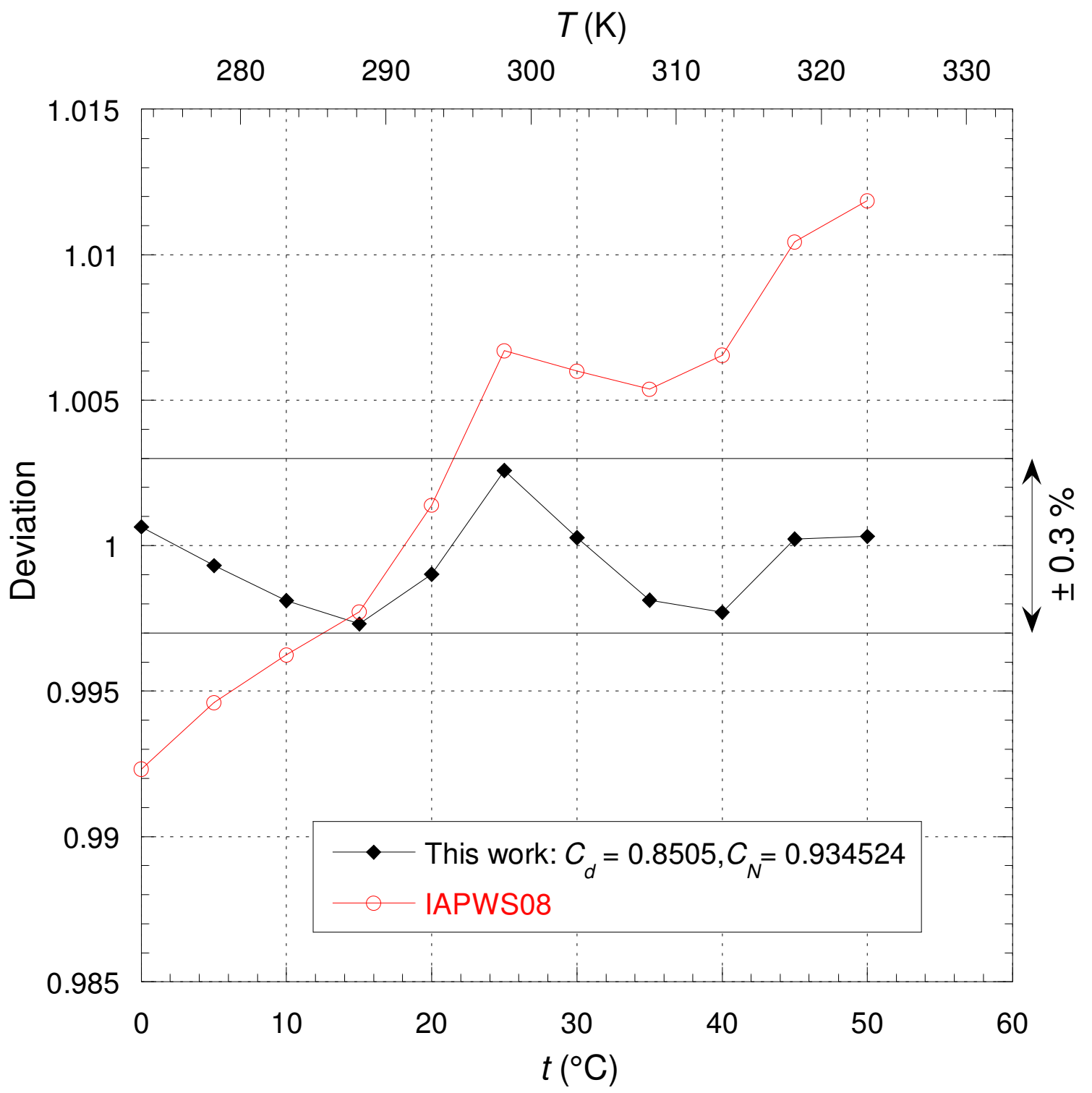

Fig. 19. Ratio of Sprung's data (from Ref. 23) with the IAPWS08 formulation (empty circles) and the present modeling (black diamonds) as a function of temperature at atmospheric pressure. The lines are eye guides.

Korson *et al.* (Ref. 24) wrote that their measurements:

> "provide more precise data than have been available before (at least as far as the temperature dependence is concerned) in the interval from 10 to 70°"

(*i.e.* in their Table II, the data given below 10 °C and above 70 °C are extrapolated from a formula and must not be considered). Fig. 20 shows as previously that these data can be very well reproduced within ±0.04% corresponding to the high accuracy of these data. It can be noticed that these data are reproduced by correcting the values of $d$ and $d_N$ by less than two thousandths. It can be seen here that the IAPWS08 formulation represents the temperature dependence quite well, but appears globally shifted in absolute value.

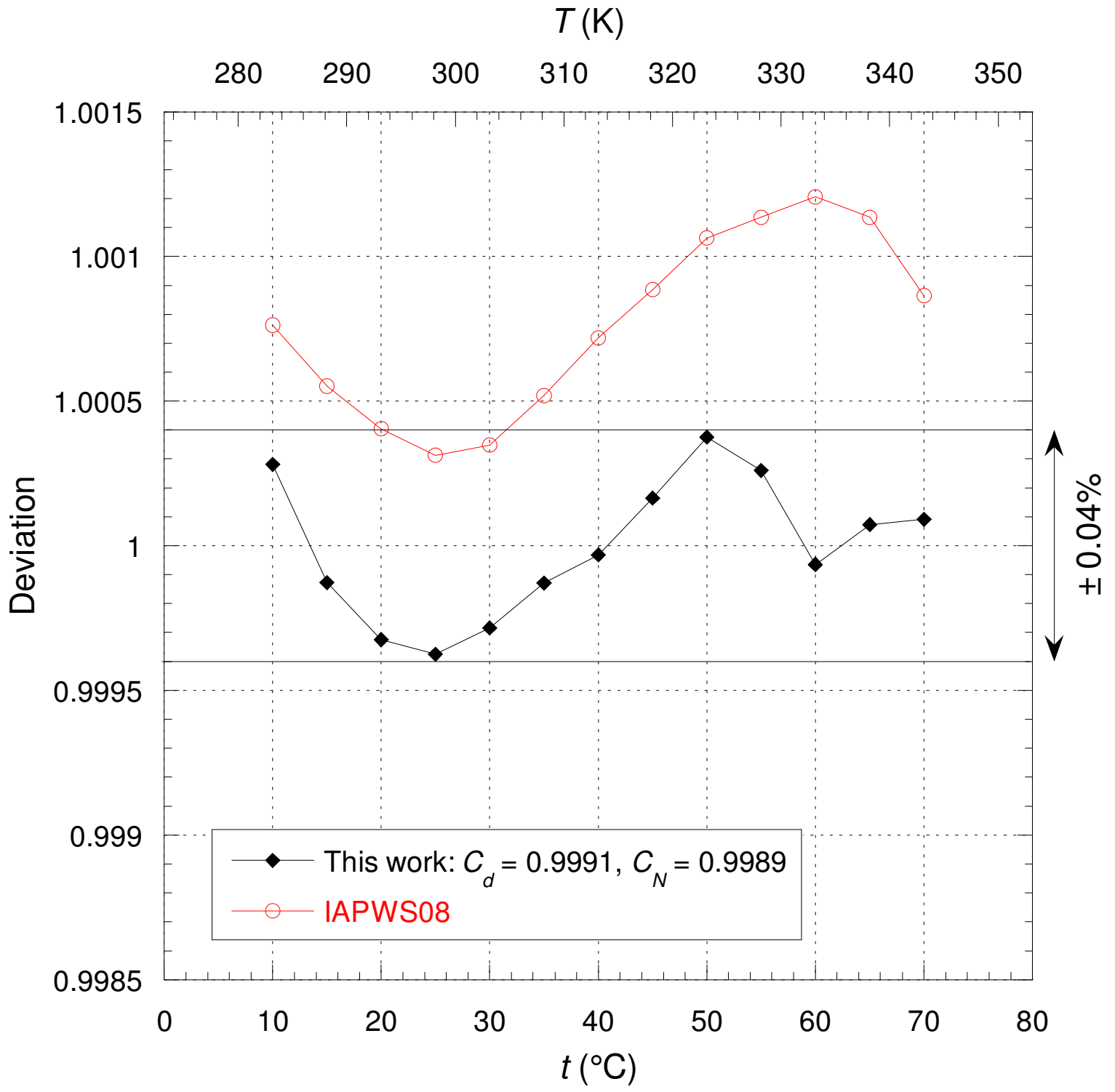

Fig. 20. Ratio of Korson *et al.*'s data (Ref. 24) with the IAPWS08 formulation (empty circles) and the present modeling (black diamonds) as a function of the temperature at atmospheric pressure. The lines are eye guides.

We will now discuss an interesting physical property that has been found by different authors who made the same experiments but with capillary tubes having different geometries. At atmospheric pressure, Thorpe *et al.*'s experiments (Ref. 23) are very representative: these authors have used two capillary tubes (they named Right and Left) having different elliptical section characteristics but the differences are very small (*i.e.* the mean radius of two tubes is close to 82 μm). The analysis (see Fig. 21 and Fig. 22) reveals that they did not obtain the same temperature dependence (i.e. the value of $C_N$ must be different). Indeed, there is a systematic deviation of the two measurements that does not correspond to a noise (i.e. the two datasets have the same uncertainties of ±0.2%). This means that the coherence volume in the two experiments is different and is slightly smaller in the Right tube than in the Left tube. In addition, the tubes were re-used to make another series of experiments but around the temperature of the maximum density of water. While the calibration method has not changed from the previous series, the difference is greater than the uncertainties between the two datasets for each tube resulting again in a very slightly different value of $C_N$ as can be seen on Fig. 21 and Fig. 22. Since the value of $C_d$ remains the same for each tube (*i.e.* the geometrical parameters of the tubes have not changed), we can see here the effect of the water "*purity*". Indeed, when doing their experiments, Thorpe *et al.* wrote that:

> "The sample of water used was distilled just before its introduction into the glischrometer from a quantity which had been repeatedly distilled in order to free it from dust. Special pains were taken in the final distillation to obtain a sufficient of air-free as well as dust-free liquid".

Very small water "purity" variations are sufficient to obtain the very small variations of the deduced $C_N$ values.

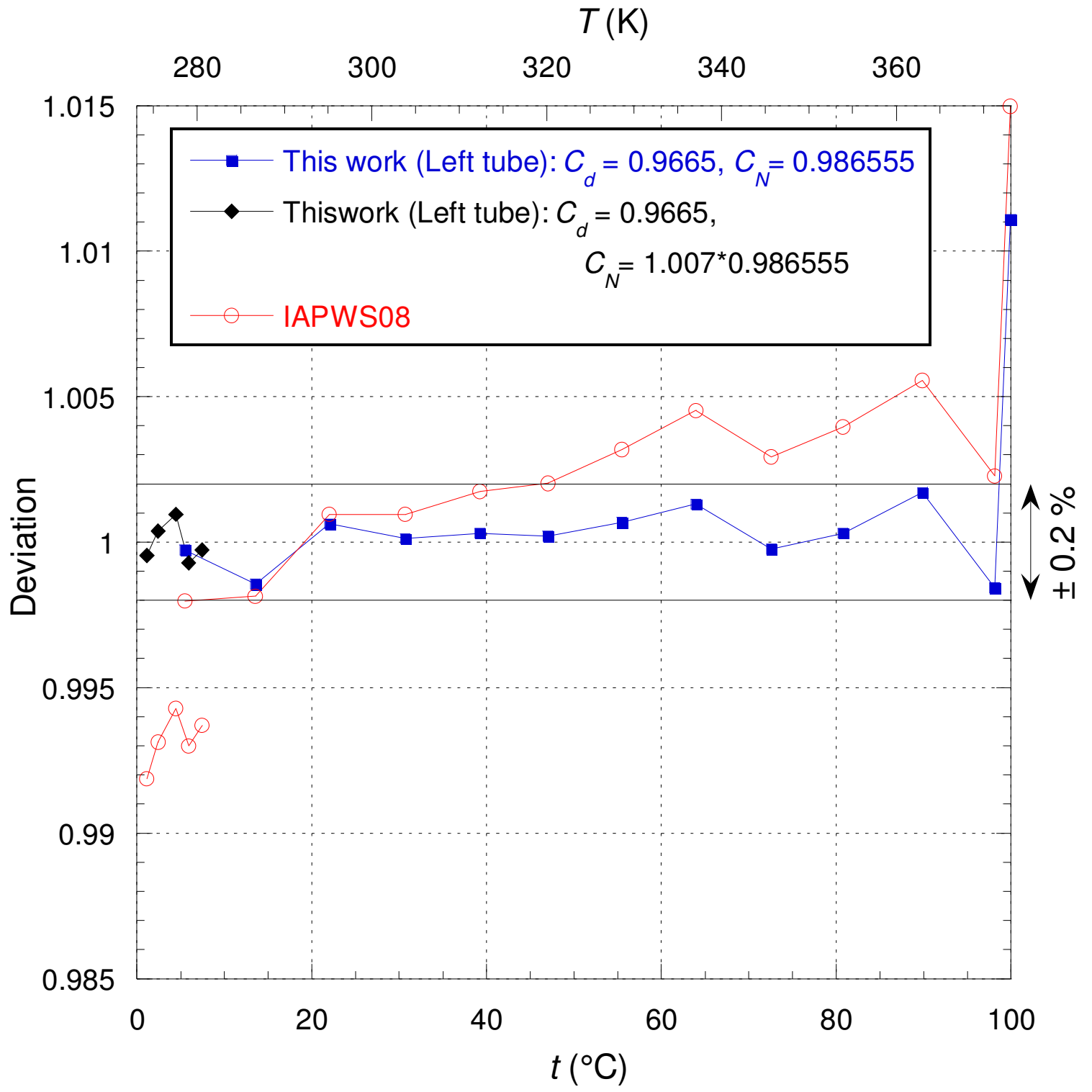

Fig. 21. Ratio of Thorpe's data (Ref. 23) with the IAPWS08 formulation (empty circles) and the present modeling (black diamonds and blue squares) as a function of the temperature at atmospheric pressure for the Left tube. The lines are eye guides.

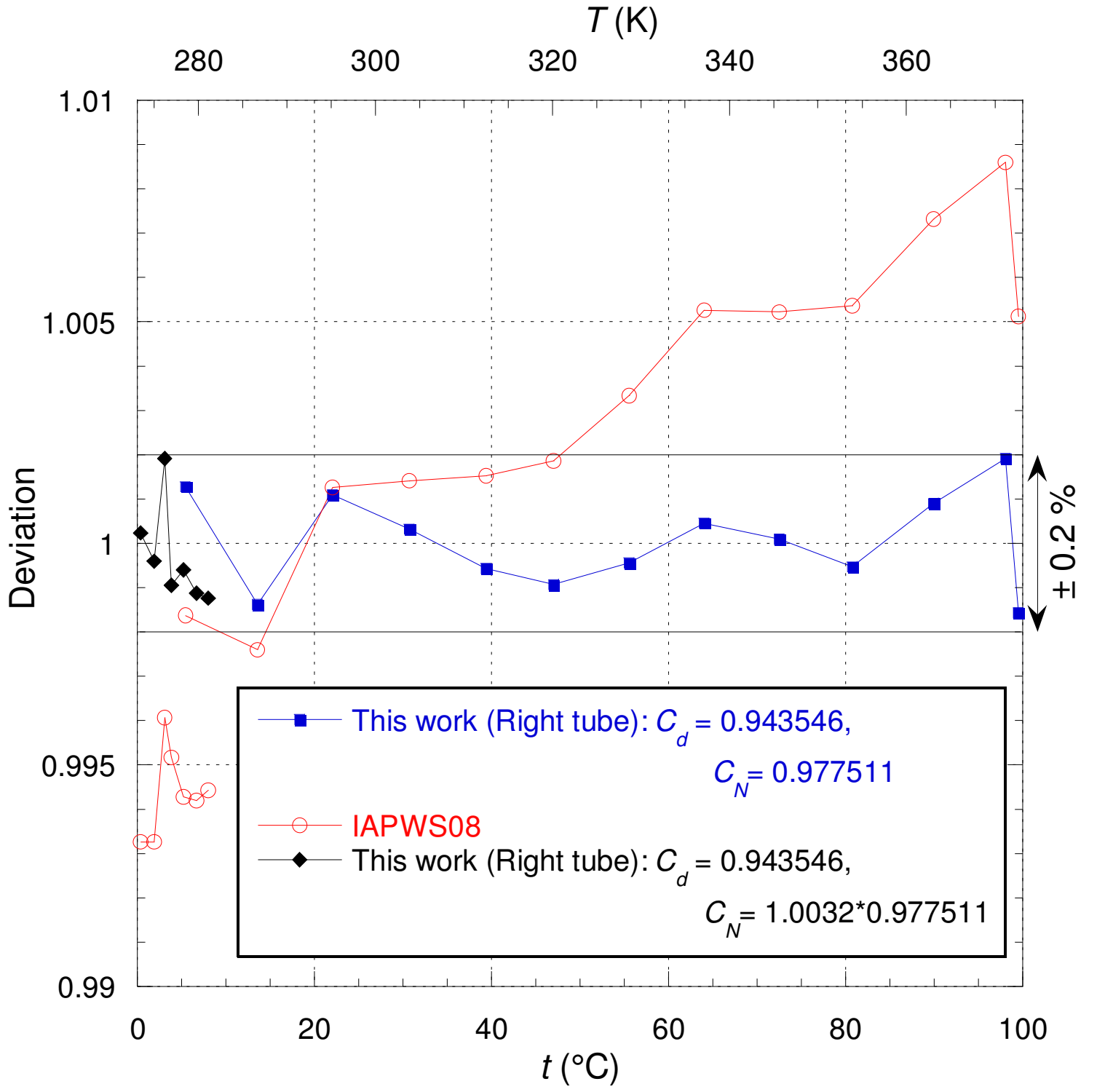

Fig. 22. Ratio of Thorpe's data (Ref. 23) with the IAPWS08 formulation (empty circles) and the present modeling (black diamonds and blue squares) as a function of the temperature at atmospheric pressure for the Right tube. The lines are eye guides.

It should be noted that the point at 99.97 °C on Fig. 21 is away from the uncertainty while the quasi-same point on Fig. 22 does not have this problem. It is likely an experimental anomaly because it corresponds to a state on SVP.
It can be seen here that the value of *d* is smaller than the one corresponding to the "virtual" experiment of the IAPWS08 formulation but this value is still higher than the tubes radius value. This shows that it is not possible to systematically identify the parameter *d* as the tube radius in this kind of experiment. However, it appears in a general way that if the radius of the tube is smaller than 100 μm then a smaller value of *d* must be imposed and *vice versa*.

Finally, we will explore the datasets in the supercooled phase at atmospheric pressure. Most of these data were taken into account in the IAPWS08 formulation but were not discussed nor presented.
In a small range of low temperatures, White *et al.* (corrected data from Ref. 25) and Eicher *et al.* (Ref. 25) have made some experiments in almost the same temperature range below 0°C which according to Eicher *et al.*:

> "[extend] the range of accurately known viscosity data from well above to well below the ice point."

Studies prior to Eicher *et al.*'s such as Hallett's (Ref. 26) have produced data down to -24 °C, but it is true that they are less accurate as we will see.
Fig. 23 and Fig. 24 show that both datasets can be reproduced by the present modeling with an uncertainty of ±0.15% in accordance with Eicher *et al.*'s experimental uncertainty of ±0.2%. It appears that the IAPWS08 formulation reproduces correctly Eicher *et al.*'s data but cannot reproduce the temperature dependence of White *et al.*'s data which is quite different. This difference results in the present modeling in coefficients $C_d$ and $C_N$ which have different values for these two datasets.
Hallett wrote about White *et al.*'s experiment that:

> "The viscosity was measured by the time required for 2.808 cm$^3$ of water to flow through a capillary tube 0.1 mm in diameter (Bingham-type viscometer) where the entire volume of liquid was cooled to the temperature of the capillary."

As we have previously discussed, it appears that the radius in White *et al.*'s experiment is 50 μm but the corresponding value for *d* is here around 86 μm.

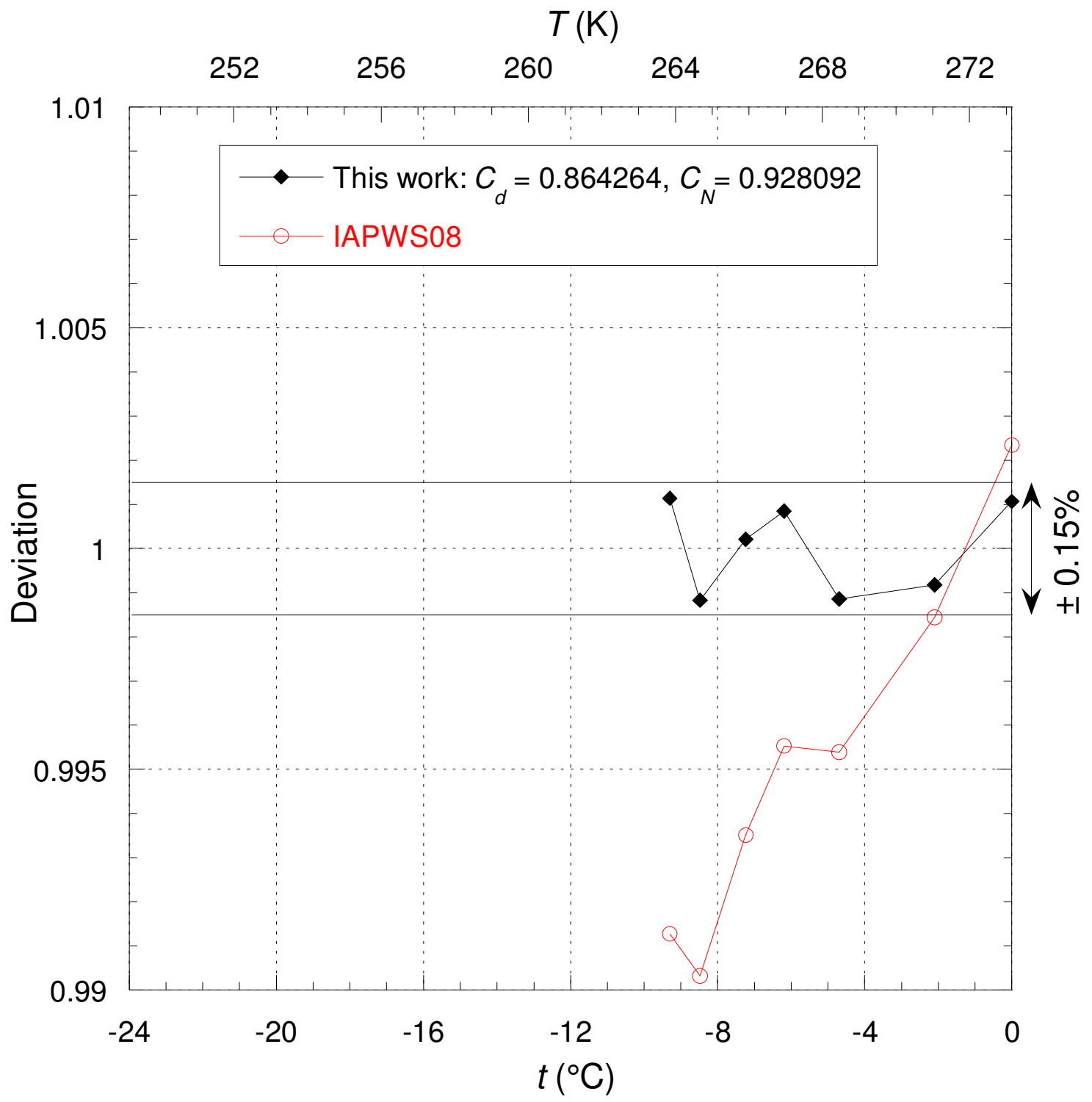

Fig. 23. Ratio of White *et al.*'s data (from Ref. 25) with the IAPWS08 formulation (empty circles) and the present modeling (black diamonds) as a function of the temperature at atmospheric pressure. The lines are eye guides.

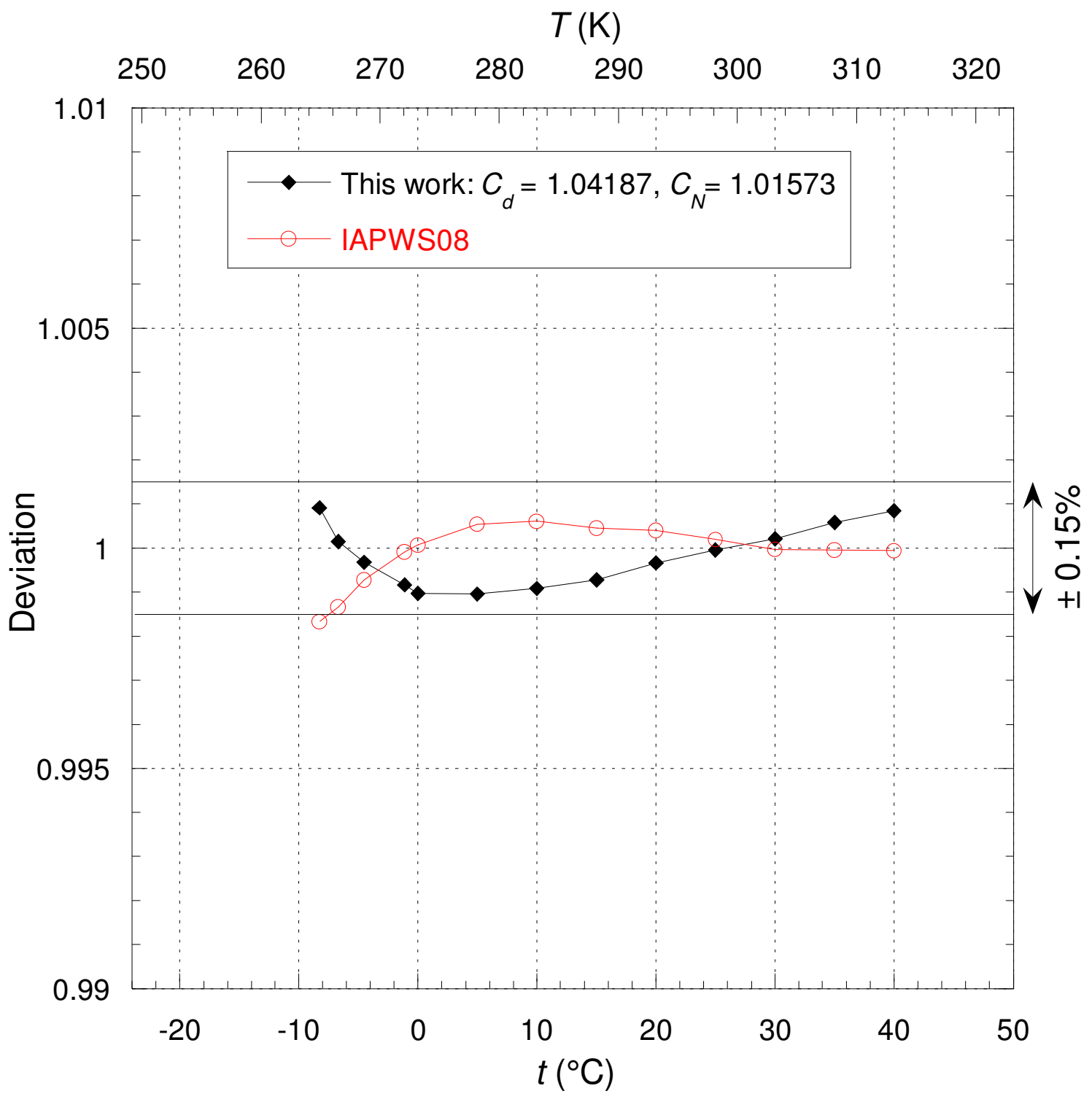

Fig. 24. Ratio of Eicher *et al.*'s data (Ref. 25) with the IAPWS08 formulation (empty circles) and the present modeling (black diamonds) as a function of the temperature at atmospheric pressure. The lines are eye guides.

For a much larger temperature range below 0 °C, there are mainly three water viscosity datasets from very different experiments; these are the data of Hallett (Ref. 26), Osipov *et al.*

(Ref. 27) and Dehaoui *et al.* (Ref. 28). The lowest temperature of -35 °C (*i.e.* 238.15 K) reached with viscosity experiments are those obtained by Osipov *et al.* but more recently the new experiment made by Dehaoui *et al.* reached almost the same lowest temperature. Fig. 25 shows that the temperature dependence is quite different for these three datasets below -10 °C (i.e. 263.15 K) and that the viscosity values from Osipov *et al.* and Dehaoui *et al.* are always higher than those of Hallet in their overlap region. Dehaoui *et al.* have proposed the following explanations for these deviations:

> "There might exist a small systematic bias between our data and ref. 26 [Hallett] at the lowest temperatures. This might originate from a temperature error. […] Hallett also writes, 'A possible defect in this technique is that the warmer water which enters the measuring system is not cooled sufficiently quickly to the bath temperature, leading to apparently smaller values of viscosity,' and then provides three pieces of evidence suggesting this effect is negligible. However, we note that he [Hallett] carried out the corresponding tests at 273.15 K [0 °C], and the effect might be larger at lower temperatures. We conclude that the small negative deviation of ref. 26 [Hallett] data from ours at low temperature is not significant. In contrast, the large positive deviation of ref. 27 [Osipov *et al.*] data from ours is far above the reported error bars. We attribute this discrepancy to a bias of the Poiseuille flow experiment of ref. 27 [Osipov *et al.*], due to electro-osmotic effects as proposed in *Materials and Methods*."

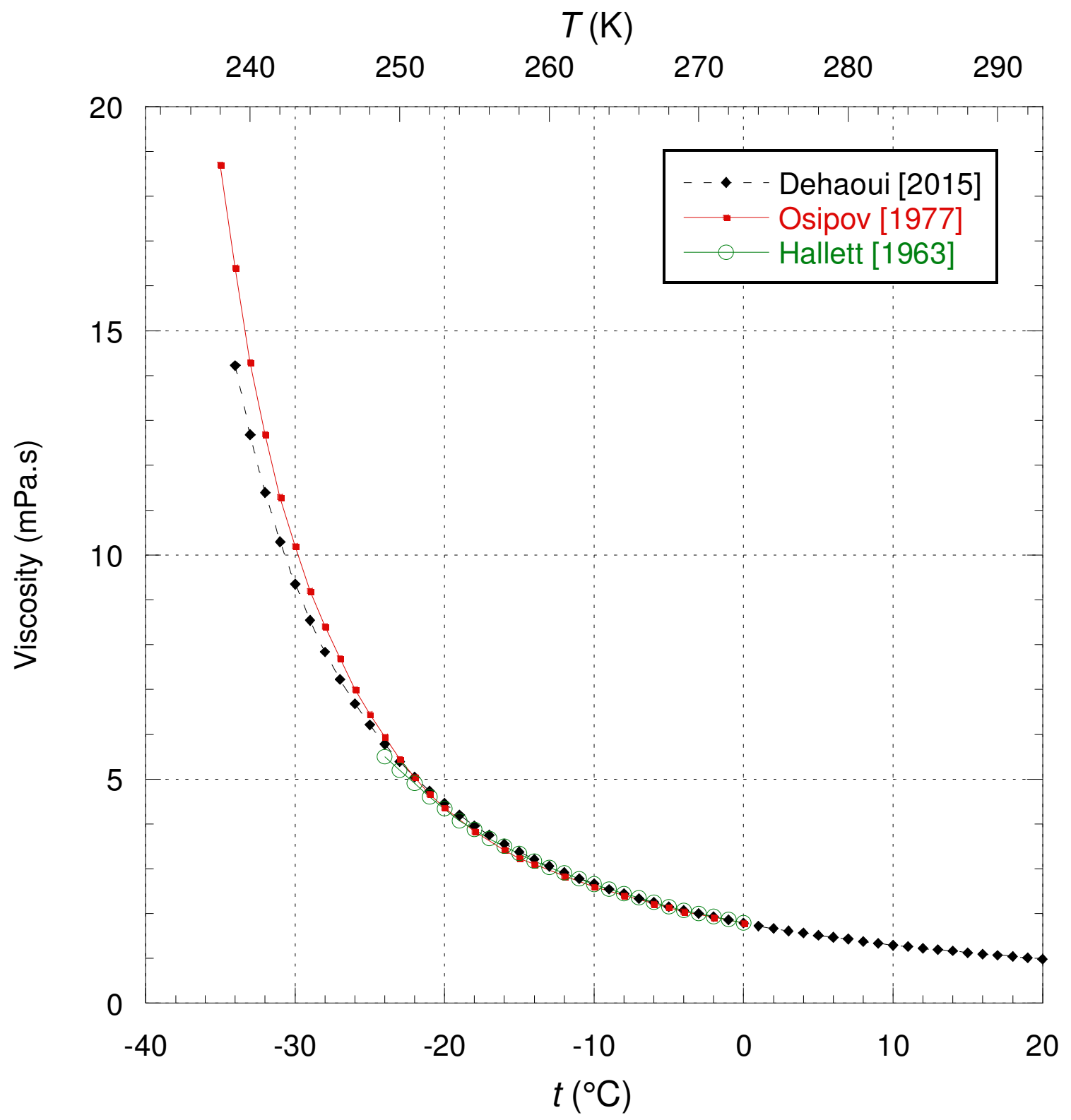

Fig. 25. Water viscosity experimental data at atmospheric pressure obtained by Dehaoui *et al.* (Ref. 28, smooth data), Osipov *et al.* (Ref. 27) and Hallett (Ref. 26). The lines are eye guides.

By considering the different experimental conditions, we will show that it is possible to account for these experimental data with their uncertainty in a consistent way (both with other viscosity data and self-diffusion coefficient data): these data can be explained by accounting for some physical effects without having to rely on experimental bias.

In Hallett's experiment (Ref. 26), the viscosity was measured by the time required for $5\times10^{-3}$ cm$^3$ of water to flow through a capillary tube of 0.2 mm in diameter (*i.e.* a radius corresponding to 100 µm). In comparison to White *et al.*'s experiment, Hallett states that:

> "In the present series of experiments, measurements of viscosity were made in an apparatus where only the capillary tube was cooled, and using a smaller volume of water."

Fig. 26 shows that the present modeling leads, whatever the value of $C_N$, to a poor description of the temperature dependence of Hallett's data; it is even worse than the description of the IAPWS08 formulation. That said, at low temperatures in both cases, both models largely overcome the experimental uncertainty.

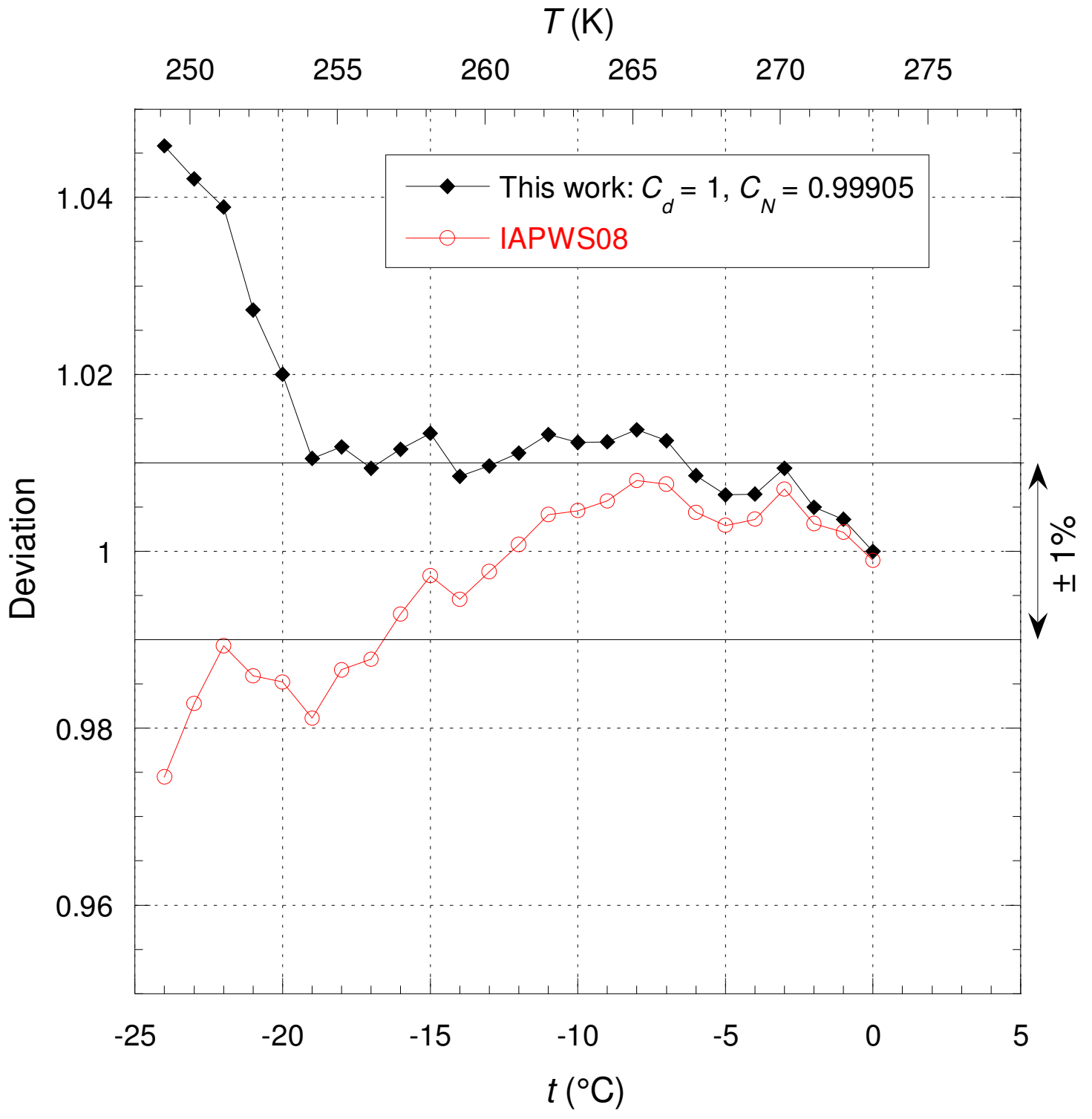

Fig. 26. Ratio of Hallett's data (Ref. 26) with the IAPWS08 formulation (empty circles) and the present modeling (black diamonds) as a function of the temperature at atmospheric pressure. The lines are eye guides.

To understand this discrepancy, it is necessary to remember that the present modeling leads to the existence of a correlation length $\xi$ given in reduced form by Eq. (10). This correlation length present throughout the sample volume also concerns the liquid layer on the tube walls. This thickness is very small for normal liquid states and its effects are negligible: for water at 0 °C, it is close to 4.35 Å. We note that this value is almost twice as high as the one determined in Appendix C for the example of Xie *et al.*'s experiment (Ref. 88) because the values of $d$ and $d_N$ are much higher here.

So to understand Hallett's data, it is enough to admit that when cooling water below 0 °C, this correlation length at the tube walls increases faster than the corresponding correlation length in the bulk. This amounts to considering for the viscosity experiment that everything happens as if the effective radius of the tube was reduced. Hallett wrote that:

"The viscosity was measured by the time taken for a fixed volume (0.01 $cm^3$) to flow through the apparatus under a constant head".

In view of this experimental procedure and the fact that Hallett wrote:

"[…] we can neglect both the kinetic energy term, and the end correction",

his data can be simply reproduced by correcting them to the power of 4 of the ratio of the hydrodynamic radius above 0 °C with its value corresponding to the chosen temperature, thus for a Poiseuille type flow:

$$\eta_{\text{tube}}(T, 1\,\text{atm}) = \left(\frac{a}{a_h(T)}\right)^4 \eta_{C_d,C_N}(T, 1\,\text{atm}) \tag{22}$$

where $\eta_{C_d,C_N}(T, 1\,\text{atm})$ represents the viscosity from the present model with the two constants defined on Fig. 26. The variation of the ratio $a_h(T)/a$ can be well represented with the following empirical function:

$$\frac{a_h(T)}{a} = 1 + \left\{-1 + 0.998968\left(1 - \exp\left(-\left|\frac{T - 233.83}{1.186}\right|^{0.6}\right)\right)\right\}\exp\left(-\left(\frac{233.83}{T}\right)^{600}\right)\exp\left(-\left(\frac{T}{T_{\text{tr}}}\right)^{35}\right) \tag{23}$$

with $a$ = 0.01 cm. Taking into account the correction of the hydrodynamic radius, Fig. 27 shows that the data can be reproduced with a deviation less than 1%, which corresponds to the experimental uncertainty (indeed the deviation is well distributed in this band). It can be noted that this kind of argument cannot be used here to explain the deviation of the IAPWS08 formulation since this would require increasing the diameter of the tube even more as it is cooled, which seems physically meaningless.

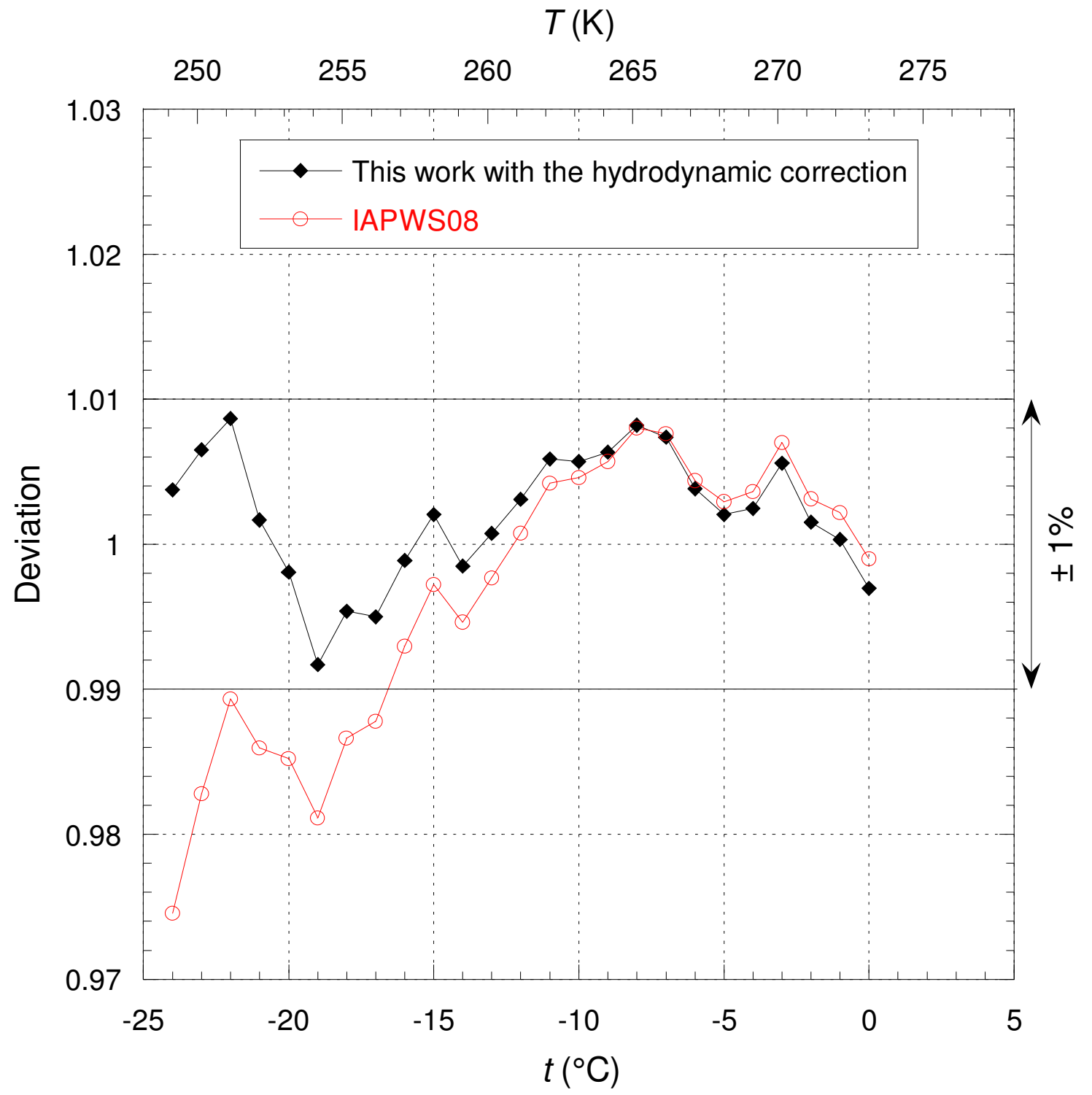

Fig. 27. Ratio of Hallett's data (Ref. 26) with the IAPWS08 formulation (empty circles) and the present modeling integrating the hydrodynamic radius correction (black diamonds) as a function of the temperature at atmospheric pressure. The lines are eye guides.

Osipov *et al.* (Ref. 27) did the same kind of experiment as Hallett's but with an even smaller volume of liquid (approximately $10^{-8}$ $cm^3$) and quartz capillary tubes with an approximate radius of 1 µm. Fig. 28 shows that it is again not possible to reproduce Osipov *et al.*'s data just by adjusting the value of the two constants $C_d$ and $C_N$: the residual deviation is still too high in both cases whatever the model.

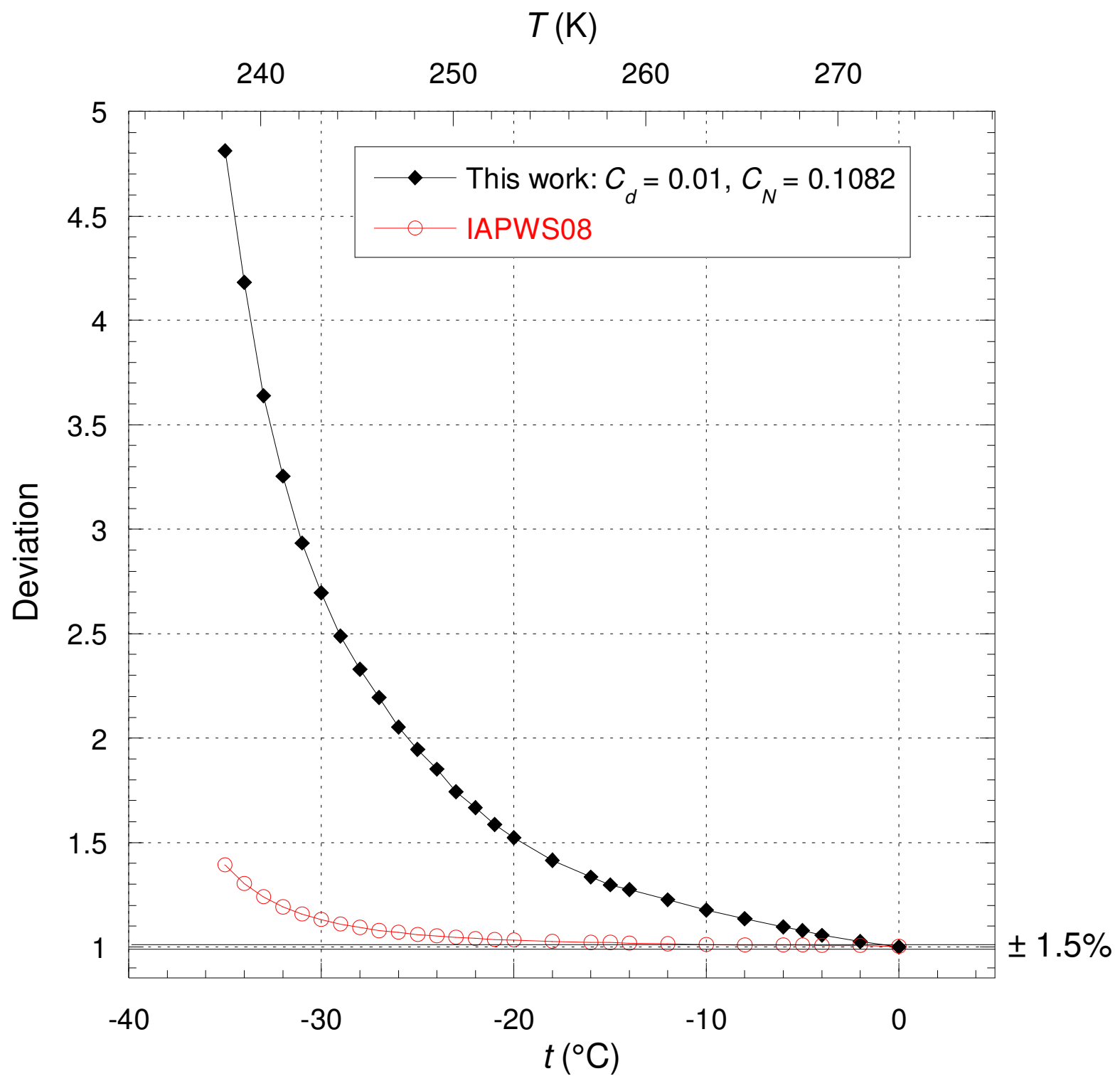

Fig. 28. Ratio of Osipov *et al.*'s data (Ref. 27) with the IAPWS08 formulation (empty circles) and the present modeling (black diamonds) as a function of the temperature at atmospheric pressure. The lines are eye guides.

For Osipov *et al.*'s dataset, the same Eq. (22) of hydrodynamic radius correction should apply except that the variation of the ratio $a_h(T)/a$ is different but it can be expressed in a very similar form to Eq. (23) such that:

$$\frac{a_h(T)}{a} = 1 + \left\{ -1 + 0.99975\left(1 - \exp\left(-\left|\frac{T-220}{16.313}\right|^{1.1694}\right)\right)\right\} \exp\left(-\left(\frac{220}{T}\right)^{600}\right) \exp\left(-\left(\frac{T}{270.56}\right)^{78.7}\right) \quad (24)$$

with $a = 10^{-4}$ cm. Concerning his data, Osipov *et al.* wrote that:

> "The accuracy of the values obtained is approximately to within 1.5% at a temperature above -30 °C for $H_2O$ […]; below these temperatures, the error may reach 2-3%".

Fig. 29 shows that with the hydrodynamic correction of the tube radius, the experimental data can be reported according to their uncertainty. Moreover, the deviation is well distributed around the unit value.

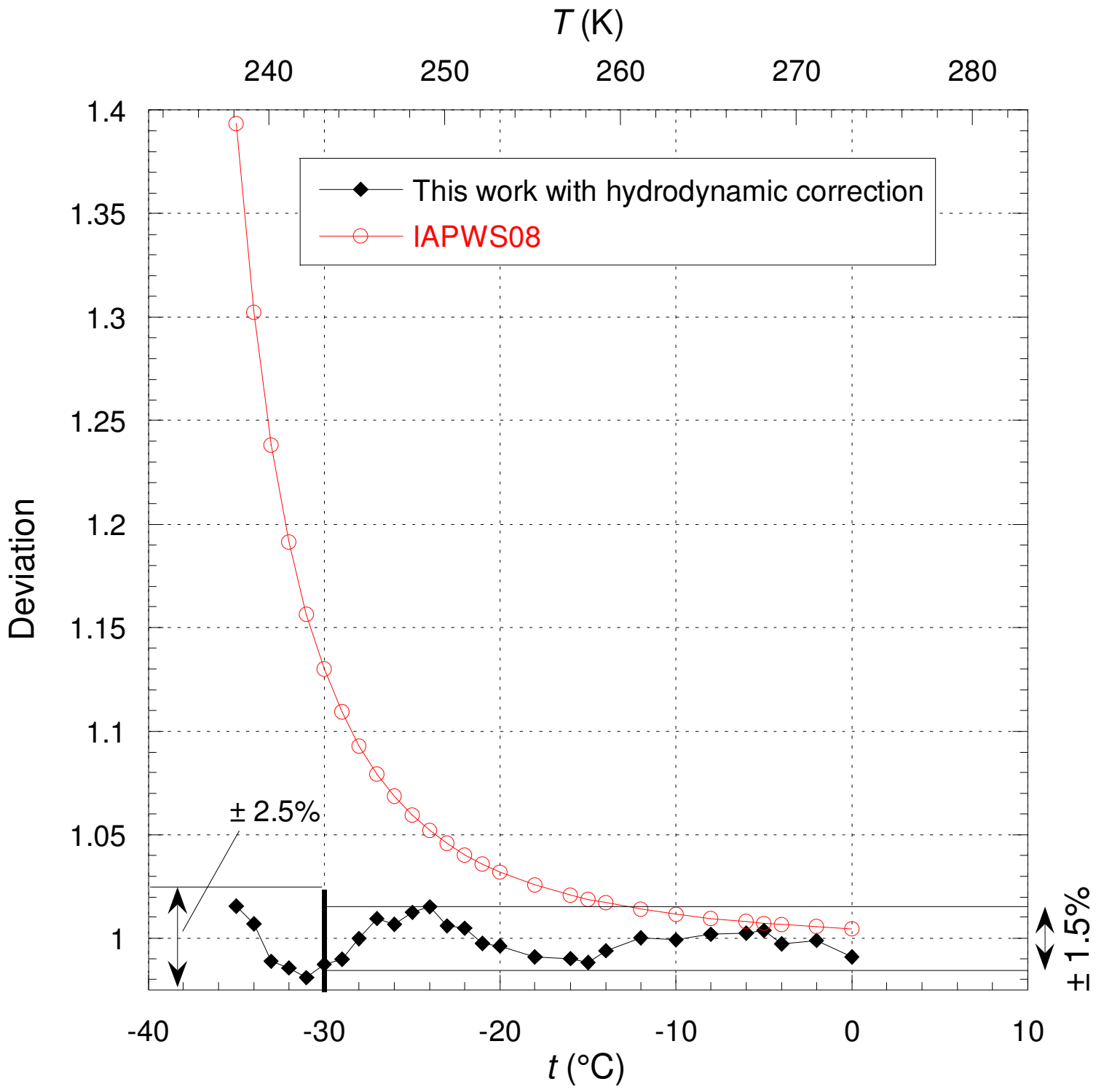

Fig. 29. Ratio of Osipov *et al.*'s data (Ref. 27) with the IAPWS08 formulation (empty circles) and the present modeling integrating the hydrodynamic radius correction (black diamonds) as a function of the temperature at atmospheric pressure. The lines are eye guides.

The experiment of Dehaoui *et al.* (Ref. 28) consists of observing the Brownian motion of polystyrene spheres having a physical radius $a$ = 175±3 nm in a capillary tube. Since the viscosity is derived from diffusion experiments, it must be considered that there is no shear and thus no gas is released (i.e. here $\eta_{Knu} = 0$). Fig. 30 shows that it is not possible below -10 °C to reproduce Dehaoui *et al.*'s data by just adjusting the value of the two constants $C_d$ and $C_N$: one more time the residual deviation at low temperature is still too high in both cases whatever the model.

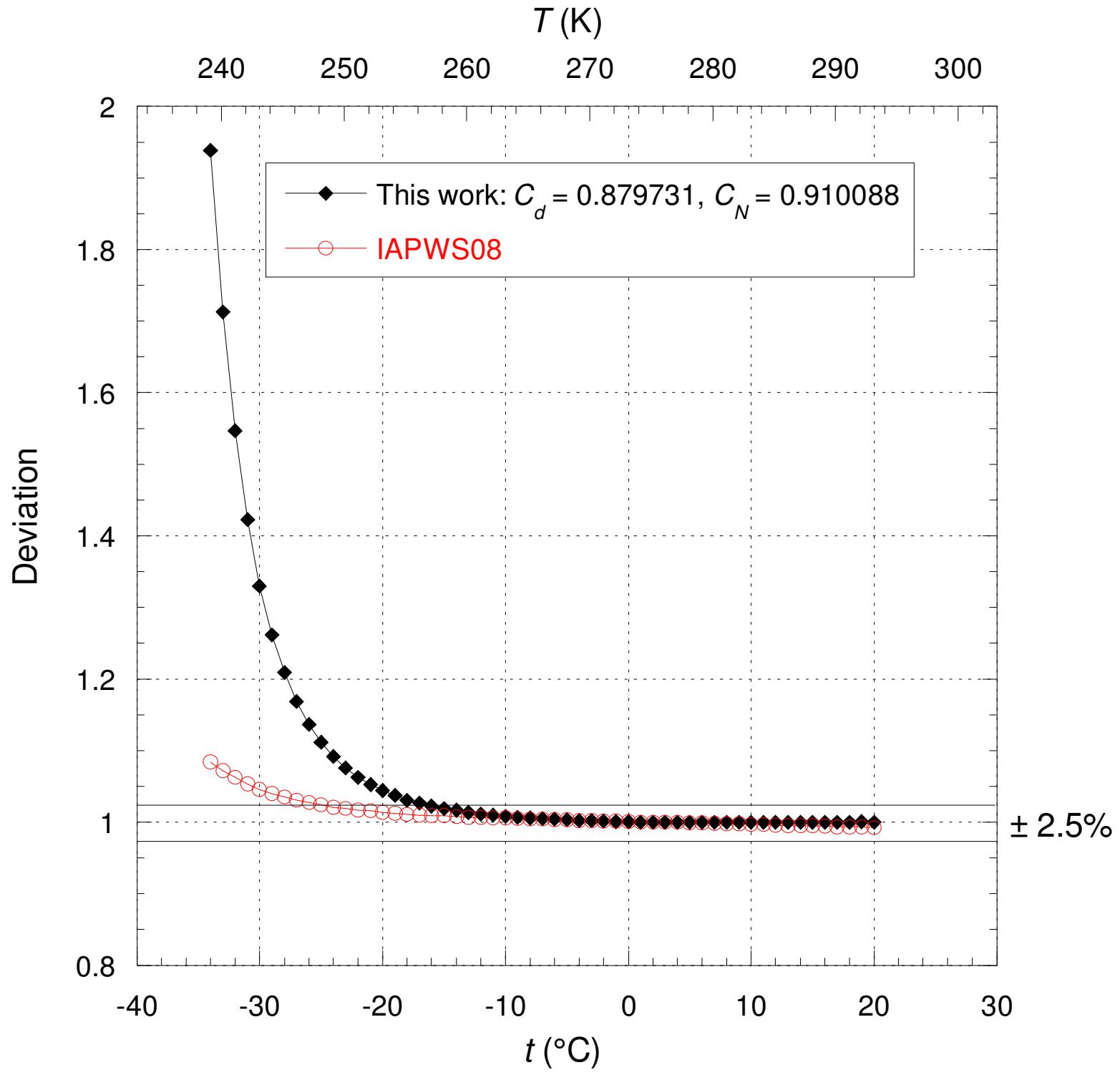

Fig. 30. Ratio of Dehaoui *et al.*'s data (Ref. 28, smoothed data) with the IAPWS08 formulation (empty circles) and the present modeling (black diamonds) as a function of temperature at atmospheric pressure. The lines are eye guides.

Dehaoui *et al.* deduced the viscosity from SE law by using a constant hydrodynamic sphere radius of approximately $a_h$ = 179.9 nm. As with the tubes of Hallett and Osipov *et al.*, it is sufficient to consider that below 0 °C the correlation length of the fluid surrounding the polystyrene spheres increases faster when the temperature decreases than the same correlation length in the bulk. This effect results in an increase in the hydrodynamic radius when the temperature decreases below 0 °C. The use of SE law implies that Dehaoui *et al.*'s data can be simply reproduced by correcting them for the ratio of the hydrodynamic radius at 0 °C with its value corresponding to the chosen temperature, thus:

$$\eta_{\text{sphere}}(T,1\,\text{atm}) = \frac{a_h(T)}{a}\eta_{C_d,C_N}(T,1\,\text{atm}) \tag{25}$$

where $\eta_{C_d,C_N}(T,1\,\text{atm})$ represents the viscosity from the present model with the two constants defined on Fig. 30. The variation of the ratio $a_h(T)/a$ can be well represented with the following empirical function:

$$\frac{a_h(T)}{a} = 1 + \exp\left(-\left(\frac{400}{T}\right)^5\right)\left(\frac{63.95}{|T-212.31|}\right)^{15}\exp\left(-\left(\frac{T}{T_{\text{tr}}}\right)^{50}\right) \tag{26}$$

with $a = 175.38\,\text{nm}$. Taking into account the correction of the hydrodynamic radius, Fig. 31 shows that the smoothed data can be reproduced with a maximum deviation of 0.5%, which is well below the experimental uncertainty between 2.3% and 2.9%. It is also much better observed in Fig. 31 that the IAPWS08 formulation does not reproduce the temperature dependence at all even for temperatures above 0 °C although the model uncertainty is lower than the experimental uncertainty.

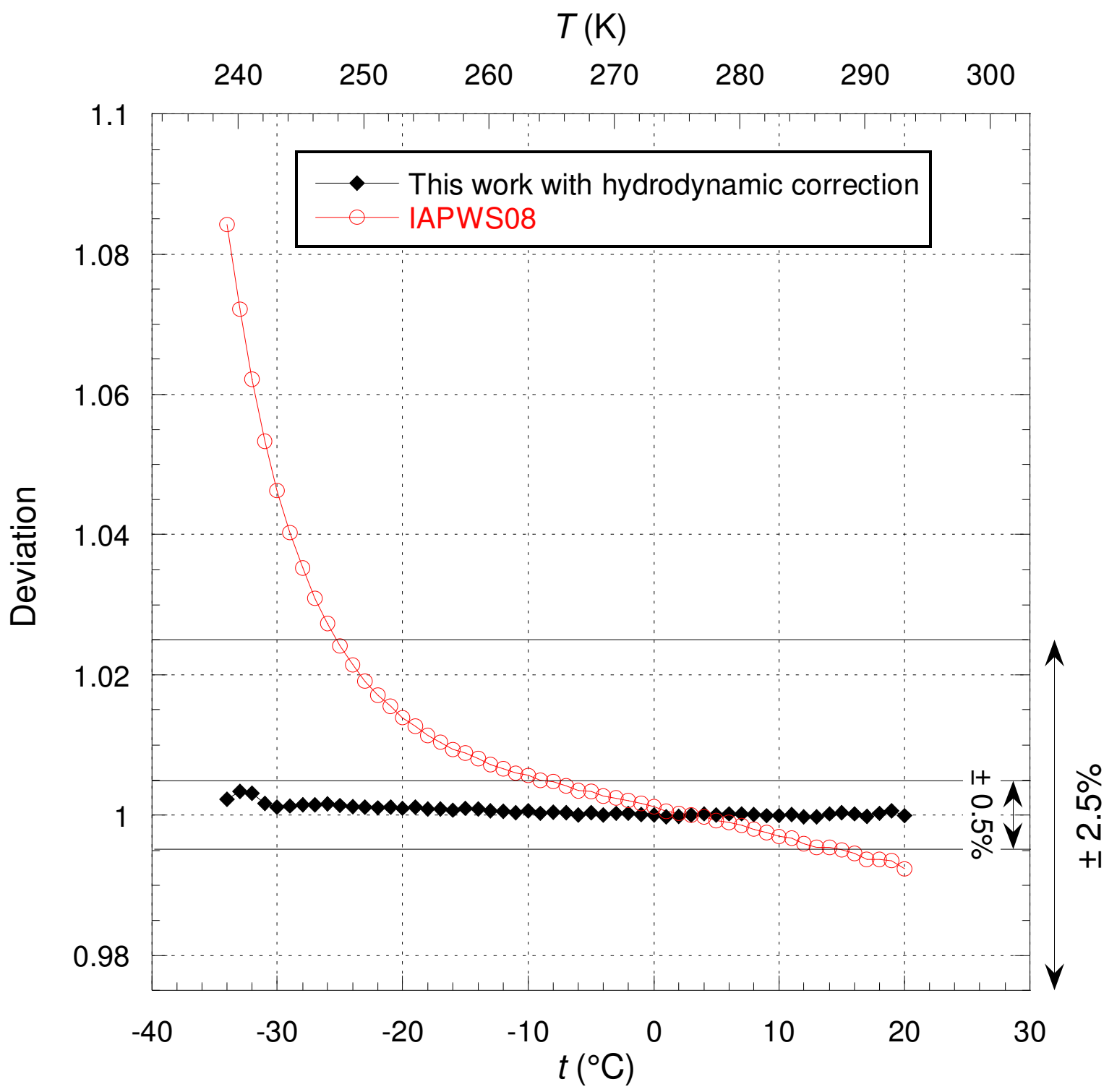

Fig. 31. Ratio of Dehaoui *et al.*'s data (smoothed data of Ref. 28) with the IAPWS08 formulation (empty circles) and the present modeling integrating the hydrodynamic radius correction (black diamonds) as a function of the temperature at atmospheric pressure. The lines are eye guides.

Fig. 32 shows the corresponding variations for the shear correlation length to the walls in relation to the theoretical length in the bulk. It appears that the temperature dependence is quite similar for Hallett and Osipov *et al.* but is very different for Dehaoui *et al.* and this can be attributed to different geometric curvatures and different materials: for Dehaoui *et al.*, the spheres are made of polystyrene and for Hallett and Osipov *et al.* the tubes are made of quartz so their affinity with water is different. In the case of tubes, it is observed that the relative increase of the shear correlation length is very strong below $T_{tr}$ as if a spontaneous "solidification" occurs at the walls because we are in a metastable phase. We will see that the present modeling is consistent with that of self-diffusion coefficients of bulk water for which these wall effects are non-existent (this point will be discussed in section 4.2.2): the assumption of no slip condition is irrelevant in the absence of flow.

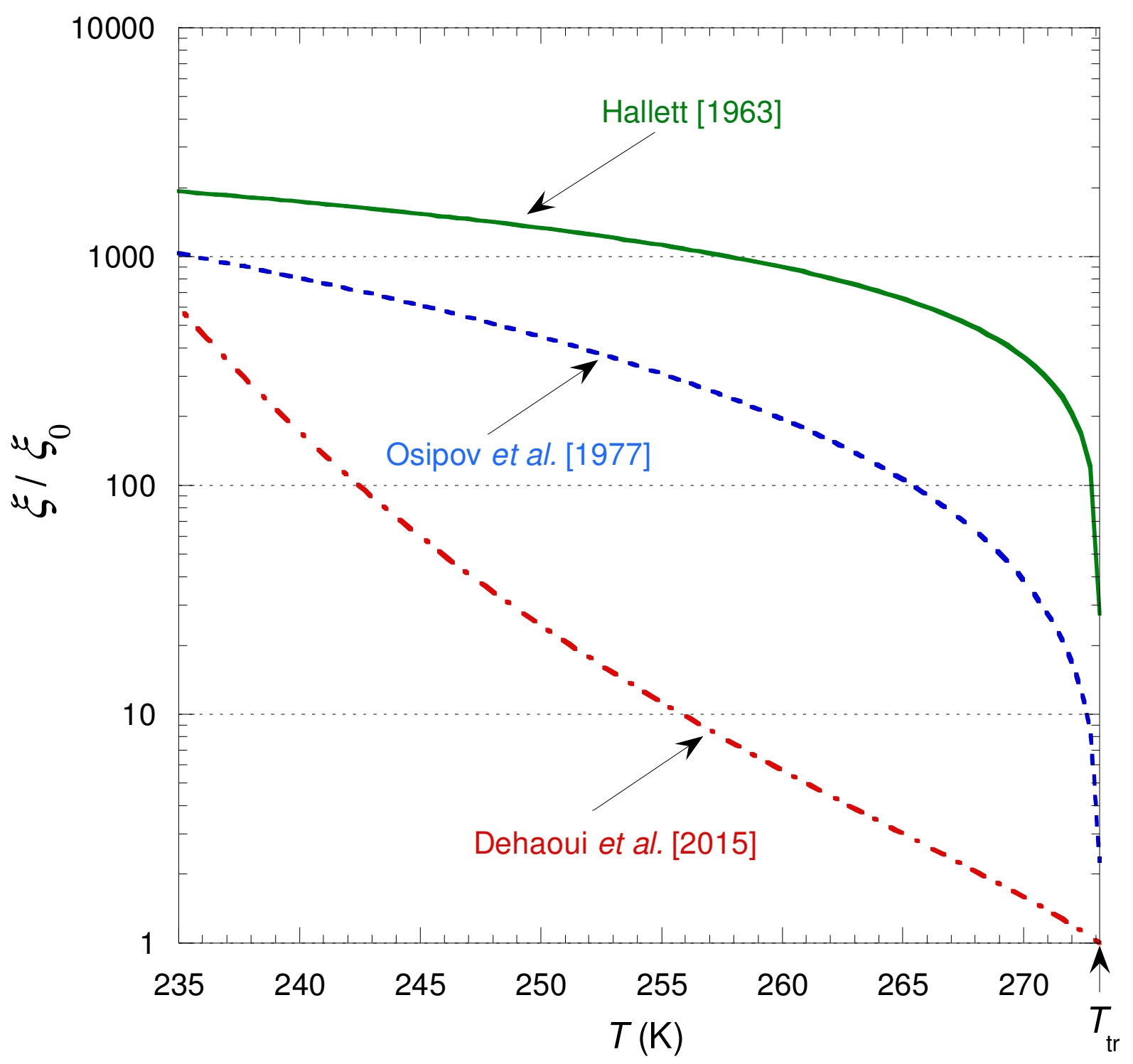

Fig. 32. Shear correlation length ratio variations in supercooled water versus the temperature for the experimental data obtained by Dehaoui *et al.*, Osipov *et al.* and Hallett (Ref. 28, 27 and 26). $\xi_0$ represents the corresponding theoretical correlation length in the bulk which is different in the three cases and $\xi$ the correlation length on the wall.

Other data corresponding to the atmospheric isobar exist but they are part of some larger dataset. They will be analyzed in the following sections as one of several elements of this dataset.

### *4.1.2. Viscosity of Water under Pressure*

We introduce this section by analyzing some experimental data on the Saturated Vapor Pressure curve (SVP). We start by discussing Korosi *et al.*'s datasets (Ref. 27) because they used two different capillary viscometers (i.e. with different tube radii, among other things): an open-type Cannon master viscometer for measuring the viscosities at atmospheric pressure (called here Apparatus I and having a tube radius of 165 µm) between 25 °C and 60 °C and a pressurized instrument (called here Apparatus II and having a tube radius of 120 µm) for the measurements between 75 °C and 150 °C and this makes the link with the previous section. Korosi *et al.* wrote that:

> "The apparatus [II] provided relative measurement, with an estimated precision of ±0.2%, on liquid water in the temperature range 75° to 150 °C".

Fig. 33 shows that the dataset corresponding to the apparatus II can be reproduced within their experimental uncertainty with the present modeling by rescaling slightly the values of $d$ and $d_N$ according to what we have already discussed in the previous section. On the other hand, it can be observed that the IAPWS08 formulation does not have the right temperature dependence at all and therefore cannot reproduce these same data with their experimental uncertainty.

No uncertainty is given regarding the apparatus I but at the end of Korosi *et al.*'s paper we find the following sentence:

> "In reviewing the paper, R. E. Manning recommended the use of [an interpolating equation which is such that] this correlation represents the experimental data within ± 0.05% average deviation".

This suggests that it is possible to represent the data of apparatus I with an uncertainty of ±0.05%. Fig. 33 shows that the present modeling reproduces these data within this deviation. On the other hand, it is somewhat borderline concerning the IAPWS08 formulation.

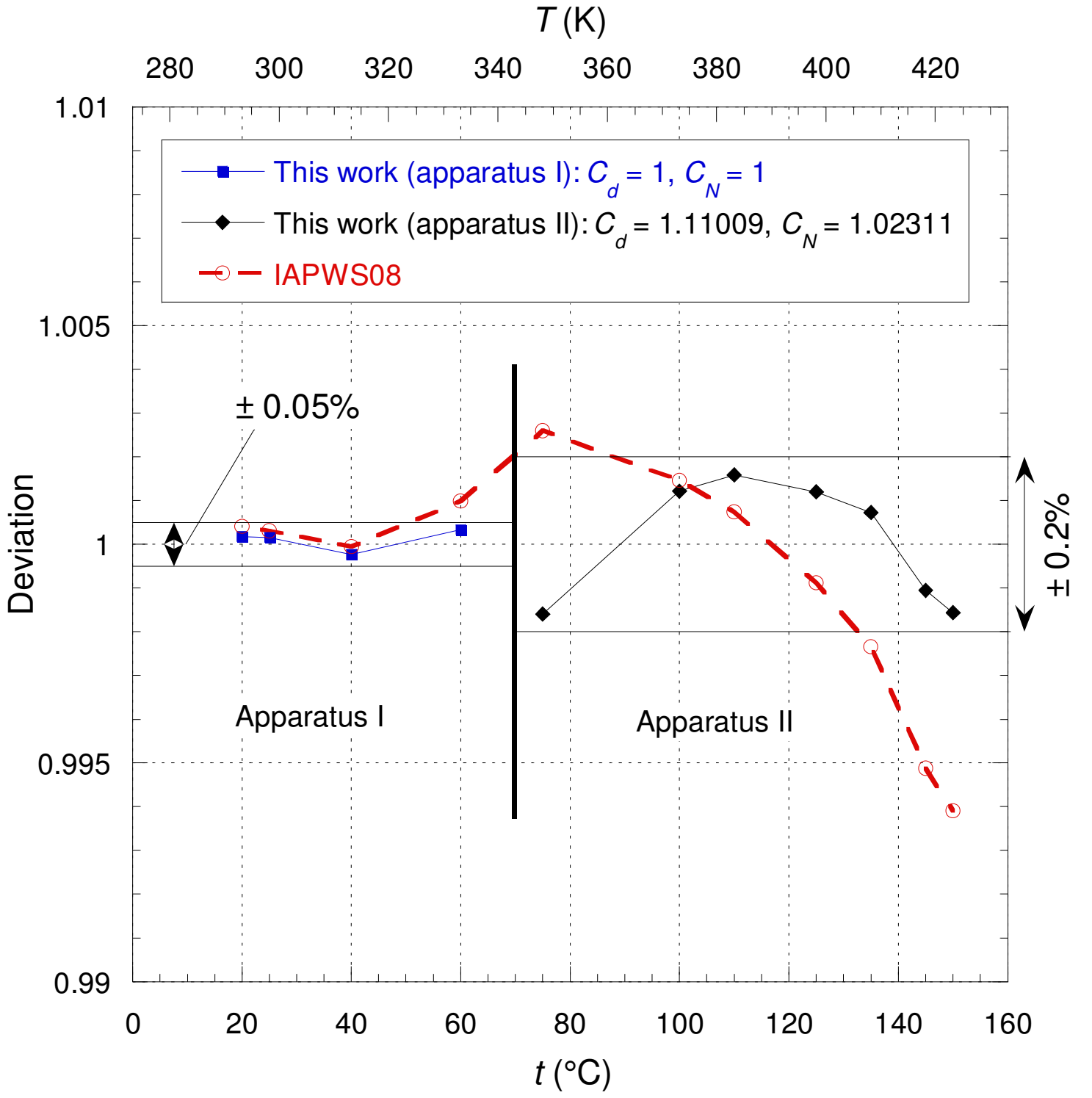

Fig. 33. Ratio of Korosi *et al.*'s data (Ref. 29) with the IAPWS08 formulation (empty circles) and the present modeling (black diamonds and blue squares) as a function of the temperature. The lines are eye guides.

Collings *et al.* (Ref. 30) have made later a similar experiment with two different capillary viscometers having two different tube radii, among other things. They wrote that:

> "[their] capillary viscometer [...] possesses some unusual features and [...] is capable of high precision".

This kind of capillary viscometer must be calibrated in order to convert flow times to viscosities. To this end, two device constants must be determined. Collings *et al.* indicated that:

> "During the calibration of this instrument, we have obtained relative values for the viscosity of water in the temperature range 1 °C to 70 °C".

And the relative values obtained for the viscosities were calibrated at 20 °C with the reference viscosity for water of 1.002 00 mPa.s.

Concerning the datasets corresponding to the first apparatus Collings *et al.* wrote that:

> "Runs 1/1 and 1/2 are results from the 0.5 mm diameter [*i.e.* a radius of 250 μm] capillary viscometer, with separate loadings of distilled water, 1/1 being carried out at atmospheric pressure, and 1/2 at the saturated vapor pressure of water".

These two datasets correspond to liquid water that has approximately the same density and temperature in both cases but not the same pressure. Since pressure is not an explicit parameter of the different models, it can be expected that there will be little effect on the present modeling parameters. Fig. 34 shows that both runs can be reported with the same $C_d$ and $C_N$ coefficients and that these data are included in the same uncertainty band. Only the hottest point on run 1/2 comes slightly out of the band, but the same trend can be observed on both runs and also with the IAPWS08 formulation. The deviation of ±0.06% is consistent with that of Korosi *et al.* for the same temperature range.

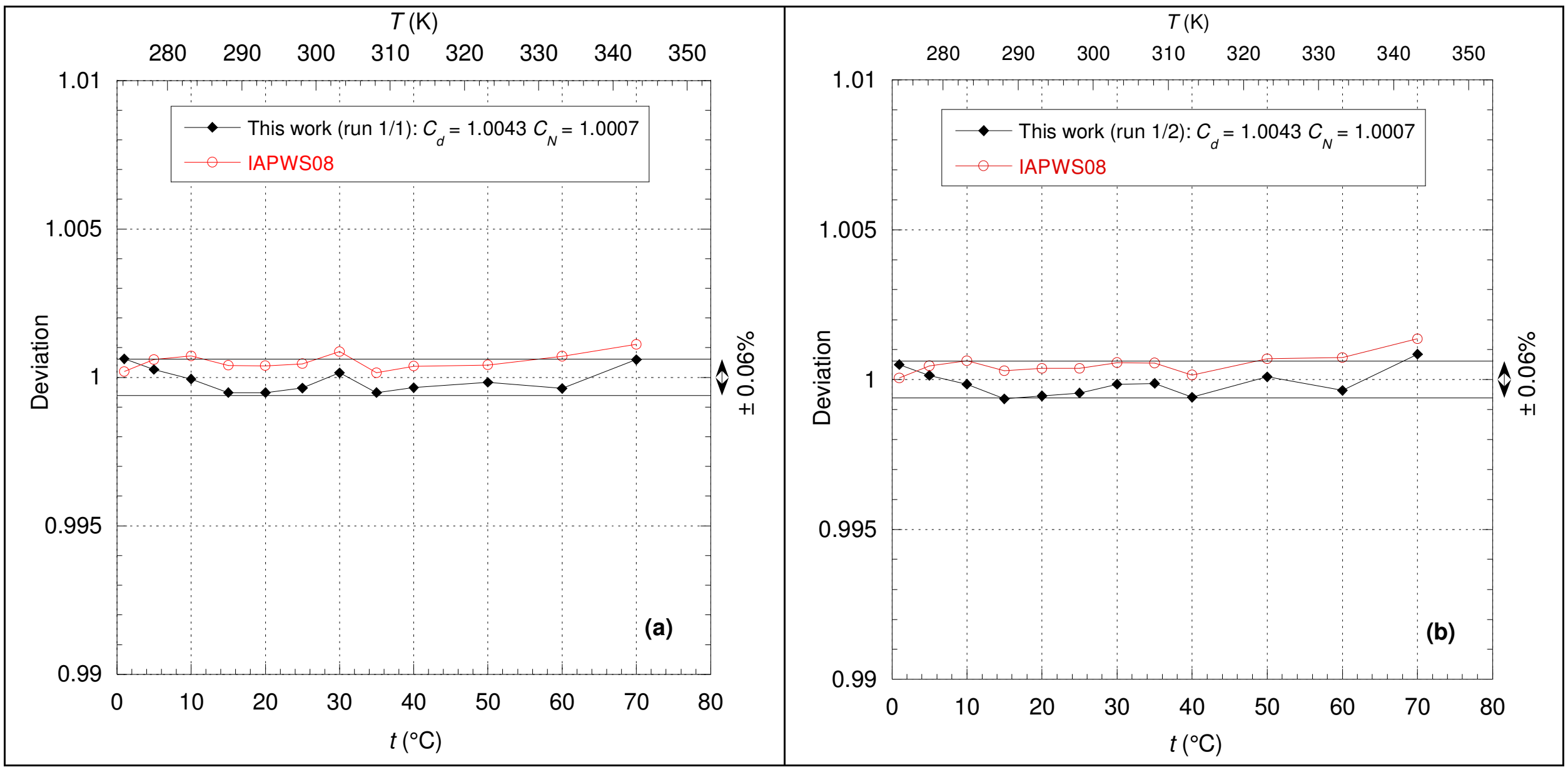

Fig. 34. Ratio of Collings *et al.*'s data (Ref. 30) with the IAPWS08 formulation (empty circles) and the present modeling for the two first runs (black diamonds) as a function of the temperature: (a) at atmospheric pressure; (b) along SVP. The lines are eye guides.

Collings *et al.* have used a second apparatus with a 0.6 mm diameter capillary tube. They wrote that:

> "the data from the 0.6 mm diameter capillary are denoted 2/1, 2/2 and 2/3".

These three runs correspond to measurements for water on SVP. It is interesting to note that the two device constants are very different between the apparatus for the runs 1/x and 2/x (x representing all the different runs). In addition, the most important device constant, named *C*, is quite different between the run 2/1 and the other two runs 2/2 and 2/3. Collings *et al.* wrote about this constant that:

> "The value of *C* is directly affected by changes in the volume of the charge and was therefore determined using the reference viscosity for water of 1.002 00 mPa.s after making the KE correction to the flow time of 20 °C."

Therefore, to account for the data of the runs 2/x, these variations must be taken into account in the calibration coefficient. So, a mean coefficient of calibration $C_{\mathrm{cal}}$ in the present modeling has been introduced. Fig. 35 shows that the data of the run 2/1 are compatible with a deviation of 0.1%, with always the same trend as for runs 1/x around 0 °C and 60 °C. By comparison we have plotted the same deviation band for the other two runs but the latter are compatible with a deviation of 0.06% as for runs 1/x. The fact that the variation in the deviation has the same shape for all runs 1/x and run 2/1 around 0 °C and 60 °C suggests that the experimental conditions have probably changed from the initial calibration in these temperature regions.

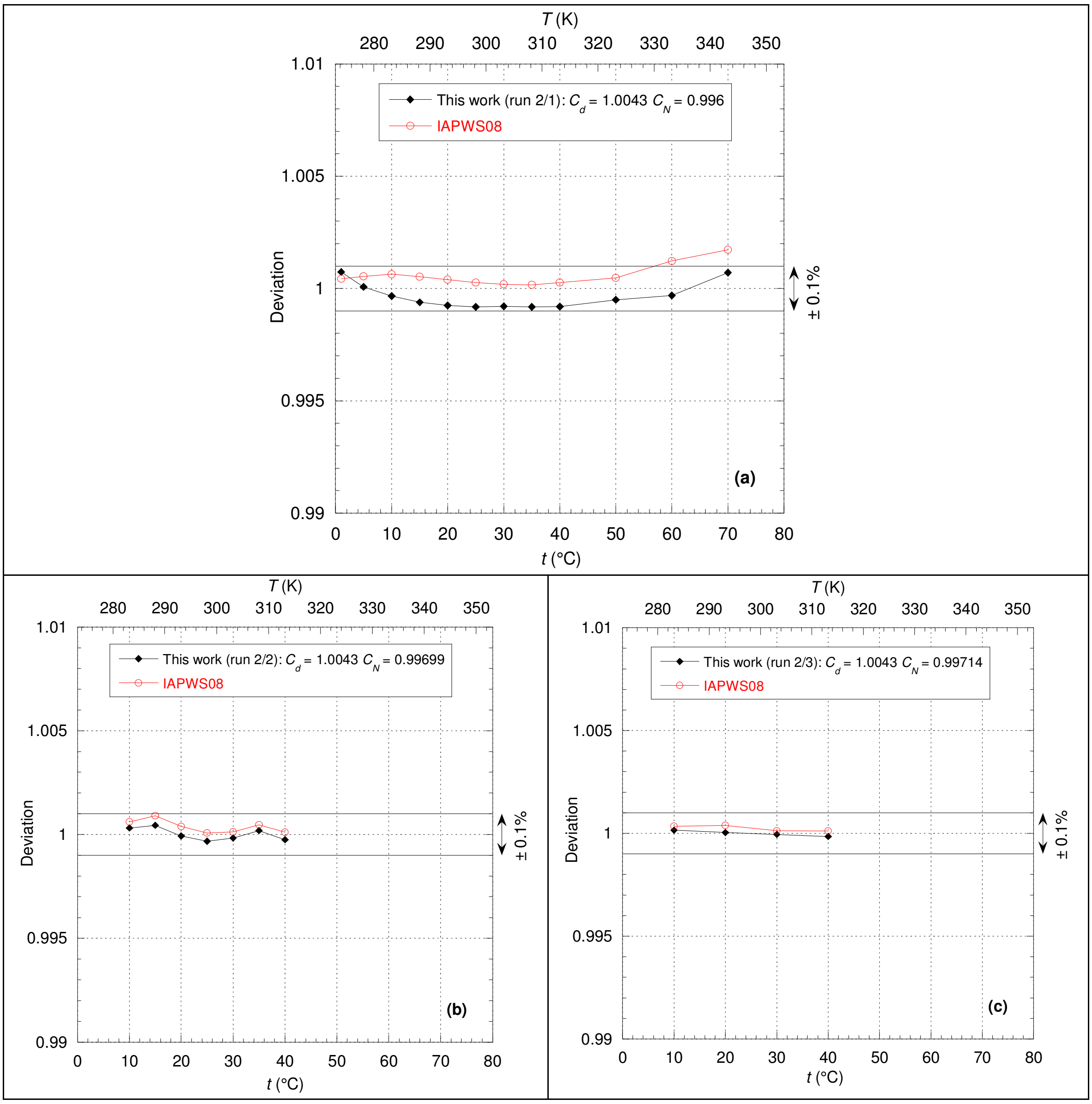

Fig. 35. Ratio of Collings *et al.*'s data (Ref. 30) with the IAPWS08 formulation (empty circles) and the present modeling for the three runs 2/x with $C_{\mathrm{cal}} = 0.997$ (black diamonds) as a function of the temperature on the liquid coexistence curve. The lines are eye guides.

It can be noted that for these three runs the coefficient $C_d$ is identical while the coefficient $C_N$ varies very slightly: this reflects the fact that the temperature dependence is slightly different for these three runs. It is interesting to note that the coefficient $C_N$ is also different between the runs 1/1 and 2/1 and therefore the temperature dependence is different although these two runs correspond to the same temperature range on SVP.

Hardy *et al.* (Ref. 31) have made later similar experiments as those of Korosi *et al.* with two different capillary viscometers having two different tube radii, among other things:

> "The Bingham viscometer used was designed for low-viscosity liquids. The capillary was 12.3 cm long, and its diameter was approximately 0.022 cm. […]
> The Ostwald viscometer was specially designed to provide for use of a capillary with 'square-cut' ends and to minimize the effect of surface tension. The capillary was about 40 cm long, and its diameter was approximately 0.04 cm."

It is also important to notice that the calibration of the instruments was done by assigning the value 1.0050 mPa.s to the viscosity of water at 20 °C. Therefore, to analyze the data, it is necessary to renormalize them with a factor $C_{cal}$ = 1.002/1.005. The Bingham viscometer was used for measurements at atmospheric pressure up to 95 °C while the Ostwald viscometer was used for measurements from 100 °C to 125 °C on SVP. As for Korosi *et al.* these data must be analyzed separately. Hardy *et al.* wrote about their measurements:

> "The limit of possible error of the determined values given in table 1 is believed to be about 0.25 percent in relation to the value 1.0050 centipoise for the viscosity of water at 20° C."

Fig. 36 shows that the atmospheric pressure data can be reproduced with the present modeling with a deviation of less than ±0.1% which is well below the experimental uncertainty. On the other hand, we observe that the data obtained on SVP have a larger deviation, in particular the point at 125 °C which seems aberrant for both models. However, Hardy *et al.* admitted the possibility of a larger error:

> "The agreement with ICT values at other temperatures is within the estimated accuracy (0.5 to 1.0%) given for the ICT values above 40° C."

It is likely that it is on the basis of this statement that Huber *et al.* (Ref. 7) attributed an overall uncertainty of 0.5% in their Table 1 for Hardy *et al.*'s data otherwise this value seems mysterious. With such a value for uncertainty, it appears that the IAPWS08 formulation accounts for most of the data.

The significant deviation of the point at 125°C probably results from the fact that to maintain the state on SVP, it is necessary to apply a pressure very close to the hydrostatic pressure applied to the charge volume.

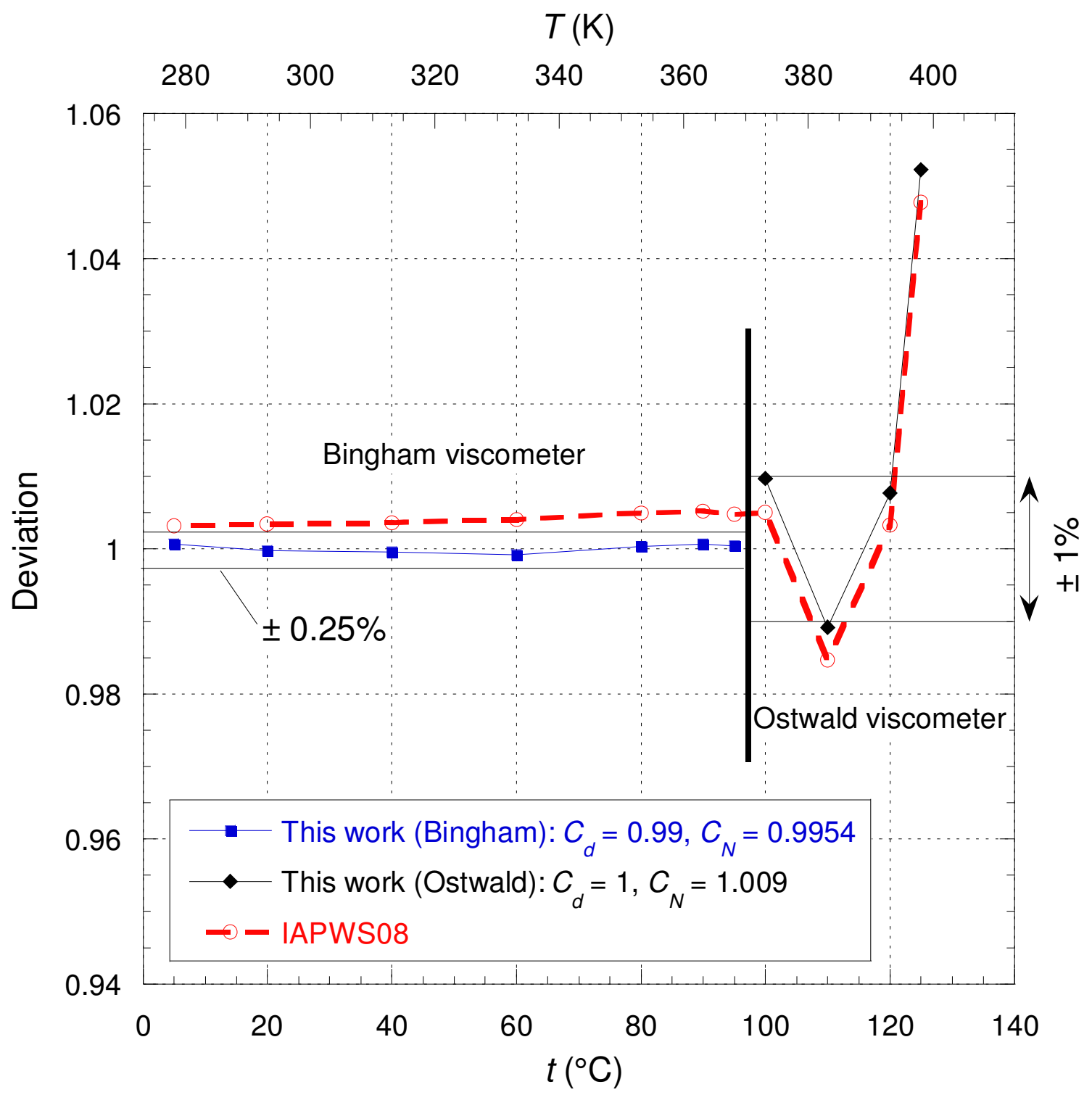

Fig. 36. Ratio of Hardy *et al.*'s data (Ref. 31) with the IAPWS08 formulation (empty circles) and the present modeling with $C_{\mathrm{cal}}$ ~ 0.997 (black diamonds and blue squares) as a function of the temperature. The lines are eye guides.

Cohen's experiment (Ref. 32) is still similar in that he used three apparatus with a capillary tube radius of 120 μm: the temperature range is quite limited, but he was able to increase the pressure up to 90 MPa. Only the datasets from apparatus I and III are complete and usable. Cohen only gives flow times that we convert here into viscosity by taking a constant of proportionality $C_{\mathrm{cal}}$ and this constant value is fixed on the data corresponding to the atmospheric isobar. Fig. 37 shows that the two models are equivalent but with two different values of the constant $C_{\mathrm{cal}}$. The experimental uncertainty is not given by Cohen for this dataset but Fig. 37 shows that the latter is around 1%.

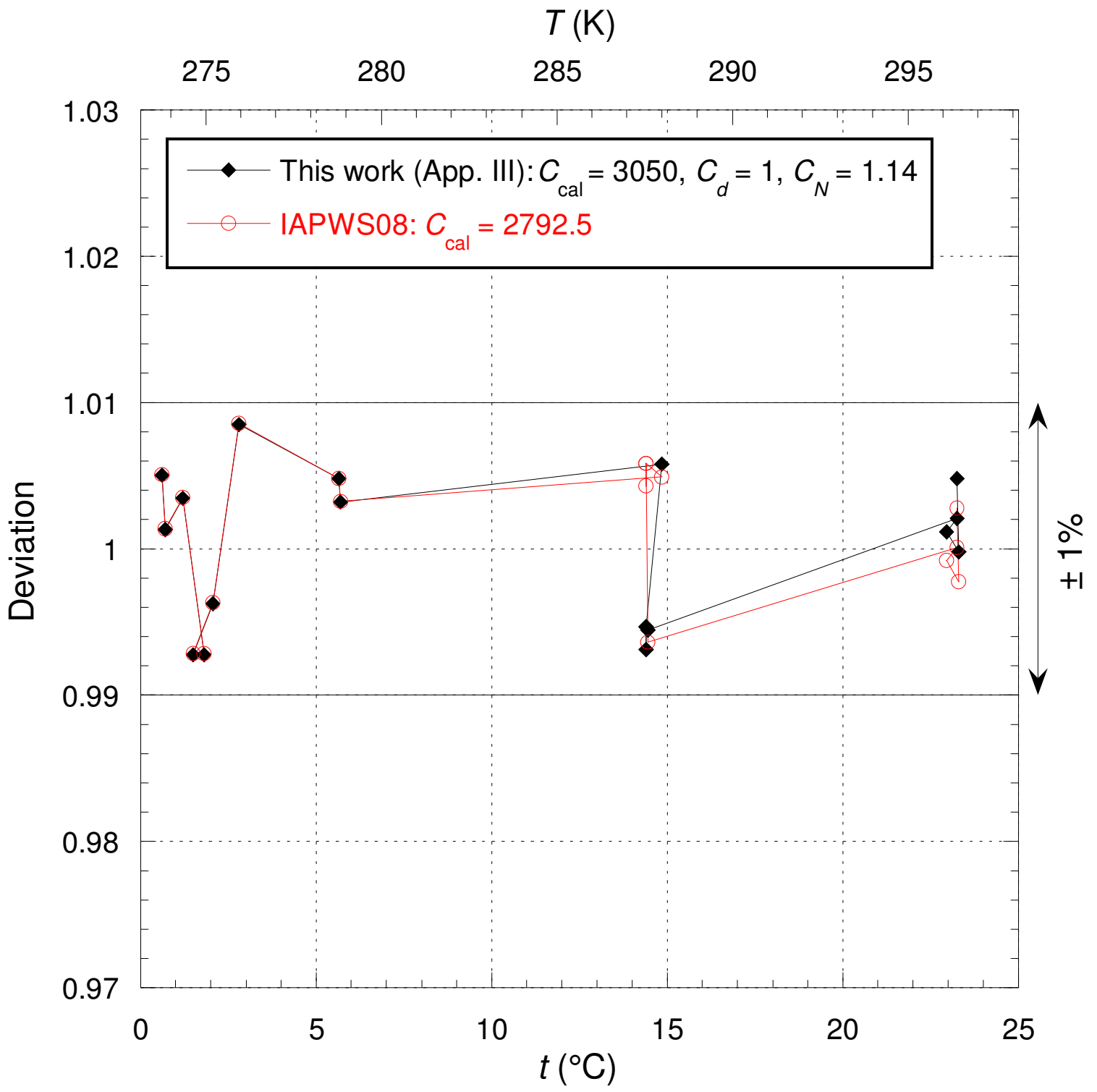

Fig. 37. Ratio of Cohen's data (Ref. 32) with the IAPWS08 formulation (empty circles) and the present modeling (black diamonds) as a function of the temperature at atmospheric pressure. The lines are eye guides.

Cohen also obtained data along quasi-isotherms with apparatus III for which he gives experimental uncertainties for each data. Fig. 38 shows that these data can be very well reproduced within their experimental uncertainties in all cases by taking into account different renormalization factors $C$ for each isotherm: it should be noted that the renormalization factors vary little in the present modeling than with the IAPWS08 formulation, which would mean that the absolute values from the IAPWS08 formulation are more shifted than those of the present modeling compared to Cohen's data. Also, it can be seen that the variations with density are slightly better reproduced with the present modeling than with the IAPWS08 formulation on the three isotherms.

For the isotherm at 15 °C (i.e. 288.15 K), Fig. 38 (c) and (d) show two datasets obtained with two different apparatus. These two sets have practically the same experimental uncertainty but it appears that the present modeling reproduces slightly better the variations with density obtained on apparatus III than the IAPWS08 formulation while these are practically identical in the case of apparatus I.

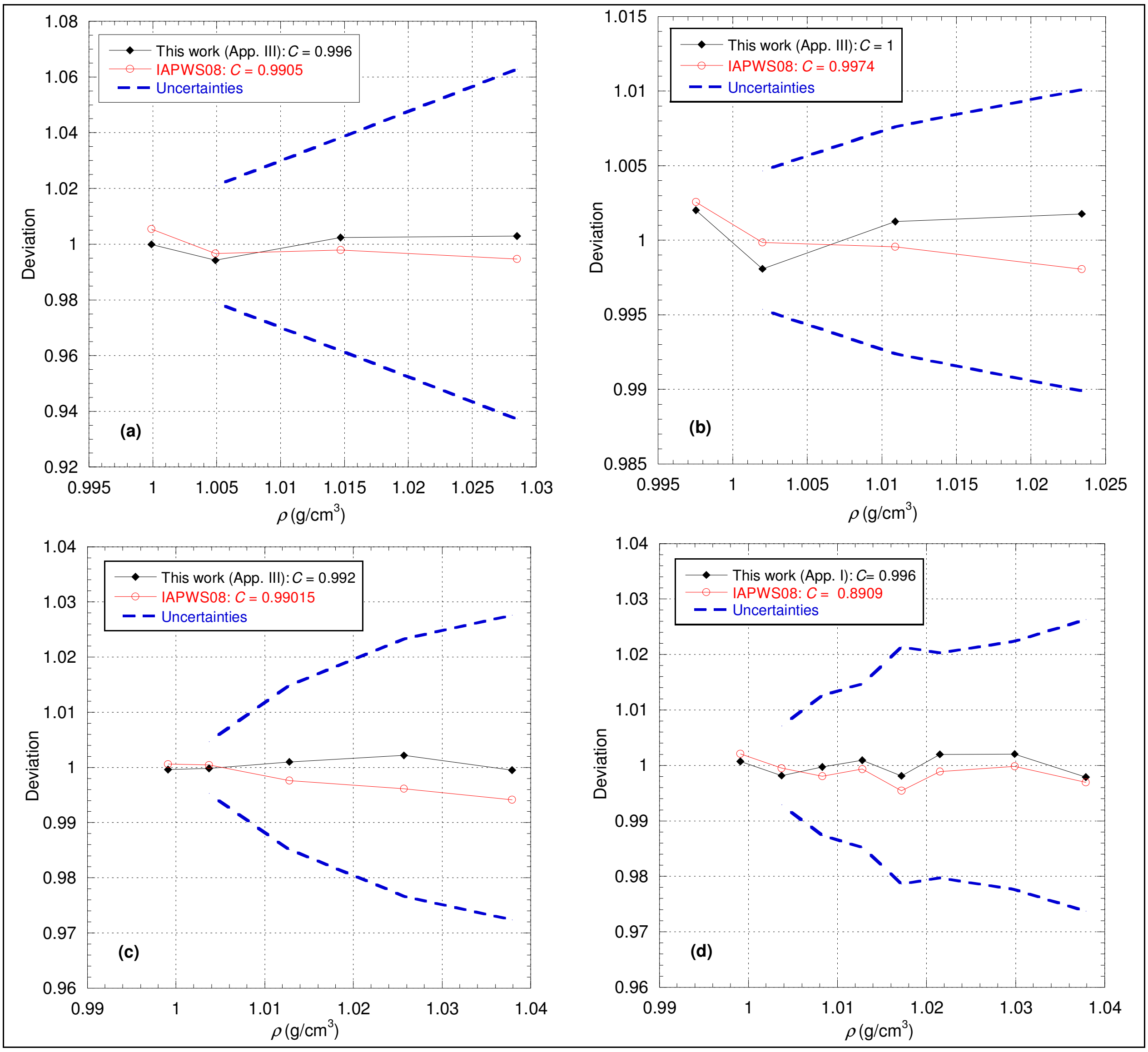

Fig. 38. Ratio of Cohen's data (Ref. 32) with the IAPWS08 formulation (empty circles) and the present modeling with $C_d$ = 1 and $C_N$ = 1.14 (black diamonds) as a function of density along quasi-isotherms: (a) ~1°C, (b) ~23°C, (c) ~15°C and (d) ~15°C. The lines are eye guides.

In section 3.2 we analyzed the smoothed data of Cappi. But in Cappi's thesis (Ref. 10), we also find the raw data which are determined in relative form, so to analyze them we must introduce a renormalization factors $C$ on each isotherm in order to compare the variations with density of the models in a relevant way. So, the value of the renormalization factors is of little interest here (i.e. it has no influence on the variations with density). In the context of the present modeling, it is useful to remember that to analyze the data from Cappi we necessarily have $C_d = C_N = 1$. To illustrate this point, we have chosen to represent only isotherms with only one run but the comments apply perfectly to all isotherms from Cappi. In section 3.2 we saw that the smoothed data were all contained in the deviation band ±0.8% with the present modeling. Fig. 39 shows that all the raw data can be represented with the present modeling in a deviation band of ±0.5% while the IAPWS08 formulation has difficulty in keeping to it. In other words, the deviation of 0.8% is an effect produced by the chosen smoothing function while the data have an experimental deviation of less than 0.5% overall. Another possible effect of the choice of the smoothing function is observed at the point of highest density on the isotherm at 20 °C (i.e. see Fig. 39(b)): indeed the value corresponding to this point given by the smoothing leads to a deviation of -1% (i.e. see Fig. 11) whereas the raw data is almost perfect with the IAPWS08 formulation and it is less than -0.3% with the present modeling. For this point we cannot exclude that it is a transcription error of the numerical value in the smoothed data. Finally, it can be seen that the IAPWS08 formulation hardly represents variations with density, particularly for the isotherm at 100 °C.

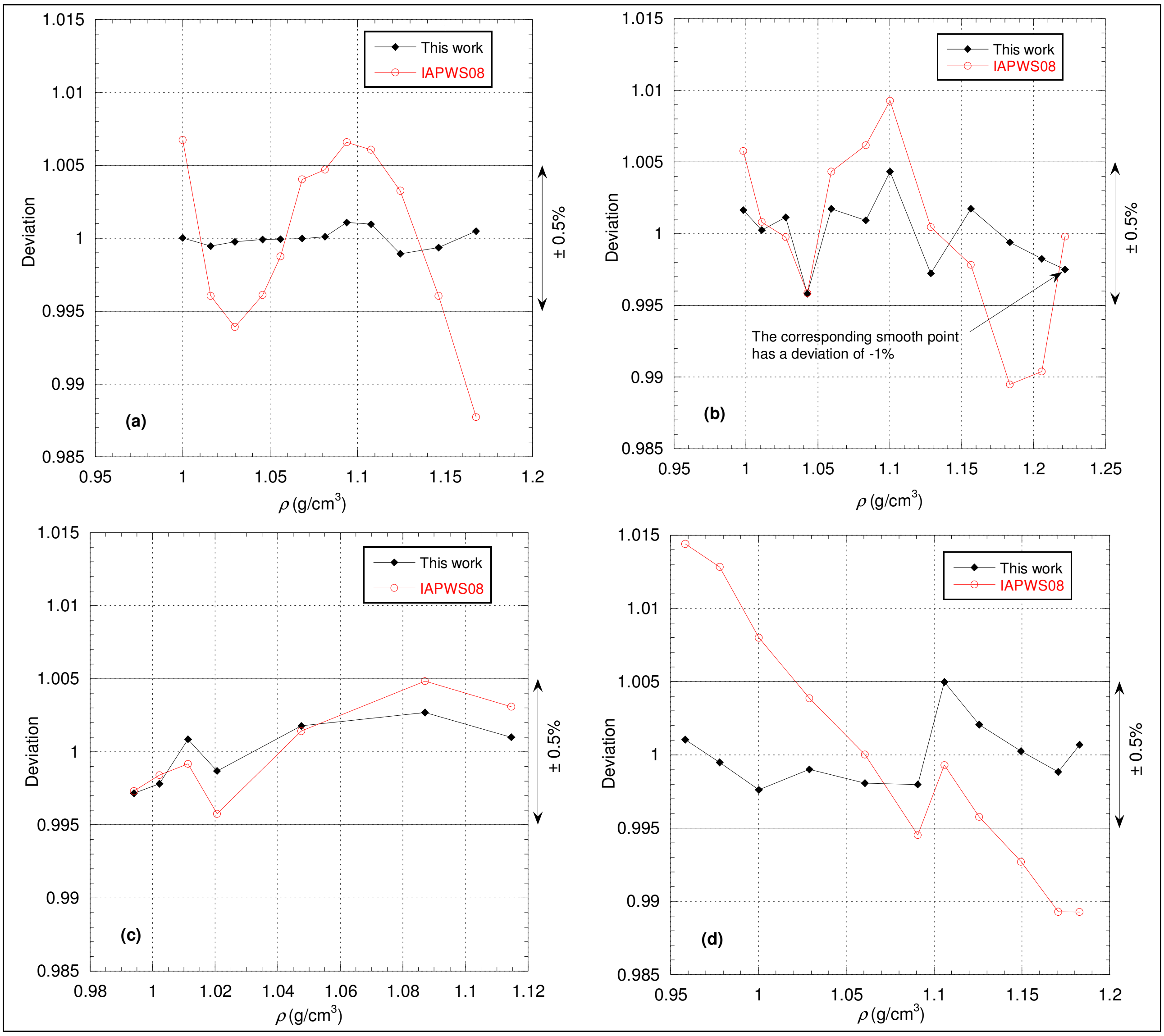

Fig. 39. Ratio of Cappi's raw data (Ref. 10) with the IAPWS08 formulation (empty circles) and the present modeling (black diamonds) as a function of the density along quasi-isotherms: (a) ~ 2.2°C, (b) ~ 20°C, (c) ~35°C and (d) ~ 100°C. The lines are eye guides.

We have observed with the datasets from Cappi the effect of the smoothing choice on the quality of data representation because in this case we compare different mathematical models. With Agaev *et al.*'s data (Ref. 33), we will see that the choice of the smoothing function can also change the structure of the data. Indeed, the work of Agaev *et al.* contains both a table of smoothed values and a plot with raw experimental data. These data are obtained from a capillary viscometer experiment so they are given in a relative form and then converted to viscosity by calibrating the device. Agaev *et al.* wrote that:

> "The error of the experimental data calculated on the basis of the general theory of errors lay within 0.5%".

Fig. 40(a) shows that smoothed data can be reproduced globally whatever the model with a deviation of 1% except for a few points on the isotherm at 150 °C (i.e. the highest temperature isotherm of this dataset). Probably the error admitted by Agaev *et al.* is a bit optimistic. Also, Huber *et al.* (Ref. 7) in their Table 9 have admitted an uncertainty of 1% for these data. In fact, the deviation of smooth data is an artifact due to the chosen smoothing procedure. Indeed, if we now observe the raw data extracted from Agaev *et al.*'s Fig. 1A, we see in Fig. 40(b) that the corresponding deviation for this same isotherm is at the opposite of the variation of the deviation on Fig. 40(a). There is therefore probably a way to smooth the 150°C isothermal data so that it can be fully included in the ±1% deviation band. Finally, it should be noted on Fig. 40(b) that the deviation of the raw data is much greater than expected. Part of this effect comes probably from their extraction on Agaev *et al.*'s Fig. 1A but it seems difficult to reduce this uncertainty to less than ±1%.

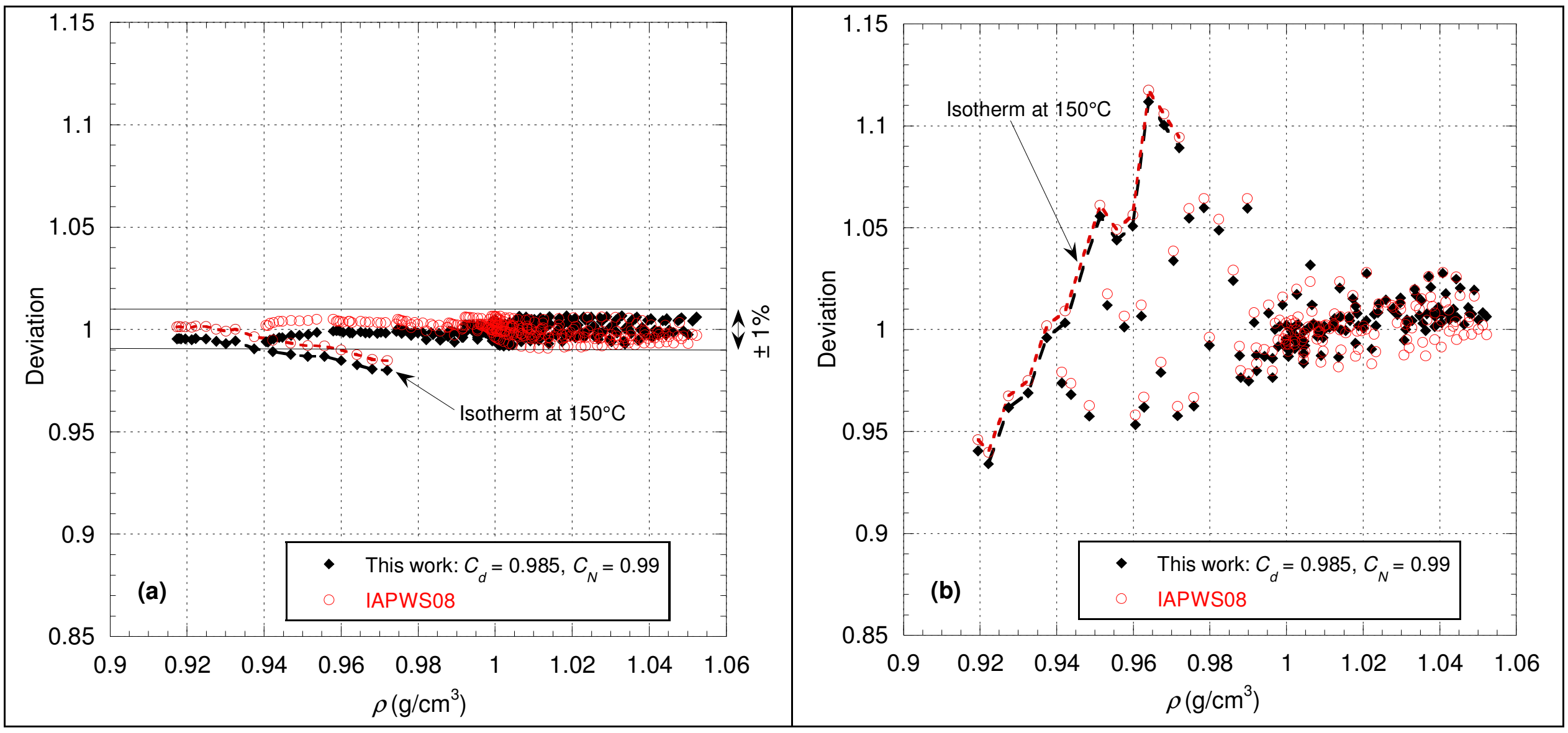

Fig. 40. Ratio of Agaev *et al.*'s data (Ref. 33) with the IAPWS08 formulation (empty circles) and the present modeling (black diamonds) as a function of the density for all water isotherms data points: (a) smoothed data, (b) raw data. The data linked with a dashed curve correspond to same isotherm at 150°C.

More recently, Abdulagatov *et al.* (Ref. 34) have obtained viscosity data also using a capillary flow technique (the inner radius of the capillary is here equal to 150.91 μm) for higher temperatures than those of Agaev *et al.* but at lower pressures. As a result, Abdulagatov *et al.*'s data are distributed over a wider density range. Concerning their data Abdulagatov *et al.* wrote:

> "Based on the detailed analysis of all sources of uncertainties likely to affect the determination of viscosity with the present apparatus, the combined maximum relative uncertainty $\delta\eta/\eta$ in measuring the viscosity was 1.5%."

Fig. 41 shows that Abdulagatov *et al.*'s data can be reproduced better than 1%. We also observe that the structure of the deviation is the same as for the IAPWS08 formulation except that our deviation is better centered on the unit value.

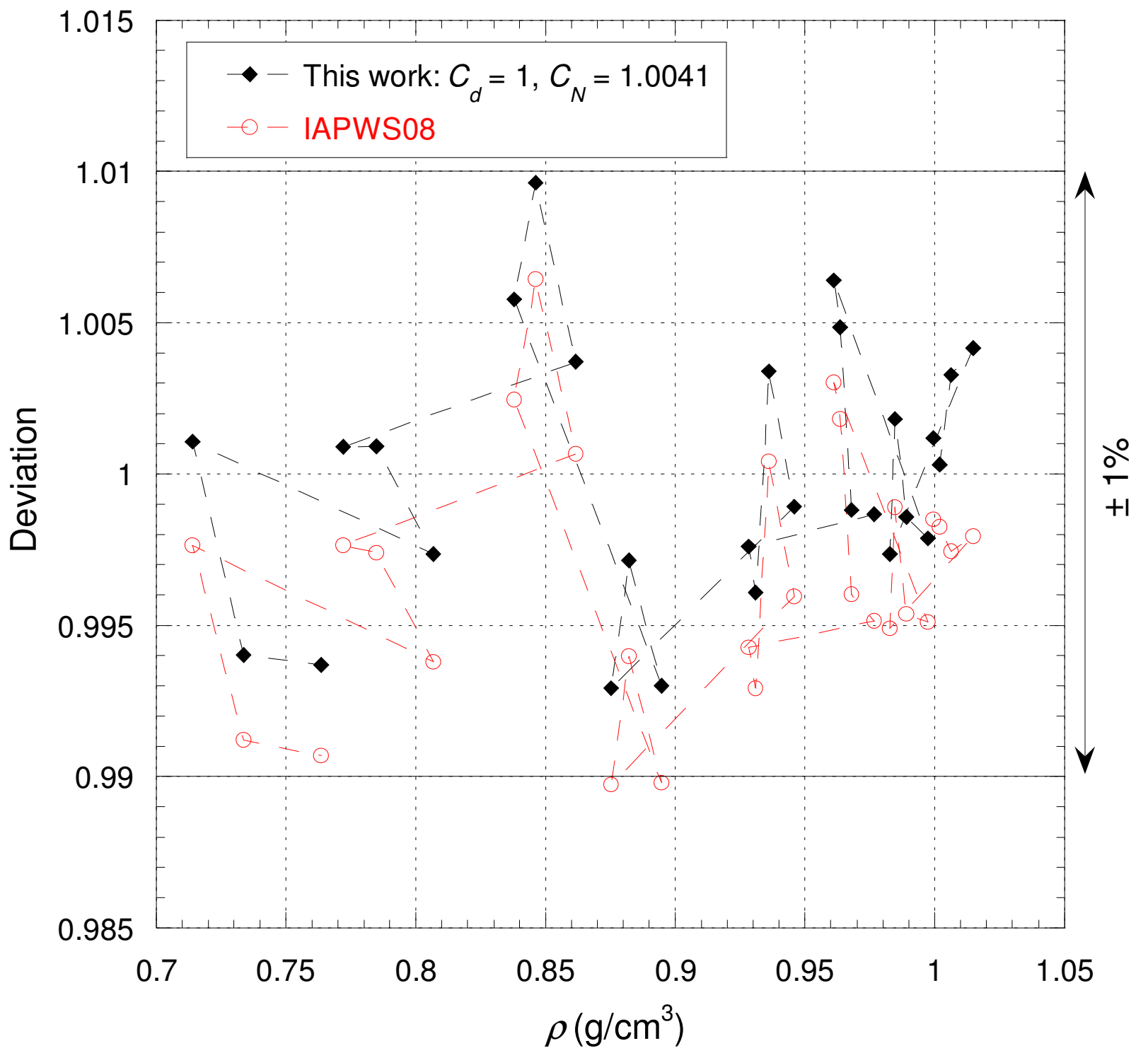

Fig. 41. Ratio of Abdugalatov *et al.*'s data (Ref. 34) with the IAPWS08 formulation (empty circles) and the present modeling (black diamonds) as a function of the density for all water isotherms data points. The dashed lines are eye guides.

Very recently a new experiment has been carried out in supercooled water by varying the pressure from 20 MPa to almost 300 MPa. These new data of Singh *et al.* (Ref. 35) are interesting because they correspond to densities higher than 1 g/cm$^3$, i.e. densities that coincide with those of Cappi. The uncertainties of these new measurements are quite large (from about ±1% up to ±5%) but they are given for each measurement point which makes it possible to represent them in the form of curves in the deviation plots. This experiment consists of measuring the time-of-flight of a slightly saline solution flowing through silica capillaries with a diameter of about 8.8 μm. The conversion of the measured time-of-flight into viscosity is essentially dominated in this experiment by a constant of proportionality between these two quantities. Thus, this experiment consists of a Poiseuille type flow in a tube but two kinds of tube characteristics were used which, according to the isobars, lead to two sets of different data named "run x". We therefore choose to represent in Fig. 42 only some characteristic isobars with the two runs. Also, for a comparison with the datasets at atmospheric pressure, we will keep here the temperature scale in Celsius degree with the symbol *t*.

We can notice on Fig. 42 that the two runs have systematically different viscosity deviations which mean that the temperature dependence is each time different as in the case of Thorpe *et al.* Fig. 42 shows that by just rescaling *d* and $d_N$ it is possible to reproduce Singh *et al.*'s data within their experimental uncertainties except for few erratic points for which the uncertainty is probably a bit too optimistic. This shows that isochors deduced from Cappi's data are correctly represented at all temperatures.

Note that the IAPWS08 formulation is perfect for the 25 °C (i.e. 298.15 K) point of run 1, which is due to the fact that this formulation was used for the calibration of Singh *et al.*'s data at this point. It can also be seen that the IAPWS08 formulation systematically exceeds the experimental deviation when the normal solidification temperature is reached and this deviation is in the same direction as for Hallett's data: the IAPWS08 formulation systematically has too strong variations.

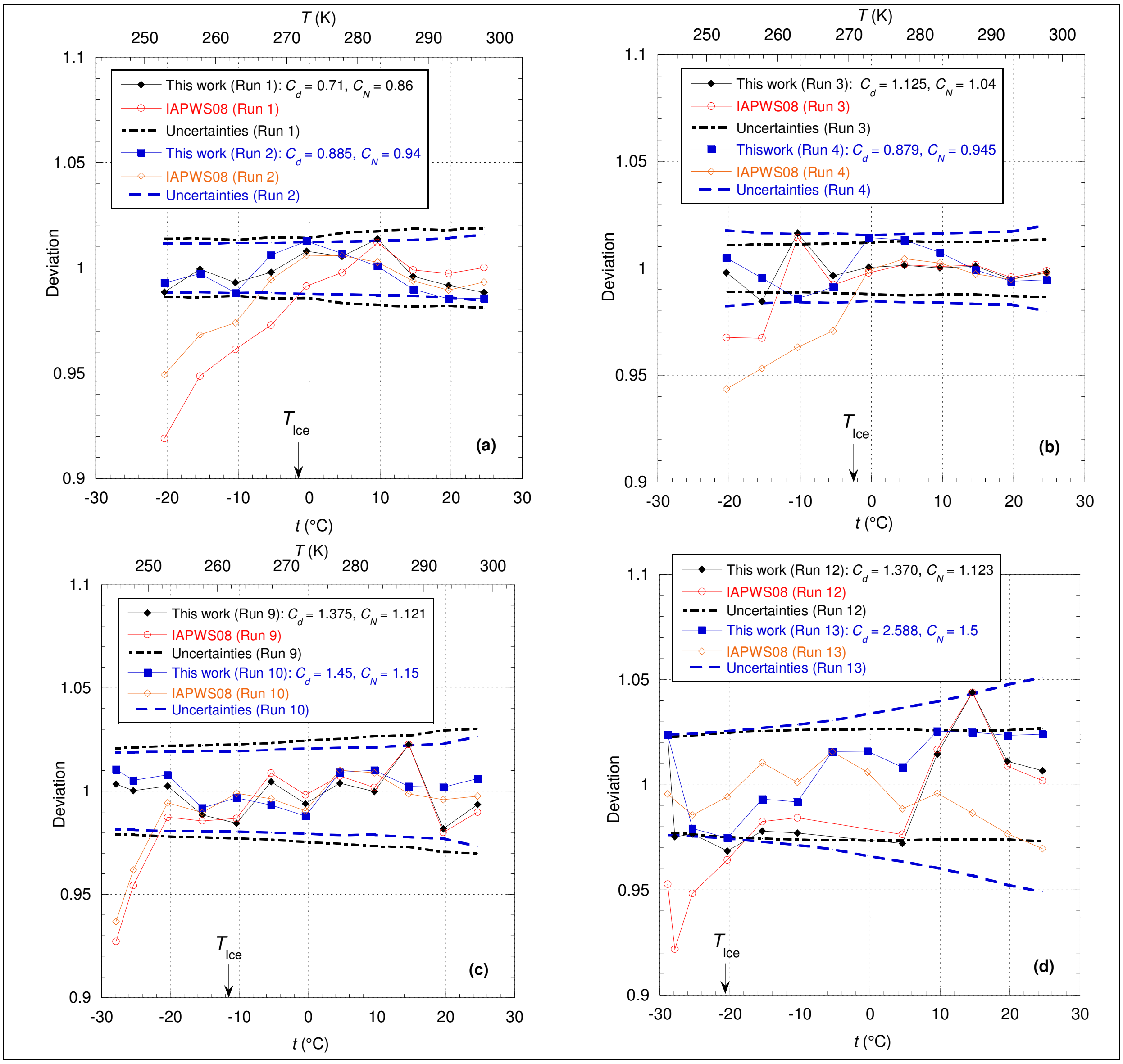

Fig. 42. Ratio of Singh *et al.*'s data (Ref. 35) with the IAPWS08 formulation (empty circles, empty diamonds) and the present modeling (black diamonds, blue squares) as a function of the temperature along different isobars: (a) 20 MPa, (b) 33 MPa, (c) 125 MPa, (d) 200 MPa. The lines are eye guides. $T_{Ice}$ represents the temperature of the melting line for the corresponding isobar.

These recent data of Singh *et al.* can be compared with Harris *et al.*'s data (Ref. 36) which are globally more accurate but correspond to a smaller temperature range below 0 °C. However, these data reach pressures of about 385 MPa. Harris *et al.*'s experiment consists of a sinker with a diameter of 6.3 mm falling into a tube of 6.5 mm of diameter. So, the shear layer in this experiment is 200 μm. It means that this experiment is equivalent to the flow in a tube of radius 100 μm. In other words, for temperatures below 0 °C the same hydrodynamic correction as Hallett must apply. Indeed, using Eq. (23), Fig. 43 shows that Harris *et al.*'s dataset can be reproduced within the experimental uncertainty estimate of ±1 % (this is also the uncertainty retained by Huber *et al.* (Ref. 7) in their Table 1).

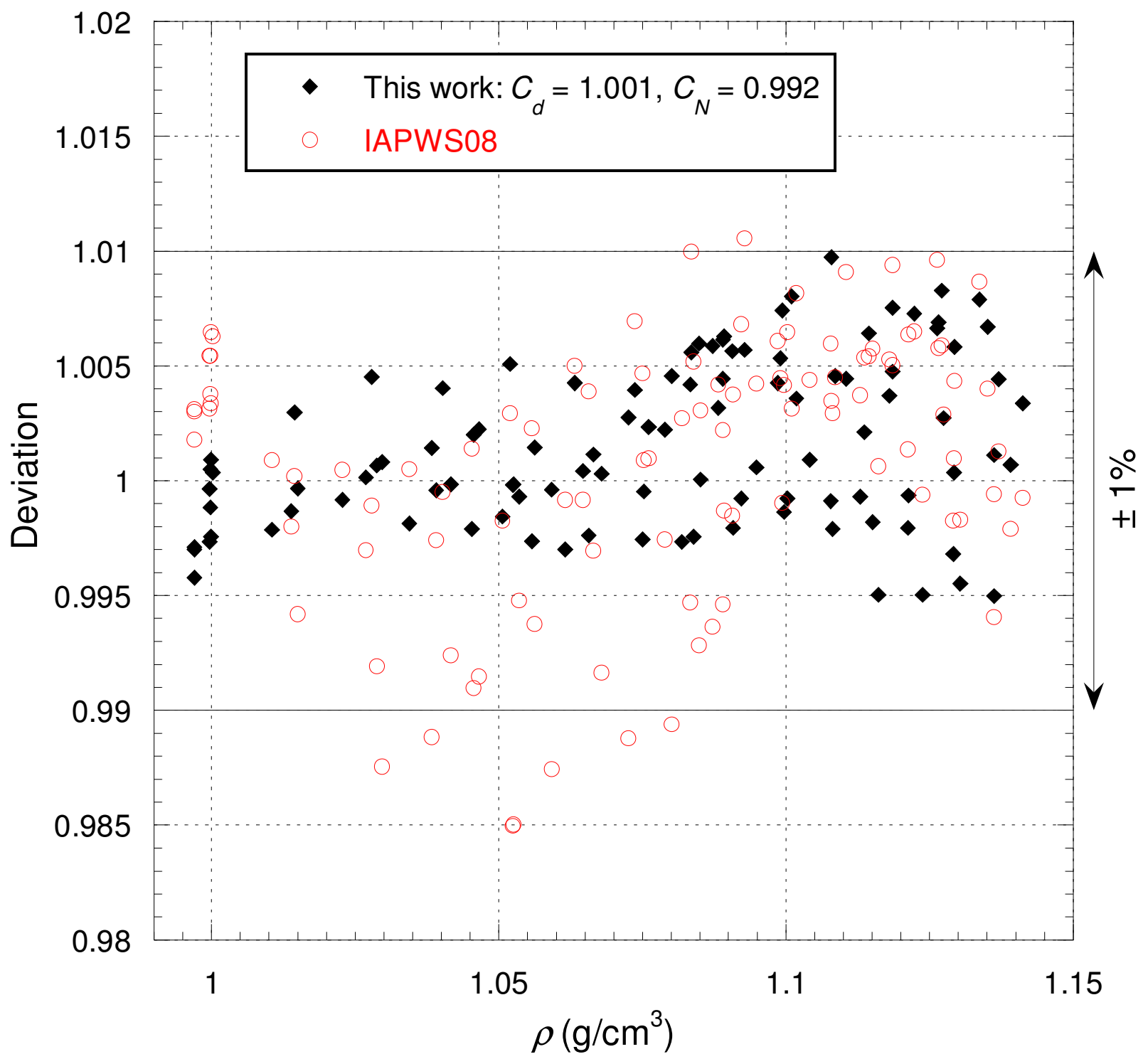

Fig. 43. Ratio of Harris *et al.*'s data (Ref. 36) with the IAPWS08 formulation (empty circles) and the present modeling integrating the hydrodynamic correction for low temperatures (black diamonds) as a function of the density.

The data of DeFries *et al.* (Ref. 37) and Först *et al.* (Ref. 38) are available approximately in the same temperature range as Harris *et al.*'s. Först *et al.* reaches higher pressures (up to 700 MPa) but with a lower accuracy. However, the experiments of DeFries *et al.* and Först *et al.* are very different from those of Harris *et al.* Indeed, these data are determined from a rolling-ball viscometer as with Abramson's experiment (Ref. 11). All these data have a common density range between 1 and 1.2 g/cm$^3$ but all these devices have different properties and geometric characteristics. Until now we have seen that to reproduce the experimental data it was enough to multiply $d_N$ by a factor but $d_N$ being an external parameter related to a particular experiment it is in the logic of the present modeling that the variations can be different from one experiment to another. Indeed, we have seen in Table 2 that the function $f_{N,\mathrm{high}}$ is the sum of two terms, one of which is "adapted" to Abramson's experiment corresponding to a rolling ball viscometer in a diamond-anvil cell. It turns out that for data obtained from other types of rolling-ball viscometers lead to slightly different variations with density of the coherence parameter $d_N$ from that of Abramson but keeps the same mathematical structure. Thus, to reproduce the data of DeFries *et al.* and Först *et al.*, the expression $f_{N,\mathrm{Abramson}}(\rho)$ must be replaced in $f_{N,\mathrm{high}}$ by the corresponding functions below:

$$f_{N,\mathrm{DeFries}}(\rho)=\frac{55.14\left(1-\exp\left(-\left|(\rho-1.1)/0.32\right|^{3}\right)\right)-70\left|\rho-1.1\right|\exp\left(-\left|(\rho-1.1)/0.01\right|^{1.1}\right)}{1+\left(\dfrac{1.1}{\rho}\right)^{1000}} \tag{27}$$

$$f_{N,\text{Först}}(\rho)=\frac{55.14\left(1-\exp\left(-\left|(\rho-1.01)/0.32\right|^{2.5}\right)\right)-83\left|\rho-1.01\right|\exp\left(-\left|(\rho-1.01)/0.066\right|^{1.1}\right)}{1+\left(\dfrac{1.01}{\rho}\right)^{1000}} \qquad (28)$$

which represent the deviation of the coherence parameter from Cappi. Fig. 44 shows the variation of the different functions and it can be seen that the increase of the coherence parameter begins at lower densities in Först *et al.* experiment than in DeFries *et al.*'s and Abramson's.

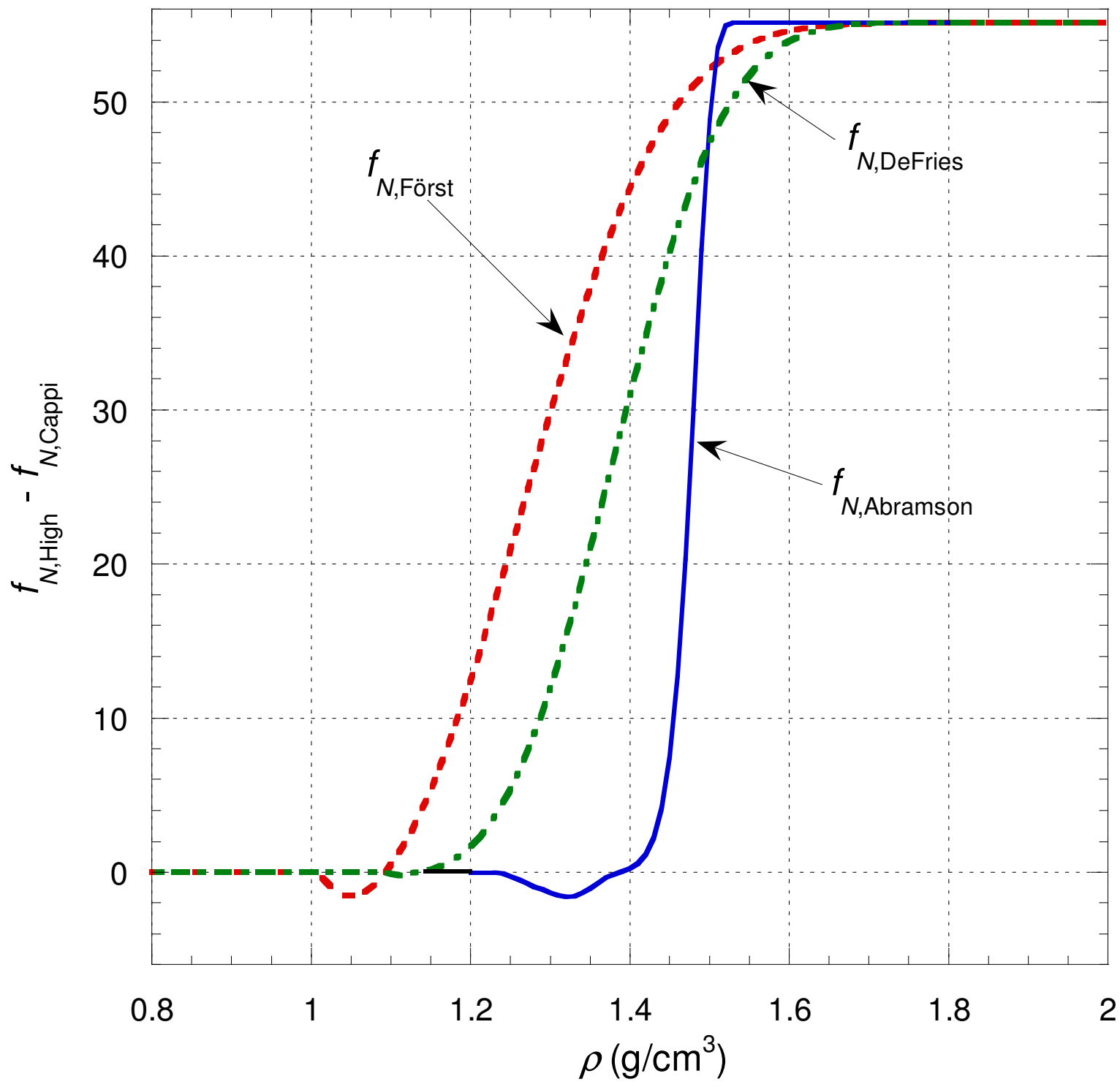

Fig. 44. Representation of the different coherence functions $f_N$ corresponding to various types of rolling ball experiments.

DeFries *et al.* (Ref. 37) wrote that:

> "The viscometer was calibrated using the 10 °C, 0-6 kbar data of Harlow. The viscosities have an estimated accuracy of ±2%".

But a few paragraphs later, it is written that:

> "Previous pressure measurements at 10 °C are also plotted. They agree with the new data within 3%, which is about the error of our measurements."

Huber *et al.* (Ref. 7) in their Table 9 have considered that Harlow's uncertainty is within 2%. This requires that the isotherm at 10°C must be centered on the unit value in the deviation plot. To do this, we have taken into account a very slight calibration factor and we can see in Fig. 45 that the deviation of the data for the isotherm at 10 °C are well centered and are practically compatible with an uncertainty of ±1%. Now taking into account this calibration factor, Fig. 45 shows that most of the data can be reproduced with the experimental uncertainty of ±2% and only two points are outside the uncertainty if the latter is considered at 3%. The two points with the greatest deviations are those corresponding to the coldest isotherms and the extreme pressure values for these isotherms.

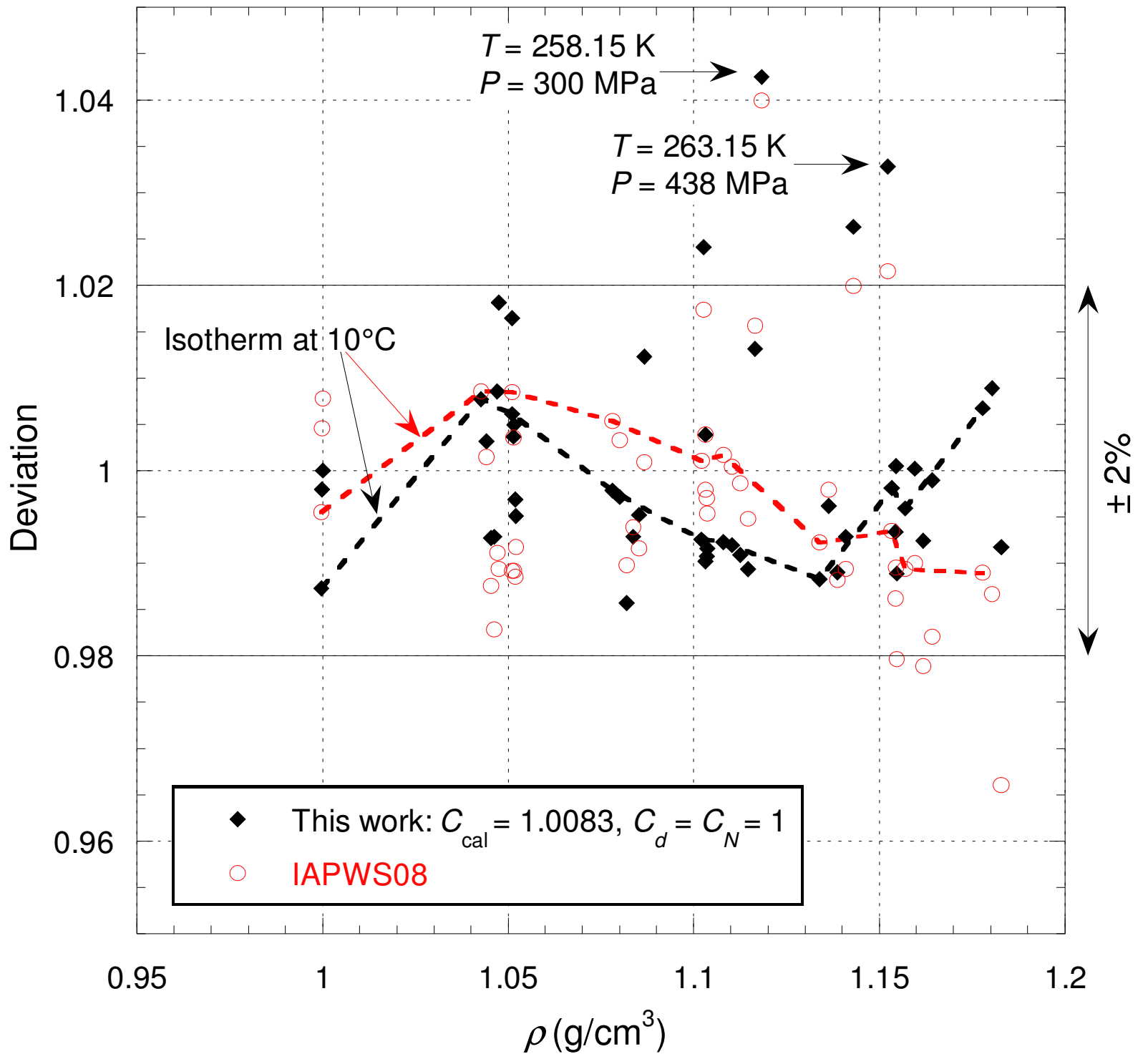

Fig. 45. Ratio of DeFries *et al.*'s data (Ref. 37) with the IAPWS08 formulation (empty circles) and the present modeling (black diamonds) using a single calibration coefficient $C_{cal}$ as a function of the density for all water isotherms data points. The dashed curves link the data corresponding to the isotherm at 10°C.

About the important deviations at high pressures of DeFries *et al.*'s data with those of Först *et al.* (Ref. 38), the latter wrote:

> "The deviation between the data of DeFries and Jonas (1977) and the new data is small for moderate pressures. With increasing pressure the relative deviation increases up to 6.4%. This may again be due to dimensional changes with pressure that cause a change in the calibration factor of their viscometer. DeFries and Jonas (1977) used data of Harlow (1967) to calibrate at high pressures and to allow for dimensional changes. Harlow used for his measurement the apparatus of Cappi (1964) with some small modifications. Therefore it is assumed that the discrepancy at higher pressures is caused by the same errors as described above."

Following the recommendations of Först *et al.*, Fig. 46 shows that by slightly modifying the calibration factor for each isotherm, all data can be reproduced now under experimental uncertainty: it can be seen that this factor must be increased on average when cooling.

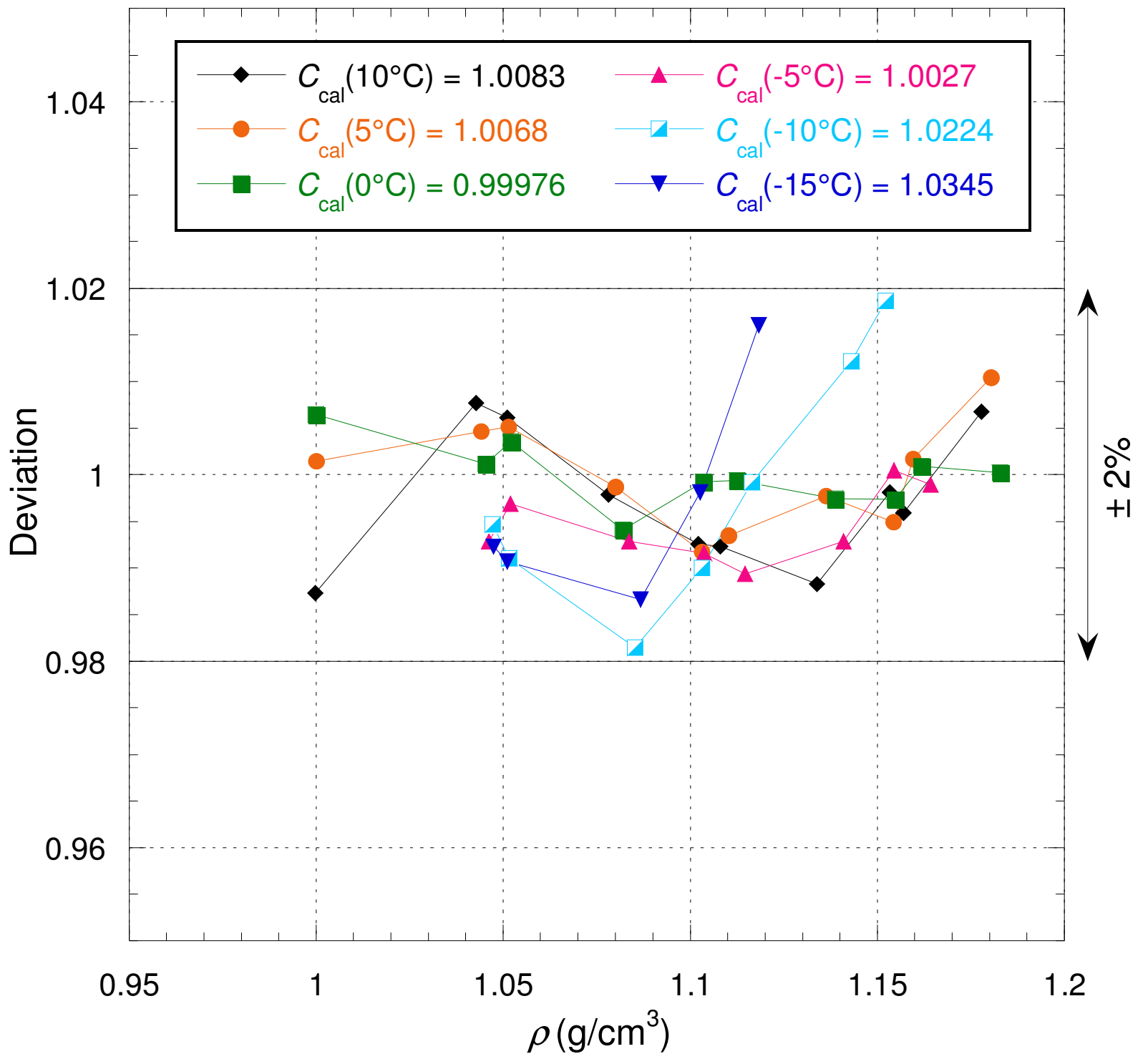

Fig. 46. Ratio of DeFries *et al.*'s data (Ref. 37) with the present modeling ($C_d = C_N = 1$) as a function of the density along each isotherm. The lines are eye guides.

Now it is interesting to analyze Först *et al.*'s data (Ref. 38) because they are also subject to criticism from different authors. In particular, Huber *et al.* (Ref. 7) wrote:

> "We did not consider the data of Först *et al.*, because they appear to have unexplained systematic deviations from other data in this region [*i.e.* subregion 1]".

Först *et al.* (Ref. 38) performed two kinds of experiments. We are interested first in the one corresponding to the "Inductive method" which consists of a rolling-ball experiment with different inclinations of the running surface. Fig. 47 shows that these data can be reproduced with an uncertainty of ±4% using Eq. (28) to describe the variation in water coherence length. It can be seen that the IAPWS08 formulation achieves a deviation of -11% at the highest densities which is not satisfactory given that Huber *et al.* (Ref. 7) in their Table 9 have considered that the uncertainty is within 2%.

As for other experiments of this type, a calibration of the measurement system is necessary, resulting in a proportionality constant between the viscosity and the rolling time. This calibration factor depends both on the temperature and pressure. Fig. 48 shows that by slightly modifying this calibration factor on each isotherm, it is now possible to reproduce all the data variations within the expected uncertainty of 2% and even better.

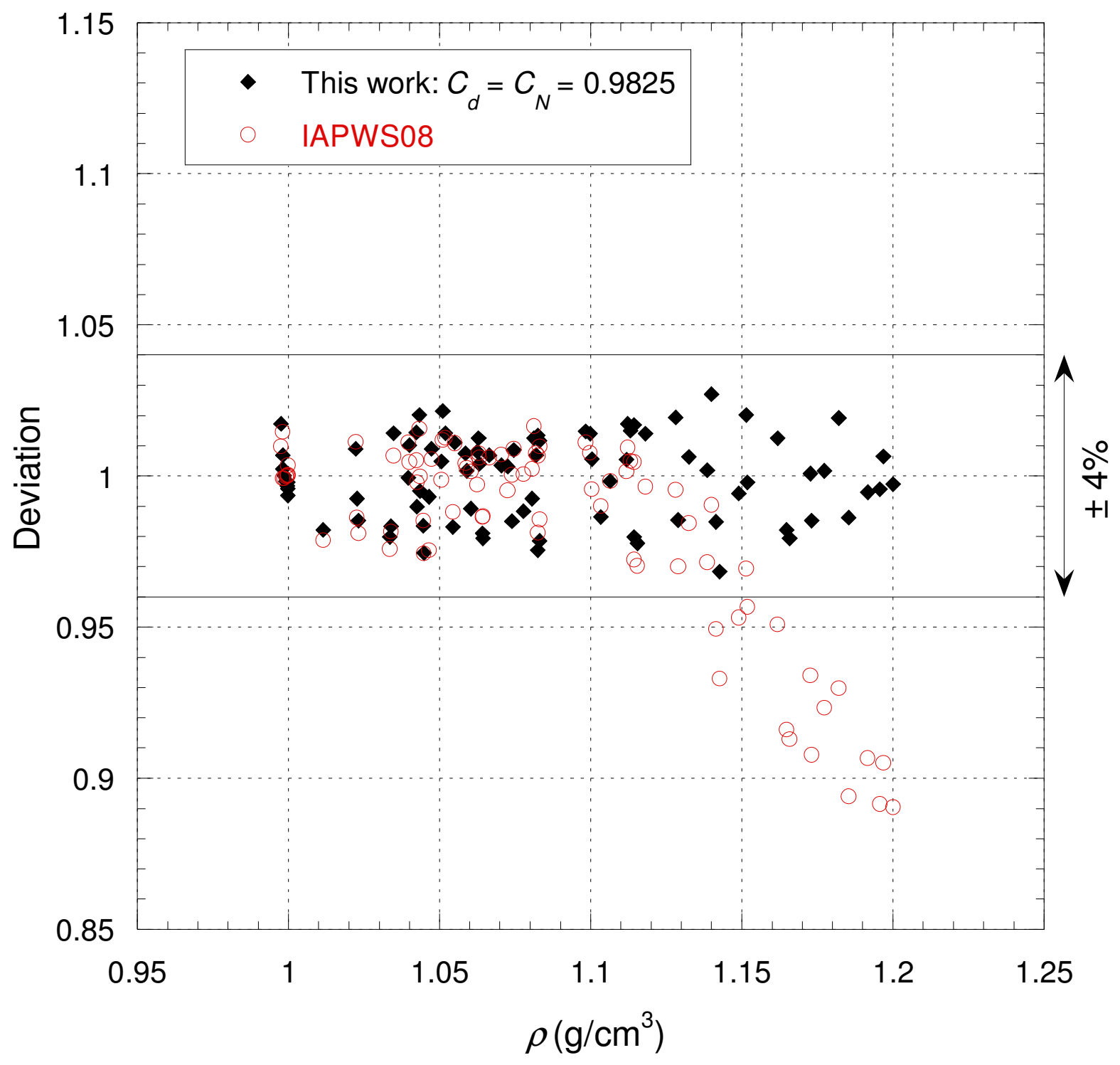

Fig. 47. Ratio of Först *et al.* “inductive” data (Ref. 38) with the IAPWS08 formulation (empty circles) and the present modeling (black diamonds) as a function of the density for all water isotherms data points.

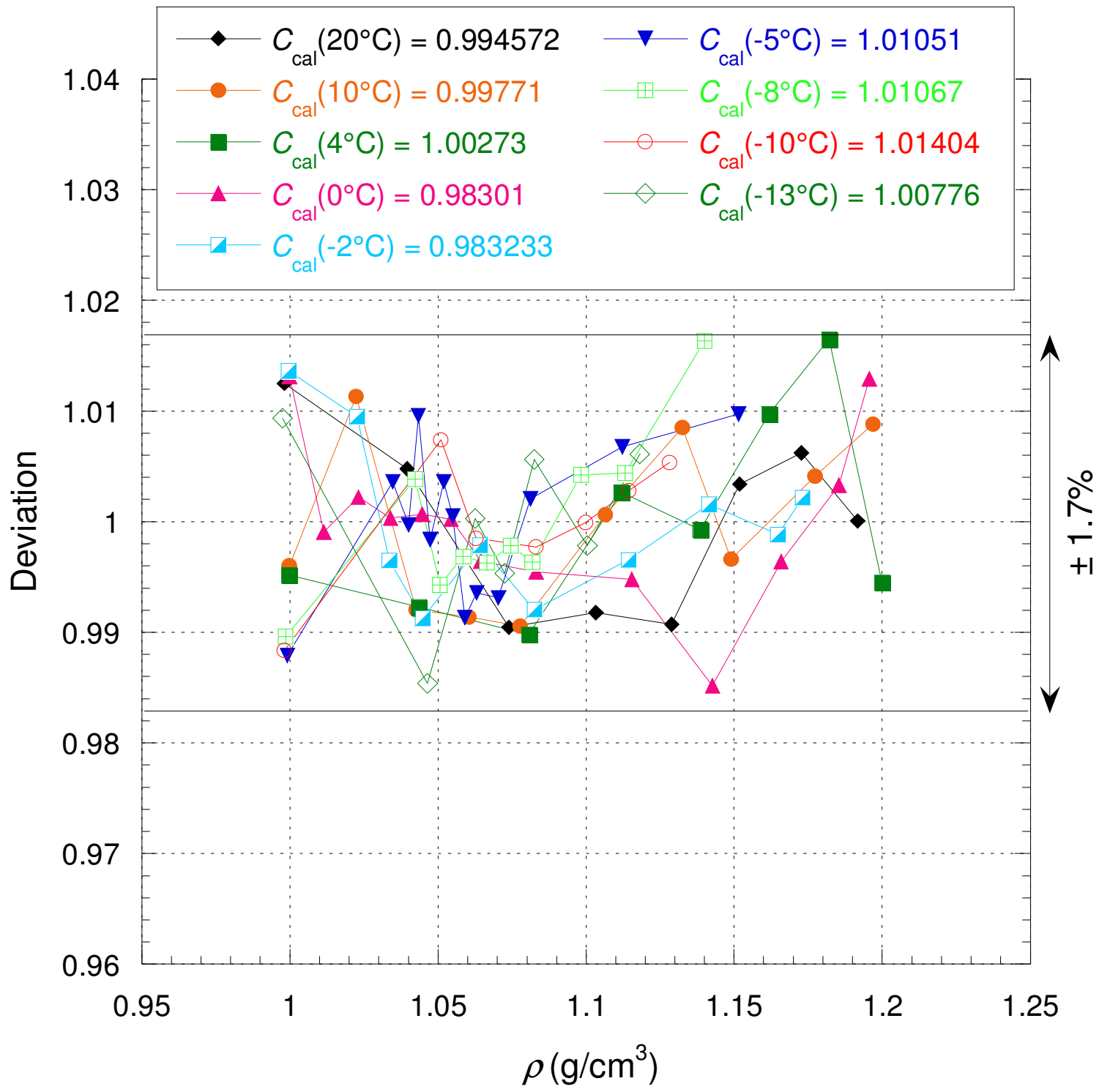

Fig. 48. Ratio of Först *et al.* “inductive” data (Ref. 38) with the IAPWS08 formulation (empty circles) and the normalized present modeling with $C_d = C_N = 0.9825$ (black diamonds) as a function of the density along each isotherm. The calibration factor $C_{\text{cal}}$ are indicated for each isotherm. The lines are eye guides.

Först *et al.* (Ref. 38) also wrote that:

> "The deviation from the data of Bett and Cappi (1965) to the data of Bridgman (1926) and the data found here is estimated to be due to insufficient consideration of the dimensional changes of the apparatus under pressure. Bett and Cappi [Ref. 12] were partly accounting for dimensional changes by considering the change in measurement distance. The maximum correction applied to the data because of dimensional changes was 0.2%. These authors did not make corrections for the change in radial dimensions, because the compressibilities of both falling body and tube of the viscometer were assumed to be equal."

In the same vein, Horne *et al.* (Ref. 39) wrote:

> "Quantitatively, however, the agreement among the values obtained by different experimentalists is poor. [...] the present results are in agreement with the recent results of Bett and Cappi [Ref. 12] however, by making many measurements over a rather narrow temperature region the present results give evidence of phenomena which escaped the notice of Bett and Cappi."

Horne *et al.* also performed an experiment with a rolling-ball type viscometer and the results they obtained are quite different, for example, from those of Cohen (Ref. 32) or Bett and Cappi (Ref. 12) (i.e. the water in Horne's experiment is more viscous than the water in Cappi's experiment). However, we have previously shown that all these datasets can be correctly reproduced in the frame of the present modeling. In other words, to understand the data of Horne *et al.* it is necessary to do the same kind of analysis as previously, as for example for Först *et al.* The density range covered by Horne *et al.* is smaller than that of DeFries *et al.* and Först *et al.*, so if there are variation effects in the coherence funtion, the corresponding $f_N$ function defined for the data analysis from Först *et al.* is the most favourable to capture these effects. Horne *et al.*'s data are given directly in relative form with respect to the viscosity at 1 atm for the isotherm considered. Therefore, for the analysis of these data, normalization factors $C$ for each isotherm must be introduced. The inverse of these factors represents the "image" of the viscosity on the atmospheric isobar therefore we can represent these factors in a parity plot like the one in Fig. 49: it is generally observed that the normalization factors of this work are closer to the bisector than those related to the formulation of IAPWS08 which means that the deviation is better centered around the unit value in our case.

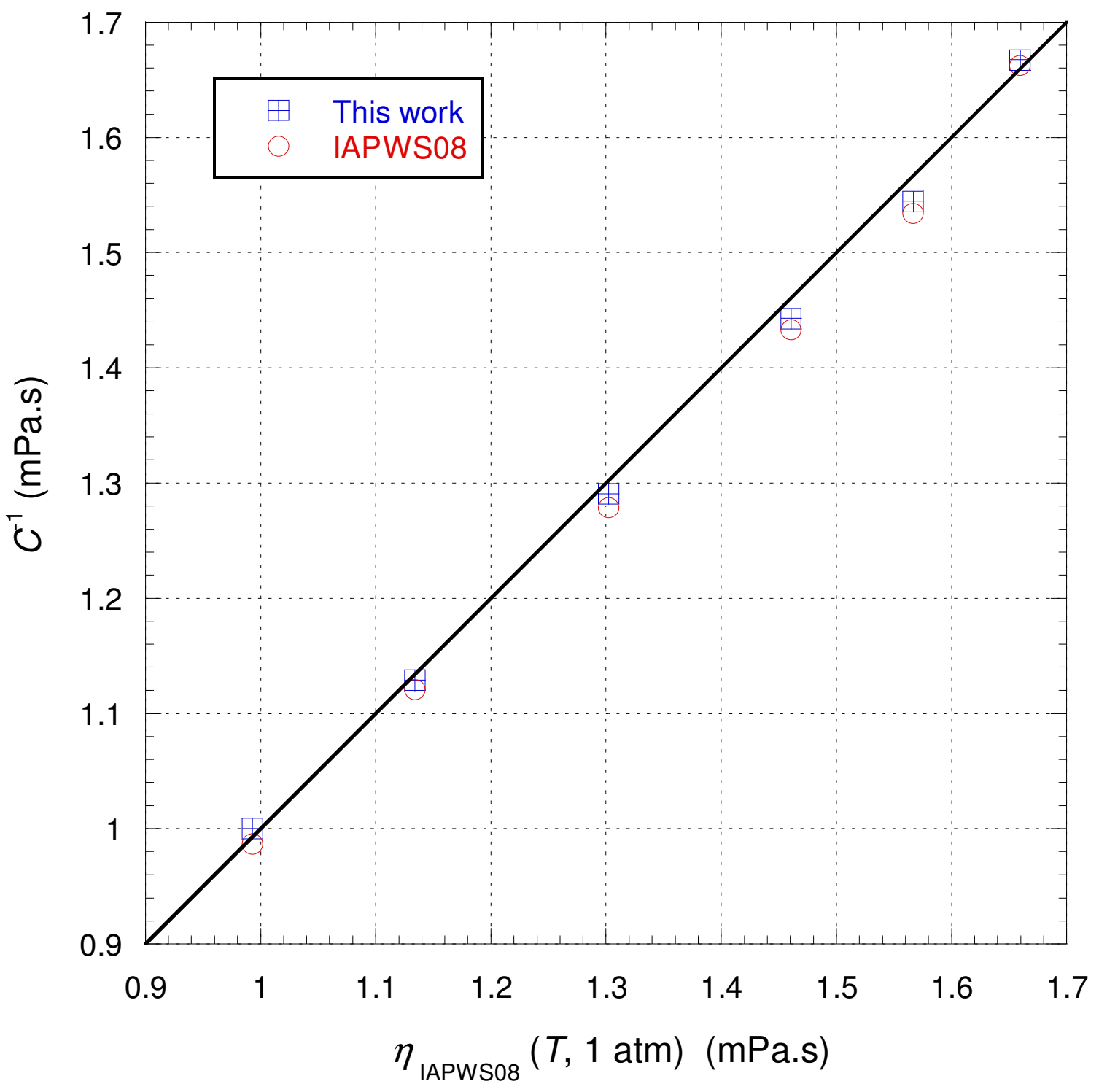

Fig. 49. Parity plot of the inverse of normalisation factors corresponding to each isotherm of the dataset from Horne *et al.* (Ref. 39).

The normalization factors do not change the possibility of reproducing more or less correctly the variations corresponding to the isotherms. Huber *et al.* (Ref. 7) in their Table 9 have considered that the uncertainty is within 2% for this dataset. Fig. 50 shows that the models are approximately equivalent in terms of uncertainty and are compatible with the experimental uncertainty, but it can be seen that the deviations are better centered on the unit value with the present modeling than with the IAPWS08 formulation.

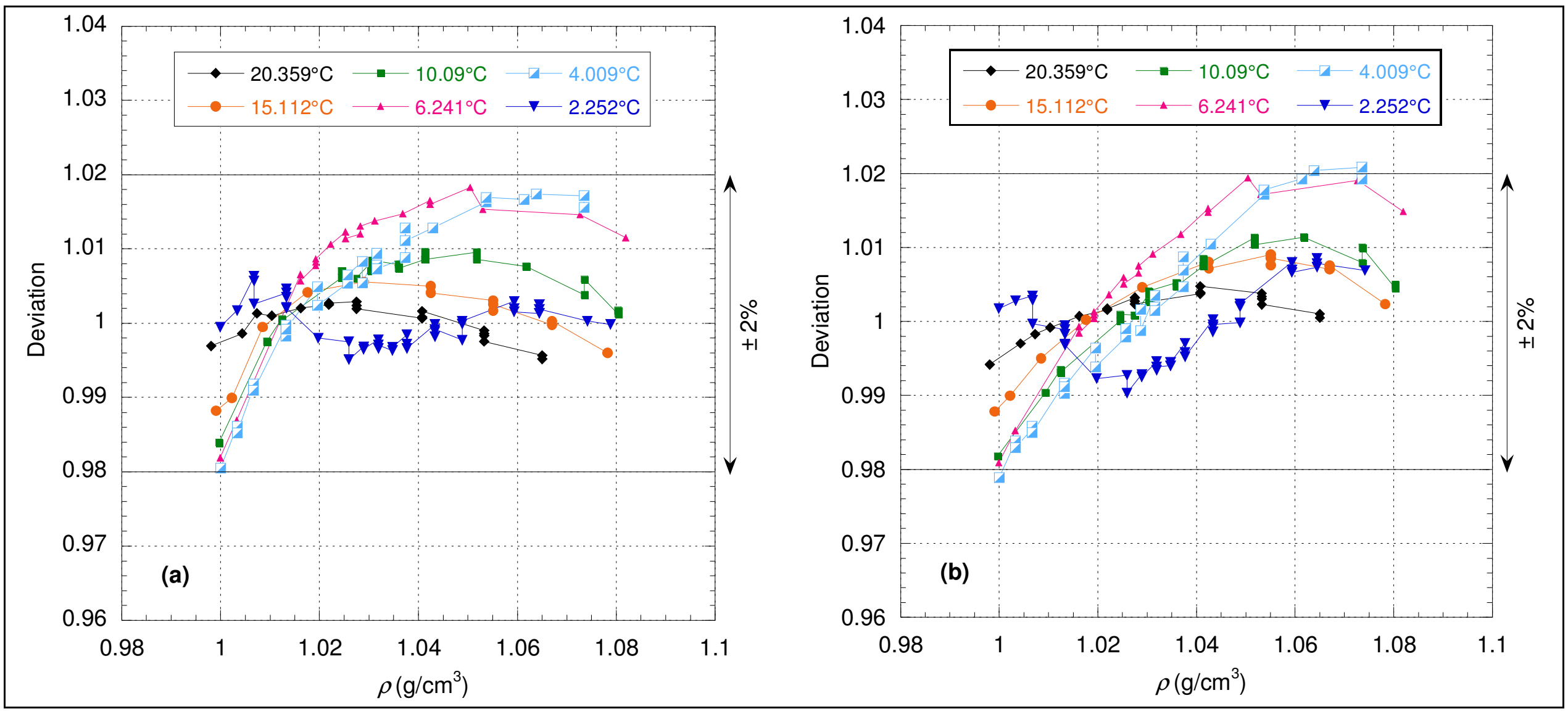

Fig. 50. Ratio of Horne *et al.*'s data (Ref. 39) as a function of the density along isotherms: (a) with present modeling using $f_{N,\text{Först}}(\rho)$ and $C_d = C_N = 0.9825$; (b) with the IAPWS08 formulation. The lines are eye guides.

Previous data of Harris *et al.*, DeFries *et al.*, Först *et al.* and Horne *et al.* correspond to overall densities above $\rho_{\mathrm{tr,Liq}}$. Fig. 44 shows that differences in coherence occur for densities greater than $\rho_{\mathrm{tr,Liq}}$ (*i.e.* in the "compressed" liquid phase). This indicates that this phenomenon of coherence modification occurs only in the compressed liquid phase.

As mentioned above, Först *et al.* conducted another experiment with an optical method to determine the pressure dependence of the viscosity and these authors explained that:

> "The key feature of this method is that no calibration is needed and the viscosity can be directly calculated according to Stokes' law provided the Reynolds number is small. The technique is based on the microscopic examination of the settling behavior of glass microspheres [*i.e.* the sphere diameter is about 50.8 µm] in a high pressure optical cell, provided the fluid is transparent both at ambient pressure and at high pressure."

But the counterpart of this optical method is that it is restricted to temperatures above 0 °C on the one hand and on the other hand:

> "the error of the optical method is quite high because of the very short falling times of the glass microspheres. The measurement times for water were about 2±3 s and therefore the relative error of the time measurement was of the magnitude of 10% and the total error about 15%."

Fig. 51 shows that both models can reproduce the data within their own experimental uncertainty. However, it can be seen that the IAPWS08 formulation is slightly better for low densities up to 1.08 g/cm$^3$ while with the present modeling the data are reproduced within less than ±5% except for the point on the isotherm at 20°C corresponding to the highest pressure. This point is also the one with the largest gap for the IAPWS08 formulation. This uncertainty of 5% is then comparable to that of Fig. 47 obtained with the inductive method which is in line with the conclusion of Först *et al.*:

> "The data obtained by both methods are found to be in agreement."

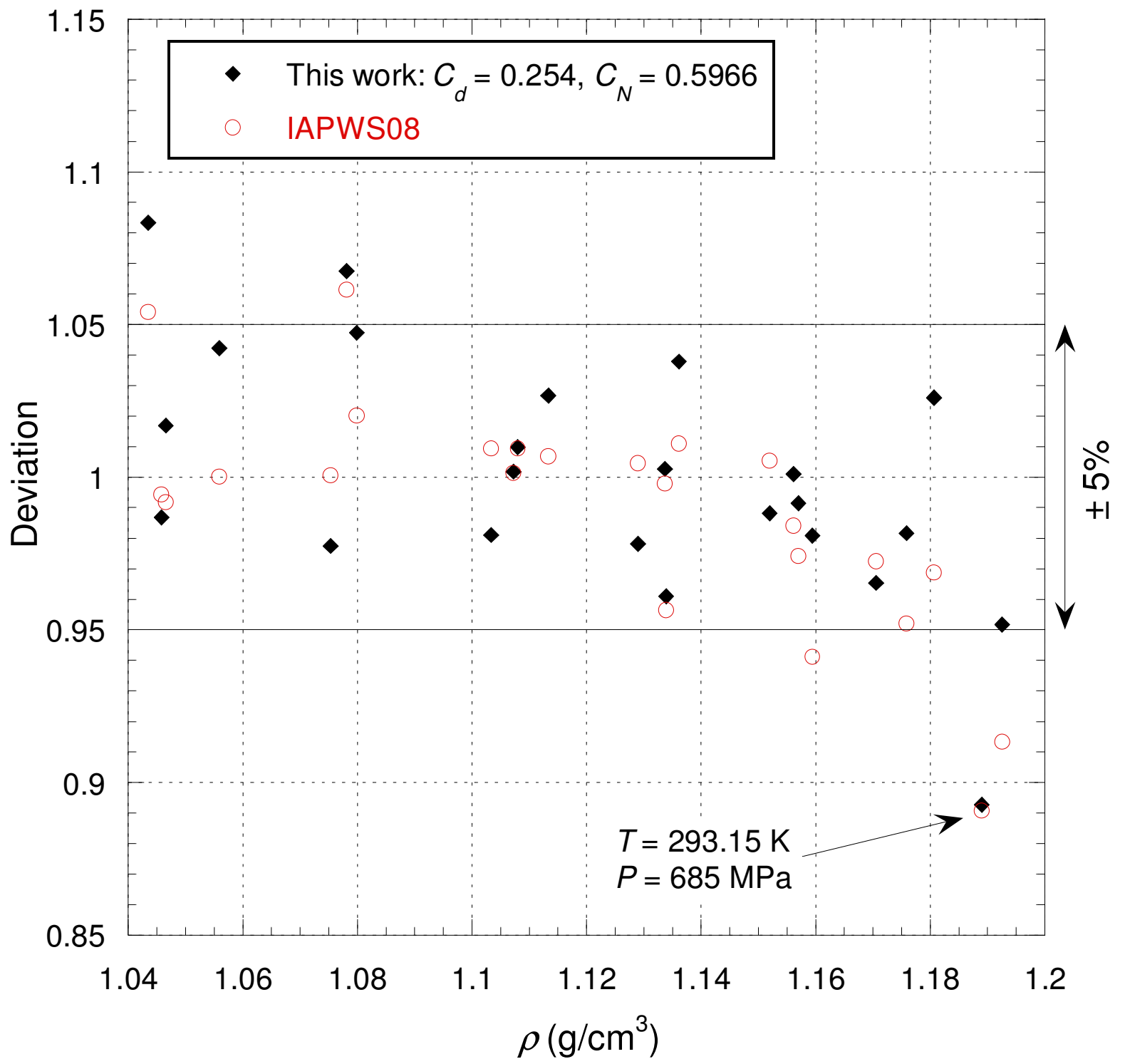

Fig. 51. Ratio of Först *et al.* "optical" data (Ref. 38) with the IAPWS08 formulation (empty circles) and the present modeling (black diamonds) as a function of the density for all isotherms data points.

Most of the low temperature and high-pressure experiments have been performed to study the liquid water anomaly which consists of observing a minimum of viscosity when increasing the pressure along an isotherm. A "usual" explanation for this phenomenon is to assume that the increase of pressure induces a deformation which reduces the strength of the hydrogen-bonded network and this reduction in cohesiveness more than compensates for the reduced void volume. To analyze this anomaly, we focus as an example on the 2.2 °C (i.e. 275.35 K) isotherm from the raw data of Cappi (i.e. see Fig. 39(a)). Fig. 52(a) shows that in our approach, this anomaly is contained in the liquid term (i.e. corresponding to Eq. (13)) while the term corresponding to the gas released is small and only increases with the pressure increase. Then Fig. 52(b) shows that the minimum of viscosity is mainly due to the variation of the transition temperature $T_t$ which consists of a competition in the ratio between the increase of $K_0^*(\rho)$ and the increase in density $\rho$: at low pressure $\rho$ increases faster than $K_0^*(\rho)$ then it is the opposite. Finally, it should be noted that the pressure corresponding to the minimum of viscosity is shifted with that corresponding to the minimum of the transition temperature.

It appears in our approach that this anomaly is not related to a variation in the cohesion of the liquid, which systematically increases with increasing pressure, but mainly to the existence of the transition temperature and its variation.

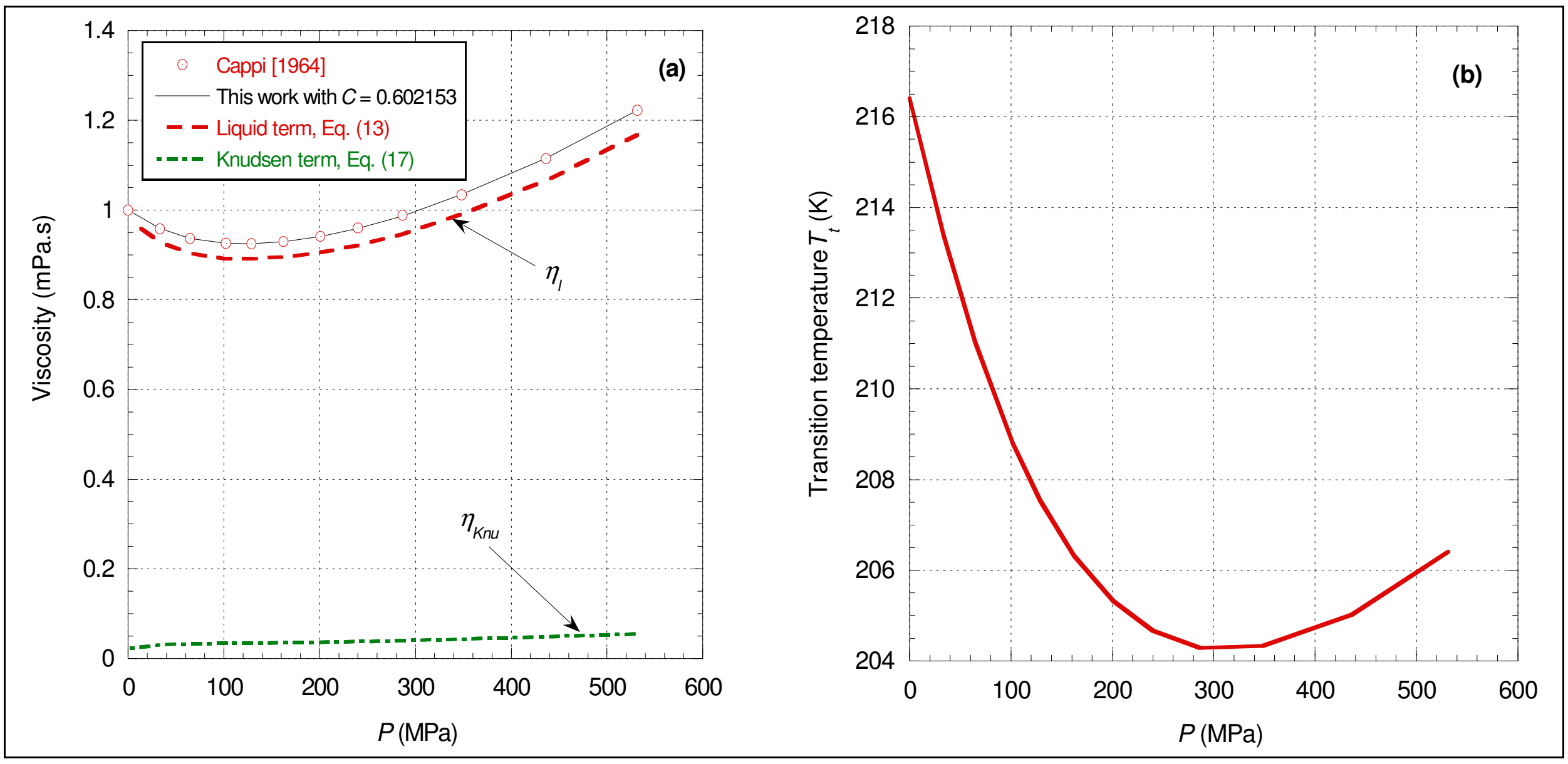

Fig. 52. Isotherm 275.35 K: (a) viscosity variations of water versus pressure of the two terms constituting Eq. (18); (b) transition temperature variation versus pressure.

Deguchi *et al.*'s experiment (Ref. 40) is the counterpart of Dehaoui *et al.*'s experiment (Ref. 28) but on the isobar corresponding to 25 MPa and for temperatures between 15°C and 300°C (i.e. the viscosity was obtained from the translational diffusion coefficient of probe particles dispersed in the medium by the SE law). Two main types of particles have been used: polystyrene latex with a mean hydrodynamic diameter $d_H$ = 214.0 nm and colloidal silica with a mean hydrodynamic diameter $d_H$ = 137.4 nm. In other words, these particles are significantly smaller than those of Dehaoui *et al.*

Deguchi *et al.* wrote that:

> "In the case of the polystyrene latex, good agreement was found between the measured and calculated values [from the IAPWS 1995 formulation, Ref. 17] up to 275 °C, but the measured value at 300 °C was slightly larger than the calculated value. […] The measured values agree with the calculated values within 6% error up to 275 °C, but the value measured at 300 °C was 27% larger than the calculated value.
>
> A similar trend was observed when the colloidal silica was used as the probe particle. Good agreement was found between the measured and calculated values up to 200 °C, and the error between the values was less than 4%. In this case, however, the measured viscosity at 251 °C was 17% smaller than the calculated value. Furthermore, strong scattering from the particles diminished at 275 °C, and a value 10% smaller than the calculated value was obtained."

Fig. 53 shows that the present modeling is globally in agreement with Deguchi *et al.*'s comment above, with the exception that the extreme values of deviation are slightly higher. On the other hand, we can see that the IAPWS08 formulation is highly shifted even if the general trend is globally identical to ours. This is because here the IAPWS08 formulation is quite different from the data of NIST (Ref. 16), as we have already mentioned in section 3. An explanation put forward by Deguchi *et al.* concerning the significant deviation at high temperature would be due to the swelling of the particles, or even coalescence in the case of polystyrene particles. These effects are therefore beyond the scope of this analysis.

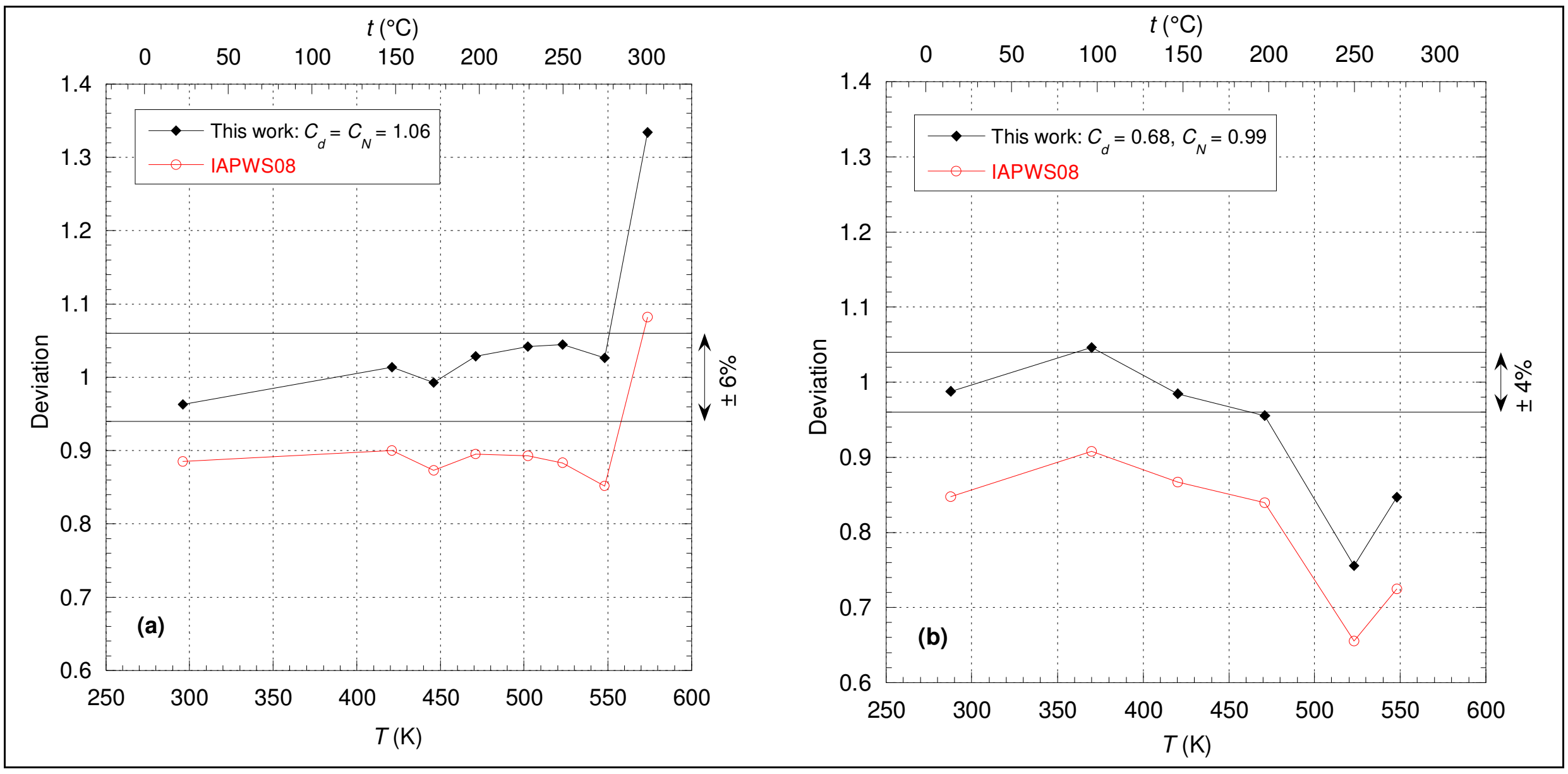

Fig. 53. Ratio of Deguchi *et al.*'s data (Ref. 40) with the IAPWS08 formulation (empty circles) and the present modeling (black diamonds) as a function of the temperature along the isobar equal to 25MPa: (a) polystyrene latex particles ($d_H$ = 214.0 nm); (b) colloidal silica particles ($d_H$ = 137.4 nm). The lines are eye guides.

To complete this section, we will analyze the data of Moszynski (Ref. 41) and Assael *et al.* (Ref. 42) which are considered the most accurate in their density and temperature range.

Regarding Moszynski's experiment, the author tells us that:

> "The measurements of viscosity of water were carried out using a steel sphere oscillating in a large container.
> In this latter case the results were evaluated using an exact solution relating the viscosity to the decrement of the oscillatory motion with appropriate corrections, for the damping due to ancillary parts of the oscillating elements, being applied. […] Thus the measurements in water may be regarded as absolute."

Fig. 54(a) shows that the deviation with the present modeling and the IAPWS08 formulation has the same general trend but they are shifted so that finally the present modeling correctly represents (i.e. within 1%) all the data on the quasi-isotherms except the one corresponding to the highest temperature while for the IAPWS08 formulation it is the opposite.
Moszynski notes that:

> "[…] a sample of fresh distilled water which, after induction into the instrument, was deaerated carefully by evacuation for a prolongated period of time, gave at 19.83 °C a value of viscosity of 1.0092 centipoises. Using a temperature coefficient of 0.024 centipoise per deg C, this yields a value of 1.0051 centipoises at 20 °C."

In other words, at 20 °C Moszynski's data have a deviation of about ±0.3% from the accepted reference value. This suggests to admit that the uncertainty of Moszynski's data is at best 0.3%. Based on this observation, Fig. 54(b) shows that by considering a renormalization coefficient *C* for each quasi-isotherm, we can reproduce Moszynski's data with an overall deviation of ±0.4% with the exception always of the isotherm with the highest temperature which seems irreducible for all models.

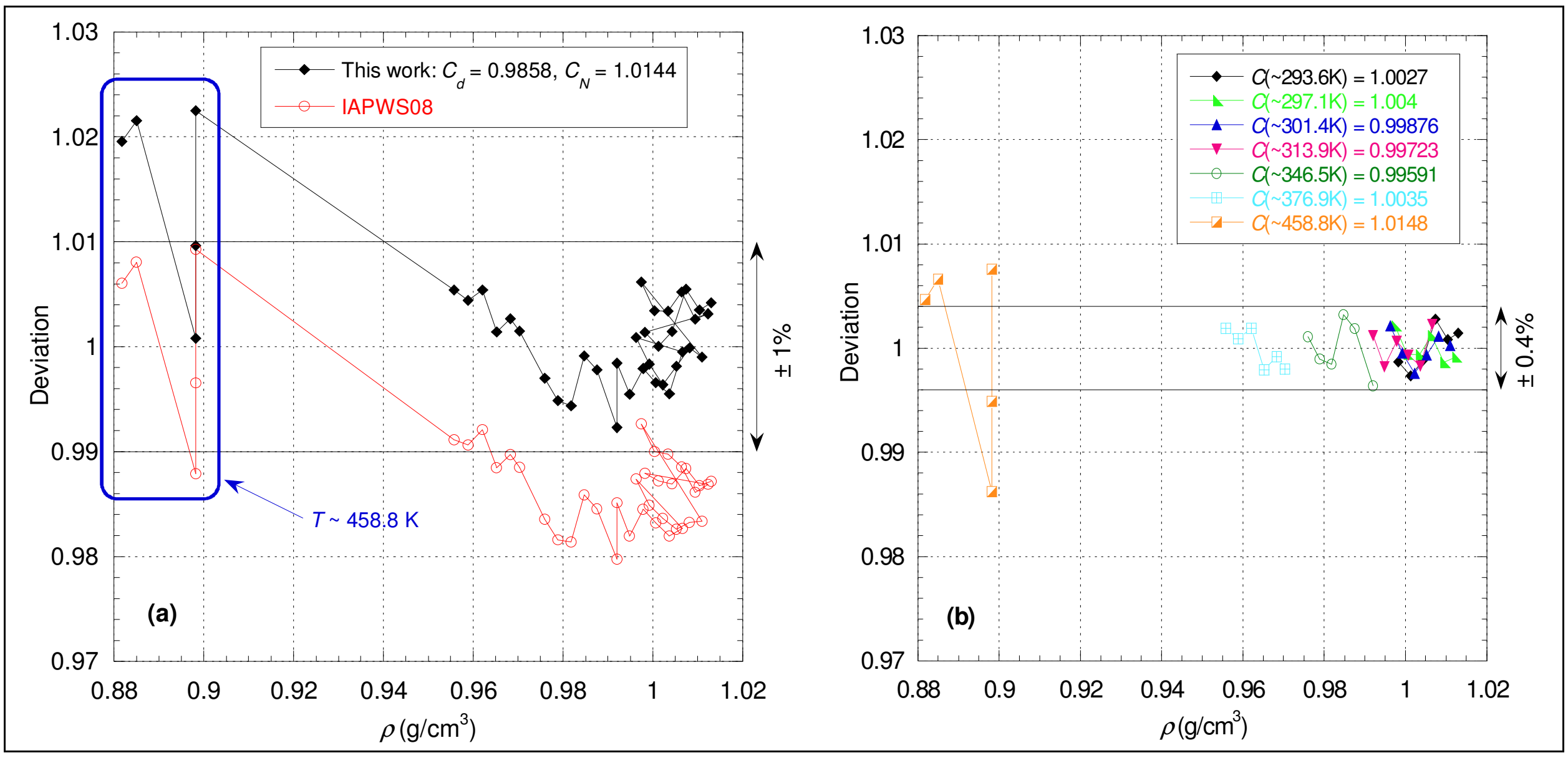

Fig. 54. (a) Ratio of Moszynski's data (Ref. 41) with the IAPWS08 formulation (empty circles) and the present modeling (black diamonds) as a function of the density for all quasi-isotherms; (b) Ratio of Moszynski's data (Ref. 41) with the present modeling ($C_d$ = 0.9858 and $C_N$ = 1.0144) rescaled as a function of the density for each quasi-isotherm. The lines are eye guides.

Assael *et al.* (Ref. 42) initially built a vibrating-wire viscometer to perform very accurate viscosity measurements on liquids hydrocarbons. This instrument was then modified to measure the viscosity of alcohols and more particularly the viscosity of water. Assael *et al.* wrote that:

> "The instrument was further calibrated, as described in detail elsewhere [Ref. 43], with respect to the viscosity of water at a pressure of 0.1 MPa and a temperature of 293.15 K for which an accurate reference value is available. With due regard to the precision of the determination of the individual quantities and the accuracy of the calibration data, it is estimated that the overall uncertainty of the present results is better than +0.5% under all conditions."

It is worth noting that the vibrating wire has a nominal diameter of 100 μm.

Fig. 55(a) shows that the two models are roughly comparable and it can be seen that the deviation for the isotherm with the highest temperature is borderline with the uncertainty estimated by Assael *et al.* However, it should be noted that Assael *et al.*'s data are obtained in relative form and that, to determine viscosity, these authors used an equation of state different from the IAPWS-95 formulation (Ref. 17) used here. Fig. 55(b) shows that by applying a small renormalization constant $C$, it is then possible to reproduce these data very precisely (i.e. within 0.2%).

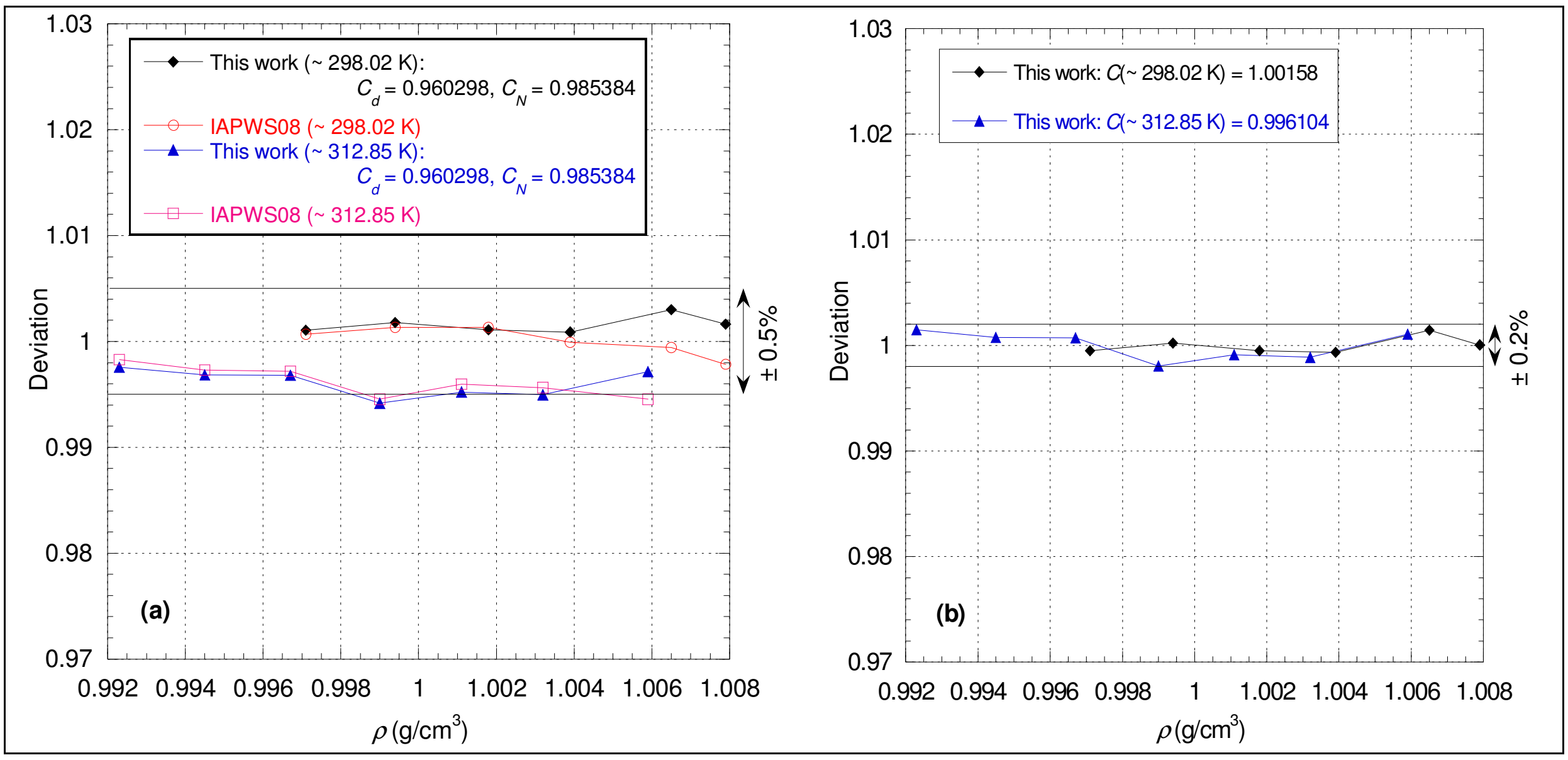

Fig. 55. Ratio of Assael *et al.*'s data (Ref. 42) with the IAPWS08 formulation (empty circles and empty squares) and the present modeling (black diamonds and blue triangles) as a function of the density: (a) for all quasi-isotherms; (b) for each quasi-isotherm with the present modeling ($C_d$ = 0.960298, $C_N$ = 0.985384) rescaled. The lines are eye guides.

To show the possible extent of the data that can be analyzed, it is interesting to apply the present modeling to the data of King *et al.* (Ref. 44) for slightly salty water. King *et al.*'s experiment is almost the same as the one made by Abramson (Ref. 11) but for one mean temperature around 23.3 °C (i.e. ~296.45 K). Therefore, the same coefficients $C_d$ and $C_N$ should be applied as for Abramson, i.e. $C_d$ = 0.554 and $C_N$ = 0.825. King *et al.*'s data are given in relative form normalized to ambient-pressure viscosity value. Taking into account a calibration factor $C_{\mathrm{cal}}$, Fig. 56 shows that with the same parameters as for Abramson's experiment, it is possible to reproduce the variation of King *et al.*'s data along a quasi-isotherm within their standard deviation.

*A priori*, the effect of salt should have an impact on the $K^*$ function, but here, given the very small amount of salt and the relatively low accuracy of the measurements, this effect cannot be observed.

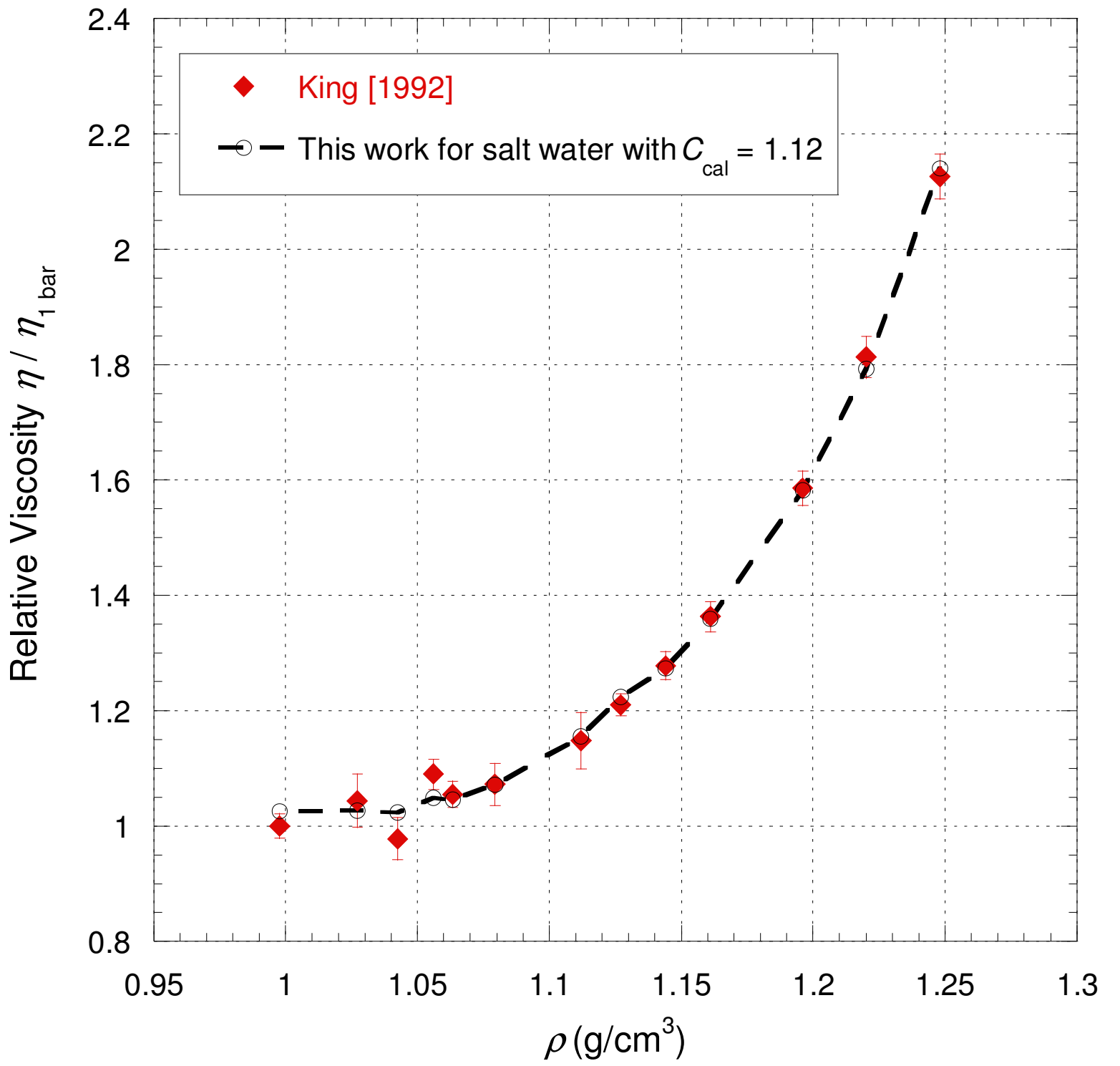

Fig. 56. Relative viscosity data obtained for salt water by King *et al.* (Ref. 44) and the present modeling with $C_d$ = 0.554 and $C_N$ = 0.825 as a function of the density along the quasi-isotherm at ~296.45 K. The lines are eye guides.

To conclude this section, we will briefly discuss a hysteresis effect (i.e. memory effect) that occurs when liquid water is compressed and then decompressed at sufficiently high pressures. This effect was first suggested by Warburg (Ref. 45) and then highlighted by Moszynski (Ref. 41) who carried out a viscosity measurement at 1 atm., then he increased the pressure till 341 atm. and then dropped back to 1 atm.; the value measured at the end is then different from that measured initially but above all the difference obtained is greater than the experimental uncertainty. In view of this observation, the data obtained by Moszynski and analyzed in Fig. 54 were carried out from high to low pressures: the analysis of these data allowed us to determine the values of the two parameters $C_d$ and $C_N$ under these experimental conditions. The values of these parameters can be used to analyze Moszynski's data in Table 1. Thus Fig. 57 shows that the present modeling with the parameters determined to analyze the data in Fig. 54 logically allows reproducing within the uncertainty the points obtained from high pressures to low pressures; it is then that the initial point is slightly outside the error bar as indicated by Moszynski.

In the present modeling, this effect can be taken into account in two different ways:

- or initially the water had a slightly higher volume of coherence than in the final state, i.e. $C_N$(initial point) = 1.0262 instead of $C_N$(final point) = 1.0144;
- or the water was in a state that when sheared it releases less gas in the initial state than in the final state.

Given the very limited data available, it is not possible to discriminate between these two possibilities. This being said, if it is the first case that occurs, then we should also have a trace of it on the self-diffusion data, whereas in the second case it is an effect that only occurs on viscosity.

Although such an effect can be taken into account by the present approach here (i.e. this effect can be understood, for example, as a creating or annealing of defects in the lattice), it appears necessary in all cases as written by Moszynski that:

> "Extensive experimentation would be required to establish the cause of this effect […]".

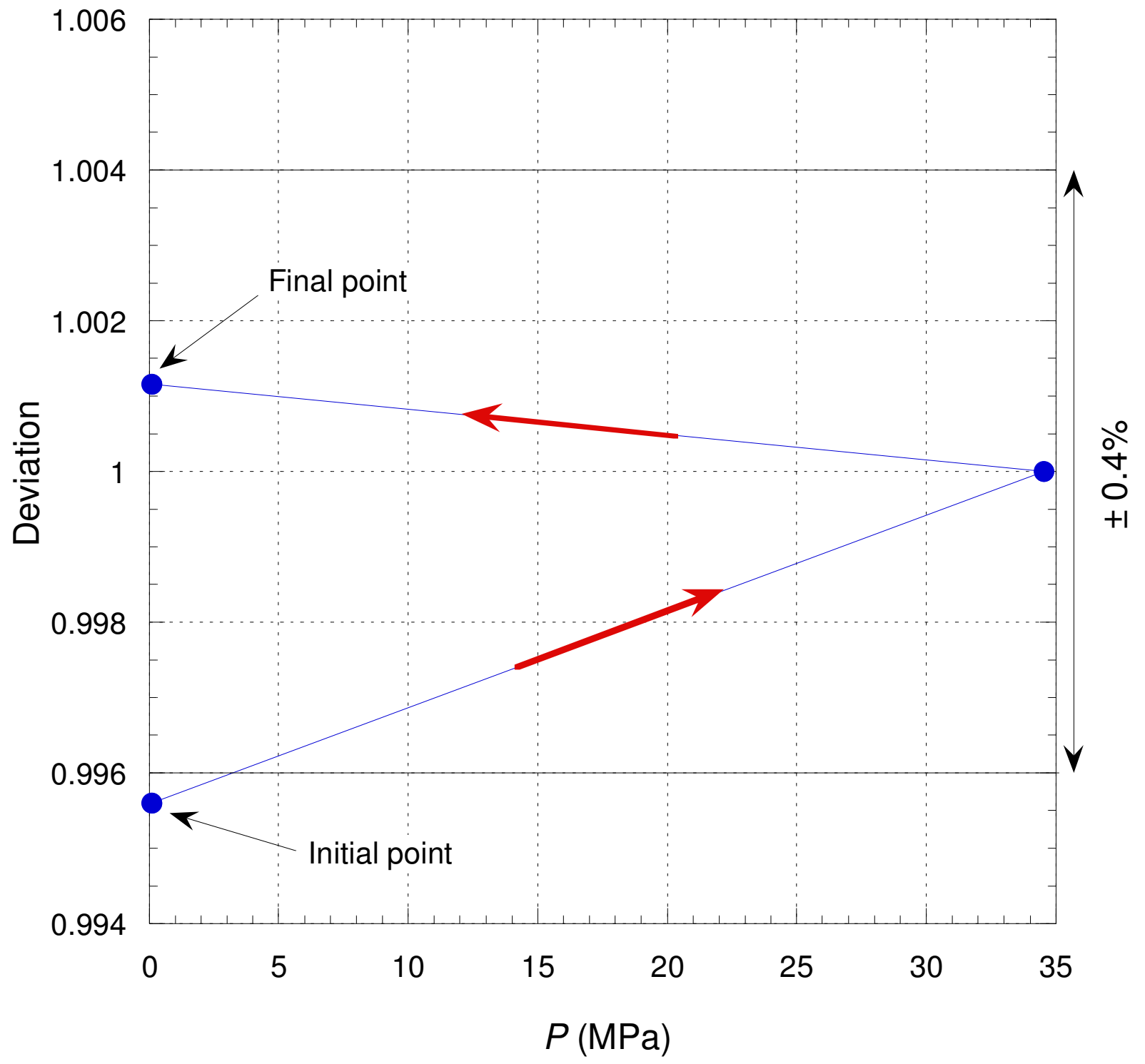

Fig. 57. Ratio of Moszynski Table 1 data (Ref. 41) with the present modeling as a function of the density for the isotherms at 20°C.

### *4.1.3. Viscosity of Gaseous Water*

As Huber *et al.* (Ref. 7) wrote that:

> "[…] the previous IAPWS formulation for the viscosity did not include the data set of Yasumoto [Ref. 46] in the analysis, and the recent data of Teske *et al.* [Ref. 47] were not available at the time, thus leaving room for improvement in this region. The present correlation was developed with both of these data sets and offers significant improvement in the representation of the data in this region."

we will start by analyzing these two datasets.

First of all, we will analyze the experimental data obtained by Yasumoto (Ref. 46) because these measurements correspond to the lowest densities reached in the vapor phase of water. Yasumoto's experiment consists in measuring the vapor viscosity by means of a tandem capillary-flow viscometer whose capillary diameters are about 1.2 mm. Yasumoto wrote that:

> "The precision of the viscosity, $\eta_2$ [*i.e.* a correction for slip is added], obtained was estimated to be less than 0.5%."

But later in the text, Yasumoto considers that formulas that represent his data within 1% are excellent or remarkable. Fig. 58 shows that the two models cannot reproduce Yasumoto's data within 0.5% but they are both compatible with an uncertainty close to 1%. However, it can be observed that the present modeling has a deviation slightly better centered on the unit value than the IAPWS08 formulation.

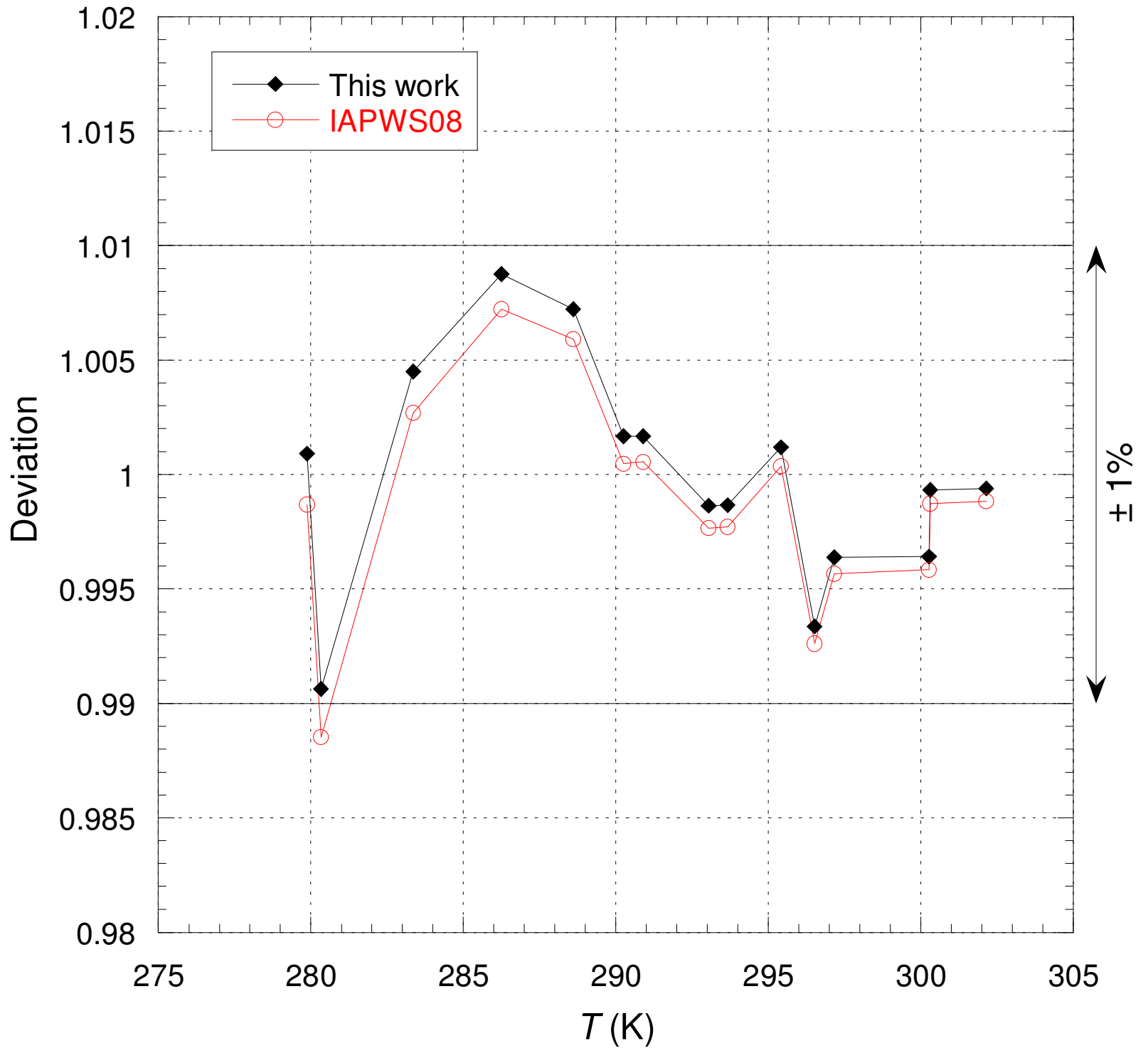

Fig. 58. Ratio of Yasumoto's data (Ref. 46) with the IAPWS08 formulation (empty circles) and the present modeling (black diamonds) as a function of the temperature. The lines are eye guides.

Yasumoto's data being those obtained with the lowest density, they therefore constitute the preferred set for observing the maximum ratio $\rho_{Knu}/\rho$. Taking for $\delta$ the tube radius of Yasumoto's experimental set-up, Fig. 59 shows that the ratio $\rho_{Knu}/\rho$ is maximum at low temperature and that for all water experimental datasets it does not exceed few thousandths. This confirms the use of Eq. (18) to process viscosity data in water.

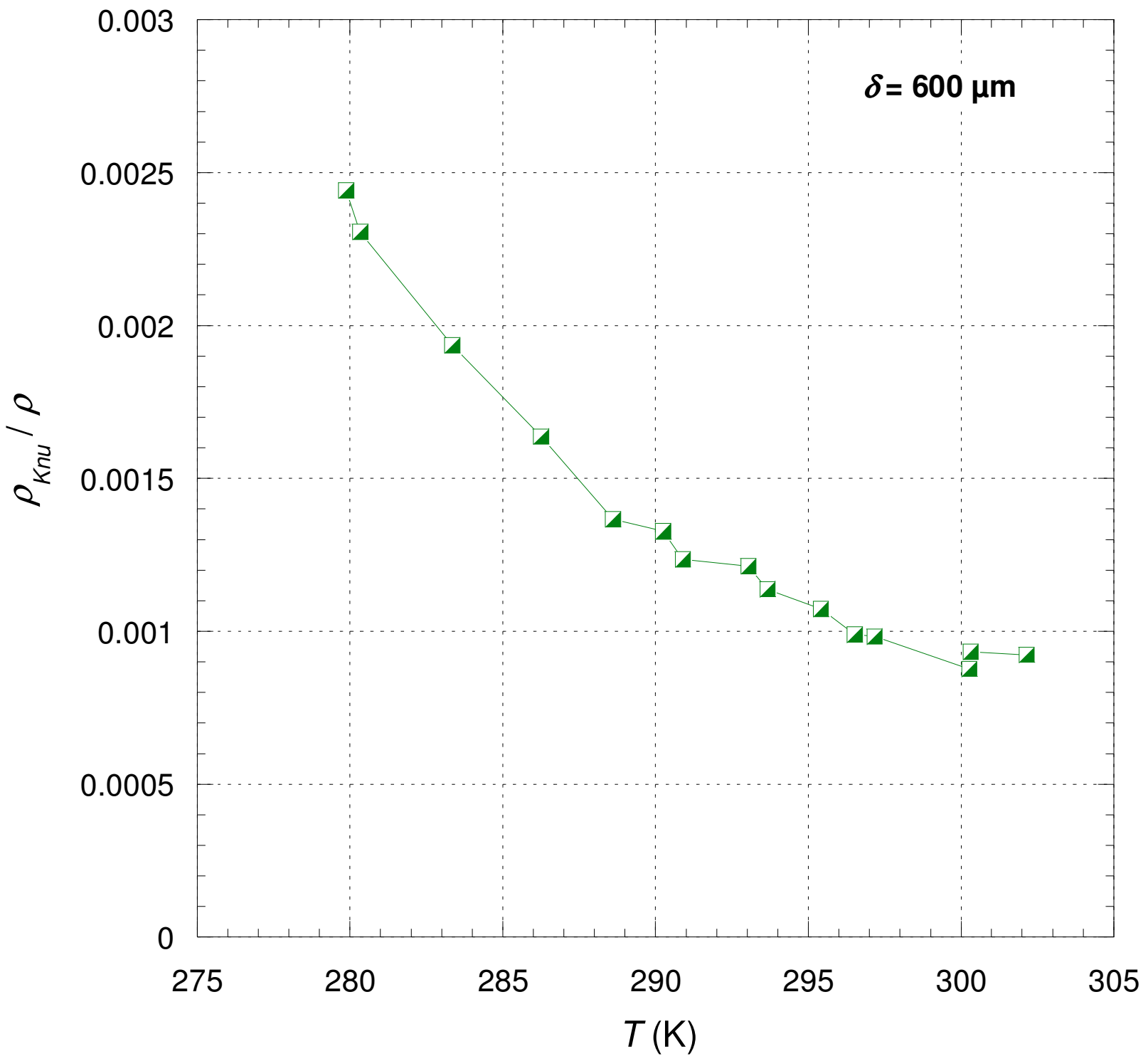

Fig. 59. Temperature variation of the density ratio $\rho_{Knu}/\rho$ corresponding to Yasumoto's dataset. The lines are eye guides.

Teske *et al.*'s data (Ref. 47) correspond to higher densities than those of Yasumoto but they are more precise and above all they are carried out along different isochors which corresponds to the essence of the construction of the present modeling. Teske *et al.* wrote that:

> "The measurements were performed using an all-quartz oscillating-disk viscometer with small gaps. [...] The uncertainty is estimated to be ± 0.2 % at ambient temperature increasing up to ± 0.3 % at higher temperatures."

Fig. 60 shows that the present modeling allows data to be reproduced within 0.4% while the IAPWS08 formulation deviates slightly at ambient temperature for the highest densities. The deviation of ±0.4% is twice too large in view of the experimental uncertainties. However, we observe that the deviation seems to be shifted above the unit value, which suggests a calibration effect. Teske *et al.* wrote that:

> "The viscometer was calibrated using a reference value by Kestin *et al.* for argon at room temperature since only one calibration point is needed in the theory of Newell in the range of moderately low densities. The performance was checked by further measurements on argon at higher temperatures up to 600 K."

This allows us to consider a renormalization of the data by rescaling very slightly each isochor by a factor *C*. Fig. 61(a) shows that the deviation can be reduced globally to ±0.2% with the exception of a few points but which are in agreement with an uncertainty of ±0.3%. Fig. 61(b) shows that the normalization coefficients are not random but increase almost linearly with density, which is significant for a device effect.

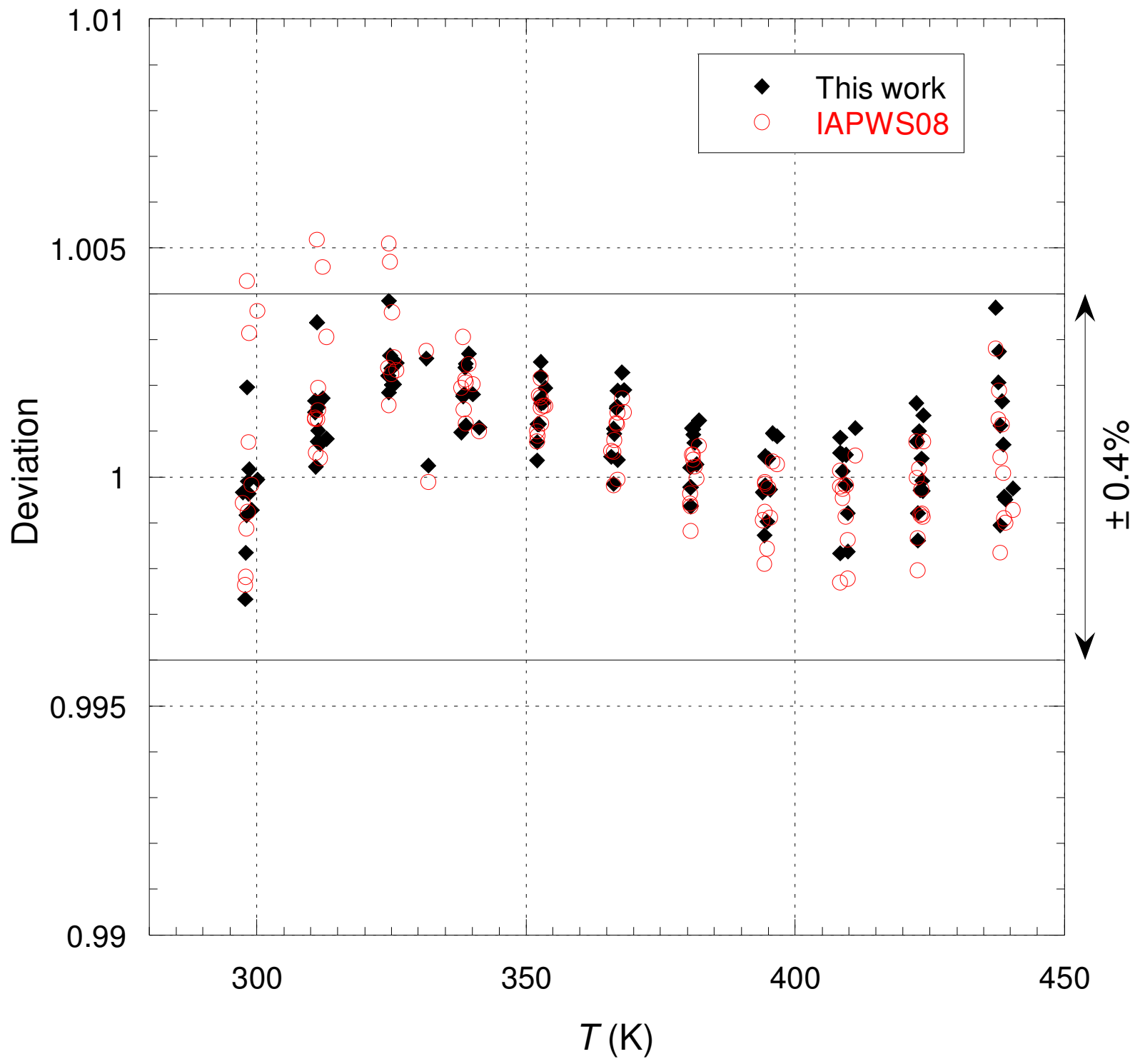

Fig. 60. Ratio of Teske *et al.*'s data (Ref. 47) with the IAPWS08 formulation (empty circles) and the present modeling (black diamonds) as a function of the temperature for all isochors data points.

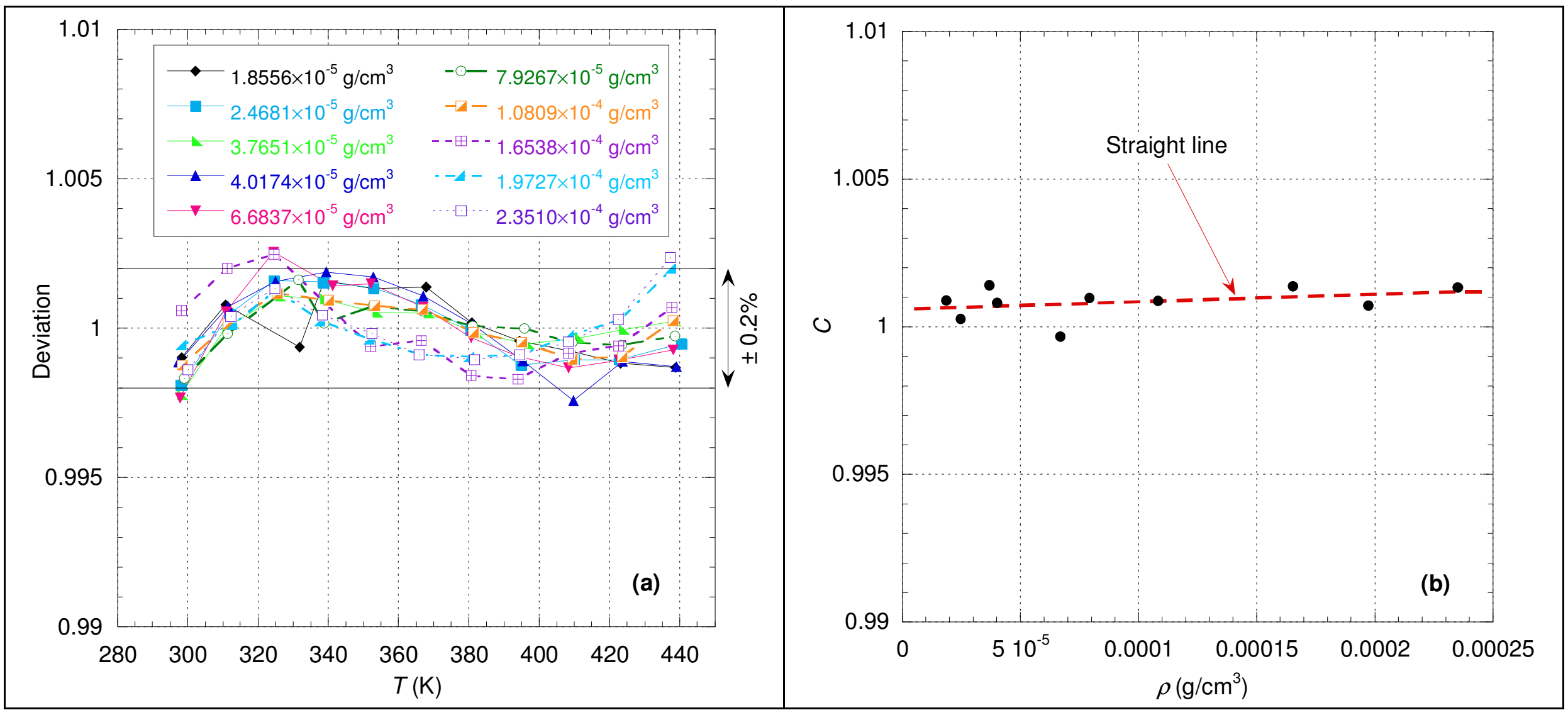

Fig. 61. (a) Ratio of Teske *et al.*'s data (Ref. 47) with the present modeling rescaled as a function of the temperature for each isochor. The lines are eye guides. (b) Renormalization factors as function of density.

Since the present modeling allows us to reproduce Teske *et al.*'s data within their uncertainty, we will take as an example the isochor with the lowest density in order to illustrate the division of Eq. (18) into two terms. Fig. 62 first shows the low-temperature extrapolation derived from the present modeling: we observe that below 100 K, viscosity is given by the liquid term while beyond this the temperature variation is entirely described by the Knudsen term. Then we observe that in the temperature range considered the liquid term is constant, equal to its limit value determined by Eq. (21), and that it represents approximately half of the viscosity value.

Fig. 62 also shows the dilut-gas limit term (i.e. Eq. (11) in Ref. 7) of the IAPWS08 formulation which corresponds to the term largely dominant on this isochor: we observe that the latter diverges around 140 K then becomes negative and therefore has no longer any physical meaning. By comparison, in the present modeling the viscosity diverges by definition at $T_t$ but for this particular isochor we have $T_t^* \approx 4\times10^{-12}$ which is more compatible with a spinodal limit than the IAPWS08 formulation.

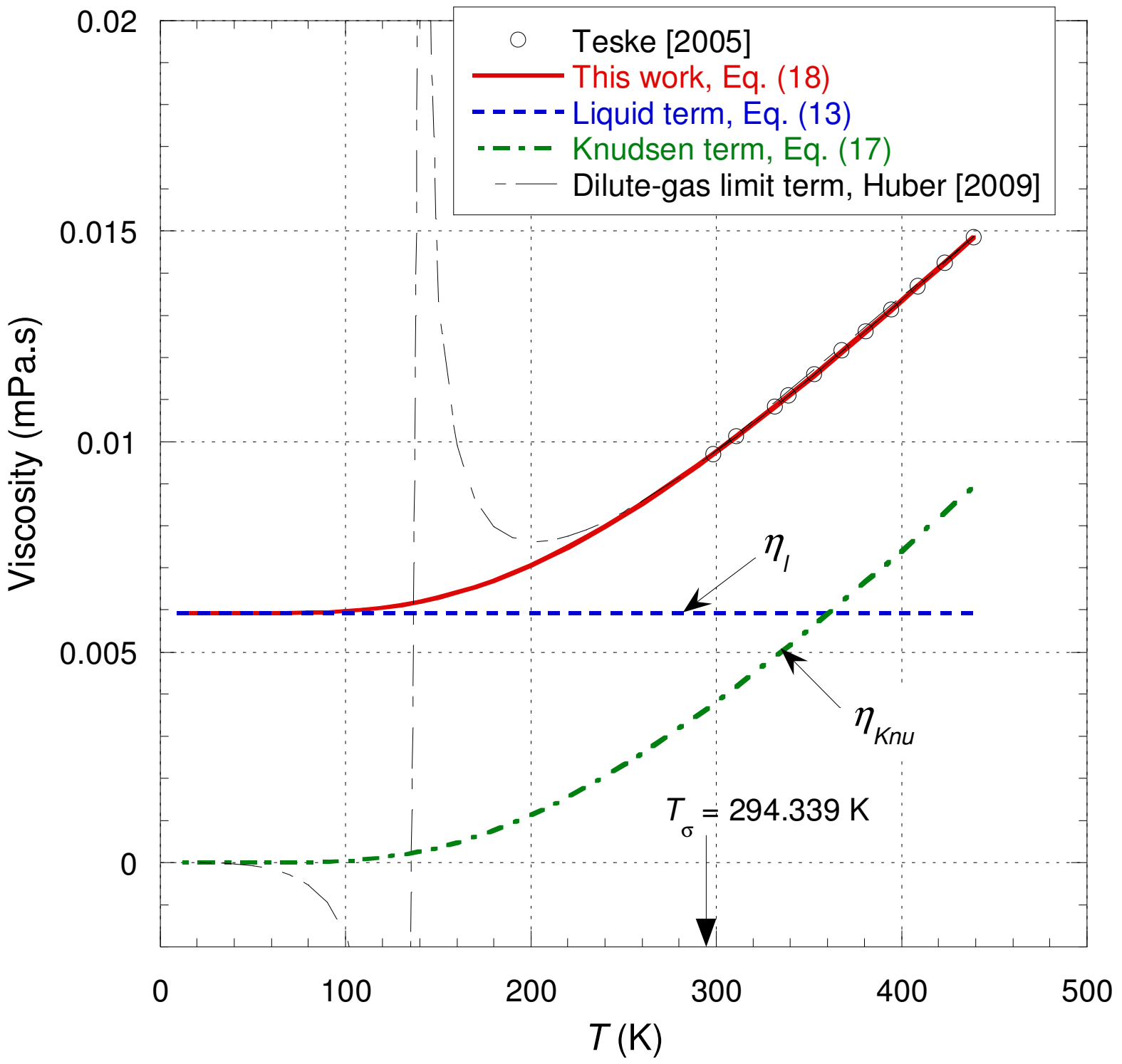

Fig. 62. Plot of the two terms constituting Eq. (18) along an isochor of viscosity corresponding to $1.8556\times10^{-5}$ g/cm$^3$ (i.e. the smallest density of Teske *et al.*'s isochor, Ref. 47) as function of the temperature. $T_\sigma$ corresponds to the temperature on the saturated vapor pressure curve for this isochor.

Shifrin (Ref. 48) made measurements with a capillary viscometer in a density range comparable to that of Teske *et al.* but for a wider temperature range while remaining on the isobar 1 kg/cm$^2$ (i.e. 0.0981 MPa). These data then allow us to partially cross-reference those of Teske *et al.* Huber *et al.* (Ref. 7) in their Table 9 have considered that the uncertainty is within 3% for this dataset. Fig. 63 shows that it is only required to renormalize the data very slightly to reproduce them with an uncertainty entirely within 3%. However, it can be observed that the IAPWS08 formulation is here also compatible with this uncertainty and has a very similar trend. In both cases it can be said that the temperature dependence is well reproduced.

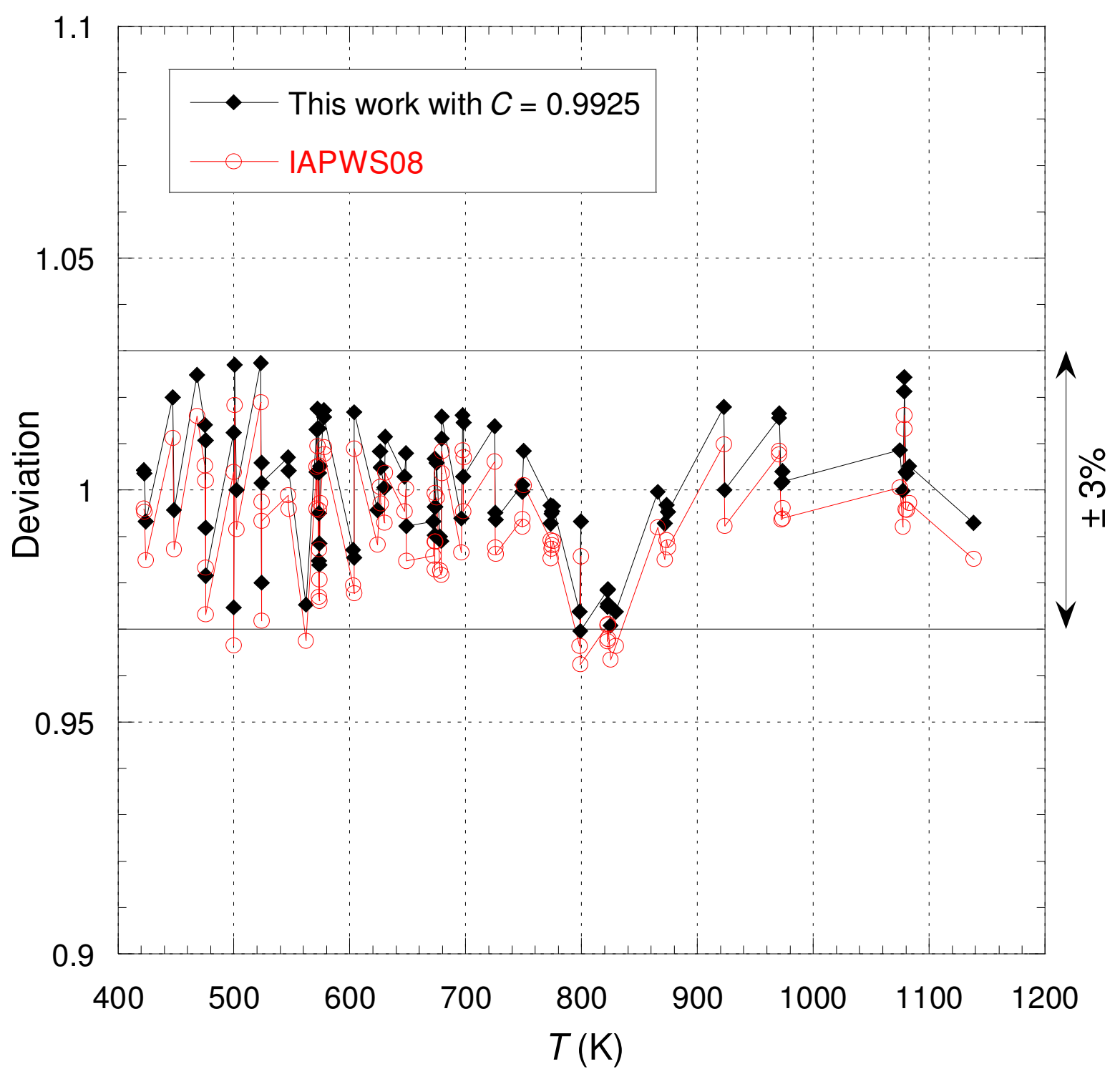

Fig. 63. Ratio of Shifrin's data (Ref. 48) with the IAPWS08 formulation (empty circles) and the present modeling rescaled (black diamonds) as a function of the temperature along the isobar at 0.0981 MPa. The lines are eye guides.

The same kind of experiment as Shifrin was performed by Bonilla *et al.* (Ref. 49) but on the atmospheric isobar and over a wider temperature range. Bonilla *et al.* provide a table with recommended values that do not correspond to the raw data but these can be found in their Fig. 1. Among the values in the recommended table, the three highest temperatures as well as the four lowest are extrapolated points and therefore they should not be considered. Fig. 64(a) shows that the raw data and the recommended values are generally equivalent, but we then preferred to analyze the raw data, which are more numerous and contain the experimental noise.

If we follow the uncertainties indicated in Huber *et al.*'s Fig. 24 (Ref. 7), then the data of Bonilla *et al.* (like those of Shifrin) are spread between subregion 4 and subregion 5, *i.e.* between 2% and 3%. Fig. 64(b) shows that by renormalizing the data (as for Shifrin's data), the deviation obtained is compatible with the uncertainty proposed by Huber *et al.* It can be seen that the IAPWS08 formulation has the same trend for the deviation until $2T_c$ but it is strongly shifted compared to the assumed uncertainty of these data. Above $2T_c$, the IAPWS08 formulation is no longer compatible with Bonilla *et al.*'s data even if these data are renormalized.

Finally, it can be noted that the renormalized data of Bonilla *et al.* are consistent with those of Shifrin throughout their common temperature range (i.e. the two temperature variations are consistent).

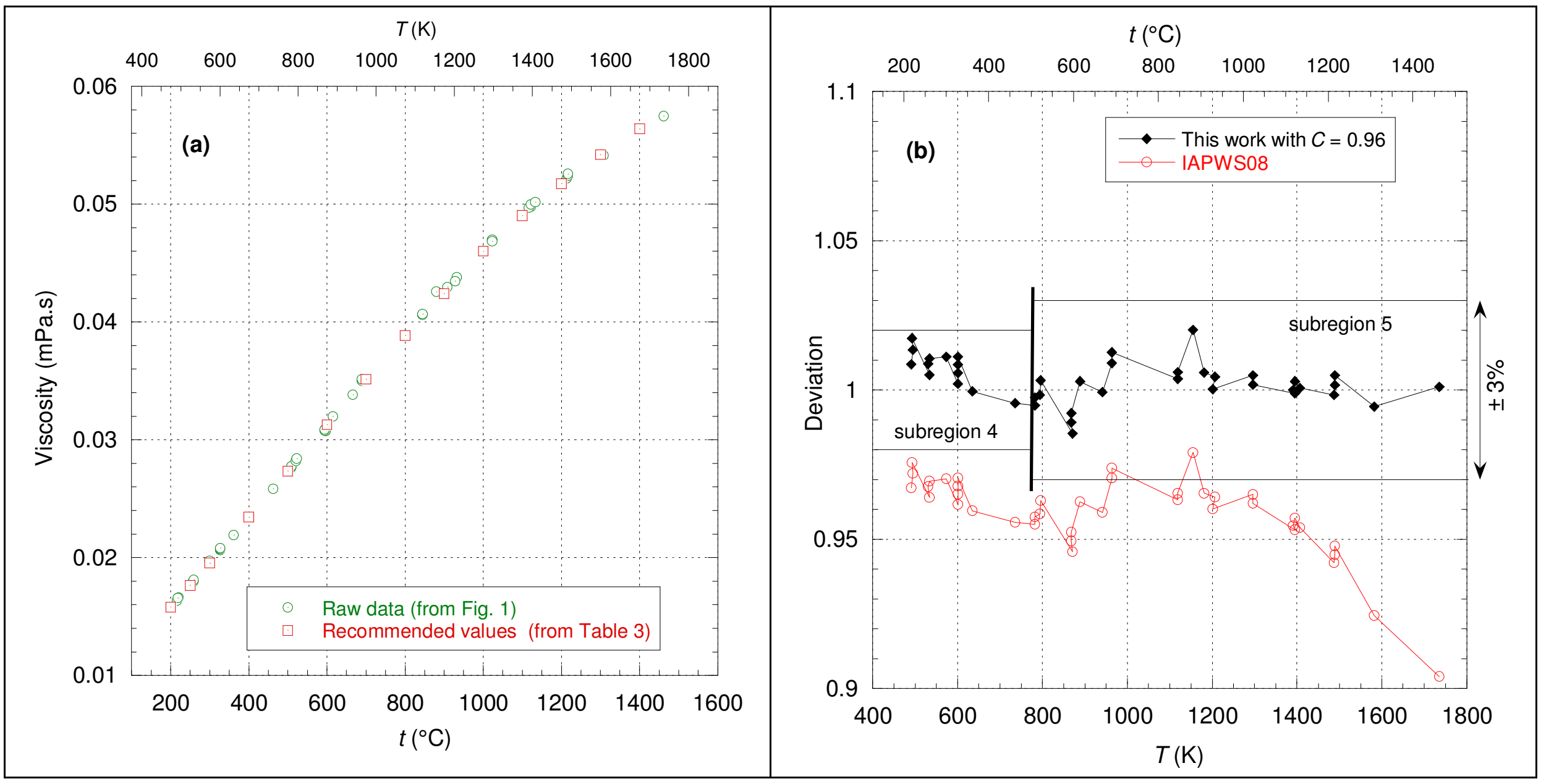

Fig. 64. (a) Bonilla *et al.* raw viscosity data and recommended values; (b) Ratio of Bonilla *et al.*'s data (Ref. 49) with the IAPWS08 formulation (empty circles) and the present modeling rescaled (black diamonds) as a function of temperature along the atmospheric isobar. The lines are eye guides.

Tanaka *et al.*'s data (Ref. 50), obtained with a falling-body viscometer, cover a wider temperature range than Shifrin's data, particularly below 400 K, but above all they allow to explore a wide range of pressure (~100 MPa). Tanaka *et al.*'s data are given in the form of two tables (i.e. Table 1 and Table 2): Table 1 contains data that are globally below the critical temperature while Table 2 contains data that are globally above the critical temperature. Both tables spread between subregion 3, subregion 4 and subregion 5, *i.e.* between 1%, 2% and 3%, according to the uncertainties indicated in Huber *et al.*'s Fig. 24 (Ref. 7).

Fig. 65(a) shows that for Tanaka *et al.*'s Table 1 dataset, the two models are almost equivalent and allow these data to be reproduced with the expected uncertainty in the gaseous phase (i.e. corresponding to subregion 4) with the exception of two points, one of which is clearly an outlier (probably due to a transcription error). The second point is at the 3% limit and belongs to subregion 5 which coincides with this uncertainty. For the data that belong to the liquid phase (i.e. to subregion 3), the uncertainty appears to be twice as large as the uncertainty expected by Huber *et al.* But considering that, for example, the dispersion of almost identical data around 300 K, it is clear that 2% reflects the experimental uncertainty of these data.

Table 2 contains two outliers omitted from this analysis. These are the points $\eta$(674.95 K, 59.271 MPa) = 0.0446 mPa.s and $\eta$(743.45 K, 98.184 MPa) = 0.0788 mPa.s. They are probably due to a transcription error. About these data Nagashima wrote in 1977 (Ref. 8):

> "The data by Mayinger [Ref. 51] and Tanaka *et al.* [Ref. 50] for $T$ > 550 °C [*i.e.* 823.15 K] were also considered reliable. […] The data [of these two authors] for $T_\sigma < T <$ 550 °C [*i.e.* $T_\sigma$ corresponding to the saturated vapor pressure temperature] show systematic deviation from those of Nagashima *et al.* and Rivkin *et al.* in the subregion where the kinematic viscosity is small. […] The existence of unfavorable flow condition can be blamed […]. For this reason, the data in these two references were included only for the liquid state and above 550 °C in the gaseous state."

More recently, Huber *et al.* (Ref. 7) wrote some similar sentences:

> "The data of Mayinger [Ref. 51] and Tanaka *et al.* [Ref. 50] both have systematic deviations from the other sets—as the pressure increases, in particular, above 30 MPa, the data of Tanaka *et al.* and of Mayinger are systematically higher than those of Nagashima and Tanishita [Ref. 52]. Nagashima [Ref. 8] noted that some of this could be caused by unfavorable flow conditions, when the kinematic viscosity is low and the flow could be turbulent, especially in the metal capillaries used, because the internal walls cannot be made satisfactorily smooth. We therefore removed the data of both Tanaka *et al.* and Mayinger from consideration for the uncertainty estimate."

Fig. 65(b) shows that the IAPWS08 formulation is in agreement with the above comment. Since Nagashima's comment, Rivkin *et al.*'s data have been renormalized in temperature (see section 3.2). Fig. 65(b) shows that by also rescaling all the data in Table 2 slightly again, the data that coincide with subregion 5 are compatible with the uncertainty expected by Huber *et al*. Now for data below 773 K, we see that the deviation is contained within 4.5% except one point that is at the limit of measurable pressures and this point is practically on the edge with subregion 2 which has an uncertainty of 7%! The distribution centered on the unit value indicates that the uncertainty simply reflects the experimental uncertainty of these data. The uncertainty may be due eventually to "unfavorable flow conditions" (as written by Nagashima) but there is no reason not to take these data into account with their own uncertainty.

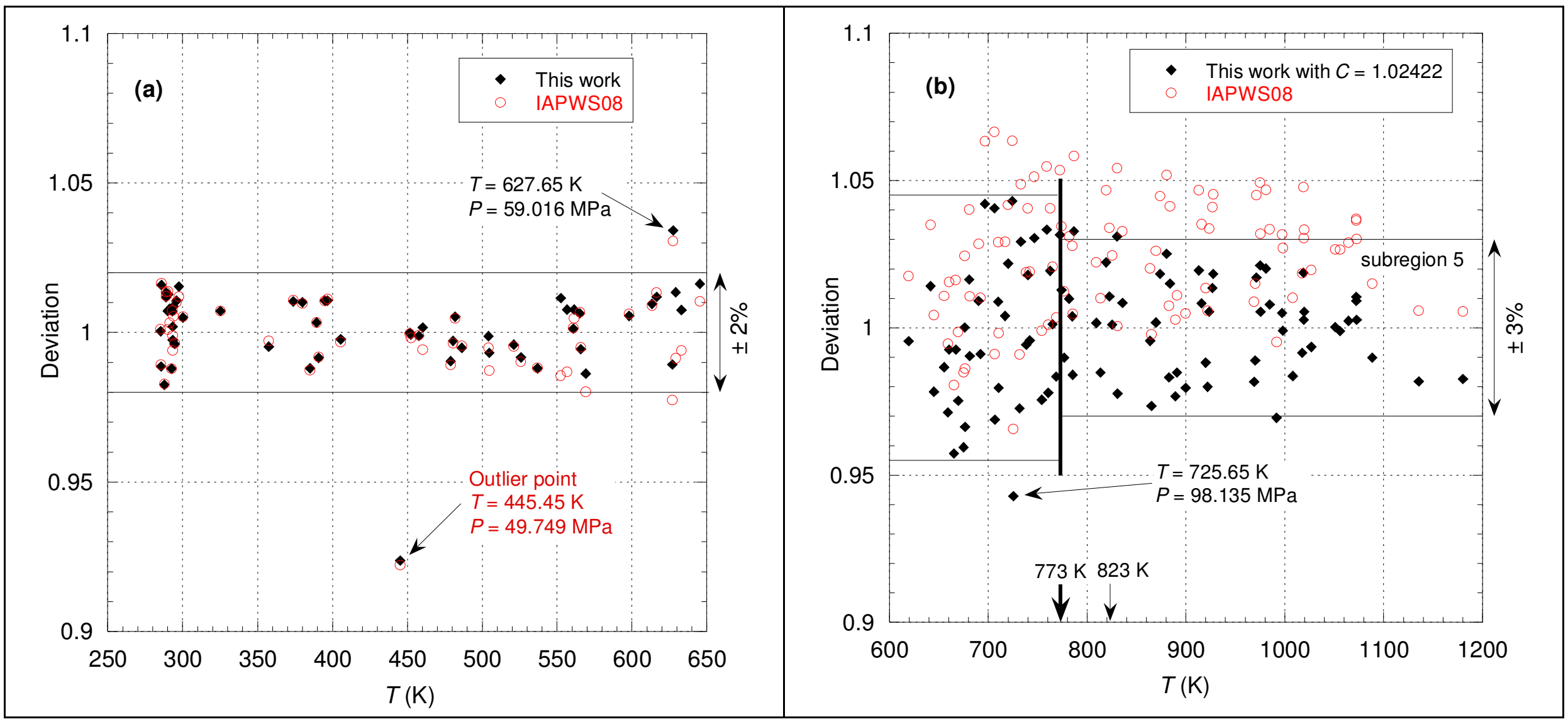

Fig. 65. Ratio of Tanaka *et al.*'s data (Ref. 50) with the IAPWS08 formulation (empty circles) and the present modeling rescaled (black diamonds) as a function of the temperature for all data points: (a) data of Table 1; (b) data of Table 2.

The comments of Nagashima and Huber *et al.* pushes us to analyze the data from Mayinger (Ref. 51) obtained with a capillary viscometer. Mayinger used two capillary diameters: ~319 µm and ~506 µm. The data with the greatest deviation are those obtained with the smallest diameter. These data are part of Table 2 of Mayinger's paper and spread between subregions 4 and 5 according to the division of Huber *et al.*'s Fig. 24 (Ref. 7) but for relatively high densities comparable to those of Fig. 65(b) from Tanaka *et al.* (Ref. 50).

First, it can be noted that Tanaka *et al.*'s Table 1 dataset corresponds to liquid water and can be reproduced with an overall deviation of ±2% which corresponds approximately to the suggested experimental uncertainty of about ±1.5%.

Tanaka *et al.*'s Table 2 dataset is distributed according to the two capillary diameters and must be therefore analyzed in two different ways. Fig. 66(a) shows that it is possible with the present modeling to reproduce the data corresponding to the smaller diameter capillary (i.e. Table 2.a) with a deviation of 10% but we observe that it is only a few particular points on the quasi-isotherm at ~665.32 K that have a very high uncertainty otherwise the data seem compatible with an uncertainty of ±5%. This is also the deviation that is obtained with Table 2.b dataset (see Fig. 66(b)). This uncertainty value is compatible, for example, with the uncertainty of the Couette constant involved in determining the pressure difference $\Delta p$ in the capillary tube.

The important deviation of the quasi-isotherm at ~665.32 K is probably due to the fact that Mayinger deduced absolute values (and not relative values) using a relation in which the value of the capillary diameter used was determined by using viscosity data for nitrogen from different experiments. Mayinger's idea was to increase the accuracy on the determination of the absolute values but since it corresponds to the fourth power of the diameter, a small error on the diameter determination leads to a large error on the absolute value. It is therefore possible to slightly renormalize each quasi-isotherm but the deviation remains essentially the same for the data of Table 2.a while the Table 2.b dataset can be reproduced with a deviation of slightly less than ±4%. Since the values of the renormalization factors allow quasi-isotherms to be centered on the unit value, it cannot be concluded that the Mayinger's data have abnormal systematic deviations with respect to other sets. Thus, apart from the quasi-isotherm at ~665.32 K, it appears possible to reproduce the variations with density of each quasi-isotherm with an uncertainty that is certainly higher than in other experiments but which is not abnormal considering the experimental conditions in which these data have been obtained.

It appears that the IAPWS08 formulation has a similar uncertainty structure to the present modeling but with a deviation of around 15%. However, it is not justified to exclude these data from the global analysis.

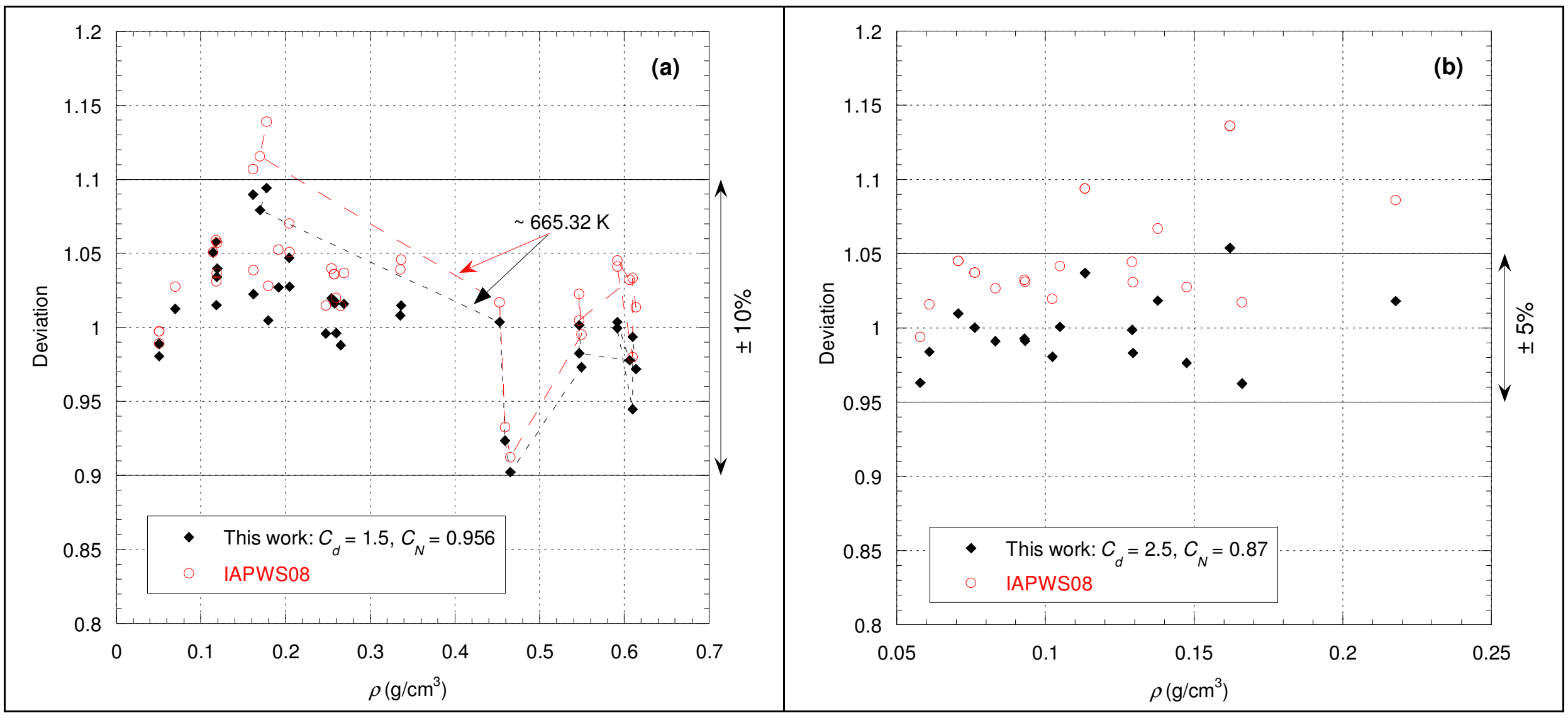

Fig. 66. Ratio of Mayinger's data (Ref. 51) with the IAPWS08 formulation (empty circles) and the present modeling (black diamonds) as a function of the density for all data points: (a) Table 2.a dataset (~319 μm

capillary diameter). The dashed lines link the data points on the quasi-isotherm ~665.32 K; (b) Table 2.b dataset (~506 μm capillary diameter).

Nagashima *et al.* (Ref. 53) have measured the viscosity of steam at subcritical pressures by using a capillary viscometer (i.e. capillary internal radius of 135 μm) along 10 quasi-isotherms from ~523.93 K to ~874.23 K. Consequently, these isotherms spread between subregions 4 and 5 in Huber *et al.*'s Fig. 24 (Ref. 7). Nagashima *et al.* wrote that:

"The estimated error of viscosity is thus about 1.5%".

Huber *et al.* considered that these data belong to a dataset whose standard deviation is 1.5% and "thus for a 95% confidence level, we obtain a 3% estimated uncertainty". Fig. 67 shows that the two models are practically equivalent and the uncertainty obtained is consistent with an uncertainty of ±3%. On the other hand, this uncertainty is not in agreement with the one expected for the points belonging to subregion 4.

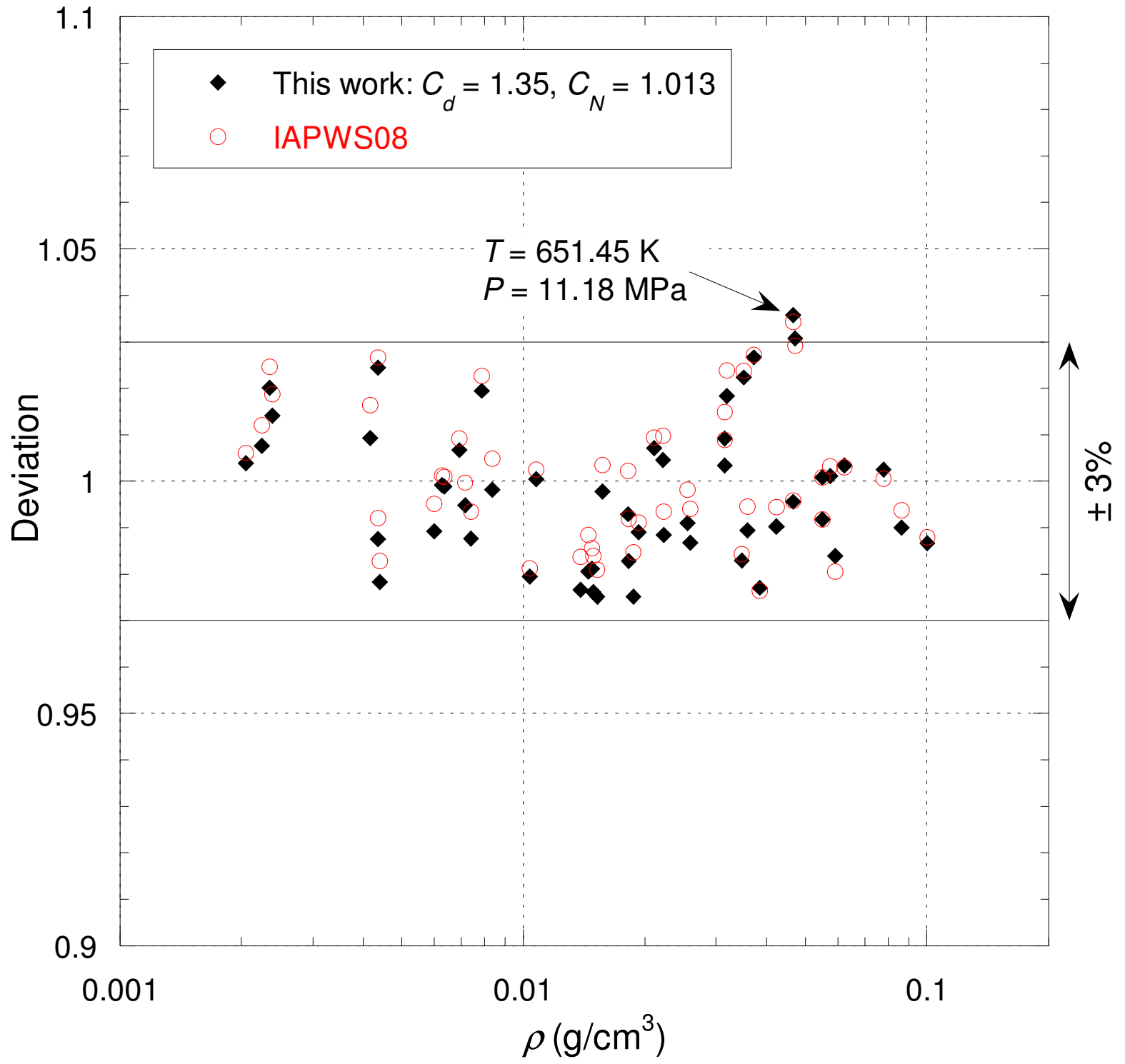

Fig. 67. Ratio of Nagashima's data (Ref. 53) with the IAPWS08 formulation (empty circles) and the present modeling (black diamonds) as a function of the density for all isotherms data points. The lines are eye guides.

Here we will use the Nagashima's low-density data along quasi-isotherms to illustrate the behavior of Eq. (19). In Fig. 68 we have chosen to represent the isotherm at 650 K for which there are 8 Nagashima's points that are close to this temperature, all in the gaseous phase. Along this isotherm with $C_d$ and $C_N$ Nagashima's parameters, the present modeling leads to the following values: $\rho_{l0}$ = 1.9×10$^{-5}$ g/cm$^3$ and $\rho_{g0}$ = 4.8×10$^{-7}$ g/cm$^3$. These two densities are significantly lower than Nagashima's lowest density along this isotherm, which is why Eq. (18) and Eq. (19) are identical to represent Nagashima's data. It can be seen in Fig. 68 that the liquid-like and gas-like terms tend quickly to zero for densities lower than $\rho_{l0}$ and $\rho_{g0}$, respectively. These densities are difficult to achieve experimentally in water but Fig. 68 can be understood as a prediction of the present modeling.

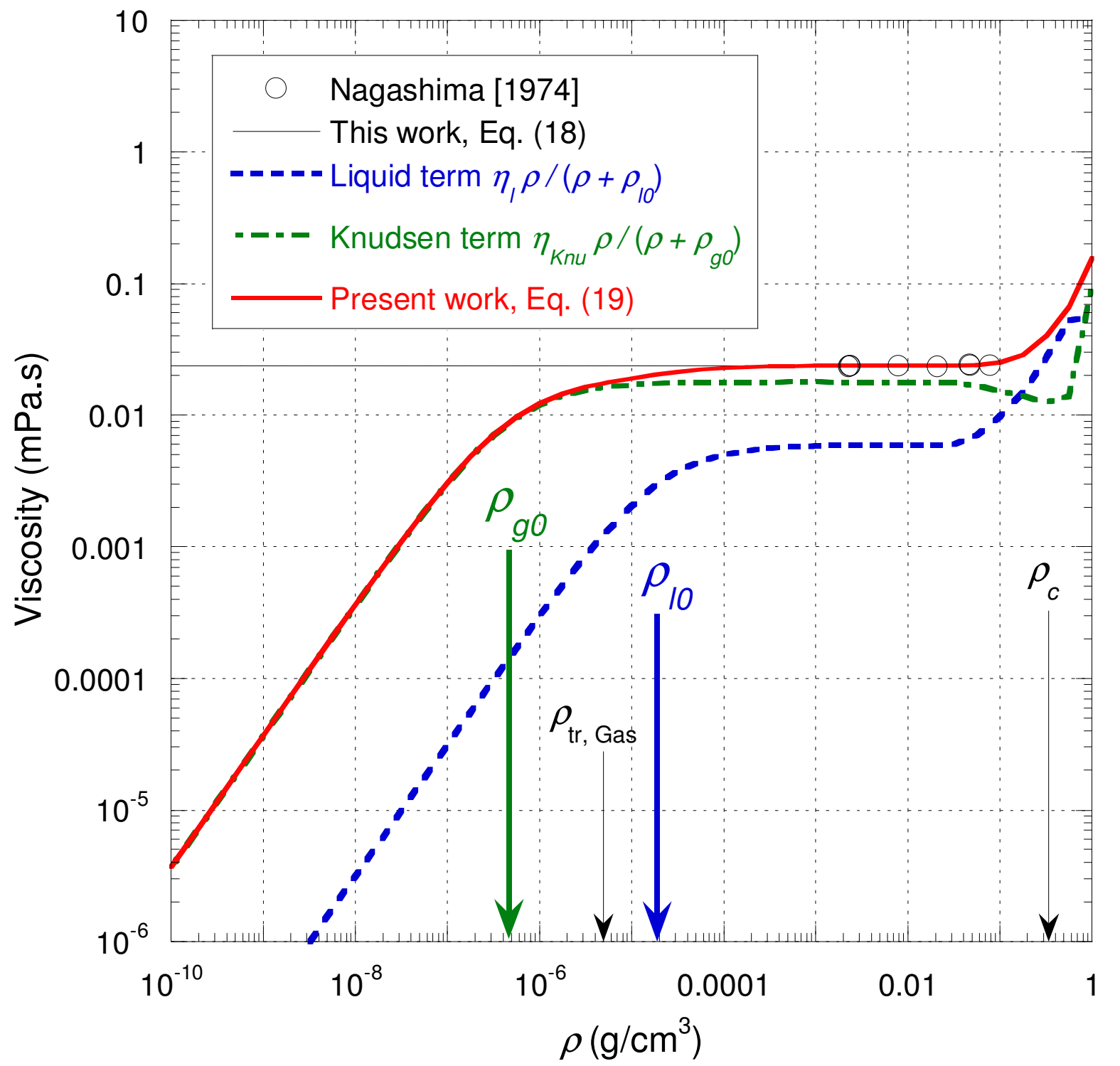

Fig. 68. Logaritmic plot of Eq. (18) and Eq. (19) along the isotherm at 650 K with Nagashima's parameters ($C_d$ = 1.35 and $C_N$ = 1.013). The black circles represent the 8 points of Nagashima's dataset which are the closest to the isotherm at 650 K. $\rho_{l0}$ represents the density for which $\left\langle u^2 \right\rangle = d^2$ and $\rho_{g0}$ represents the density for which $\rho = \rho_{Knu}$ along this particular isotherm.

In section 4.1.2 we have analyzed Moszynski's data (Ref. 41) in the liquid phase. In the same paper, the author also made measurements in steam using an oscillating-disk viscometer. The dataset consists of 6 quasi-isotherms from ~410.88 K to ~539.93 K. Therefore, all these isotherms belong to subregion 4 of Huber *et al.* (Ref. 7) corresponding to an uncertainty of ±2%. Fig. 69 shows that the present modeling makes it possible to represent these data with the expected uncertainty for subregion 4 and even better since all the data are included in ±1.5%. On the other hand, it can be seen that the IAPWS08 formulation does not satisfy its own uncertainty.
Now Moszynski wrote that:

> "If the possibility of relative movement of the disk and plates, and other errors, are accounted for one arrives at an estimated absolute accuracy of 0.7-1 per cent".

Fig. 70(a) shows that it is possible to achieve this precision if we accept that each isotherm must be slightly renormalized with a constant factor (i.e. renormalization of the calibration constant). In this figure we observe that the variation *versus* density of each quasi-isotherm is then correctly reproduced. Fig. 70(b) shows that the renormalization factor varies almost linearly, which corresponds to a systematic drift of the device constant with temperature.

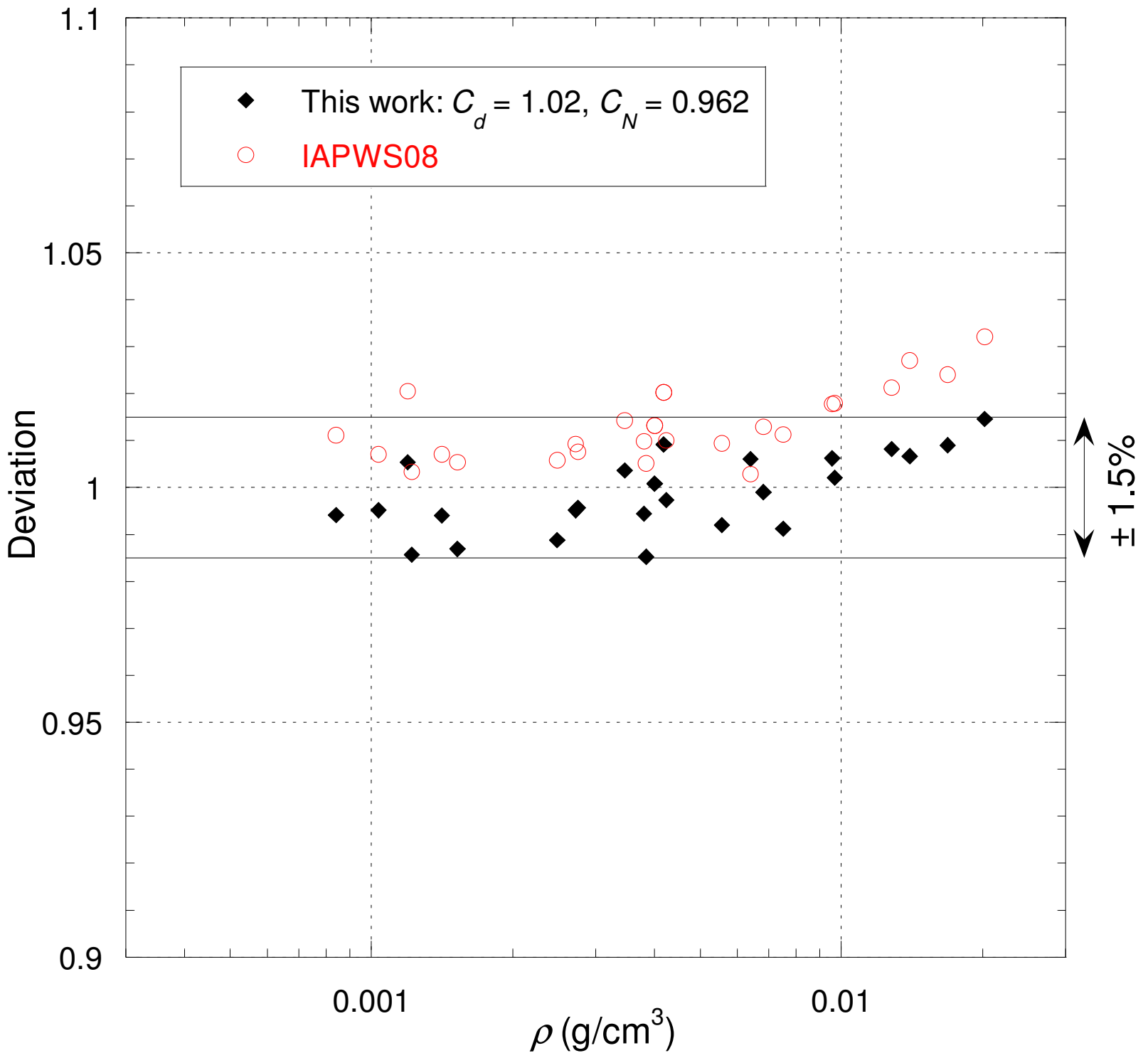

Fig. 69. Ratio of Moszynski's data (Ref. 41) with the IAPWS08 formulation (empty circles) and the present modeling (black diamonds) as a function of the density for all quasi-isotherms data points.

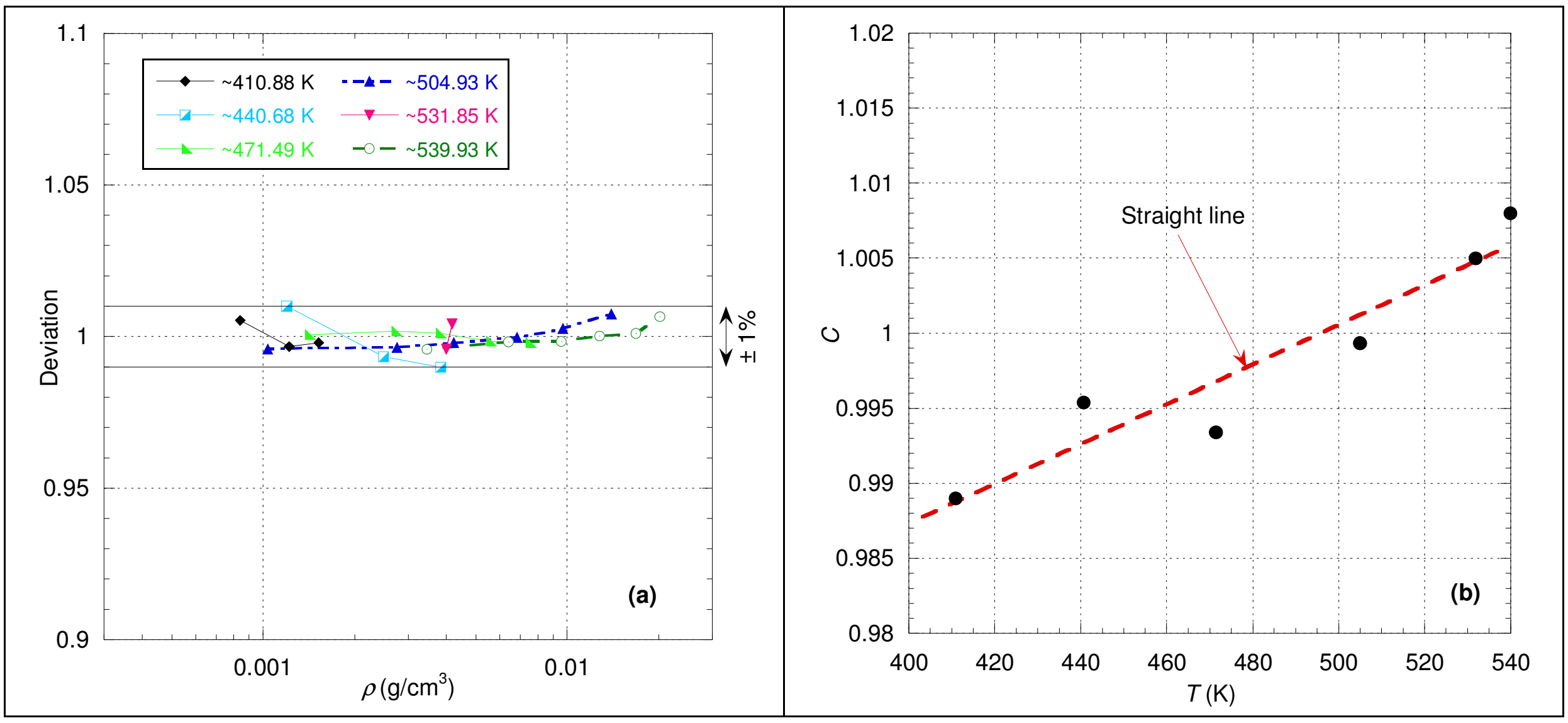

Fig. 70. (a) Ratio of Moszynski's data (Ref. 41) with the present modeling ($C_d$ = 1.02, $C_N$ = 0.962) rescaled as a function of the density for each quasi-isotherm. The lines are eye guides. (b) Renormalization factors as function of temperature.

To conclude this section, we will analyze two datasets from Kestin *et al.* (Ref. 54 and 55) which are considered very accurate and have been used as a reference for a number of studies. These two datasets correspond to measurements in steam using an oscillating-disk viscometer with two different characteristics of the suspension system. So, these measurements are similar to the one made by Moszynski (Ref. 41). Therefore, these two datasets belong to subregion 4 of Huber *et al.* (Ref. 7) corresponding to an uncertainty of ±2%.

We start by analyzing the data from Ref. 54 which corresponds to 4 quasi-isotherms from ~410.94 K to ~507.46 K. These isotherms are similar to those of Moszynski but cover lower pressures (i.e. slightly lower densities). Fig. 71 shows that the two models are practically equivalent and here they allow these data to be reproduced with the expected uncertainty for subregion 4.

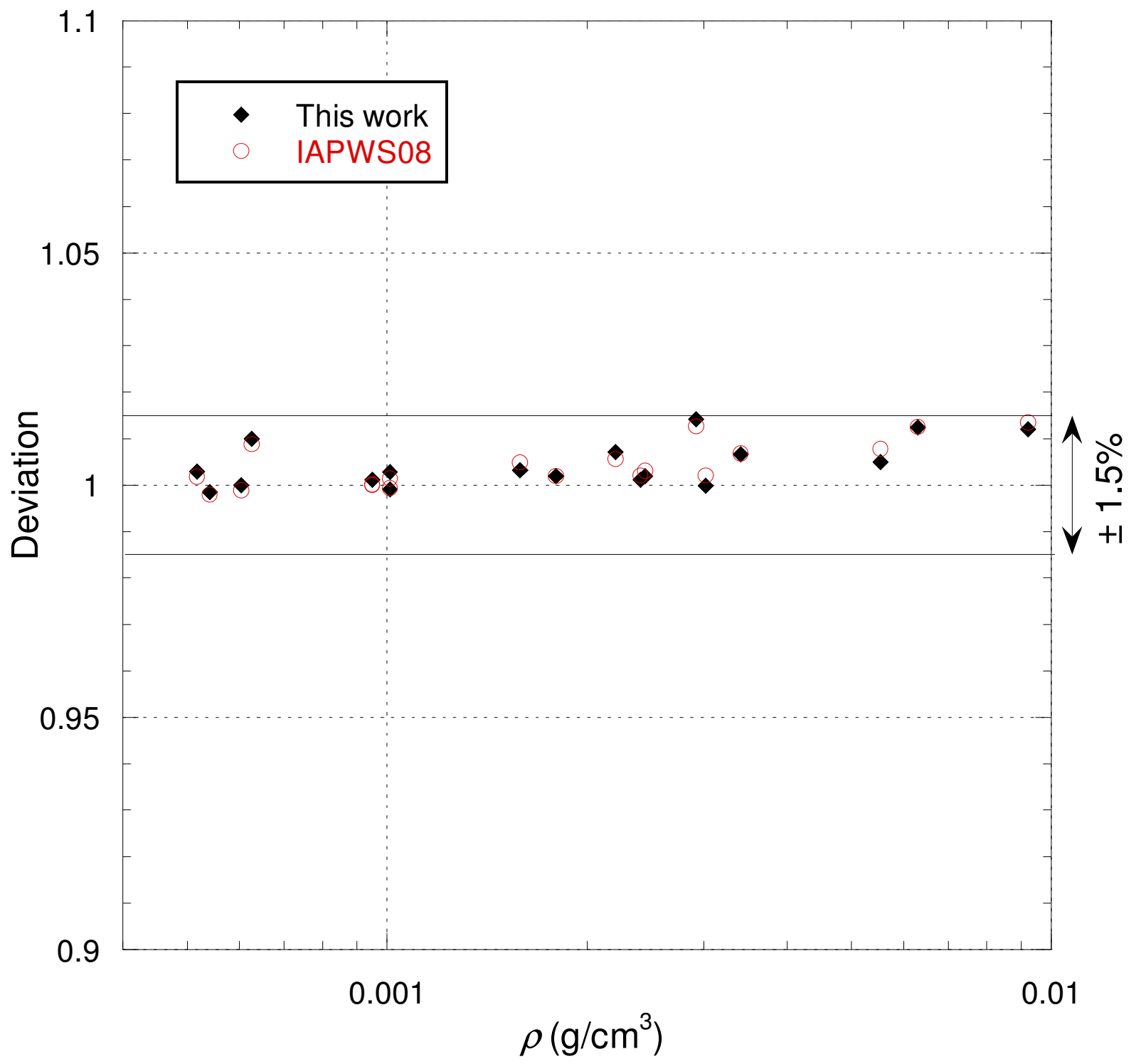

Fig. 71. Ratio of Kestin *et al.*'s data (Ref. 54) with the IAPWS08 formulation (empty circles) and the present modeling (black diamonds) as a function of the density for all quasi-isotherms data points.

As we have already shown in several previous examples, by renormalizing each quasi-isotherm with a constant $C$, we observe on Fig. 72(a) that the deviation of each quasi-isotherm is much lower and more in agreement with the uncertainty expressed by the authors, but above all we observe that the variation according to density is then well reproduced by the present modeling. Fig. 72(b) shows that the renormalization factor is constant until about 440 K and then increases almost linearly, as in the similar experiments of Moszynski (Ref. 41), reflecting a small drift of the device constant with temperature.

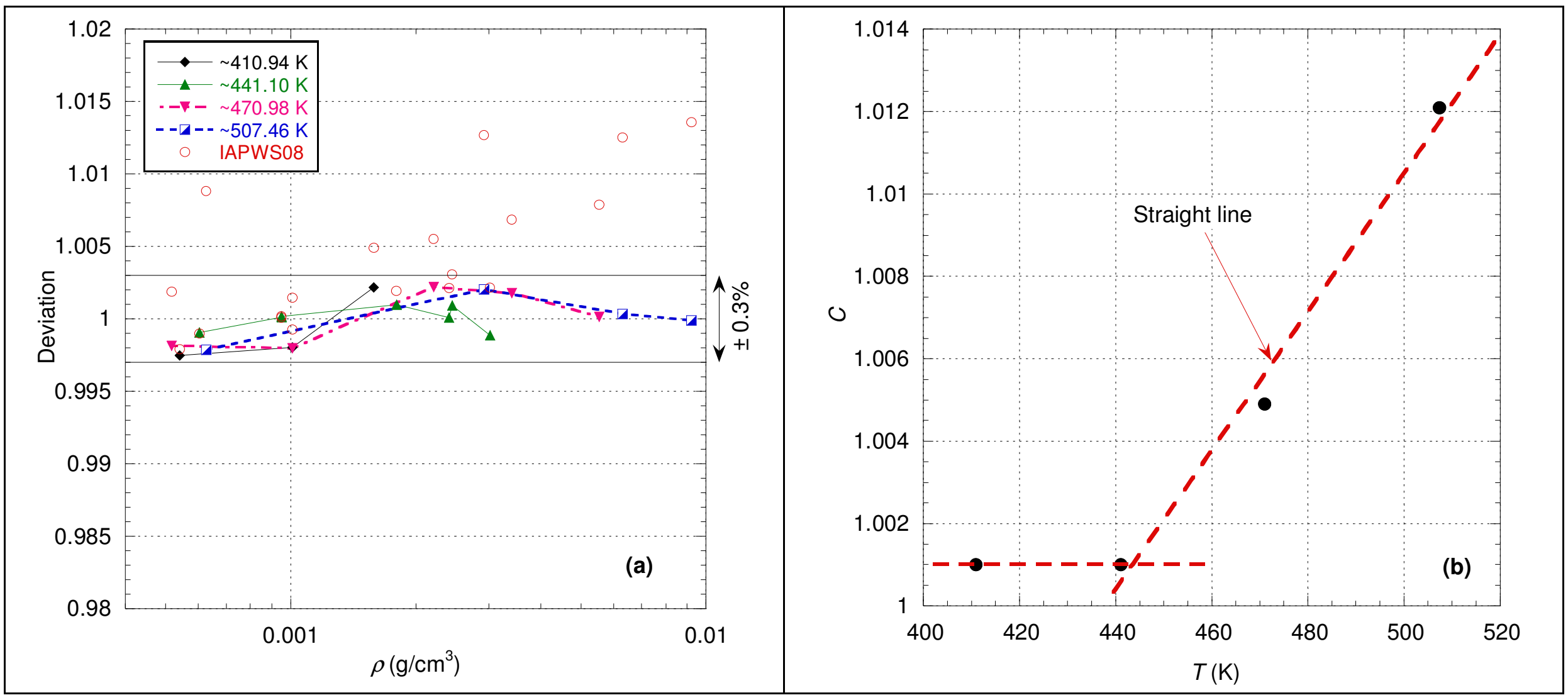

Fig. 72. (a) Ratio of Kestin *et al.*'s data (Ref. 54) with the IAPWS08 formulation (empty circles) and with the present modeling rescaled as a function of the density for each quasi-isotherm. The lines are eye guides. (b) Renormalization factors as function of temperature.

The second dataset from Ref. 55 is considered by the authors as:

> "[…] the most successful series of measurements of the viscosity of superheated steam up to 275 deg C yet performed at Brown University".

This dataset contains measurements along 8 quasi-isotherms from ~421.9 K to ~546.6 K. This set covers a wider pressure range than the previous one (i.e. it reaches slightly higher densities). Fig. 73 shows that the two models are practically equivalent and that these data can be reproduced with the expected uncertainty corresponding to subregion 4, although this is somewhat borderline for the IAPWS08 formulation.

Now Huber *et al.* (Ref. 7) in their Table 9 have considered that the uncertainty is within 0.8% for this dataset. Fig. 74(a) shows once again that by renormalizing each quasi-isotherm with a constant $C$, we can reproduce all the variations versus density of each quasi-isotherm with the uncertainty proposed by Huber *et al.* and even better since all quasi-isotherms, except the one corresponding to the highest temperature, can be reproduced with an uncertainty of ±0.3%. Fig. 74(b) shows that the renormalization factor again varies in a quasi-linear way as a function of temperature (as with other experiments with oscillating-disks), but here we observe that for two very close temperatures (see the rounded rectangles in blue) there is a rather significant drift of the device constant with temperature.

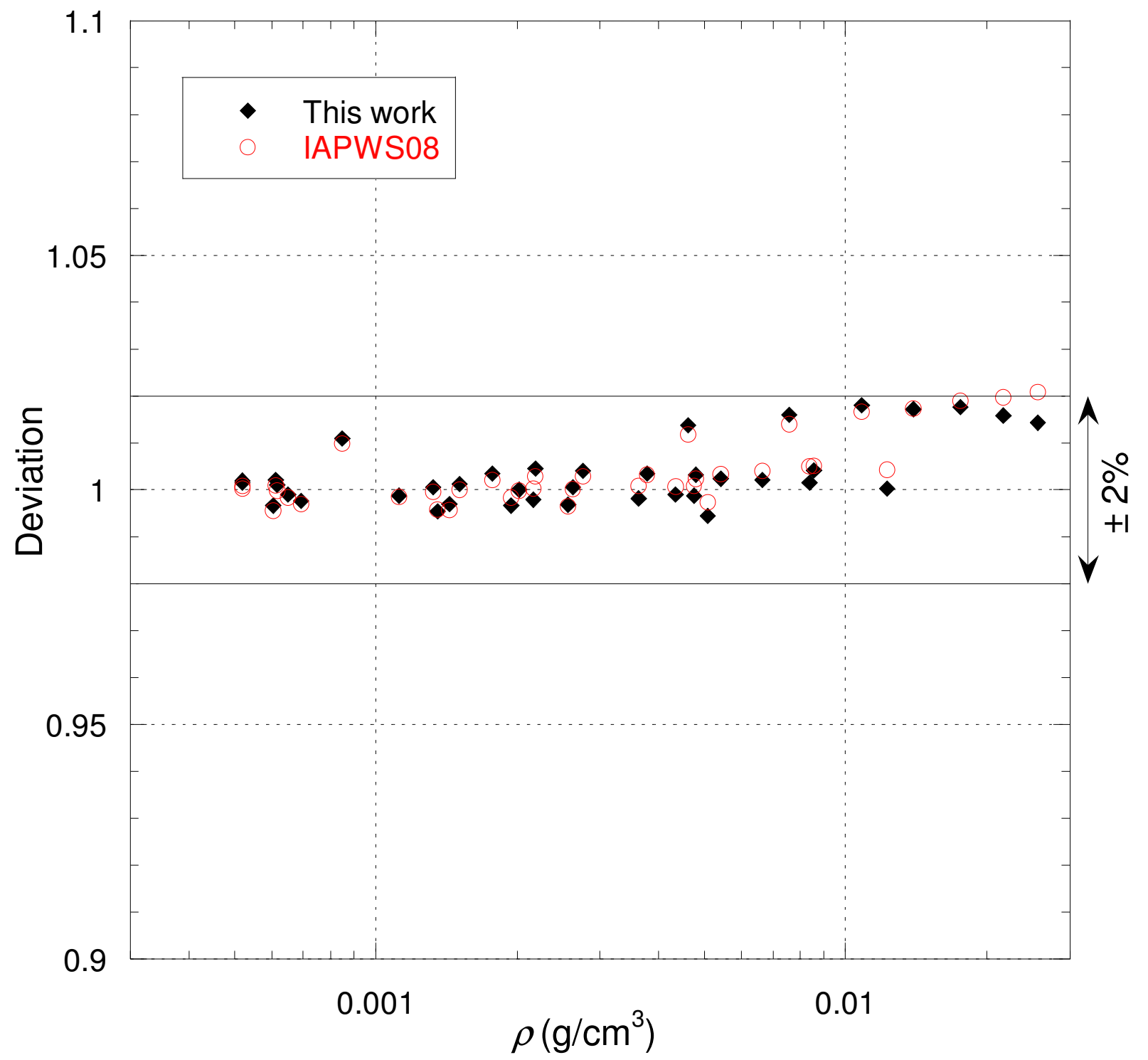

Fig. 73. Ratio of Kestin *et al.*'s data (Ref. 55) with the IAPWS08 formulation (empty circles) and the present modeling (black diamonds) as a function of the density for all quasi-isotherms data points.

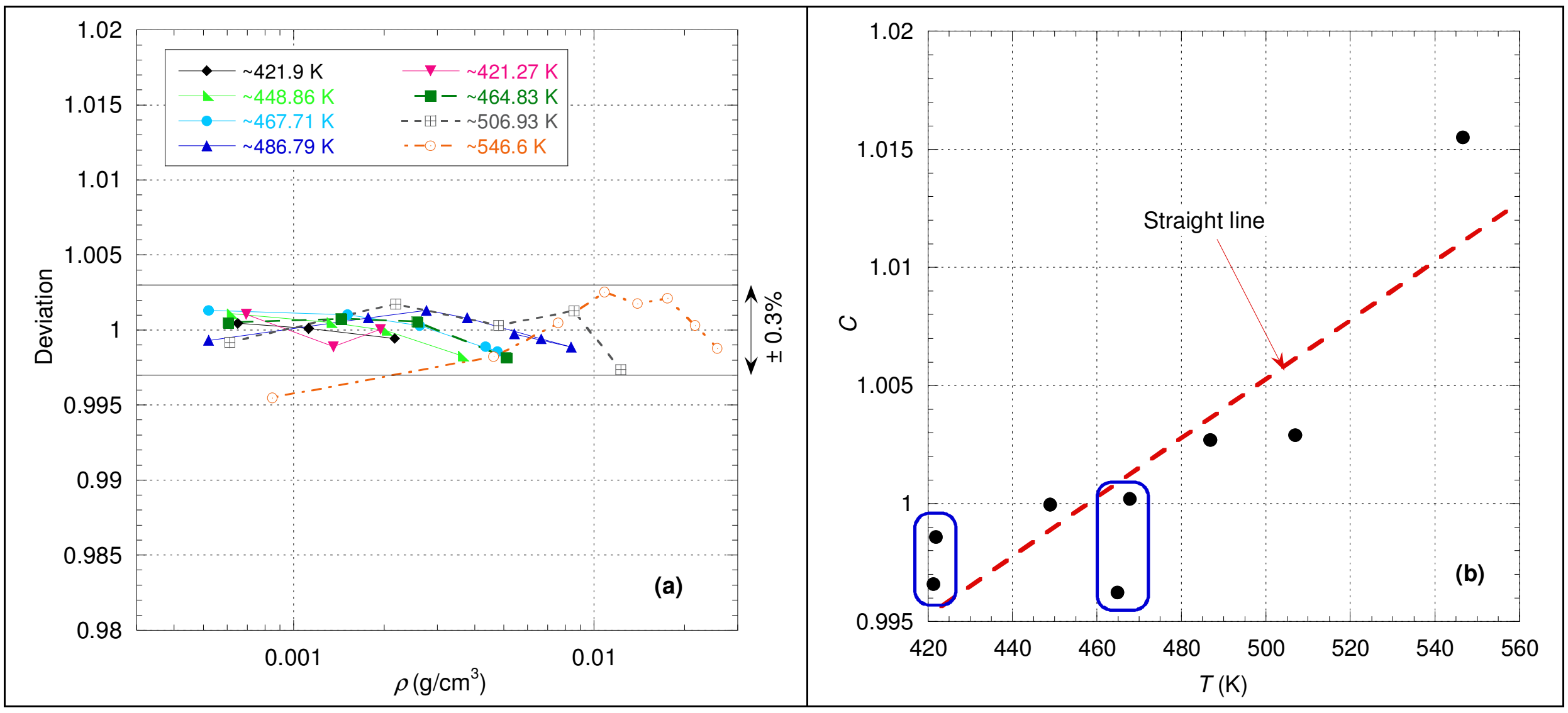

Fig. 74. (a) Ratio of Kestin *et al.*'s data (Ref. 55) with the present modeling rescaled as a function of the density for each quasi-isotherm. The lines are eye guides. (b) Renormalization factors as function of the temperature.

Now Moszynski and Kestin *et al.* experiments have been carried out in particular to study the water steam anomaly which consists in observing a decrease in viscosity as a function of pressure along isotherms when approaching SVP. Moszynski (Ref. 41) wrote:

> "No definite explanation of the observed anomalous behavior of steam in the low-pressure range can be offered at the present time. One may speculate, however, that an agglomeration of molecules near the saturation curve may be responsible for the decrease of viscosity with pressure".

Taking as an example the quasi-isotherm at 504.93 K from Moszynski, Fig. 75 shows that in the context of the present modeling, the viscosity decreases when the pressure in the steam increases comes from a "significant" viscosity decrease of the released gas (i.e. there is less gas released) while the liquid term (i.e. corresponding to Eq. (13)) remains substantially constant. So, this is in line with Moszynski's comment.

This anomaly being related to the gas released by the flow, it is interesting to note that it should therefore not occur on the self-diffusion coefficient.

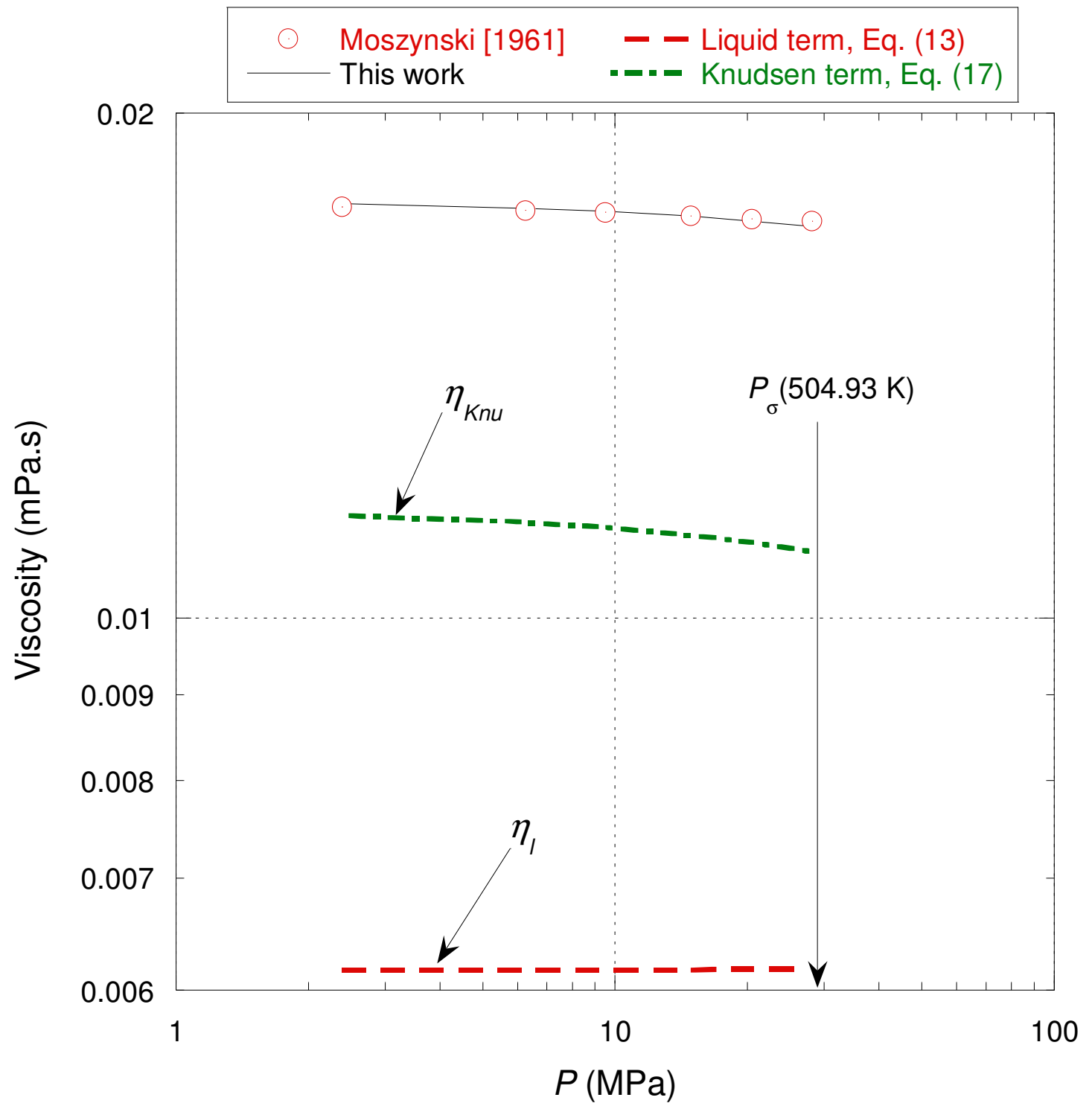

Fig. 75. Logarithmic plot of the two terms constituting Eq. (18) for the quasi-isotherm at ~504.93 K as function of the pressure in steam. $P_\sigma$ represents the saturated vapor pressure value corresponding to the quasi-isotherm.

## 4.2. Self-Diffusion Coefficient

Self-diffusion processes are linked to molecular movements in the same way as viscous processes, so there has been a longstanding attempt to link these two quantities together. Indeed, many theoretical models, either based on molecular theories or empirical relationships, have been developed to the present day; the hydrodynamic equation of Stokes–Einstein (SE) being the most popular. For example, a detailed review can be found in Ref. 56. Despite all these approaches, Suárez-Iglesias *et al.* (Ref. 56) wrote that:

> "Regarding self-diffusion coefficients, there is no predictive way of obtaining these, except for dilute gases."

Contrary to viscous phenomena, self-diffusion processes are physical processes that are usually measured in thermodynamic equilibrium systems. As a result, this property is generally more difficult to measure accurately than the viscosity observed in out-of-equilibrium systems. Thus, there is generally less data on self-diffusion coefficients in the literature than there are on viscosity. To overcome certain difficulties, tracers (e.g. isotopes) are used and it is accepted that their diffusion coefficient variations with some parameters (e.g. density, temperature) are representative of the diffusion coefficient variations of atoms/molecules specific to the medium studied. The other issue of using tracers was described by Wang (Ref. 57) as follows:

> "Liquid water has many properties that are characteristic of associated liquids. The anomalous temperature dependence of density and heat capacity of water have stimulated numerous suggestions that liquid water at room temperatures is a mixture of various species of associated molecules, e.g., hydrol $H_2O$, dihydrol $(H_2O)_2$, trihydrol $(H_2O)_3$, etc. Recently, Eucken explained these anomalous properties of liquid water quantitatively by assuming that water is an ideal solution of molecules containing 1, 2, 4 and 8 $H_2O$ each.
> On the other hand, Raman spectra and X-ray data showed that no definite associated molecules exist, that the structure of liquid water is best represented by a three-dimensional network of pliable branched chains of tetrahedra, and liquid water actually has a quasi-crystalline structure as first suggested by Bernal and Fowler. This quasi-crystalline structure of liquid water seems to be the more favored picture recently. It is of interest to see whether we can decide from examination of self-diffusion data in the present modeling which of the above outlined two alternative structures of liquid water is the correct one."

The difference between an equilibrium system and an out-of-equilibrium system is expressed in the present modeling by the fact that the self-diffusion coefficient described by Eq. (12) is linked only to the liquid term of the viscosity equation in the form of a SE-type law such that: $\dfrac{k_B T}{\eta_l D_t} = \dfrac{1}{3}\left(\dfrac{2\pi}{q_c}\right)$. But $\dfrac{k_B T}{\eta D_t}$ differs globally from a SE-type law. As a result, some "anomalies" are specific to viscosity and are not observed on the self-diffusion coefficient. However, as with the viscosity data, the data of the different authors, and their variations, do not fit with each other, although the uncertainty is generally much greater than for the viscosity data. This is essentially what Kisel'nick *et al.* (Ref. 58) wrote in 1974:

> "There is a considerable amount of data available at present dealing with the self-diffusion of water molecules. Most of this, however, was obtained at room temperature, and the spread of values is so great that the problem of establishing a true coefficient of self-diffusion is still unsolved."

Therefore, the same analysis as for the viscosity will be done in the following sections for the self-diffusion coefficient of water.

### *4.2.1. Modeling of the Cut-off Wave-Vector Modulus $q_c$*

For the description of viscosity, the parameter $q_{c0}$ has been used as a characteristic length scale but the parameter $q_c$ does not directly contribute in the description of the viscosity. On the other hand, this parameter plays a fundamental role in the description of the self-diffusion coefficient. Indeed, the cut-off wave-vector modulus $q_c$ is a fundamental intrinsic property of the fluid as the parameter $K^*$ and we have $q_c \approx q_{c0}(\rho)$ for the normal liquid states. According to Eq. (20), the function $f_{q_c}(\rho, T)$ describes the deviation of $q_c$ from $q_{c0}$ and its expression is now reported in Table 4 (this function is also included in the calculation programs of Ref. 15).

| *Name* | *formula* |
|---|---|
| $f_{q_c}(\rho,T)$ | $f_{q_c}(\rho,T)=f_{q_c,\mathrm{SVP}}(T)\times f_{q_c,1\mathrm{atm}}(T)\times f_{q_c,\mathrm{Supercrit}}(T)\times\begin{cases} f_{q_c,Gas}(\rho,T) & \text{if } \rho<\rho_{q_c} \\ f_{q_c,\mathrm{Liq}}(\rho,T) & \text{if } \rho\ge\rho_{q_c}\end{cases}$ <br> with <br> $f_{q_c,\mathrm{SVP}}(T)=1+0.2776\left(1-\exp\left(-\left\|\frac{T-362.15}{218}\right\|^{1.2}\right)\right)\exp\left(-\left(\frac{89}{T-T_{\mathrm{tr}}}\right)^{100}\right)$, <br> $f_{q_c,1\mathrm{atm}}(T)=1-\frac{\exp\left(-(T/T_{\mathrm{tr}})^{10}\right)}{1+(T/234.25)^{55}}$, <br> $f_{q_c,\mathrm{Supercrit}}(T)=1+\frac{0.32116\exp\left(-\left(\frac{T-613.05}{200.4}\right)^{6}\right)+0.076108\exp\left(-\left(\frac{T}{T_c}\right)^{2.8}\right)}{1+(T_c/T)^{150}}$, <br> $f_{q_c,\mathrm{Liq}}(\rho,T)=f_{1q_c}(T)+(1-f_{1q_c}(T))\exp\left(-f_{2q_c}(T)\left\|1-\frac{\rho}{\rho_{q_c}(T)}\right\|^{2.5}\right)$, <br> $f_{1q_c}(T)=1+\frac{\dfrac{0.2}{1+\left(\frac{334}{T}\right)^{2}\left(\frac{T-334}{165.7}\right)^{2}}-\dfrac{0.57161}{1+\left(\frac{525}{T}\right)^{1.1124}\left(\frac{T-525}{225}\right)^{1.444}}}{1+\left(\frac{T_{\mathrm{tr}}}{T}\right)^{1000}}$, <br> $f_{2q_c}(T)=0.87197\exp\left(\left(\frac{659.04}{T}\right)^{0.13747}\left\|\frac{T-659.04}{112.02}\right\|^{1.3522}\right)+\frac{2.29}{1+\left(\frac{T-488}{23}\right)^{4}}$, <br> $f_{q_c,\mathrm{Gas}}(\rho,T)=1+\alpha_{q_c,\mathrm{Gas}}(T)\exp\left(-\frac{\rho}{\rho_{q_c}(T)}\right)\left(1-\frac{\rho}{\rho_{q_c}(T)}\right)\left(\frac{\rho_c}{\rho}\right)^{\frac{4}{3}}$, <br> $\alpha_{q_c,\mathrm{Gas}}=0.312\exp\left(-\left(\frac{T}{2700}\right)^{3}\right)-0.07629\exp\left(-\left(\frac{T-729.57}{80.652}\right)^{2}\right)$ $-0.044268\exp\left(-\left(\frac{T-589.14}{150.57}\right)^{2}\right)+0.0153\exp\left(-\left(\frac{T-736}{30.54}\right)^{2}\right)$ $-0.01606\exp\left(-\left(\frac{T-817.92}{32.8}\right)^{2}\right)$ <br> where <br> $\rho_{q_c}(T)=\begin{cases}\rho_{1\mathrm{atm}}(T) & \text{if } T<T_{\mathrm{tr}} \\ \rho_{\mathrm{SVP}}(T) & \text{if } T_{\mathrm{tr}}\le T<T_{\mathrm{c}} \\ \rho_{\mathrm{Supercrit}}(T) & \text{if } T\ge T_{\mathrm{c}}\end{cases}$ <br> See Appendix E for the expression of $\rho_{\mathrm{Supercrit}}(T)$ and Appendix D for |
| | the expression of $\rho_{1\text{atm}}(T)$. |

Table 4. The cut-off wave-vector modulus function: $T$ in K and $\rho$ in g/cm$^3$. $T_{\text{tr}}$ = 273.16 K represents the triple point temperature, $\rho_{\text{SVP}}$ represents the liquid density along the saturated vapor pressure line and $\rho_{\text{Supercrit}}$ represents the density along the Frenkel/Widom line. We notice that both functions $f_{q_c,\text{Liq}}(\rho,T)$ and $f_{q_c,\text{Gas}}(\rho,T)$ are equal to 1 for $\rho = \rho_{q_c}$ whatever the temperature. These formulas must not be extrapolated above 2000 K.

Similarly to other parameters of Table 1 to 3, we preferred to define this function by part in order to have fairly simple mathematical forms. The connection line chosen for these functions is the liquid coexistence curve (i.e. SVP curve) below the critical isotherm and above the Frenkel/Widom line (determined from the isochoric heat capacity peaks of the IAPWS-95 formulation, Ref. 17) which is in continuity with the SVP curve as can be seen on Fig. 76. This connection line (i.e. the red curve) separates dense states with liquid-like behavior from low-density states with gas-like behavior. In Fig. 76 we have also positioned some characteristic isotherms corresponding to experimental data to show where the data are located at the highest densities and temperatures. Therefore, beyond these highest densities and temperatures the $f_{q_c}$ function must be considered as possible extrapolation.

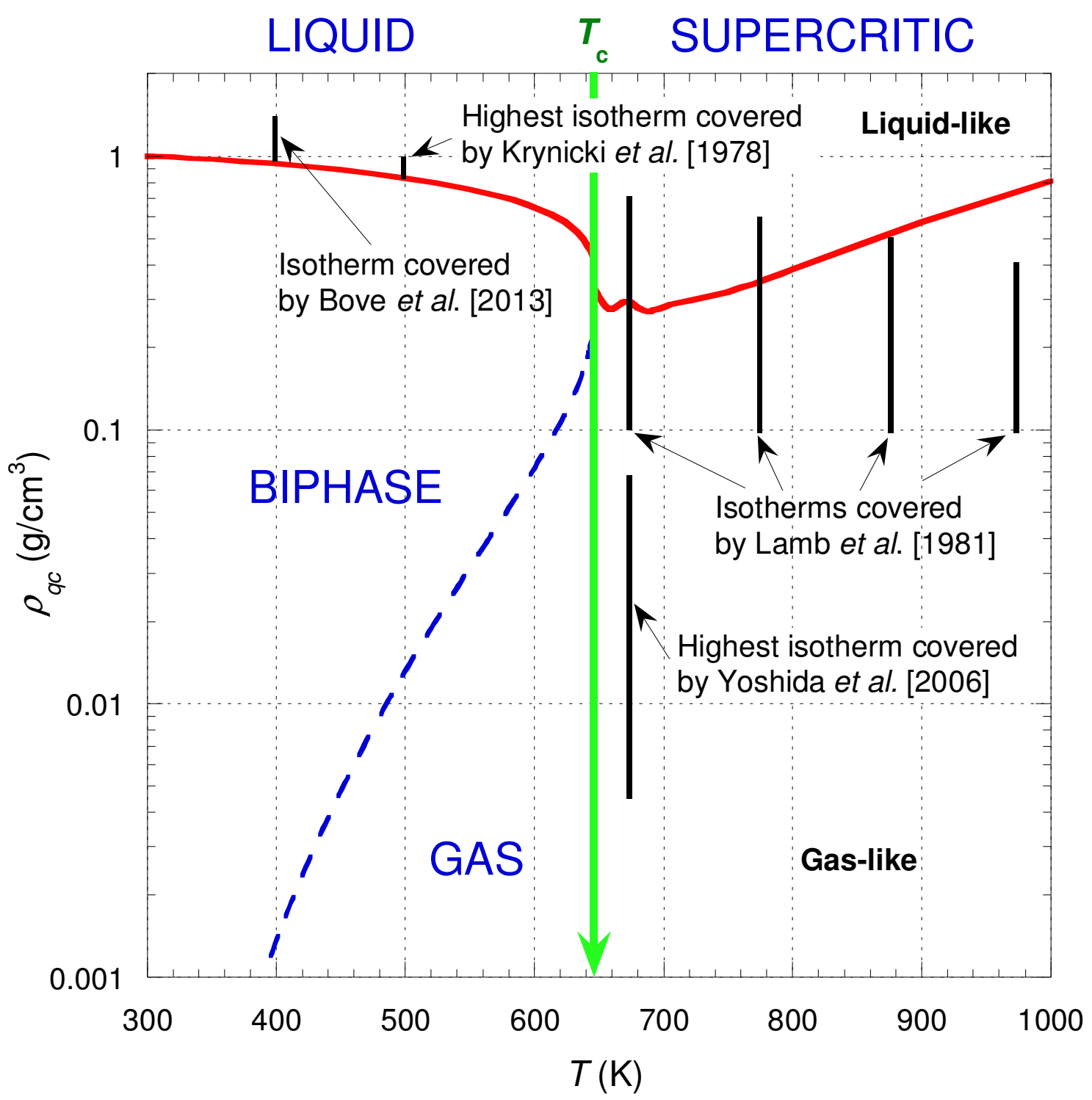

Fig. 76. Representation of the connection line (i.e. red curve) for the expressions of Table 4 in a density-temperature phase diagram. The blue dashed curve represents the vapor coexistence line.

It is interesting to have a look at the variations of the function $f_{q_c}(\rho,T)$. First of all, we observe that $f_{q_c}(\rho,T) \approx 1$ for liquid states between 250 K and 330 K whatever the density (i.e. the basic units correspond to the unit cells). Below 250 K when the temperature decreases it appears that the cut-off wave-vector modulus $q_c$ decreases very quickly, in other words the basic units have larger and larger sizes, which seems to go in the direction of intuition. Now beyond $T_{tr}$, we observe that the states close to SVP have a larger cut-off wave-vector modulus $q_c$ than $q_{c0}$ and the former decreases with increasing density on an isotherm. In other words, close to SVP the basic units are smaller than those expected by $q_{c0}^{-1}$ and then when the density is increased on an isotherm they become greater. We also observe that there is a transition on the critical isotherm so that in the supercritical domain $f_{q_c}(\rho,T)$ is always greater than 1 but remains very close to 1 when moving away from the Frenkel/Widom line at a sufficiently high temperature.

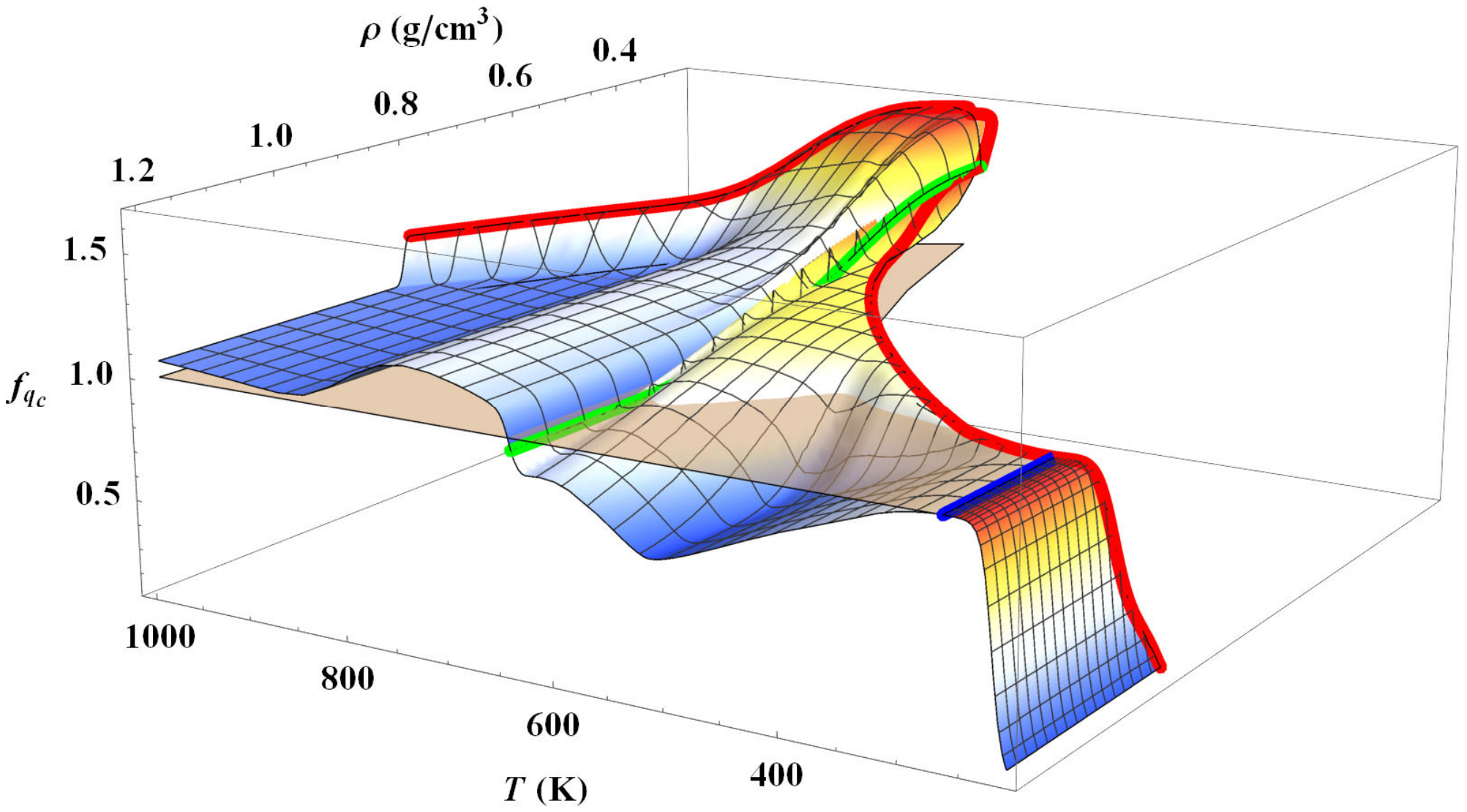

Fig. 77. Plot of the function $f_{q_c}(\rho,T)$ in the liquid-like domain. The red thick curve represents the connection line, the blue thick curve represents the isotherm at $T_{tr}$ and the green thick curve represents the critical isotherm. The light-colored horizontal plane represents $f_{q_c}(\rho,T)=1$.

In the gas-like density region, Fig. 78 shows that $f_{q_c}(\rho,T)$ is generally greater than 1 and systematically increases when the density decreases on each isotherm, which simply means that the number of modes increases strongly when the density decreases, in other words, the disorder increases accordingly.

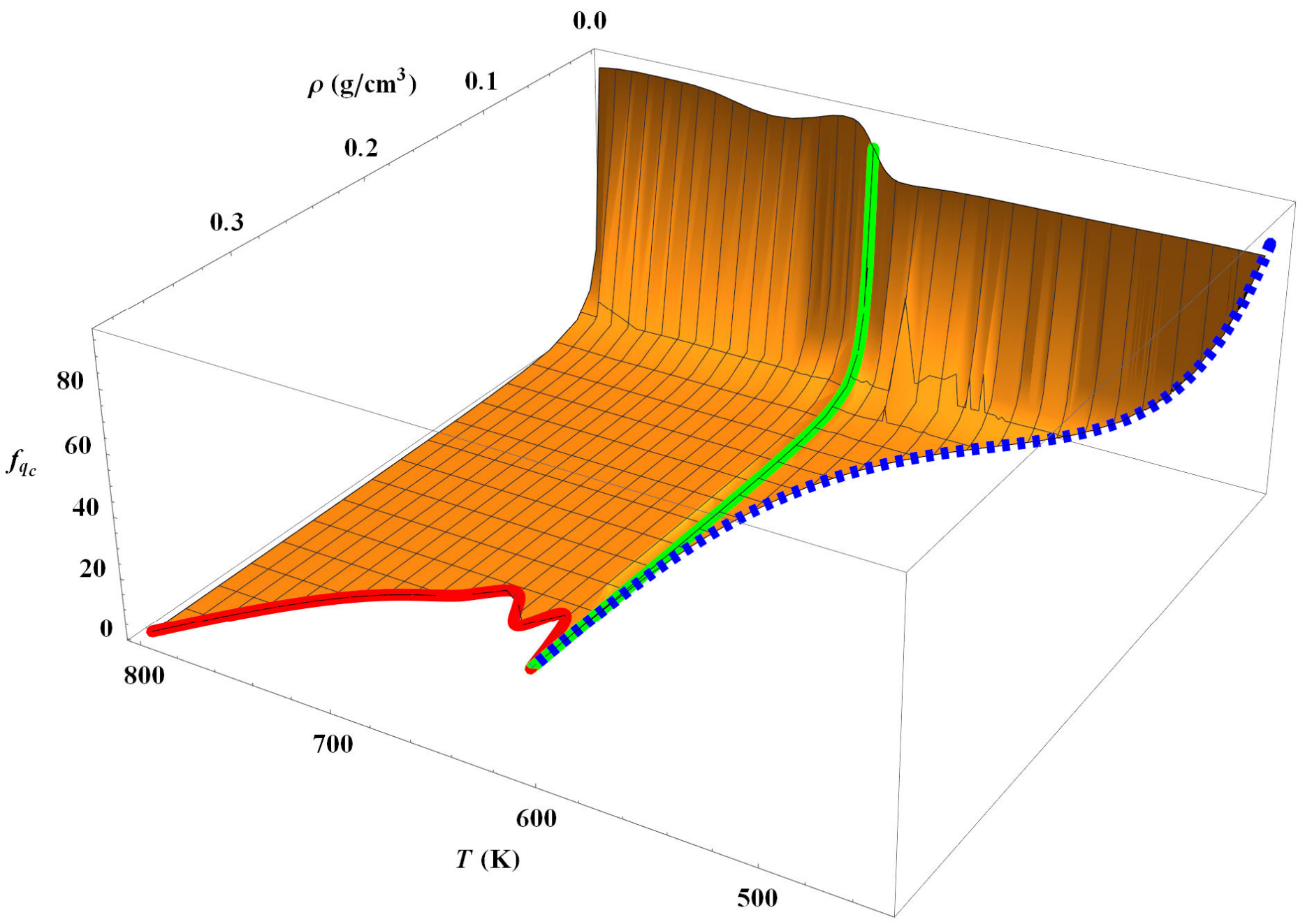

Fig. 78. Plot of the function $f_{q_c}(\rho, T)$ in the gas-like density region. The red thick curve represents the Frenkel/Widom line (i.e. a part of the connection line), the dashed blue curve represents the vapor coexistence curve and the green thick curve represents the critical isotherm.

It is important to note that in Eq. (12) it is the ratio $q_c/d$ which operates. As $d$ is not known *a priori*, it must be considered that $q_c$ is known to just one multiplicative constant. On the other hand, the variations of $q_c$ defined in Table 4 are intrinsic to the medium and do not depend on a particular set-up. It thus appears that the value of $d$ leads to setting the absolute value of the self-diffusion coefficient. Consequently, the calibration effects will be referenced on this value without there being in general any way of discerning the physical aspect of the renormalization.

#### *4.2.2. Self-Diffusion of Liquid Water at Atmospheric Pressure*

As for viscosity, many water self-diffusion coefficient data under one atmosphere exist in the literature, and as for viscosity, there is a disagreement over the numerical values between the different sources. Thus Trappeniers *et al.* (Ref. 59) wrote for the data at 25 °C only:

> "It is evident from table 1, however, that there exists a wide range of values for $D_t(H_2O)$, at about 25°C, as determined by spin echo as well as by various tracer methods."

The generally accepted explanation for this situation is that there are systematic errors in experiments whose origin is not controlled. The explanation of this dispersion of values and especially of experimental variations in the context of the present modeling is quite different. Indeed, as for the viscosity, we will see that these experimental variations can be reproduced by taking into account the characteristics of the experimental devices, characteristics which are essentially symbolized in the present modeling by the parameters $d$ and $d_N$. Note that the values of $d$ and $d_N$ are different from the corresponding value used to represent the data of the IAPWS08 formulation because the experimental devices are quite different (i.e. have different geometric characteristics) from those corresponding to the various viscometers.

Three methods of measurements are used in the references analyzed here, namely the use of molecules labeled with isotopes (radioactive or not tracers, e.g. deuterium, tritium, $^{18}O$), NMR methods in the presence of magnetic field gradients and quasielastic neutron scattering (QENS) experiments. The QENS diffusion data are strongly model dependent so we will start by analyzing the data from NMR experiments because they are more characteristic of the self-diffusion of water as Holz *et al.* (Ref. 60) wrote:

> "The main advantages of PFG NMR diffusion methods over conventional isotopic tracer methods are: fast measurements, small sample volumes, determination of 'true' self-diffusion coefficients (which are not influenced by interfering isotope effects) and an easy application to a wide pressure and temperature range."

Kisel'nick *et al.* (Ref. 58) have measured the self-diffusion coefficient of pure water using a spin-echo NMR method at pressures of 1-7500 kg/cm$^2$ and at temperatures of 4-60 °C. Here we will only analyze the data at 1 kg/cm$^2$ (i.e. 0.0980665 MPa) corresponding to the data points on the vertical axis in Fig. 1 of Ref. 58. We will discuss the pressure data in section 4.2.3.

Fig. 79(a) shows that it is possible to reproduce Kisel'nick *et al.*'s data at 0.0980665 MPa within an uncertainty of 2% which is consistent with the SE-type relationship as these authors concluded:

> "It is clear from Table 1 that the value of $D_t\eta/T$ remains constant within the limits of experimental error. This indicates that the radius of the diffusing particle is unchanged over this particular range of temperatures and pressures. The calculated radius of a water molecule is 1.42 Å is close to the generally accepted value."

It should be noted that this conclusion was established by using the 4π factor for slipping boundary limit in the SE-law instead of the more usual 6π factor for sticking boundary limit. The latter is used when the diffusing particles are larger than the medium particles while when the particles are approximately the same size, the 4π factor is preferred. In the present modeling it is the 6π factor that should be favoured since the basic units of the theory are always larger than a single molecule in the normal liquid phase. This being said, the SE-type relationship of the present modeling leads to replacing $6\pi a$ ($a$ being the hydrodynamic radius of the diffusing particle) with $(2/3)\pi q_c^{-1}$ if the viscosity $\eta$ can be approximated by $\eta_l$.

Along this isobar for the temperature range considered, we have $q_c \approx q_{c0}(\rho) \approx Const$ so $D_t\eta_l/T \approx Const$. In this temperature range we can refer to Korson *et al.*'s viscosity data (Ref. 24) and we observe on Fig. 79(b) that the gas-like term is in absolute value very small compared to the liquid-like term and its variation with temperature is low. So as a first rather rough approximation, we can say that Kisel'nick *et al.*'s conclusion is reasonable along the isobar at 0.0980665 MPa for the corresponding temperature range.

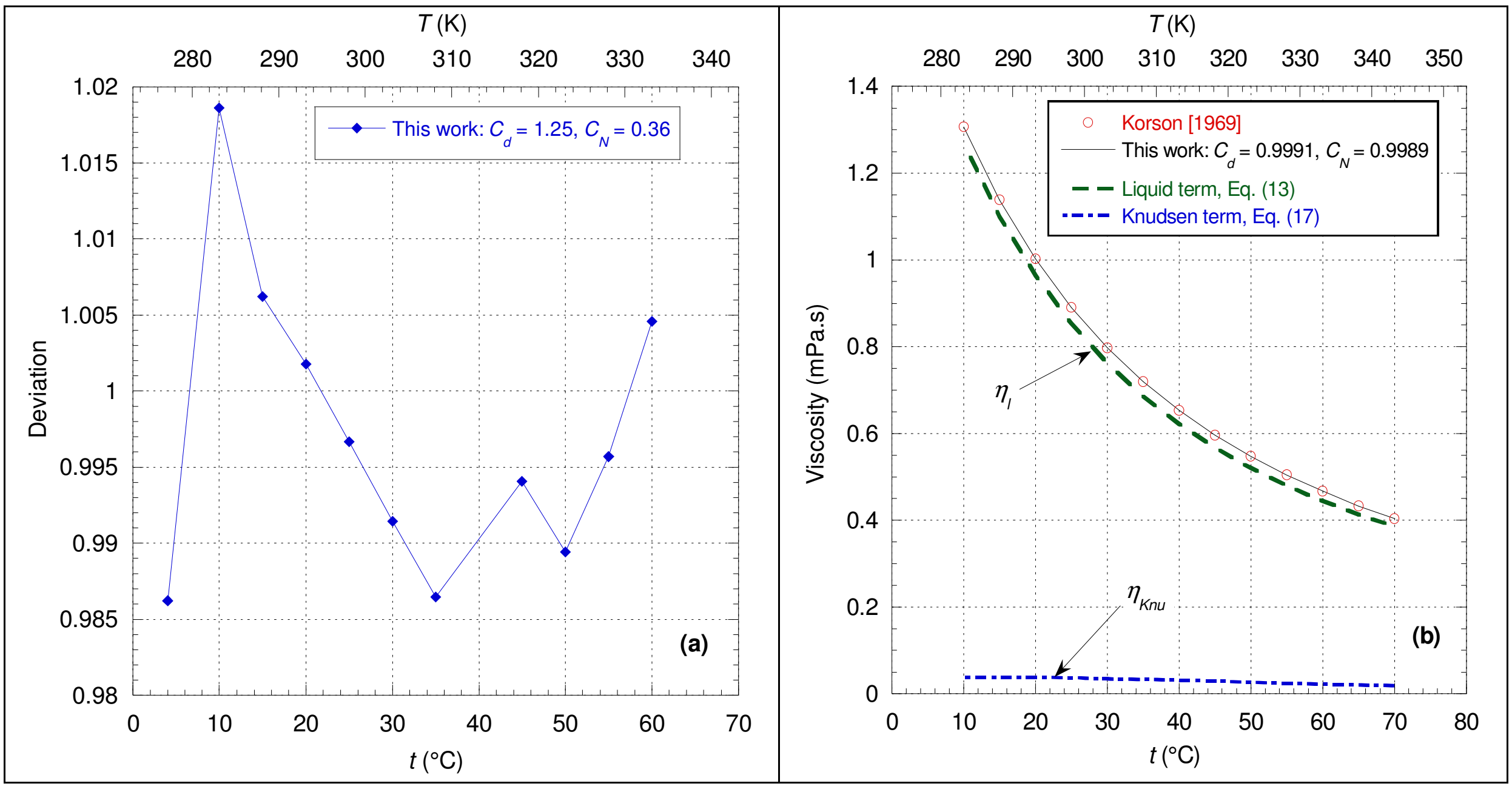

Fig. 79. (a) Ratio of Kisel'nick *et al.*'s data (Ref. 58) with the present modeling as a function of the temperature along the isobar equal to 0.0980665 MPa. The lines are eye guides. (b) Viscosity data of Korson *et al.* (Ref. 24) and the corresponding viscous terms in the present modeling.

More recent and more accurate data than Kisel'nick *et al.* were made by Holz *et al.* (Ref. 60). Fig. 80 shows that the deviation here is divided by two and it is now within 1% according to the uncertainty suggested by Holz *et al*. It is observed that, although different, the values of $C_d$ and $C_N$ are comparable to the values of these parameters for Kisel'nick *et al*.

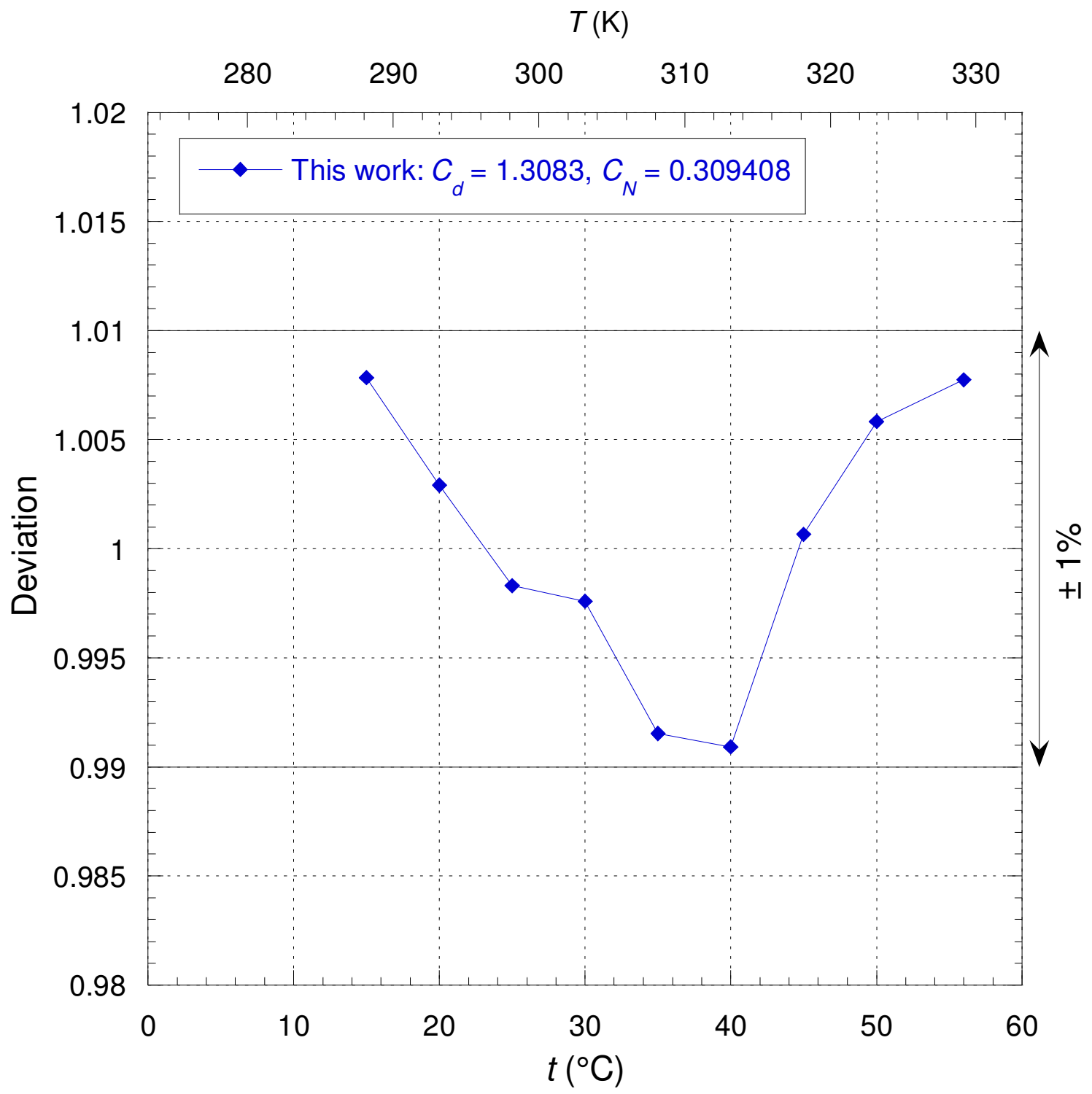

Fig. 80. Ratio of Holz *et al.*'s data (Ref. 60) with the present modeling as a function of the temperature along the atmospheric isobar. The lines are eye guides.

The same year as Holz *et al.* (Ref. 60), Tofts *et al.* (Ref. 61) have made measurements in a very narrow temperature range but with high accuracy. Each given value is an average of 3 or 4 measurements at the same temperature and the authors wrote that:

> "Total uncertainty in estimates of diffusion coefficient were in the range 1.4–2.7%."

Fig. 81 shows that the present modeling allows reproducing the data within 0.2% which corresponds not to the uncertainty but to the reproducibility of each data point.
It should be noted that between the three experiments of Kisel'nick *et al.*, Holz *et al.* and Tofts *et al.* the coefficients $C_N$ are slightly different, which means that the temperature dependence is different in these three cases.

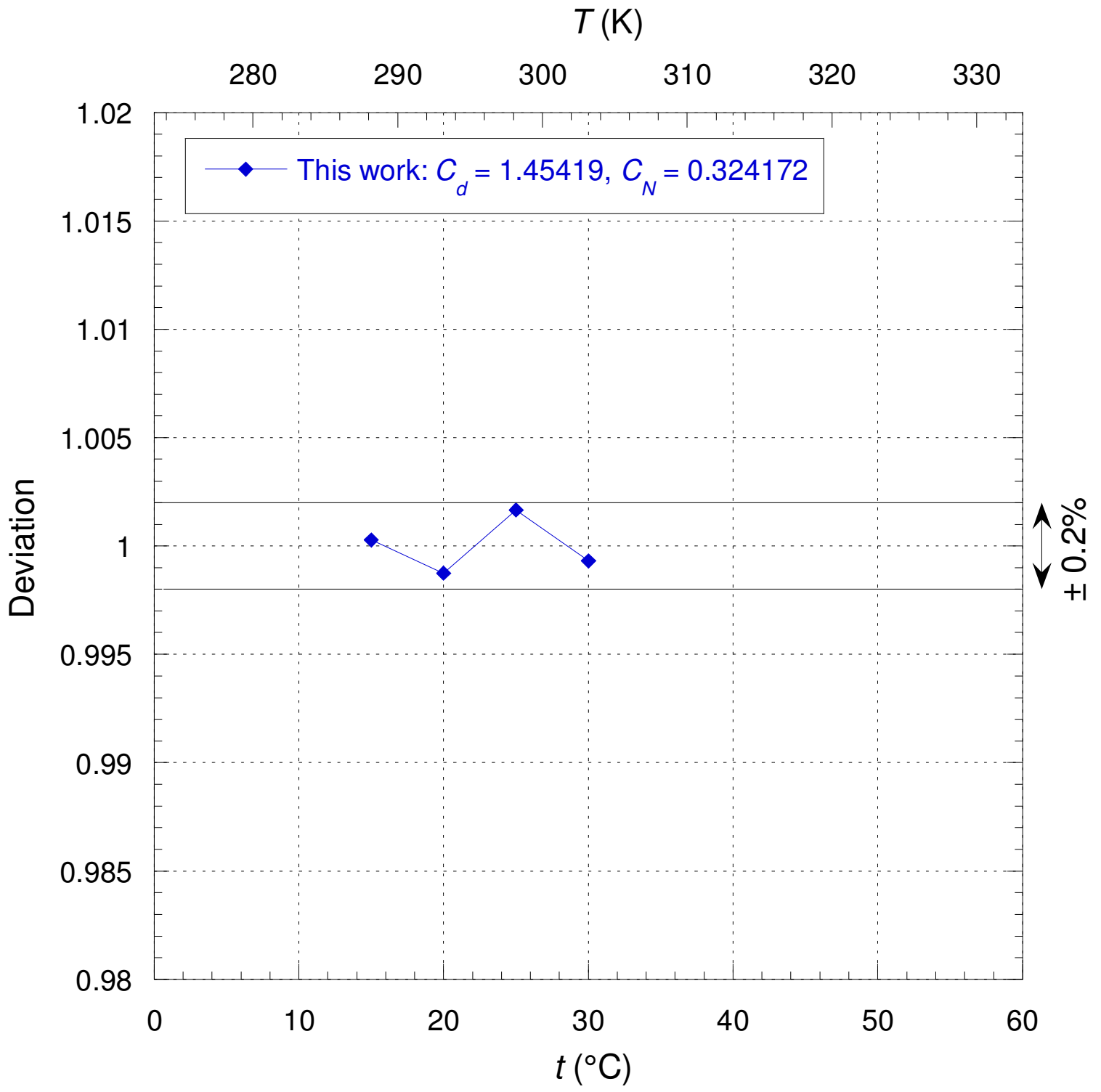

Fig. 81. Ratio of Tofts *et al.*'s data (Ref. 61) with the present modeling as a function of the temperature along the atmospheric isobar. The lines are eye guides.

Most of other NMR experiments focus on the supercooled phase. To carry out these experiments it is generally necessary to use tubes of diameters smaller than those in the normal liquid phase in order to avoid crystallization. In this context, Gillen *et al.*'s experiment (Ref. 62) is interesting since these authors made two kind of experiments using different diameters of the tubes to reach -31 °C (i.e. 242.15 K) as they wrote it:

> "Standard 5-mm-o.d. NMR tubes containing triply distilled water were used above -12°C. Between -12 and -19°C, five 1-mm melting-point tubes, carefully chosen for supercooling propensity, were packed in the 5 mm tube. For supercooling beyond -19°C, smaller diameter tubes were required and prepared by distillation into freshly drawn capillaries."

Therefore, data above -12°C should be treated differently from data below. Fig. 82 shows that taking into account both kinds of experimental characteristics, it is possible to reproduce the data (i.e. blue diamonds) with an accuracy comparable to that of Holz *et al.* (Ref. 60) above -12 °C. On the other hand, the data (i.e. black triangles) corresponding to the lowest temperatures have a very strong deviation that continues to grow with the decrease in temperature. This observation has also been mentioned by Easteal *et al.* (Ref. 63) who proposed the following renormalization:

> "Work covering the lower regime (242-298 K) has been reported by Gillen and others using n.m.r. techniques. Their values above 273.15 K are on average 3.18% lower than the accepted values of Mills [Ref. 68]. It thus seems reasonable to scale Gillen's data by this factor."

However, this deviation is not acceptable in view of the experimental uncertainty given by the authors:

> "Errors in self-diffusion coefficients reported in Table 1 are estimated to be ±5% (95% confidence limits)."

This deviation can be explained by the fact that in this experiment all the tubes and the gradient coil are cooled and therefore this is sufficient to admit that probably the field gradient does not remain constant as a function of temperature. The importance of keeping the coil gradient at constant temperature will be discussed again later with the analysis of Prielmeier *et al.*'s data (Ref. 65). If this effect is taken into account, then the formula for determining the self-diffusion coefficient should be corrected as follows:

$$D_{t,\mathrm{Gillen}}(T,1\,\mathrm{atm}) = f_{g,\mathrm{Gillen}}(T)^2\, D_t(T,1\,\mathrm{atm}) \tag{29}$$

because it is the square of the field gradient that is involved and so that here:

$$f_{g,\mathrm{Gillen}}(T > 242.5\,\mathrm{K}) = \left[1\text{-}0.07243\exp\left(-\left(\frac{T}{246.42}\right)^{200}\right)\right]\exp\left(-\left(\frac{238}{T}\right)^{213}\right) \tag{30}$$

Eq. (30) reveals that the effect of decreasing temperature is to reduce the gradient coil calibration constant with respect to the estimated one. Taking this correction into account, Fig. 82 shows that the data (i.e. red circle) can now be reproduced within the experimental uncertainty.

It can be noted that we logically obtain that $C_d$ and $C_N$ are smaller for smaller tube diameters but we observe that the temperature dependence of the two kinds of experiment is very different above and below 12°C.

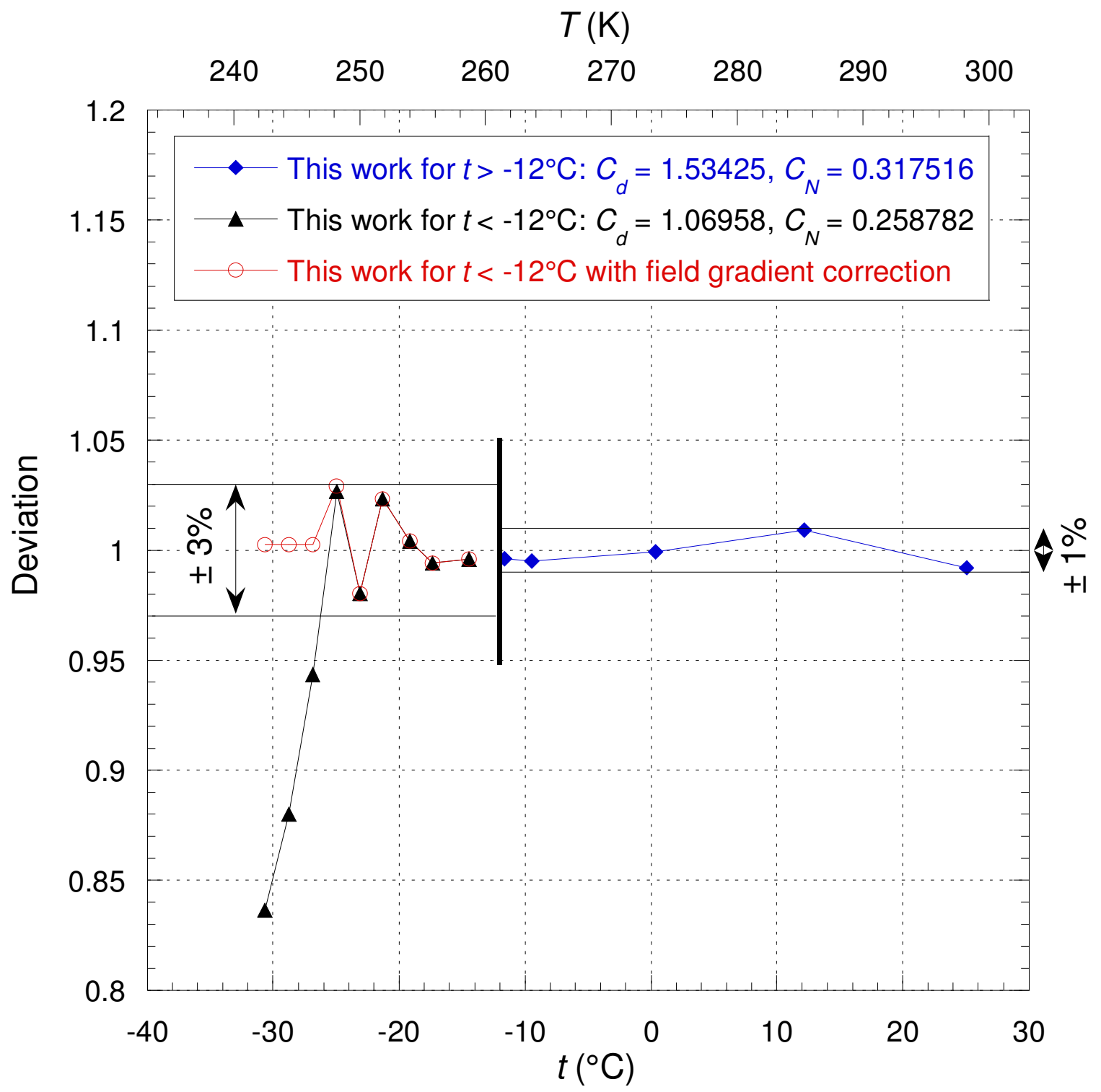

Fig. 82. Ratio of Gillen *et al.*'s data (Ref. 62) with the present modeling as a function of the temperature along the atmospheric isobar. The lines are eye guides.

In order to overcome crystallization effects such as those reported by Gillen *et al.*, Price *et al.* (Ref. 64) used a multitude of small diameter tubes (i.e. internal diameter of 130 µm) in order to have enough signal. By using a pulsed-gradient spin-echo NMR method, they could reach little lower temperatures than in Gillen *et al.*'s experiment but their data represent an average of physical phenomena for all the tubes. Price *et al.* adopted the following approach for the calibration of the field gradient factor:

> "It should be noted that in our data we defined the diffusion coefficient at 298.15 K to be $2.30\times10^{-9}$ m$^2$ s$^{-1}$. This value is the most well-defined value of the self-diffusion coefficient of water. This was the basis for the calibration of $g$."

Fig. 83 shows that the present modeling allows reproducing Price *et al.*'s data only above -10 °C with a relatively high uncertainty compared to the uncertainty of ±1% suggested by these authors. For temperatures below -10 °C, the same kind of deviation is observed as with Gillen *et al.*'s data. It is therefore very likely that the calibration constant of the field gradient varies more strongly with temperature than assumed. Then assuming the same kind of correction as that given by Eq. (29), we find here:

$$f_{g,\mathrm{Price}}(T > 237.8\ \mathrm{K}) = 1 - \frac{0.9775}{1+\left(\dfrac{T}{233.82}\right)^{104}} - 0.0225\exp\left(-\left(\frac{T}{267}\right)^{200}\right) \tag{31}$$

Again, Eq. (31) reveals that the effect of decreasing temperature is to reduce the calibration constant with respect to the estimated one. Fig. 83 shows that it is now possible to reproduce the data (i.e. red circle) within a uniform uncertainty of 3%. Given that, the deviation is well centered on the unit value, this 3% band can be considered as the “real” experimental uncertainty.
It also appears that the coefficient $C_N$ is much smaller than for Gillen *et al.*’s data which indicates that the temperature dependence is significantly different when using a great number of tubes.

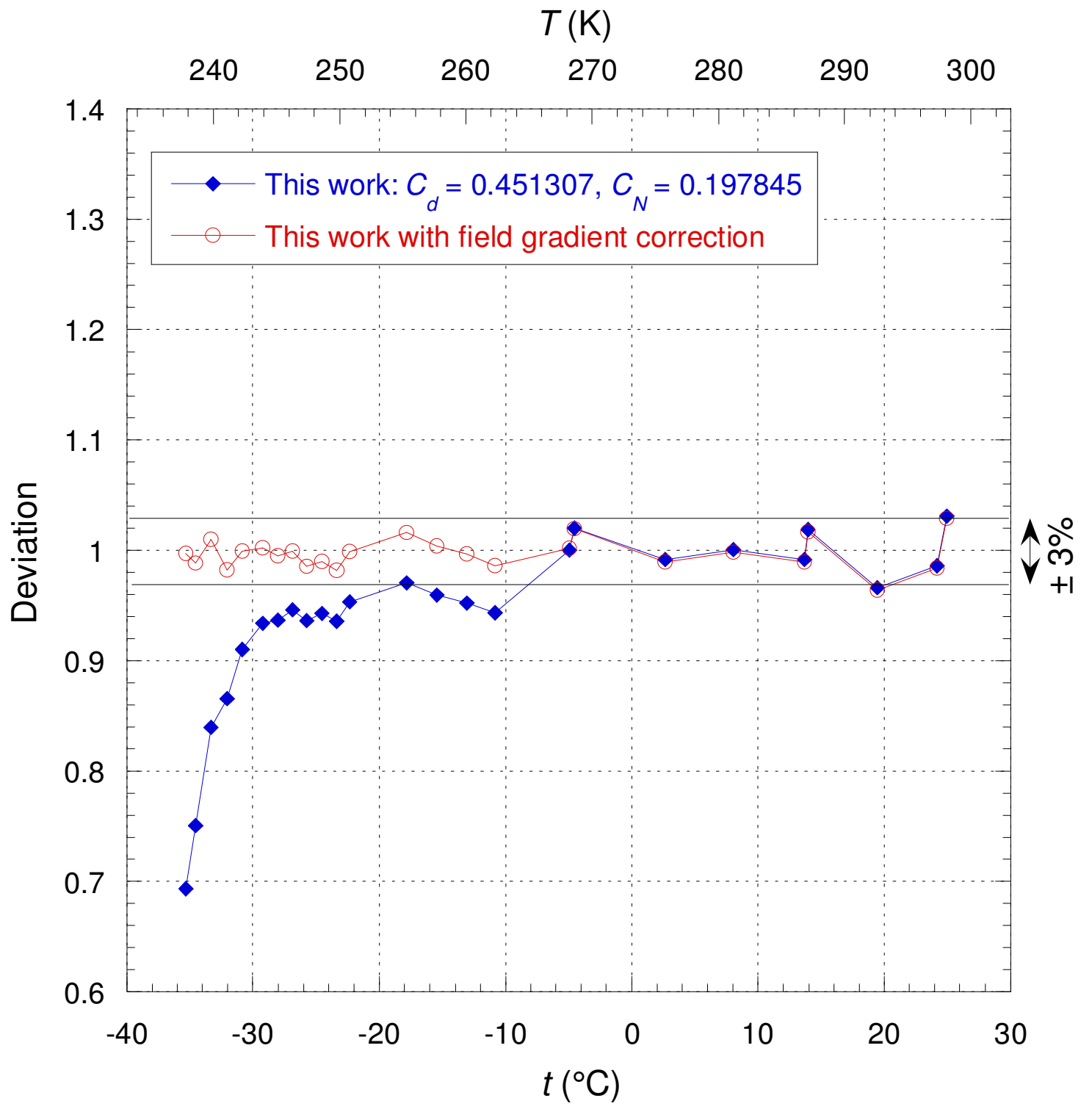

Fig. 83. Ratio of Price *et al.*’s data (Ref. 64) with the present modeling as a function of the temperature along the atmospheric isobar. The lines are eye guides.

Prielmeier *et al.* (Ref. 65) have measured the self-diffusion coefficient in supercooled water with the NMR spin-echo technique at pressures up to 400 MPa and temperatures down to 203 K. These data will therefore be analyzed in detail in section 4.2.3. However, we will now analyze the data along the 0.1 MPa isobar because the experimental design of Prielmeier *et al.* is such that the gradient coil is maintained at a constant temperature and therefore it is expected that there will be no temperature effect on the parameter *g*. Harris *et al.* (Ref. 66) have particularly emphasized the advantage of such a system:

> “Lüdemann’s apparatus uses the etched capillary cell with gradient coils mounted outside the thermostat in which the cell is mounted (Prielmeier, 1988; Lang and Lüdemann, 1991). The calibration constant is therefore independent of the state of the sample.”

In the Prielmeier *et al.*’s paper there are two types of data: raw data in the form of points on their Fig. 2 and smoothed data in their Table 1; these two datasets do not contain the same number of data points. Here we will focus only on the raw data. Along the 0.1 MPa isobar the smallest temperature reached is only 252 K (i.e. -21.15 °C), therefore if there is an effect of the temperature dependence of the field gradient it will not be as pronounced as in the experiments of Gillen *et al.* and Price *et al.* but still it should be “visible”. Prielmeier *et al.* wrote about their measurements:

"The self diffusion coefficients for $H_2O$ are judged reliable to ±3% ."

Fig. 84 shows that the present modeling makes it possible to reproduce these data very precisely (i.e. well below the experimental uncertainty) without any correction being considered.
We find here that the coefficients $C_d$ and $C_N$ are larger than for Price *et al.*'s experiment, which is consistent with the fact that the internal diameter of the tubes for low pressures is here equal to 200 μm, i.e. the tubes diameter is larger than that of Price *et al.*

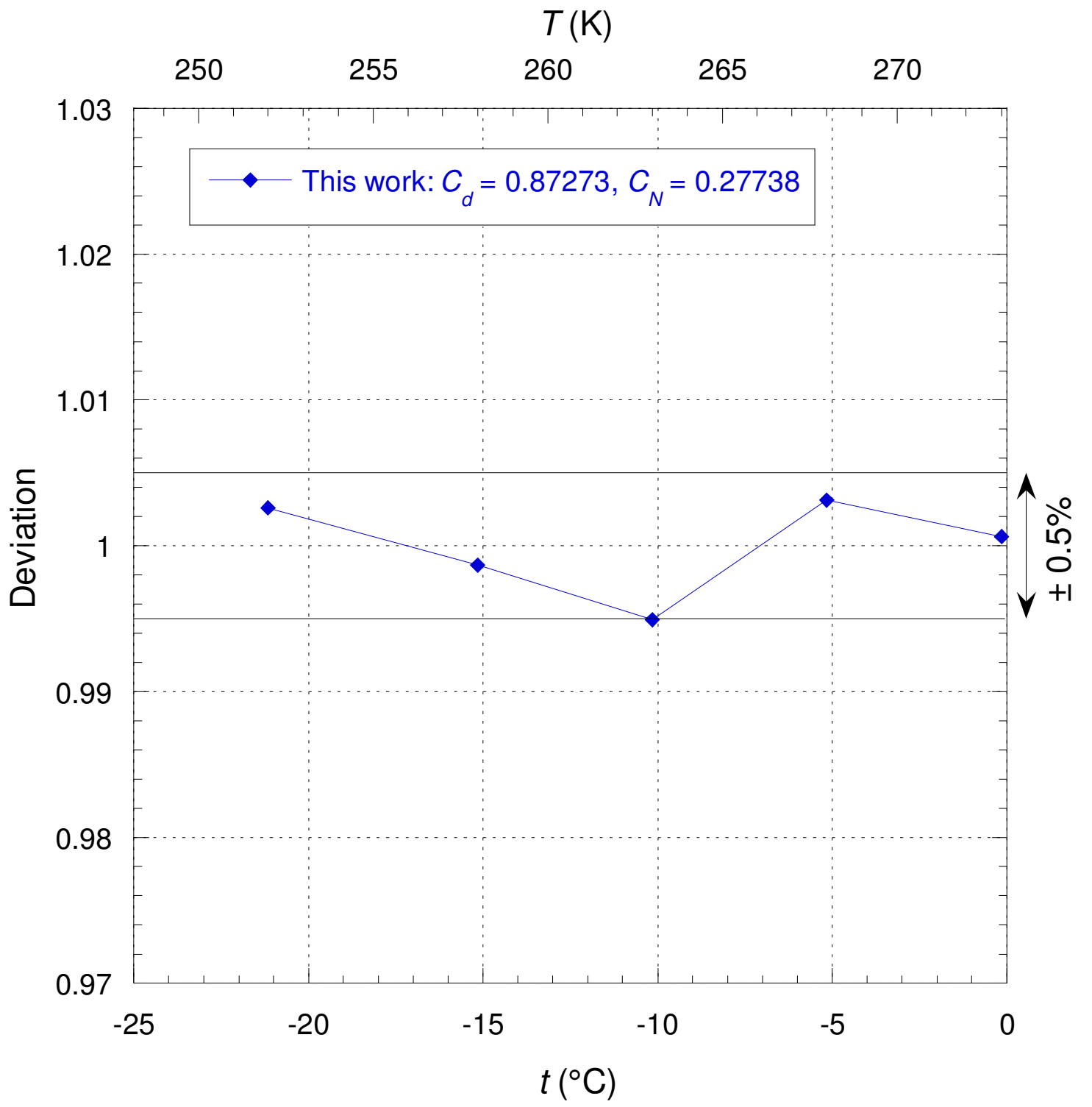

Fig. 84. Ratio of Prielmeier *et al.*'s raw data (Ref. 65) with the present modeling as a function of the temperature along the 0.1 MPa isobar. The lines are eye guides.

To avoid the problem of calibrating the field gradient parameter, it is interesting to analyze the QENS data of Qvist *et al.* (Ref. 67) which covers the same negative temperature range as that of Prielmeier *et al.* The data analyzed here are those presented in Table III of the Qvist *et al.*'s paper: given the uncertainty about the last digit, the relative error of these data is about ±0.5%. Fig. 85 shows that the present modeling makes it possible to reproduce these data as well as possible with a deviation of ±1.3%, which seems *a priori* too large in view of the experimental uncertainty estimated by the authors. However, if we refer to the comment of Qvist *et al.*, this deviation seems reasonable:

> "If our QENS analysis is sound, it must produce $D_t$ values consistent with these more direct measurements. This is indeed the case (Fig. 7, Table III); the average absolute deviation at the six temperatures is merely 1.6% [with the data from Price *et al.* and Mills]."

Now it should be realized that QENS data are highly model dependent (in absolute value and for variations). In particular, the value of the diffusion coefficient deduced depends on the apparent jump parameter $d_{jump}$ which, when left free as here, varies in opposite way with temperature compared to what is expected, namely $d_{jump}$ decreases when temperature increases. The authors have tried to justify this opposite variation by using some MD simulations but the physical interpretation of this parameter remains uncertain. It is easy to understand that if this parameter had been set at least to a constant value (instead of being a free temperature dependant parameter), the variation in the diffusion coefficient would be slightly different but especially the relative error on the diffusion coefficient would be higher. In other words, the relative error of 0.5% must be understood as a minimum value. Taking into account the various approximations in the neutron modeling, a relative error close to 1% is probably more appropriate.

Assuming that the uncertainty is compatible with that obtained in Fig. 85, it can be seen that the parameters $C_d$ and $C_N$ have values very close to those of Prielmeier *et al.* rather than those of Price *et al.* This seems consistent with the fact that the tubes of Qvist *et al.* have an internal diameter of 300 μm comparable to those of Prielmeier *et al.*

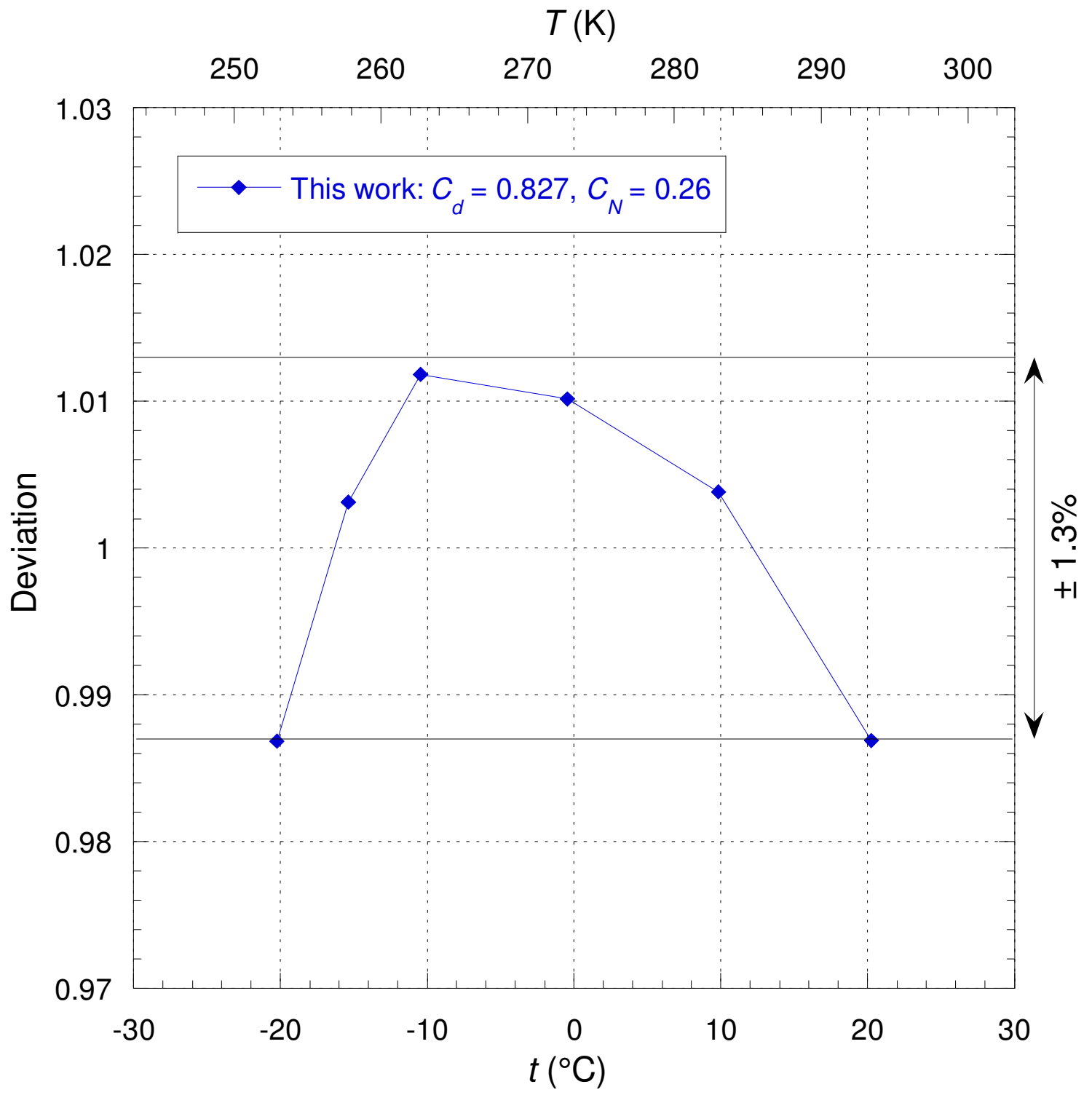

Fig. 85. Ratio of Qvist *et al.*'s QENS data (Ref. 67) with the present modeling as a function of the temperature along the atmospheric isobar. The lines are eye guides.

In the case of viscosity data in the supercooled phase we saw that it was necessary to correct the latter due to the shear correlation length increase at the walls. However, in the case of self-diffusion coefficient data, no similar effect is observed since it is a phenomenon in the bulk that is observed with no shear stress, implying that the no slip condition is irrelevant.

We will now analyze some experiments using isotopes such as tritiated water (HTO) and $H_2O^{18}$ tracer in normal water. It is accepted in the literature that the diffusion coefficients derived from these isotopic tracer methods have similar variations to those of "normal" water (i.e. $H_2O^{16}$) self-diffusion coefficient. In particular it is currently admitted that the diffusion coefficient of $H_2O^{18}$ closely approximates the "true" self-diffusion coefficient of normal water.

Mills (Ref. 68) performed experiments with various isotopic forms of water in order to better understand the self-diffusion coefficients dependence on the mass of tracers. He deduced from the rates of variation with concentration of the diffusion coefficient an extrapolation for the self-diffusion of normal water. The two determinations of Mills for HTO and for normal water allow fixing the parameters for the diffusion coefficient of HTO in normal water. In addition, the data of Mills have been used as a reference by many authors, so the analysis of this data will enable us to analyze the other experiments in a coherent way. Mills wrote about his measurements:

> “The overall reproducibility as shown in Table I was at least of order ±0.2%. There always remains of course the possibility of some unrecognized systematic error.”

First of all, Fig. 86 shows that Mills’ data can be reproduced with an uncertainty of ±0.5% except for one point, which is consistent with the reproducibility of each point estimated by Mills. It is observed that the value of the parameter $C_N$ lies between that of Price *et al.* and that of Gillen *et al.*, in other words, temperature dependence is still different here from other experiments. But this value of $C_N$ must remain the same for all Mills’ data that can be considered as they are obtained with the same device. To reproduce with the same uncertainty as for normal water the diffusion coefficient variation for HTO considering the same values of the parameters $C_d$ and $C_N$, one must assume that HTO modifies the cohesion and the size of the basic units in the system. In the present modeling this is translated by introducing a proportionality constant on $K$ and on $q_c$, respectively named $C_K$ and $C_q$. It is found that HTO induces a slight increase of the cohesion (i.e. $C_K > 1$) and increases the size of the basic units (i.e. $C_q < 1$ that is $q_c$ is smaller).

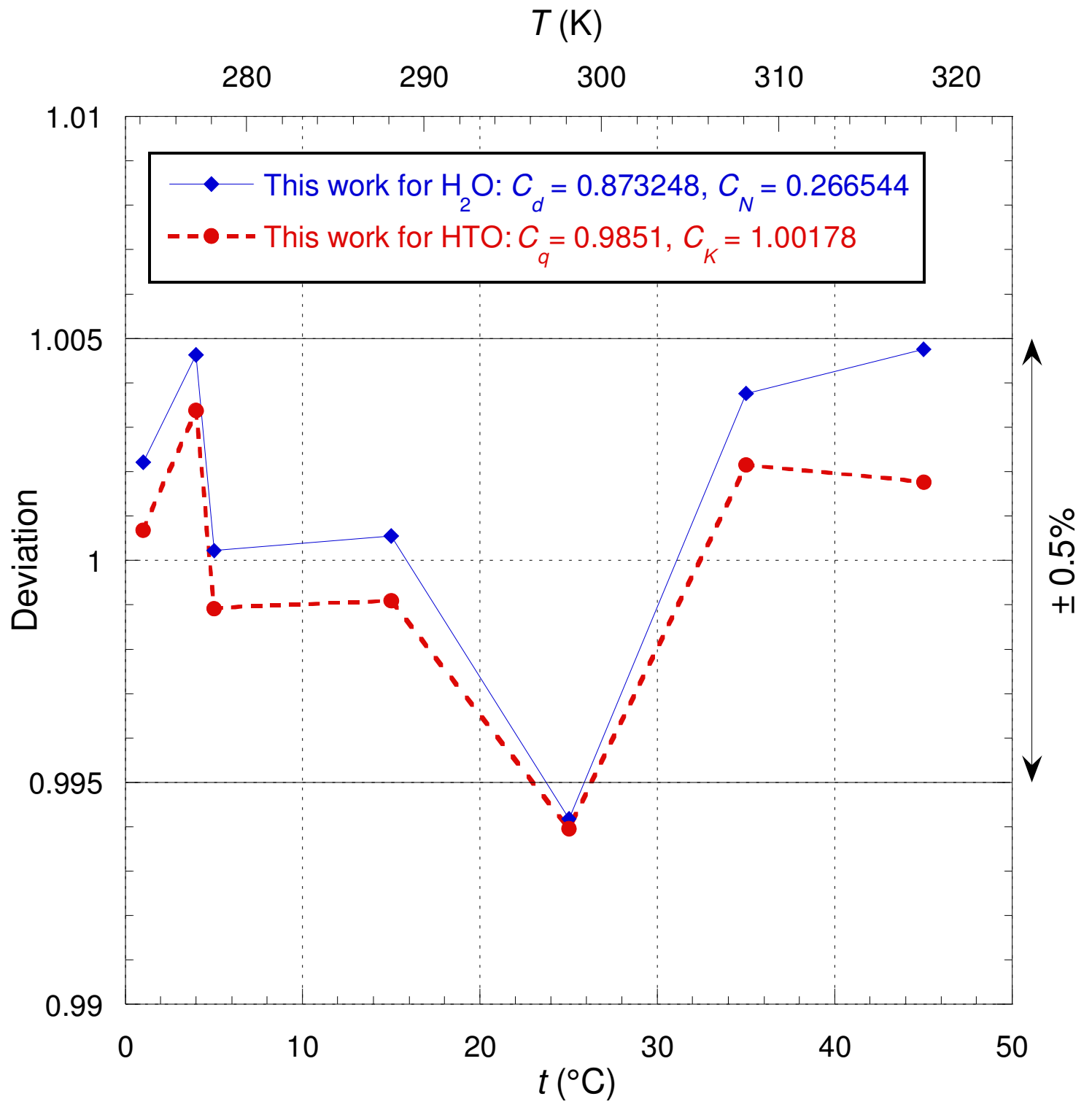

Fig. 86. Ratio of Mills’ data for $H_2O$ and HTO tracer (Ref. 68) with the present modeling as a function of the temperature along the atmospheric isobar. The lines are eye guides.

Pruppacher’s experiment (Ref. 69) is interesting because the temperature range explored extends far enough into the supercooled phase and because he used two different instruments as Gillen *et al.* depending on whether the temperature is below or above 0 °C as he explained:

> "At temperatures warmer than 0°C a 10-cm-long water column was used. At temperatures below 0°C the water column was made increasingly shorter in order to prevent the supercooled water from freezing."

Therefore, as for Gillen *et al.*, the data below and above 0 °C should be analyzed separately. Pruppacher's experimental data for the self-diffusion coefficient of water are listed as a function of temperature in his Table III. Pruppacher considered that his values of the diffusion coefficient are deduced with a constant absolute error at any temperature, which can be represented in a form of two curves in the deviation plot.

For the analysis of Pruppacher's data, we have to use the values of the constants $C_K$ and $C_q$ for HTO determined for Mills. Fig. 87 shows that Pruppacher's data can be reproduced within their experimental uncertainties. This indicates a consistency of all these results contrary to what Price *et al.* (Ref. 64) suggested:

> "In 1972, Pruppacher measured the water diffusion down to 248 K at ambient pressure using tritium as a tracer. However, the low reliability of the results of Pruppacher at room temperature places some doubt on the accuracy of his measurements at low temperature."

We observe that the temperature dependence is slightly different between the two kinds of experiments (i.e. $C_N$ is slightly different) while the coefficient $C_d$ is the same in both cases and the resulting value of $d$ is consistent with the following Pruppacher's comment:

> "Several experiments were performed with glass tubes instead of polyethylene tubes, and polyethylene tubes of 0.114 and 0.269 cm instead of 0.178 cm, without detecting any difference in the results."

The coefficients $C_d$ and $C_N$ are quite different from those of Mills but the analysis shows that these data are consistent with those of Mills.

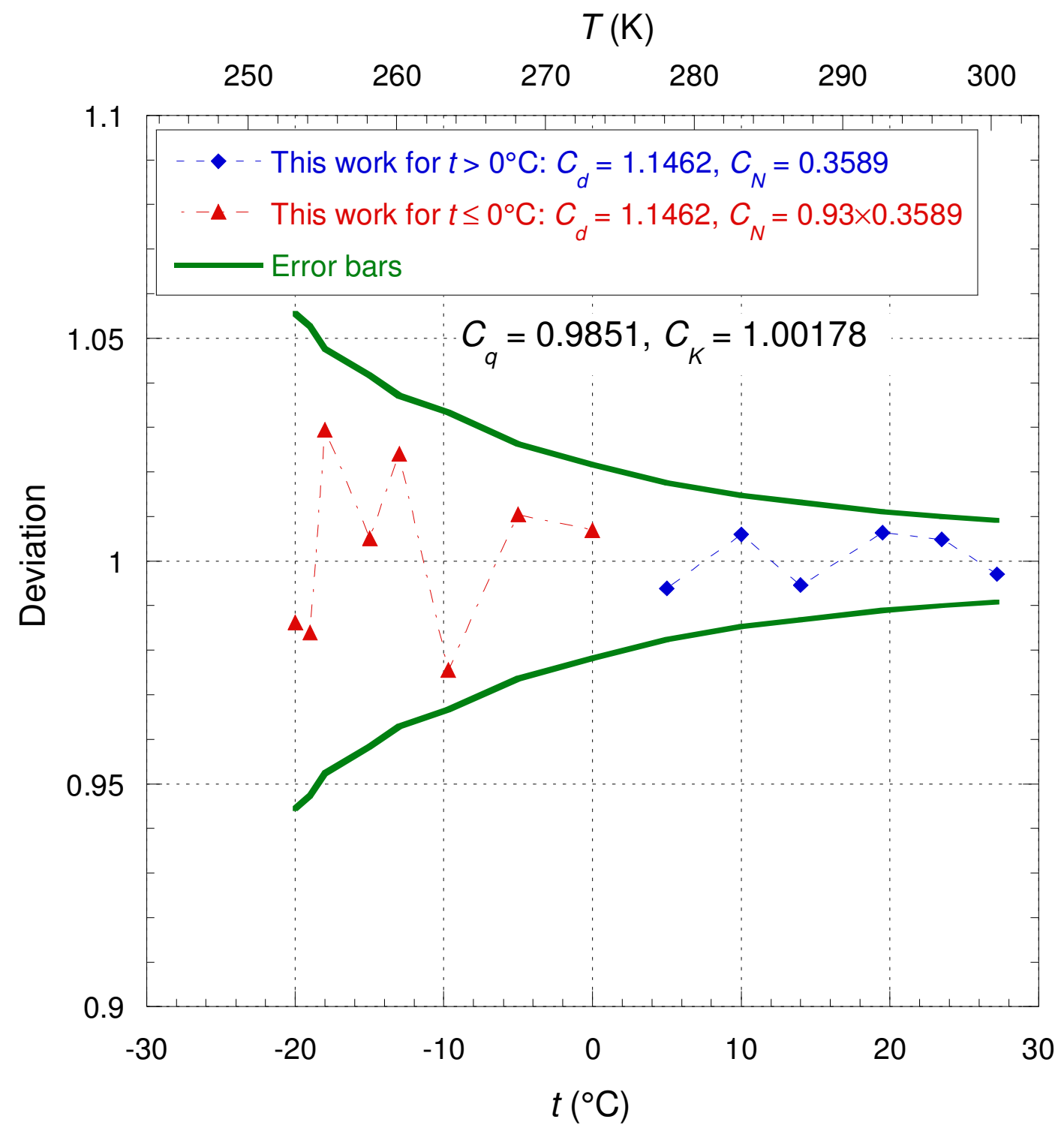

Fig. 87. Ratio of Pruppacher's data for HTO tracer (Ref. 69) with the present modeling as a function of the temperature along the atmospheric isobar. The two constants' values of $C_K$ and $C_q$ used are those determined for Mills' data (see Fig. 86). The dashed lines are eye guides.

We will now analyze Pruppacher's following comment:

> "At temperatures warmer than 0°C the values for $\eta$ listed, are those experimentally determined by Hardy and Cottingham [Ref. 31]. At temperatures colder than 0°C the $\eta$ values are taken from the experimental results of Hallett [Ref. 26]. From these data the quantity $D_t\eta/T$ was computed. As seen from Table III, Column 4, $D_t\eta/T$ is a constant to within 3% over the temperature range investigated […]".

For temperatures above 0 °C we have already shown with Kisel'nick *et al.*'s data that admitting $D_t\eta/T = Const$ was a rough but reasonable approximation. Let's observe what happens then for the supercooled phase. Since Pruppacher used Hallett's data, the liquid-like and gas-like terms corresponding to the representation of Hallett's viscosity data are shown in Fig. 88(a): it is observed that the approximation $\eta \approx \eta_l$ is even better than for Korson *et al.*'s data (see Fig. 79(b)). On the other hand, unlike the normal liquid phase, Fig. 77 shows that the function $f_{q_c}(\rho,T)$ decreases rapidly in the supercooled phase, so *a priori* one can no longer make the approximation $q_c \approx q_{c0}(\rho)$. However, Fig. 88(b) shows that the relative deviation of the variation of the function $f_{q_c}(\rho,T)$ along the atmospheric isobar is at most close to 1% in the temperature range corresponding to Pruppacher's data. Considering that the density varies by a maximum of 0.6% over this same temperature range, it can be said that Pruppacher's remark is accurate with the uncertainty margin of 3% but otherwise it is false in absolute terms.

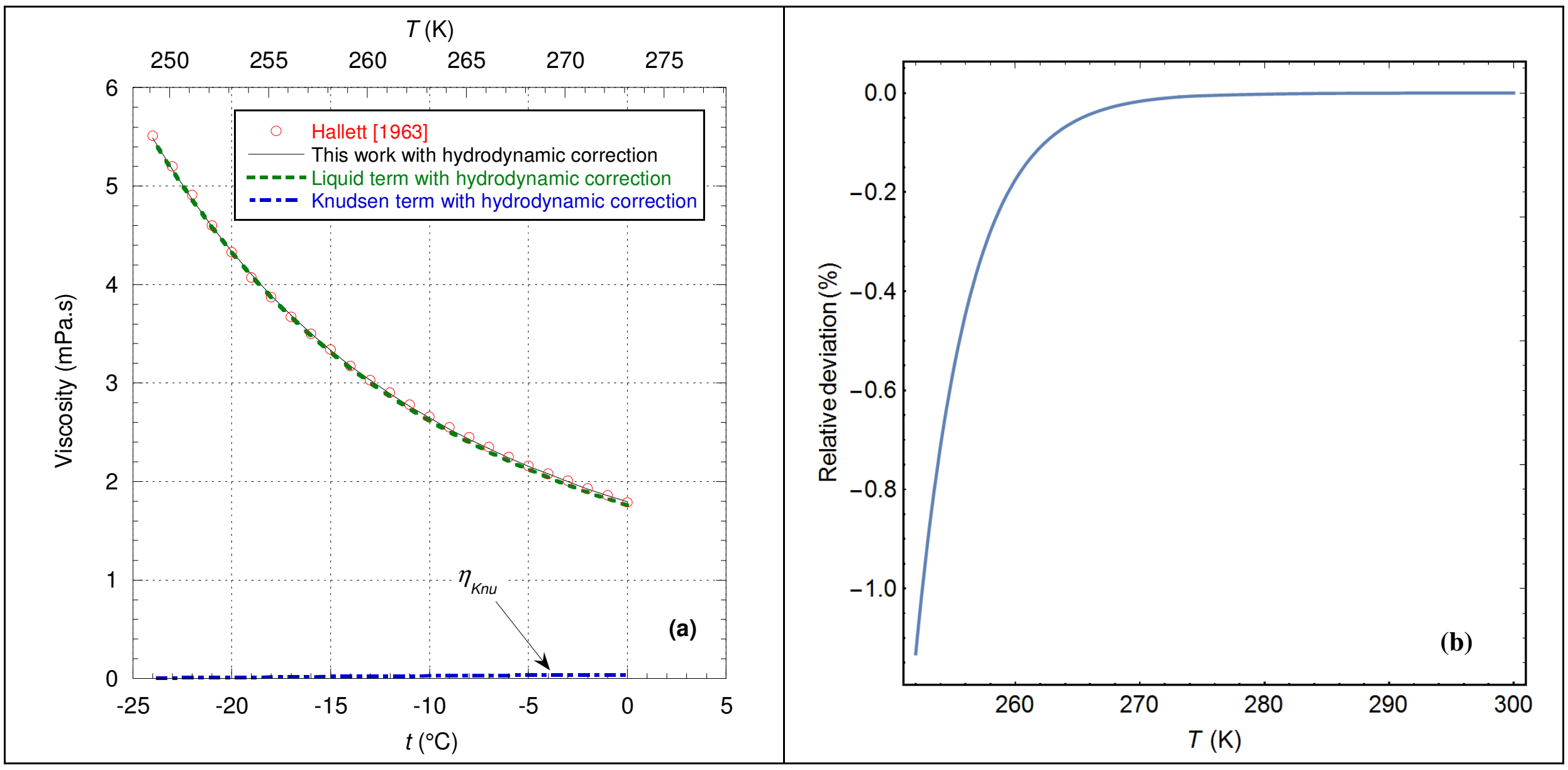

Fig. 88. (a) Water viscosity data of Hallett (Ref. 26) and the corresponding viscous terms in the present modeling. (b) Percentage deviation of the function $f_{q_c}(\rho,T)$ from the unit value along the atmospheric isobar.

Since the $H_2O^{18}$ tracer has diffusion properties quasi-identical to normal water (i.e. $C_q = C_K = 1.0$), we make an analysis comparable to that of Mills to describe the datasets of Wang *et al.* (Ref. 70 and 71) and Easteal *et al.* (Ref. 63 and 72).

Wang *et al.* carried out two different experiments between the years 1951 and 1953: i.e. in the first case the capillary tubes were identical with an internal radius of 250 µm and 4 cm long, while in the second they used different tubes with radii varying between 120 µm and 130 µm and lengths ranging from 1.94 cm to 4.79 cm. In both cases Wang *et al.* indicated error bars for each of the measurement points which can be translated in the form of two curves in the deviation plot. The diffusion coefficients are deduced from the solution of the one-dimensional diffusion equation.

Fig. 89(a) shows that the present modeling allows $H_2O^{18}$ data to be reproduced within an uncertainty of 2% compatible with the error bars. On the other hand, we can see on Fig. 89(b) that the deviation is much larger and less consistent with the error bars than previously. This greater dispersion is explained by the multiplicity of geometric characteristics of the tubes used. However, it can be seen that the parameters $d$ and $d_N$ are smaller when the tubes are smaller and it appears, in particular for the 1951 data, that the temperature dependence is quite different from other kinds of data.

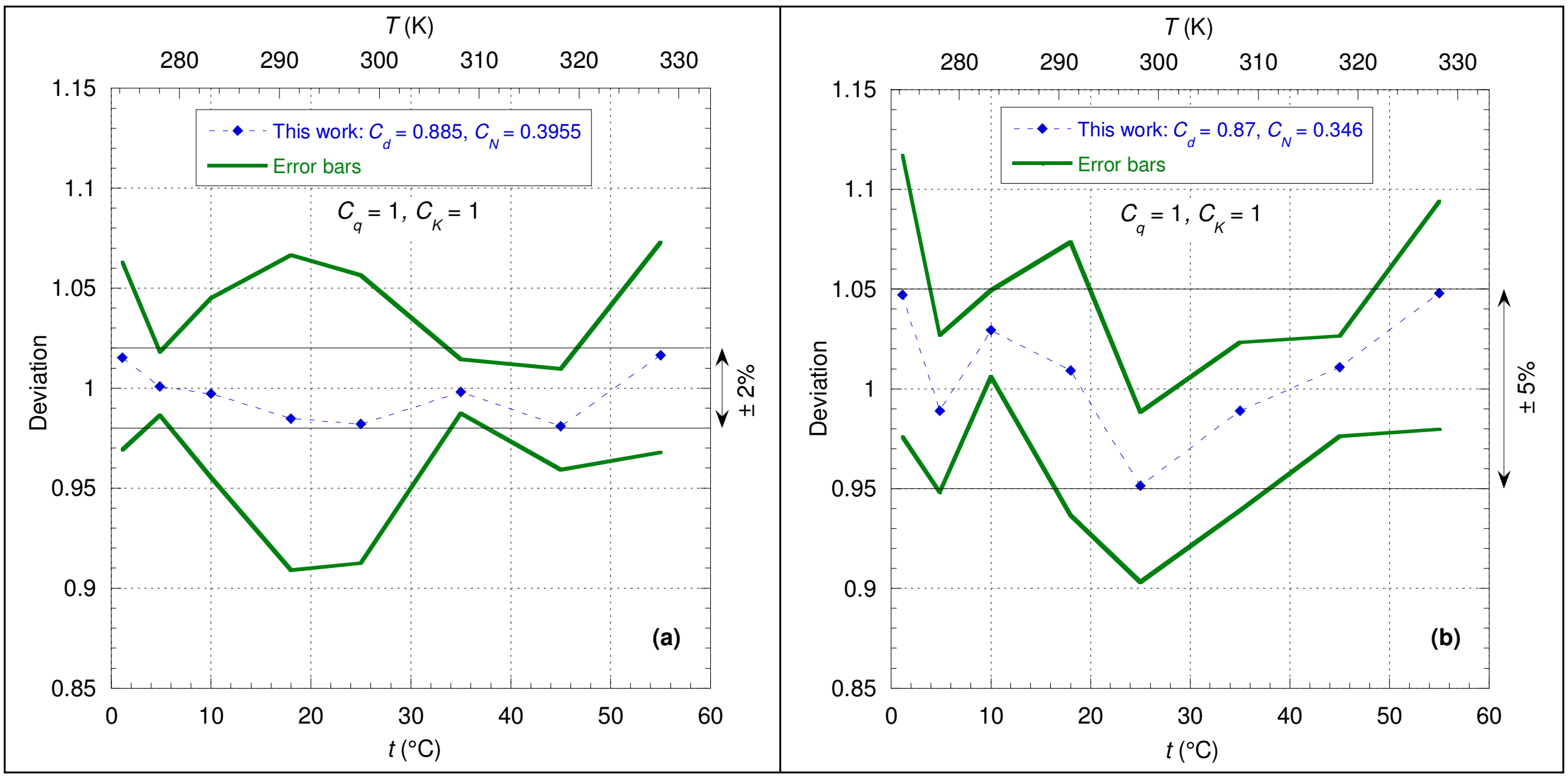

Fig. 89. Ratio of Wang *et al.*'s data for $H_2O^{18}$ tracer with the present modeling as a function of the temperature along the atmospheric isobar: (a) Table I from Ref. 70; (b) Table II from Ref. 71. The dashed lines are eye guides.

As Wang *et al.,* Easteal *et al.* (Ref. 63 and 72) carried out two different experiments between 1984 and 1989 but concerning the 1984 data, Easteal *et al.* wrote in 1989:

> "The previously obtained value [at 323.15 K] [Ref. 72] was some 3% lower, outside experimental uncertainty, and inadequate temperature control or insufficient stirring of the diaphragm cell compartments may be the cause of the discrepancy."

Therefore, we will not analyze the 1984 dataset.
An interesting feature of the 1989 paper is that Easteal *et al.* (Ref. 63) measured both HTO and $H_2O^{18}$ diffusion in the same cell. About their data, Easteal *et al.* wrote that:

> "The overall accuracy is estimated to be ±0.6-8% for the HTO results and ±0.2% for $H_2O^{18}$ or better at the lower temperatures."

Fig. 90 shows that the present modeling allows for reproducing HTO and $H_2O^{18}$ data with uncertainties compatible with those given by Easteal *et al.* Indeed, although the latter are higher than the estimated values, Easteal *et al.* have fitted their data by least squares with a logarithmic function and they have deduced that:

> "The maximum deviation of the fitted curve from the experimental data is ±1.0% $^3$HHO [i.e. HTO] values and ±0.6% for the $H_2O^{18}$ data. […] The two equations of the form of eqn (1) describe the diffusion behavior of the two tracer species in water over essentially the whole liquid range at atmospheric pressure to ±1.0%."

In other words, the best fit of Easteal *et al.* is equivalent to or even worse than the deviation obtained with the present modeling.

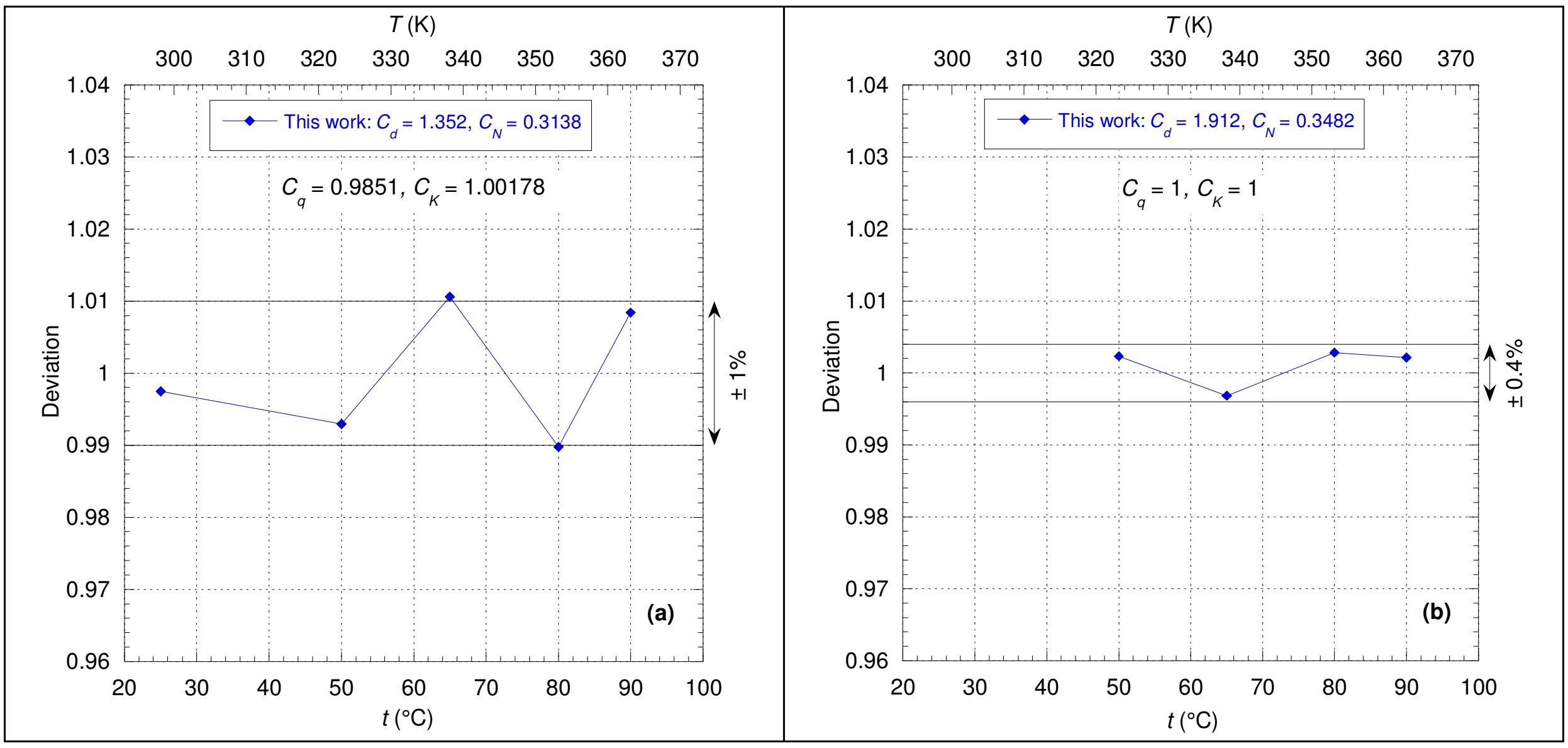

Fig. 90. Ratio of Easteal *et al.*'s data (Ref. 63) with the present modeling as a function of the temperature along the atmospheric isobar: (a) for HTO; (b) for $H_2O^{18}$. The lines are eye guides.

Easteal *et al.* then deduced from their measurements the following remark:

> "It may be seen that the percentage difference between $D_t(^3HH0)$ and $D_t(H_2O^{18})$ is not constant with changing temperature, although the molecular masses of $^3HHO$ [i.e. HTO] and $H_2O^{18}$ are the same. This indicates differences in the diffusional behavior of the two species with temperature due to effects of the position of the labelled atoms on intermolecular interactions."

With the present modeling, Easteal *et al.*'s remark can be written as follows: the temperature dependence is indeed different for these two tracers, in particular the coherence volume is different (i.e. $C_N$ is different): HTO generates larger basic units than with $H_2O^{18}$ but the coherence volume is smaller.
Finally, we observe that for $H_2O^{18}$ the $C_N$ coefficient of Easteal *et al.* is quite close to the $C_N$ coefficient of Wang *et al.* for the data corresponding to Fig. 89(b) while for HTO these coefficients are quite different between the different authors, in other words the temperature dependences differ significantly, hence the feeling of a large discrepancy between the authors.

Other types of tracers such as HDO have also been used but we do not present their analysis here because it does not bring anything more to our subject and these data can be reproduced just as well by the present modeling. From all these measurements Wang (Ref. 57) concluded in 1951 that:

> "Thus the present result supports the quasi-crystalline structure of liquid water, but contradicts the assumption that liquid water, at room temperatures, is a mixture of various definite species of associated water molecules in thermal equilibrium with each other."

For the present modeling, Wang's conclusion cannot be applied since the objects of the present theory can be interpreted as associated molecules that move randomly on a lattice and thus form a "structured" medium. Therefore, the model here includes the two characteristic properties that Wang tried to separate.

### *4.2.3. Self-Diffusion of Water under Pressure*

As with viscosity, we will start by analyzing the data on the saturated vapor pressure curve (SVP). Most of the data on SVP were obtained by NMR experiments. As Krynicki *et al.* (Ref. 73) wrote:

> "There also exist numerous measurements of $D_t$ for liquid water under its saturated vapor pressure (SVP) but only a few covered wide temperature ranges."

we will only analyze the few data that cover a wide temperature range.

Krynicki *et al.* (Ref. 73) performed measurements of the self-diffusion coefficient of normal water by NMR for temperatures between 275.2 and 498.2 K and at pressures up to 175 MPa. Data under pressure other than SVP will be analyzed later in the section. The work of Krynicki *et al.* contains both a table of smoothed values (i.e. Table 1) and a figure with raw experimental data (i.e. Fig. 1). Only the raw data will be analyzed in order not to be affected by the chosen smoothing method (i.e. relation (13)). Concerning their data, Krynicki *et al.* wrote that:

> "The general reproducibility of the results is estimated to be better than ±5%. The random error in measuring $D_t$, caused by the scatter of points on a graph corresponding to relation (l), increases from ≈1.5% at 298 K to 4% at 498 K."

Fig. 91 shows that the data can be reproduced on SVP with the estimated uncertainty. The large deviation of Krynicki *et al.*'s dataset was commented in the following words by Yoshida *et al.* (Ref. 75):

> "The $D_t$ values for $^1H_2O$ by Krynicki *et al.* [Ref. 73] are ~20% larger; the difference exceeds their experimental uncertainty (±5%). The difference can be ascribed to the convection effect in their apparatus."

For the data along SVP, the commentary of Yoshida *et al.* is exaggerated and we will see that this is also the case for the analysis of data under pressure although the deviation actually exceeds 5%.

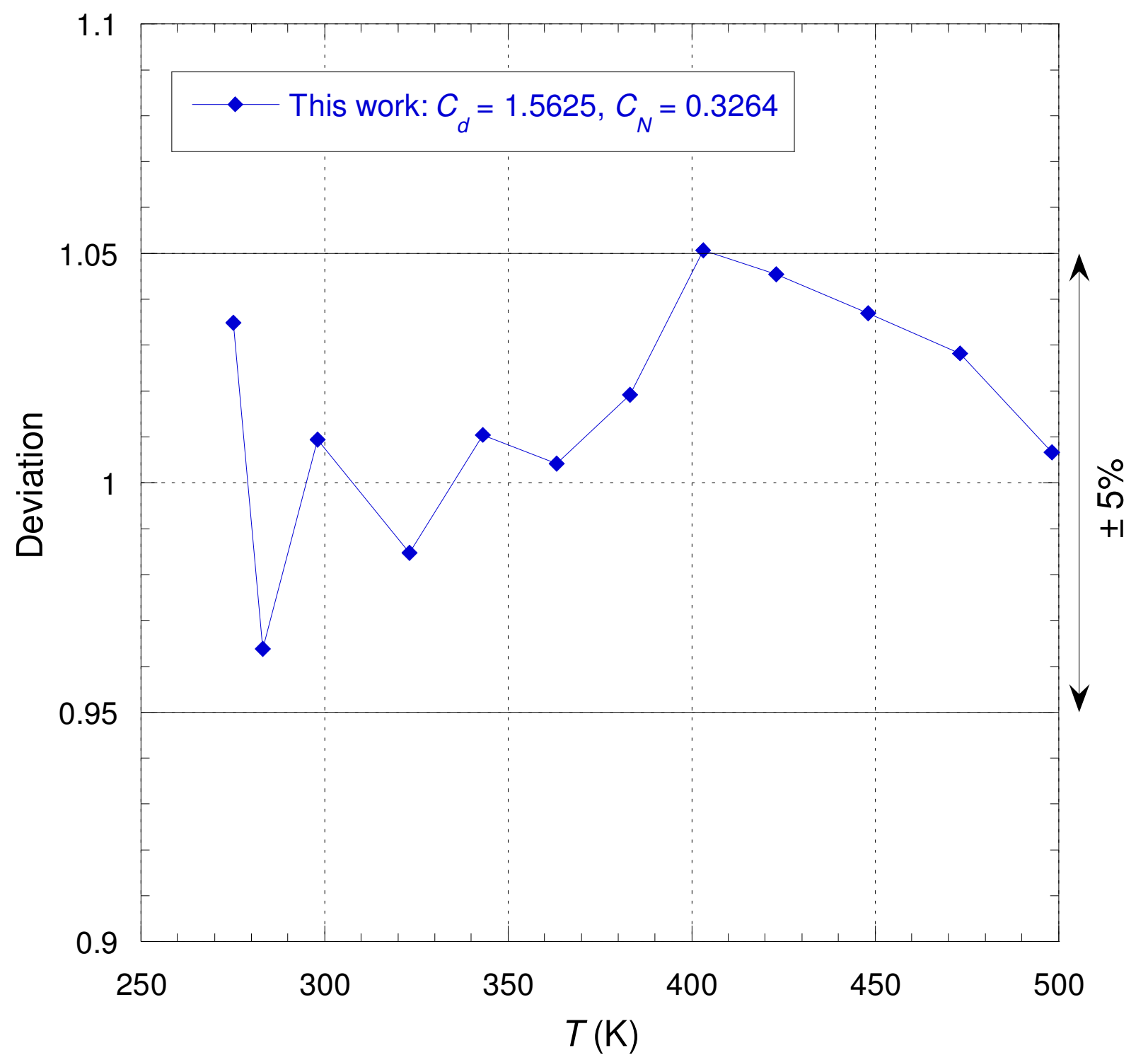

Fig. 91. Ratio of Krynicki *et al.*'s data (Ref. 73) with the present modeling as a function of the temperature along SVP. The lines are eye guides.

The analysis made by Krynicki *et al.* regarding their data leads them to write the following comment:

> "Using our values of $D_t$ from table 1 [i.e the smoothed data] and the literature viscosity data [e.g. Ref. 12], we find that at constant pressure $[D_t\eta/T] = (6.9 \pm 0.4) \times 10^{-15}$ N K$^{-1}$, i.e., $[D_t\eta/T]$ is constant within the limits of the experimental error."

Along SVP, only Korosi *et al.*'s data (Ref. 29) obtained with apparatus II are high enough in temperature to make a correct analysis (see Fig. 33). Fig. 92(a) shows that it is still acceptable to make the approximation $\eta \approx \eta_l$ along SVP at least up to 430 K so the quantity $D_t\eta/T$ depends mainly on the variation of $q_c$: the density decreases by about 16% between 283 K and 498 K (i.e. $q_{c0}(\rho)$ decreases by about 2.5%) but the function $f_{q_c}(\rho,T)$ increases by about 12% over the relevant temperature range as can be seen on Fig. 92(b). In the above quotation the experimental error considered is about 6% and the actual variations of $q_c$ are approximatively in accordance with this limit. In other words, Krynicki *et al.*'s comment is not false, but it is a very rough approximation.

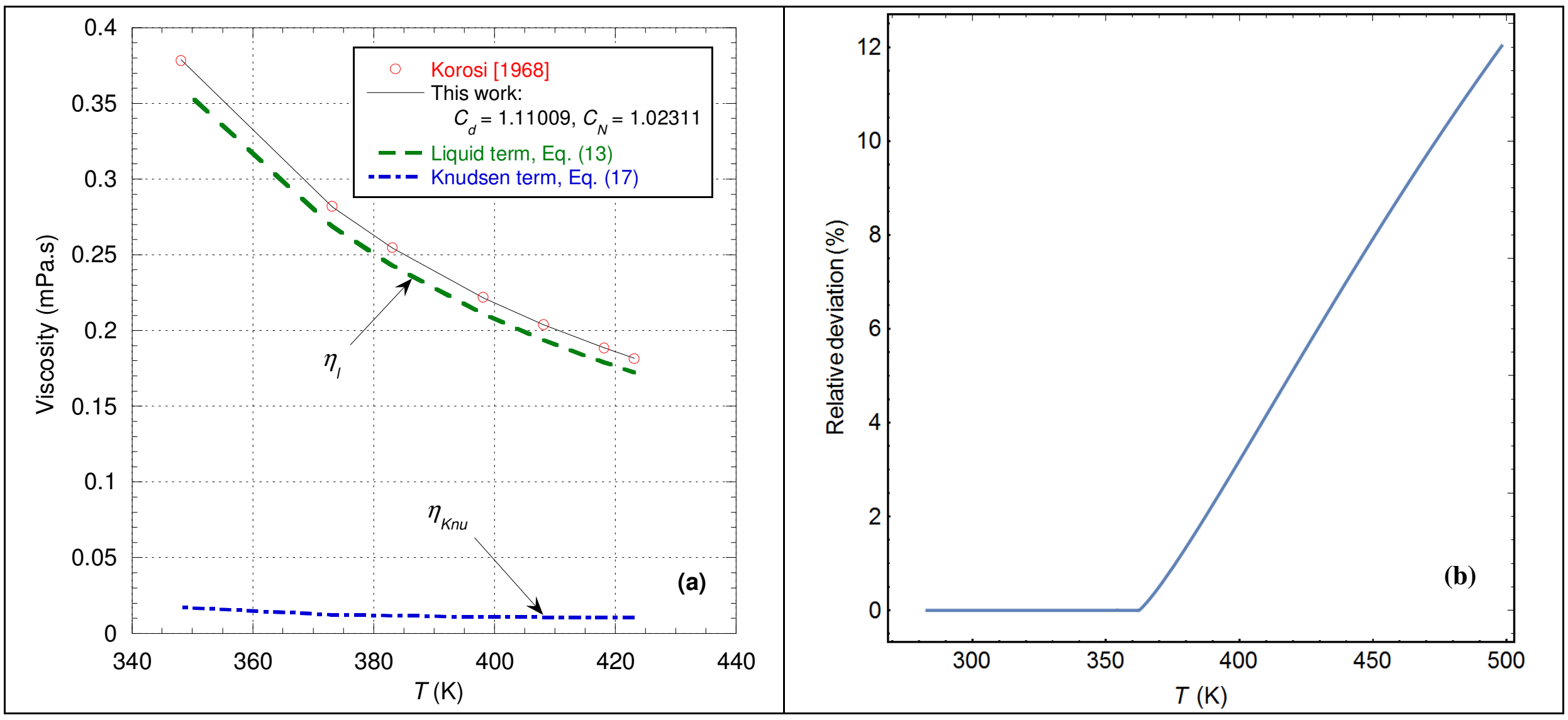

Fig. 92. (a) Water viscosity data of Korosi *et al.* (Ref. 29) and the corresponding viscous terms in the present modeling. (b) Percentage deviation of the function $f_{q_c}(\rho,T)$ from the unit value along SVP.

Simpson *et al.* (Ref. 74) NMR experiment covers a much narrower temperature range than those of Krynicki *et al.* For their experiment Simpson *et al.* used Pyrex tubes having 7 mm in length and an inner diameter of 3 mm. Simpson *et al.* give their results in the form of a table (i.e. Table I) but the values contained in this table are smoothed values as the authors indicate:

> "Five possible smooth curves have been drawn through the $D_t$ data [...]. The mean of the values read from these curves at each 5° interval is shown in Table I. The mean deviation for the values read from the $D_t$ curves is 3% [...]."

Fig. 93 shows that the present modeling makes it possible to reproduce the smoothed data of Simpson *et al.* with the expected uncertainty, but we can observe the smoothing effect, which produces a "beautiful" oscillation here that has no relation to noise on the raw data.
We note that the value of the $C_N$ coefficient is quite different from that of the Krynicki *et al.*'s data, and therefore the temperature dependencies are quite different, but it is difficult to conclude here on the relevance of this result because of the smoothing effect that is influencing the result.

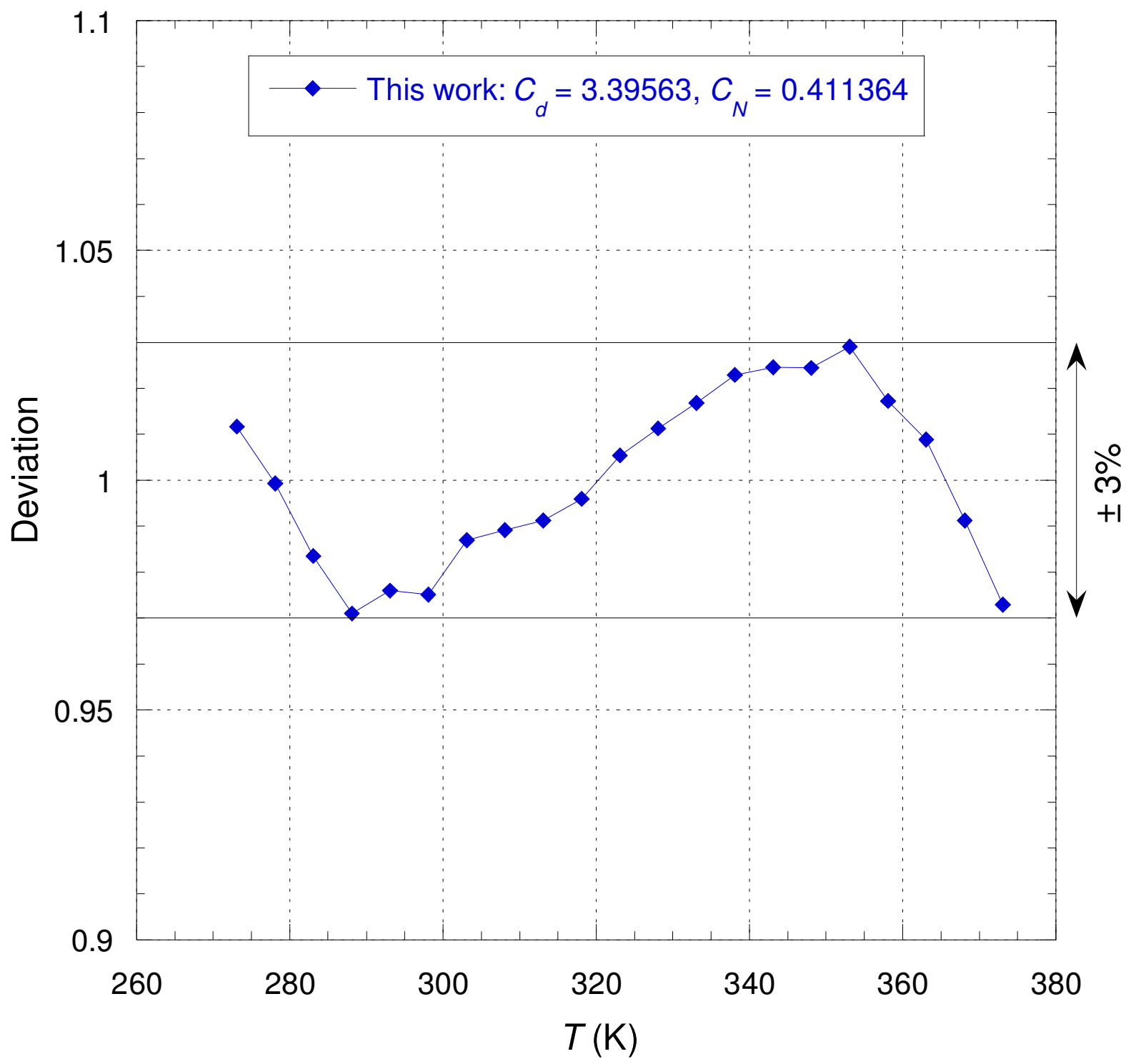

Fig. 93. Ratio of Simpson *et al.*'s smoothed data (Ref. 74) with the present modeling as a function of the temperature along SVP. The lines are eye guides.

The most accurate data are those of Yoshida *et al.* (Ref. 75) also obtained from a NMR experiment. Yoshida *et al.*'s data cover almost the entire temperature range corresponding to SVP curve but above all they are raw data and therefore more relevant than Simpson *et al.*'s data to the model analysis presented here. Yoshida *et al.* used quartz tubes having an inner diameter of 2.5 mm and concerning their measurements they wrote that:

> "The uncertainty of $D_t$ is ±0.5% in repeated measurements for a single sample and the total uncertainty is ±1% as shown in Table I."

Fig. 94 shows that the data can be reproduced with the expected uncertainty and, above all, it is observed that the deviation is well centered on the unit value. It appears that the value of the coefficient $C_N$ is very different from Krynicki *et al.* and Simpson *et al.*, so the temperature dependence differs significantly, hence the impression of a great disparity between the data.
With the analysis of their data, Yoshida *et al.* concluded that to satisfy the SE-type law, it was necessary to consider the factor 4π (i.e. corresponding to a slip boundary condition) and not not the factor 6π (i.e. corresponding to a stick boundary condition), which is consistent with Kisel'nick *et al.*'s comment in section 4.2.2. But in addition, to satisfy the SE-type law, Yoshida *et al.* have deduced that the hydrodynamic radius of a water molecule must vary so that:

> "In contrast to the previous results, the [hydrodynamic radius] *R* is not constant but dramatically increases with decreasing density above $\rho_n$=0.92(<150 °C), and the increasing rate becomes much weaker at the higher temperatures. The *R* increases from 1.6 to 2.0 Å as the density $\rho_n$ decreases from 1.00 (30 °C) to 0.80 (250 °C). The increase of the effective hydrodynamic radius can be interpreted as an indication of the strong effect of the short-range attractions between the solute (water)-solvent (water)."

In the framework of the present modeling, the interpretation is quite different and even goes in the opposite direction to that of Yoshida *et al.*: indeed, until about 360 K, we have $q_c \approx q_{c0}(\rho)$ which decreases slightly as the temperature increases along SVP, so the basic units of the theory grow slightly larger as for Yoshida *et al.* but above 360 K, the function $f_{q_c}(\rho, T_\sigma)$ increases faster than the decrease of $q_{c0}$, therefore the basic units of the theory decrease in size with the increase in temperature as one would expect for increasingly "gaseous" behavior.

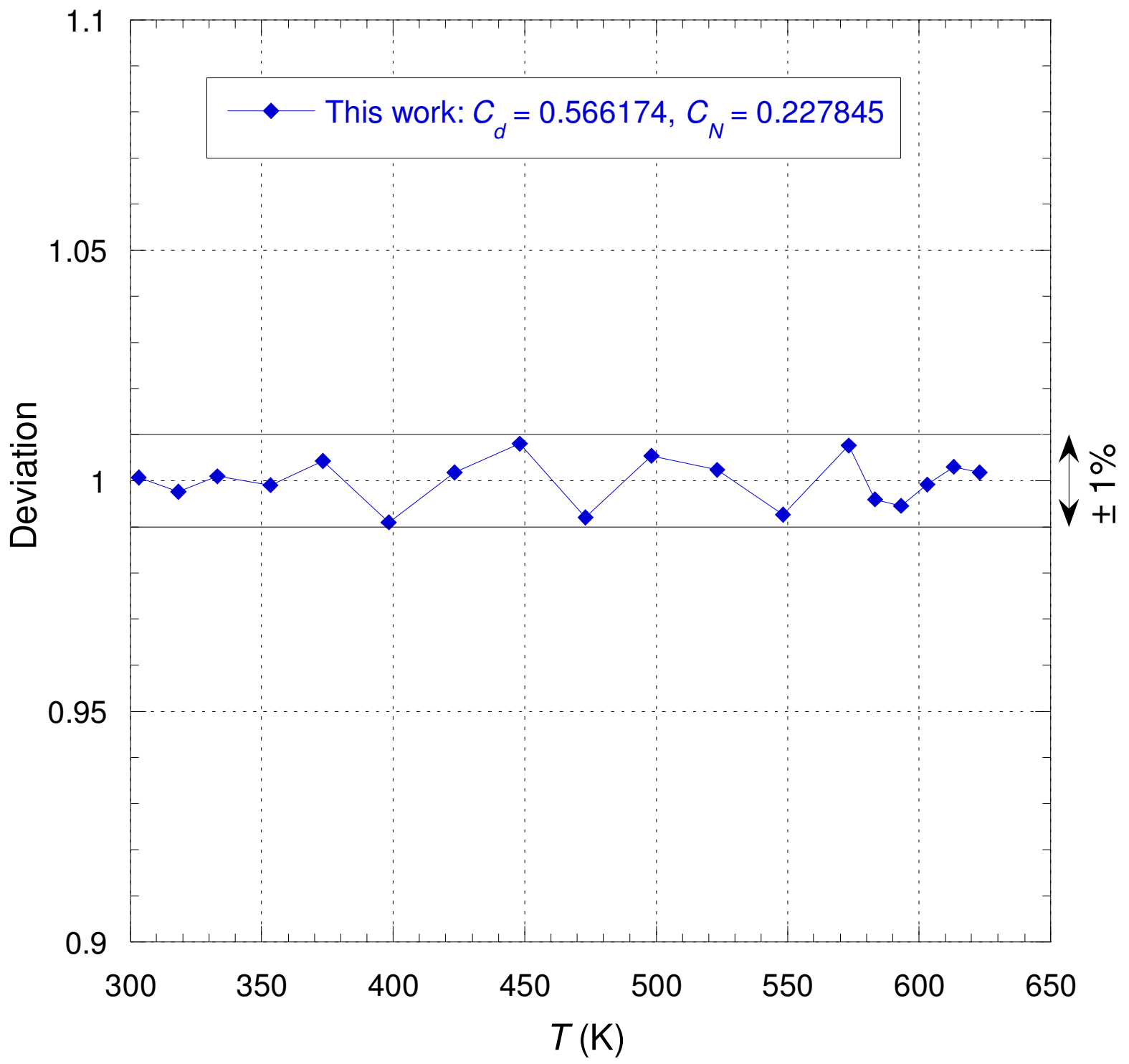

Fig. 94. Ratio of Yoshida *et al.*'s data (Ref. 75) with the present modeling as a function of the temperature along SVP. The lines are eye guides.

Hausser *et al.* (Ref. 76) also performed NMR-type experiments along SVP in a temperature range comparable to that of Yoshida *et al.* but with 10 times less accuracy as reported by the latter:

"The $D_t$ values for $^1H_2O$ by Hausser *et al.* are larger by ~10% (their uncertainty) than the present over the entire temperature range."

With an uncertainty as large as ±10%, the present modeling can reproduce these data but this does not provide anything more for the analysis.

We will now analyze the pressure datasets outside SVP. Indeed, there are diffusion coefficient data in the literature from different kinds of experimental methods but as Kisel'nick *et al.* (Ref. 58) rightly writes, we find that:

> “The data on the self-diffusion of water molecules at pressures above atmospheric is very scarce.”

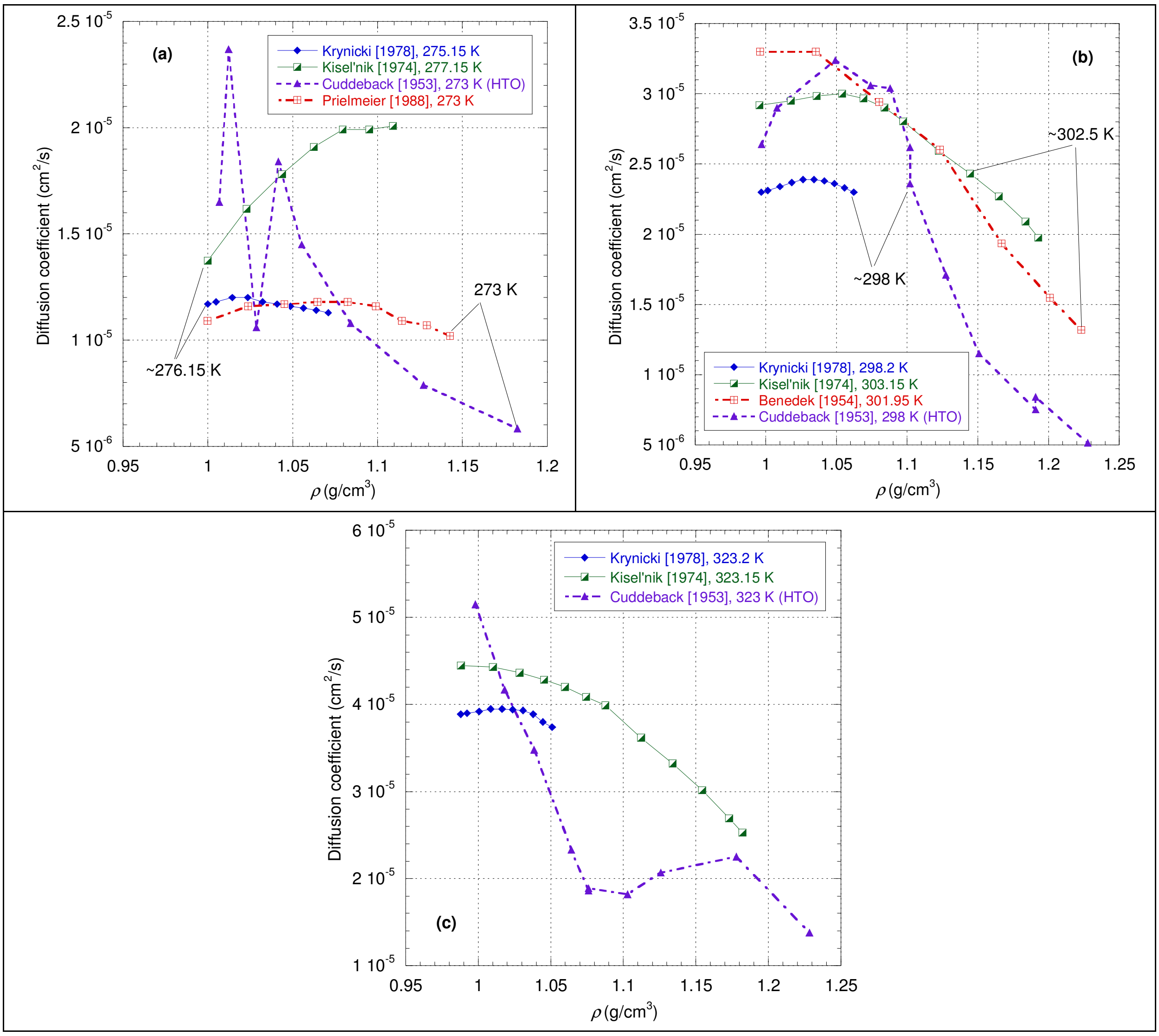

Fig. 95. Diffusion coefficients in water from different sources (Ref. 58, 73, 77 and 78) along 3 quasi-isotherms as function of the density: (a) ~275 K; (b) ~300 K; (c) ~323 K. The lines are eye guides.

Fig. 95 shows very clearly for three quasi-isotherms the very large discrepancy between the data. We observe that it is not only a shift problem but the variations are very different: it seems, for example, unthinkable to account for both the variation (i.e. in the opposite direction) of the data of Krynicki *et al*. (Ref. 73) and Kisel'nick *et al.* (Ref. 58) around 276 K; similarly the variation of Benedek *et al.*’s data (Ref. 77) seem incompatible with those of Krynicki *et al*. (Ref. 73) or Kisel'nick *et al.* (Ref. 58). Cuddeback *et al.*’s data (Ref. 78) concern HTO diffusion in water. Variations of these data appear to be systematically different from NMR water diffusion data. However, we saw in section 4.2.2 that at atmospheric pressure, HTO diffusion data were not so different from that of water. Cuddeback *et al.*’s data (Ref. 78) are therefore incompatible with these data at atmospheric pressure. Woolf (Ref. 79) has already written the following remark about these data:

> “The data of Cuddeback *et al.* [Ref. 78] for tritiated water are obviously wrong and the results of Kisel’nik *et al.* [Ref. 58] appear to be too high. […]

> The higher values of $D_t$ reported [by Kisel'nik *et al.*] probably are due in part to an error in the calibration procedure but the disagreement with our results in the pressure dependence must be due to additional features in the experimental procedures."

Therefore, on the examples given in Fig. 95, only the data from Krynicki *et al.* (Ref. 73) and Prielmeier *et al.* (Ref. 65) seem consistent with the different theoretical approaches. Prielmeier *et al.*'s data correspond to high density data entirely in the supercooled phase while those of Krynicki *et al.* have slightly lower densities but are entirely in the normal liquid phase. These two datasets are quite complementary and both correspond to NMR-type experiments.

Let's start by analyzing the low temperature Prielmeier *et al.*'s data (Ref. 65). As already noted, two types of data are present in Prielmeier *et al.*'s paper: raw data in the form of points on their Fig. 2 and smoothed data in their Table 1, and these two datasets do not contain the same number of data points. The data that belong to the 0.1 MPa isobar are also included in this analysis. We have also previously reported that the uncertainty associated with Prielmeier *et al.*'s data is estimated by the latter at ±3%. Fig. 96 shows the representation of the two datasets with the present modeling: we observe that the choice of the smoothing function is not judicious because it biases the data at high density and low temperature; in particular on the isotherm at 208.5 K, either there is an error in the transcription of the numerical values or the function defined to represent these data is badly chosen. In the case of raw data, it appears that it is possible to reproduce the latter with an uncertainty of ±5% which is clearly too high in relation to the estimated experimental uncertainty.

However, Prielmeier *et al.* indicated that:

> "The determination of $D_t$ was most conveniently accomplished by increasing $g$ while keeping all other variables constant. The coil constant was determined using the known self diffusion coefficient of water at ambient pressure and 298 K obtained by Mills [Ref. 68] which are generally acknowledged to be the most reliable values. The coil constant was controlled with a redetermination of $D_t$ for benzene."

This makes it possible to consider a renormalization of each isotherm by a factor $C$. Fig. 97(a) shows that by slightly renormalizing the isotherms corresponding to the raw data, these ones fall within a range of ±3% overall, with the exception of a few points. However, for the latter, since the corresponding smoothed point is in the band, this means that the uncertainty on these points can be understood as greater than 3%. In the same way, the smoothed data fall globally within this ±3% band with the exception of a few points, some of which belong to the isotherm at 208.5 K. We can see by observing the renormalization factors on Fig. 97(b) that the isotherm at 208.5 K of the smoothed data is problematic. The evolution of the renormalization factors with temperature shows that they fluctuate essentially around an average value and this fluctuation reflects the experimental noise of the device. In addition, we can see on Fig. 97(b) that the smoothing method has a shifting effect on the renormalization constant.

It should be noted that the parameters $C_d$ and $C_N$ are in all cases identical to those used to describe the data on the 0.1 MPa isobar (see Fig. 84). The $C_d$ coefficient represents an average of the different tubes used. Indeed, Prielmeier *et al.* mentioned that:

"The inner diameters varied between 200 and 70 μm. The tubes with the larger id. routinely withstood pressures up to 200 MPa, while a significant portion of the narrower tubes could be pressurized to 400 MPa."

It would therefore have been appropriate for the analysis to separate the data according to the tube used to obtain greater accuracy, but this is impossible to do in the absence of more detail.

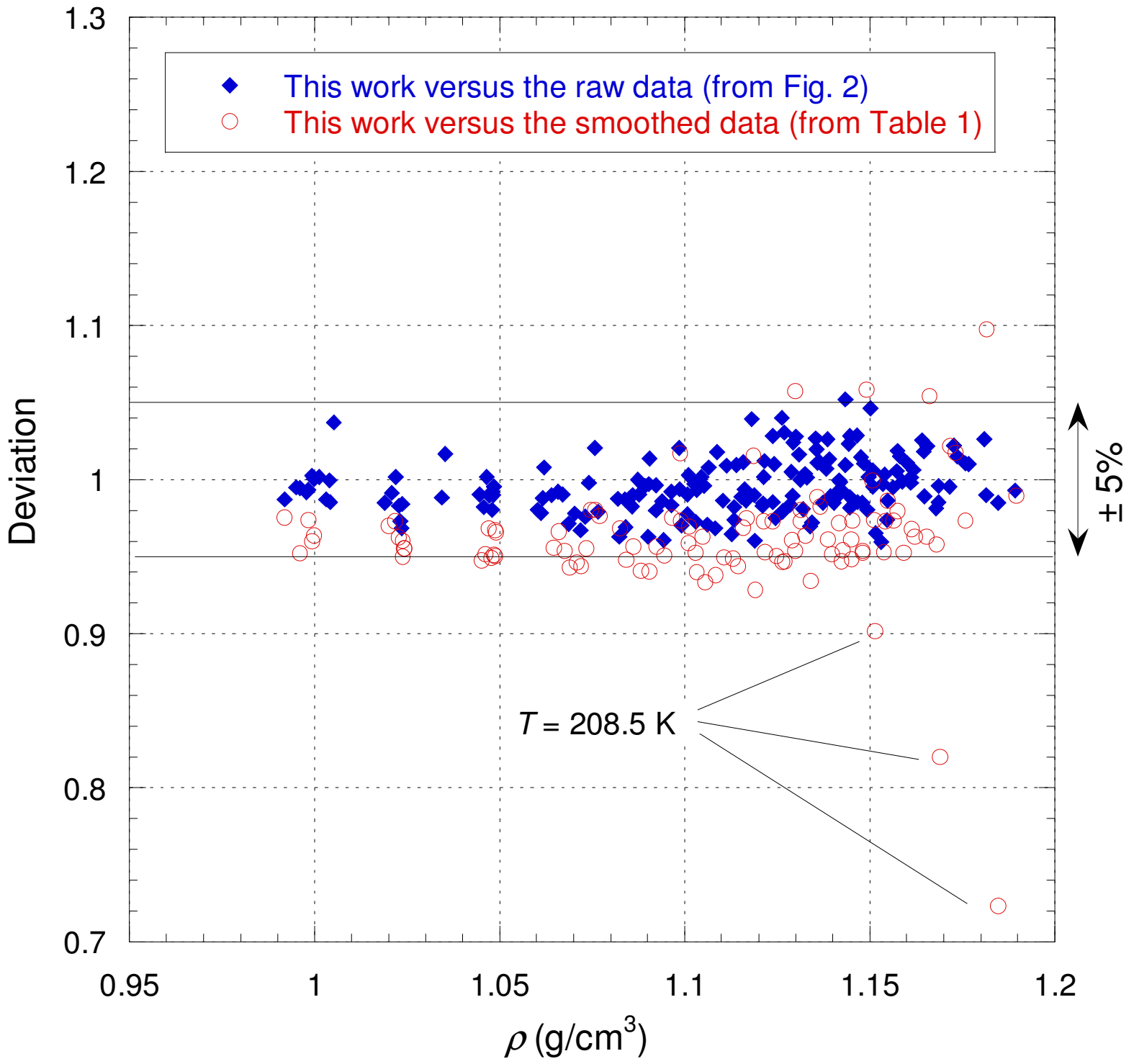

Fig. 96. Ratio of Prielmeier *et al.*'s data (Ref. 65) with the present modeling ($C_d$ = 0.87273, $C_N$ = 0.27738) as a function of the density for all the isotherms.

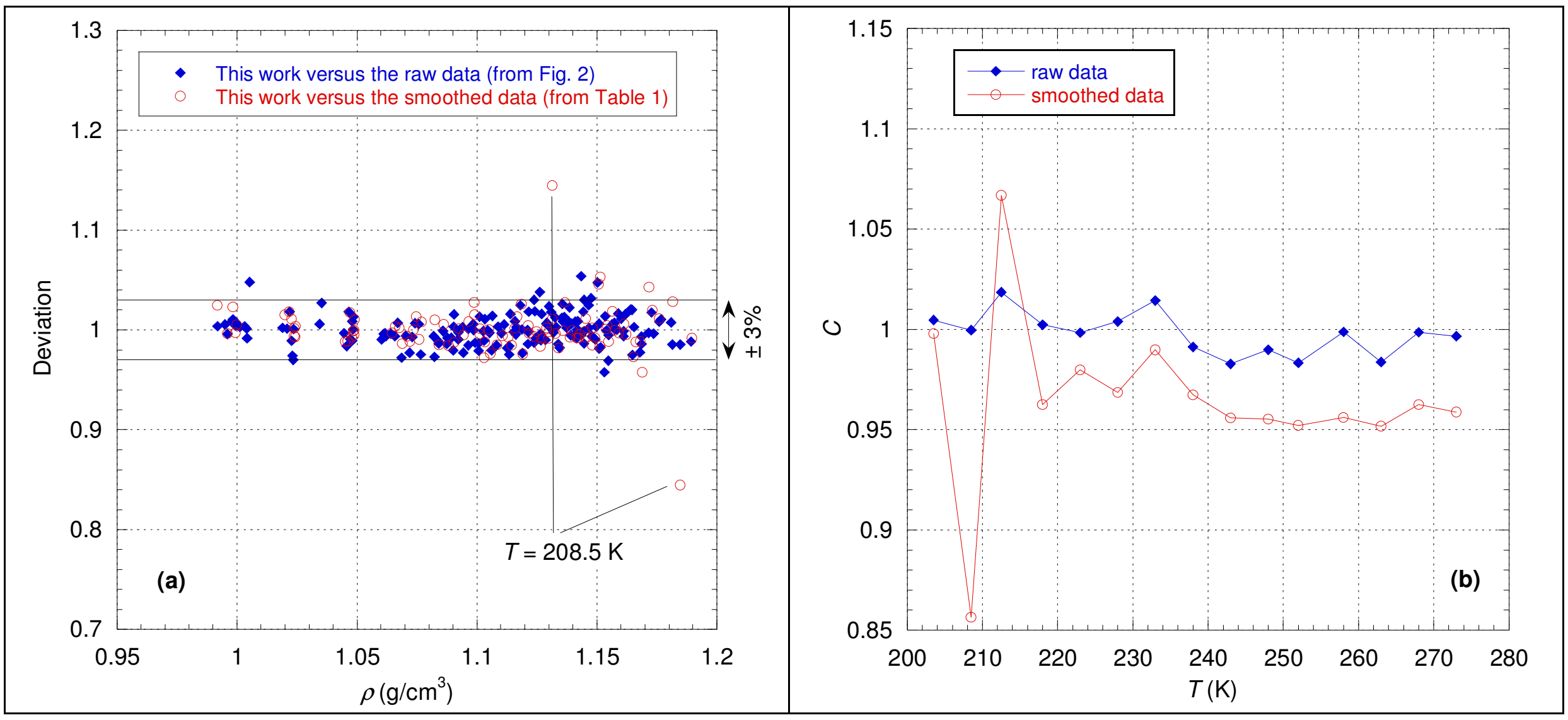

Fig. 97. (a) Ratio of Prielmeier *et al.*'s data (Ref. 65) renormalized with the present modeling ($C_d$ = 0.87273, $C_N$ = 0.27738) as a function of the density for all the isotherms. (b) Renormalization factors as function of the temperature. The lines are eye guides.

Angell *et al.* (Ref. 80) performed an NMR experiment similar to that of Prielmeier *et al.* but covers a lower density region. The diameter of the tubes used by Angell *et al.* (i.e. 300 µm inner diameter) is slightly larger than the largest Prielmeier *et al.*'s tubes. Regarding the accuracy of their data, Angell *et al.* wrote that:

> "We estimate that the ucertainty in the diffusion constants arising from the necessity to extrapolate the coil calibration to low temperatures is ± 3% at - 5°C, increasing to ± 6% at - 20°C."

Fig. 98 shows that the data can be reproduced exactly with the uncertainty given by Angell *et al.*, i.e. there is a drift when the temperature decreases. However, this drift was identified by Angell *et al.* as being due to:

> "The gradient coil was calibrated against the HTO tracer diffusion data of Woolf [e.g. Ref. 79] and Mills [Ref. 68] at several temperatures between 2 and 20°C and several pressures between 1 and 2000 bar. A small and nearly linear dependence upon pressure and temperature was observed for the gradient coil calibration; this dependence is probably attributable to thermal expansion and compression of the Teflon coil former."

This drift can therefore be corrected by renormalizing each isotherm by a factor $C$. Fig. 99(a) shows that under these conditions all isotherms fall within the ±3% band but what is more important is that the density dependence of each of these isotherms is now well reproduced, which means that the drift of $g$ with pressure is much lower than with temperature. Since the renormalization factors $C$ represent the drift of $g$, they can be represented as a function $f_{g,\mathrm{Angell}}(T)$ as defined by Eq. (29) but for isotherms. This function is shown in Fig. 99(b) and is compatible with a linear drift of $g$ as indicated by Angell *et al.* Fig. 99(b) shows here that the effect of temperature is to slightly increase the gradient coil calibration constant $g$ with respect to the estimated one at the lowest temperatures.

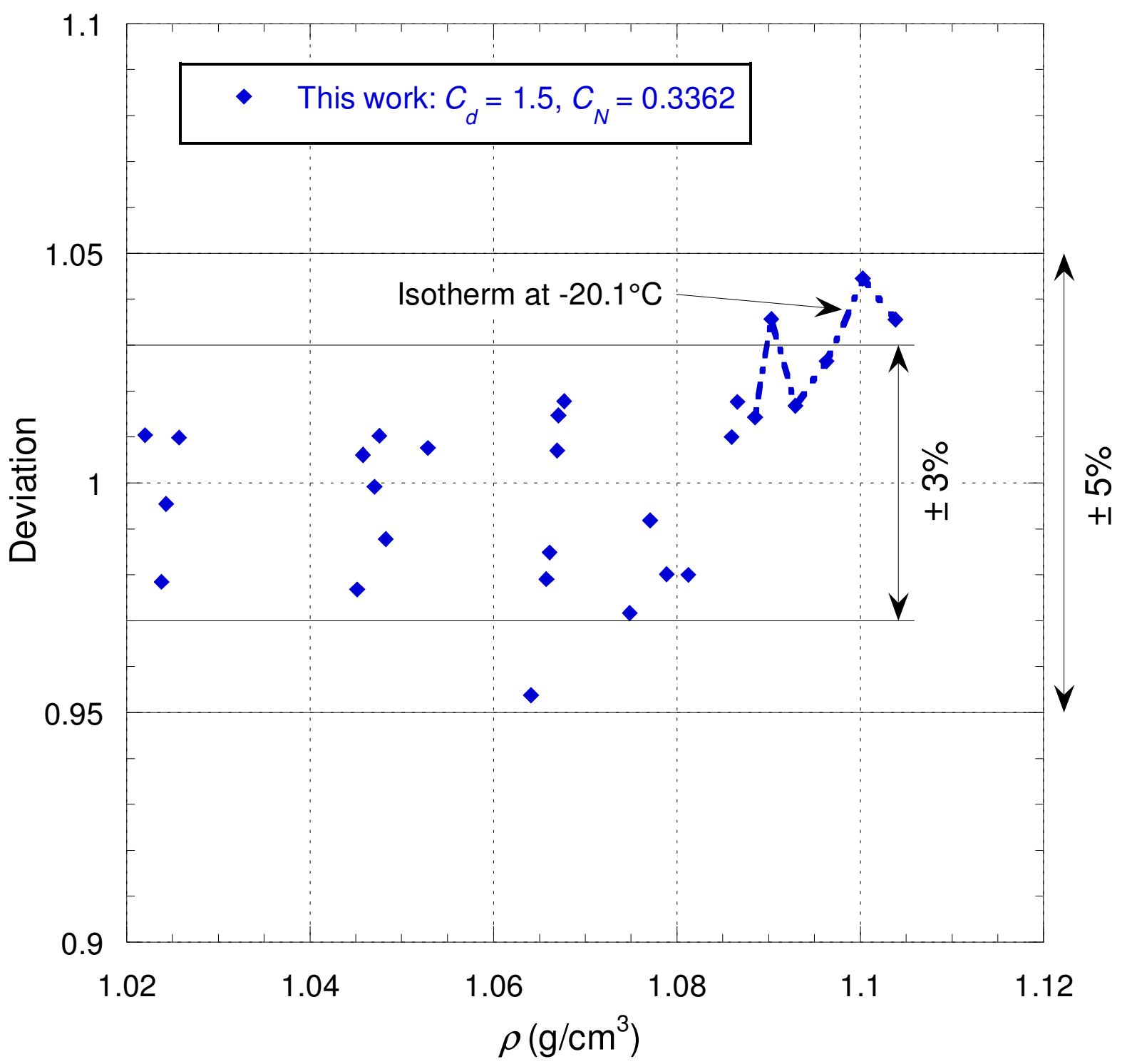

Fig. 98. Ratio of Angell *et al.*'s data (Ref. 80) with the present modeling as a function of the density for all the isotherms.

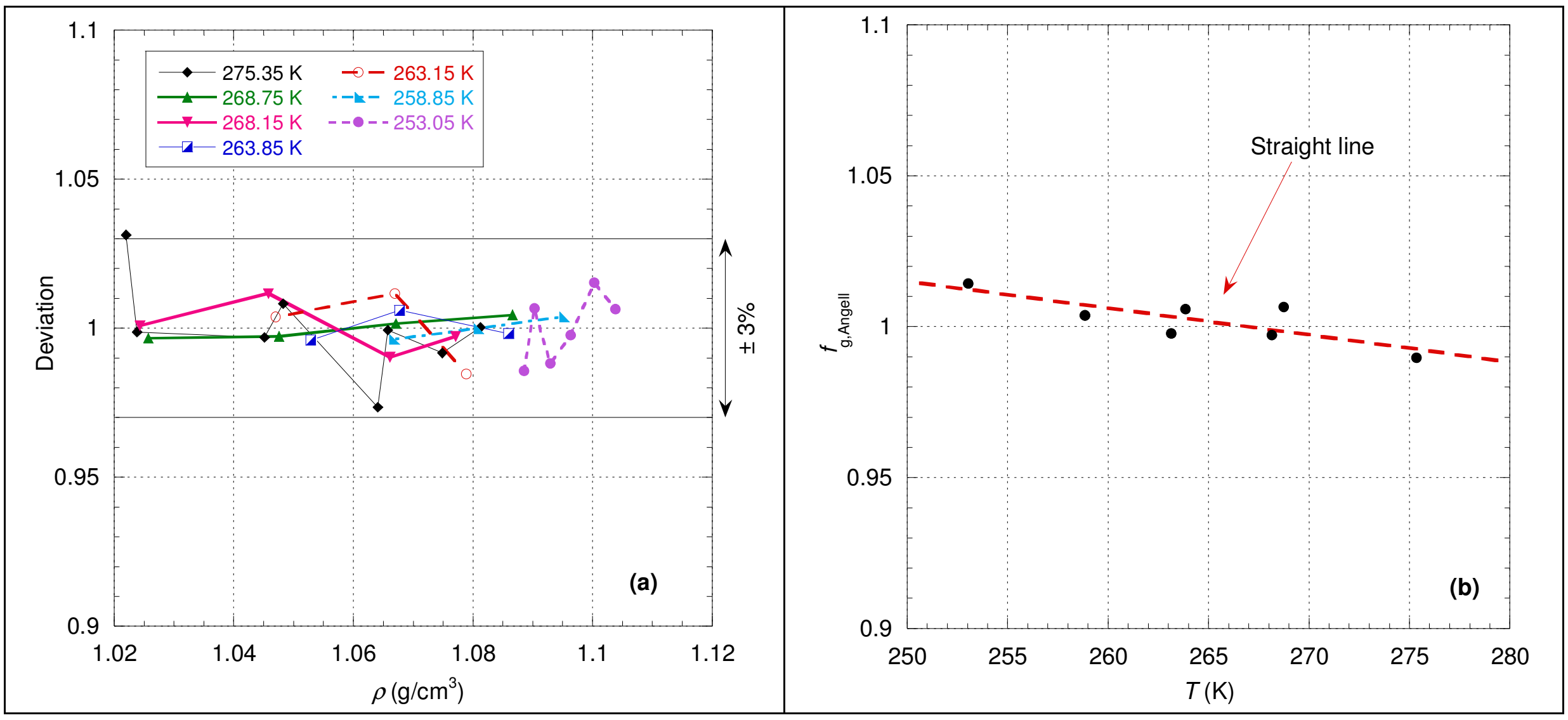

Fig. 99. (a) Ratio of Angell *et al.*'s data (Ref. 80) with the present modeling ($C_d$ = 1.5, $C_N$ = 0.3362) as a function of the density for each isotherm. The lines are eye guides. (b) Renormalization factors showing the drift as a function of the temperature of the gradient coil calibration constant $g$.

The same kind of experiment as Angell *et al.* (Ref. 80) in a similar temperature and pressure range was performed by Harris *et al.* (Ref. 66). The data given by Harris *et al.* in their Table 1 are only quasi-isotherms and the authors wrote about these data that:

"Results obtained with this cell had an uncertainty of ±(3 to 7)%, […]".

Fig. 100 shows the deviation obtained for each of the quasi-isotherms. On the one hand, it can be observed that variations with density are correctly reproduced and on the other hand that a progressive shift according to the temperature of the quasi-isotherms is observed and more particularly for the highest temperatures. The shift for the three of the largest temperatures can be easily understood by referring to the calibration method described by the authors:

> "The gradient coil was calibrated using the reference values for the self-diffusion coefficient of water established at 0.1 MPa by Mills (1973)."

Harris *et al*. used the extrapolated Mills' data for water (Ref. 68) to calibrate their device at atmospheric pressure at temperatures above 274 K only since there are no Mills' data below this temperature. Since Mills' experiment is different from Harris *et al*.'s, the shift must be corrected by a renormalization factor $C$ that counterbalances the apparatus constant imposed at least for the three highest temperatures. Fig. 101(a) shows that taking into account renormalization factors, uncertainty is then within ±2%, which is in agreement with the experimental uncertainty estimated by Harris *et al*. In addition, Fig. 101(b) shows that the renormalization factors tend to increase linearly when the temperature decreases as for other similar experiments.

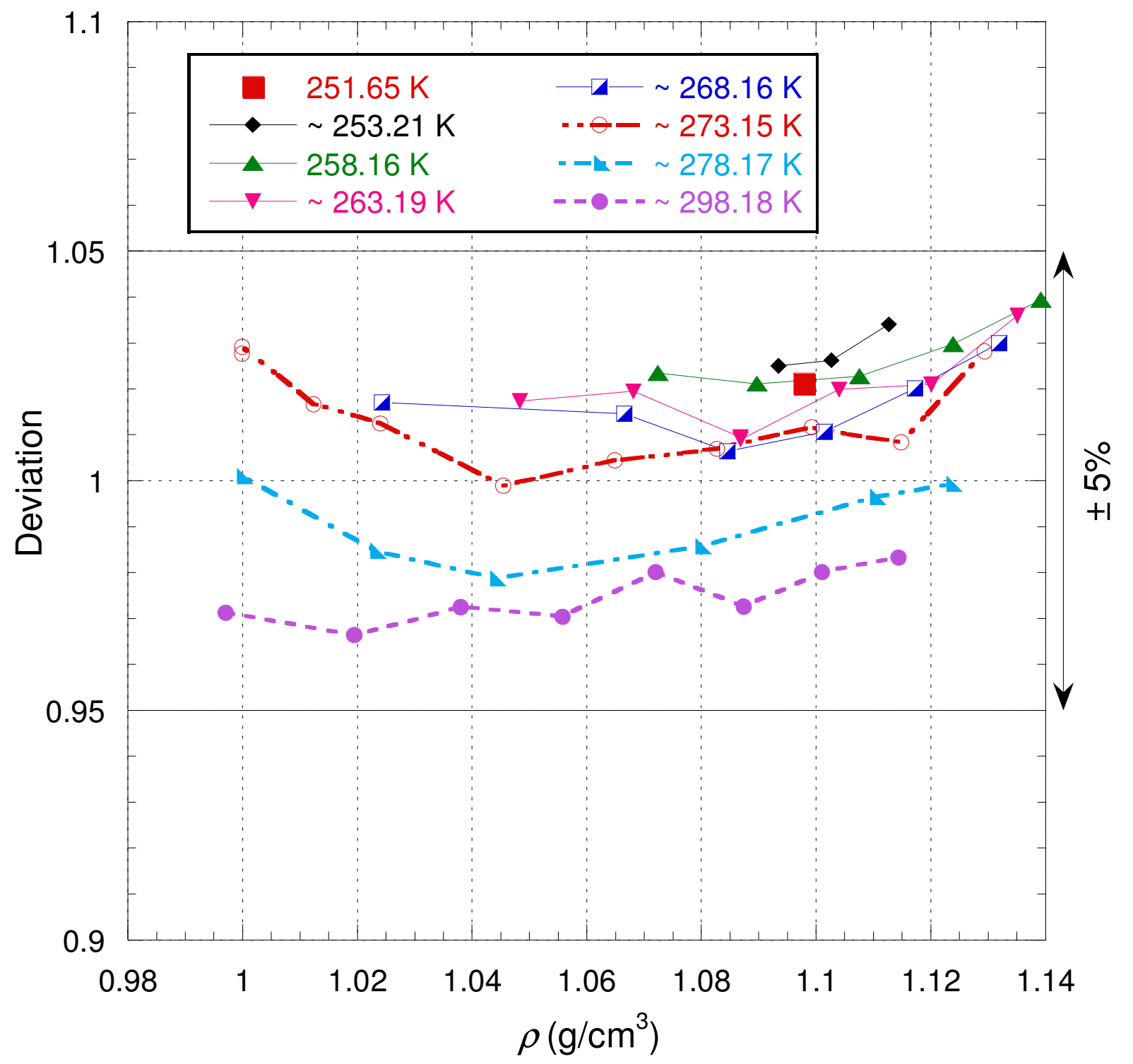

Fig. 100. Ratio of Harris *et al*.'s data (Ref. 66) with the present modeling ($C_d$ = 1.824, $C_N$ = 0.3565) as a function of the density for each isotherm. The lines are eye guides.

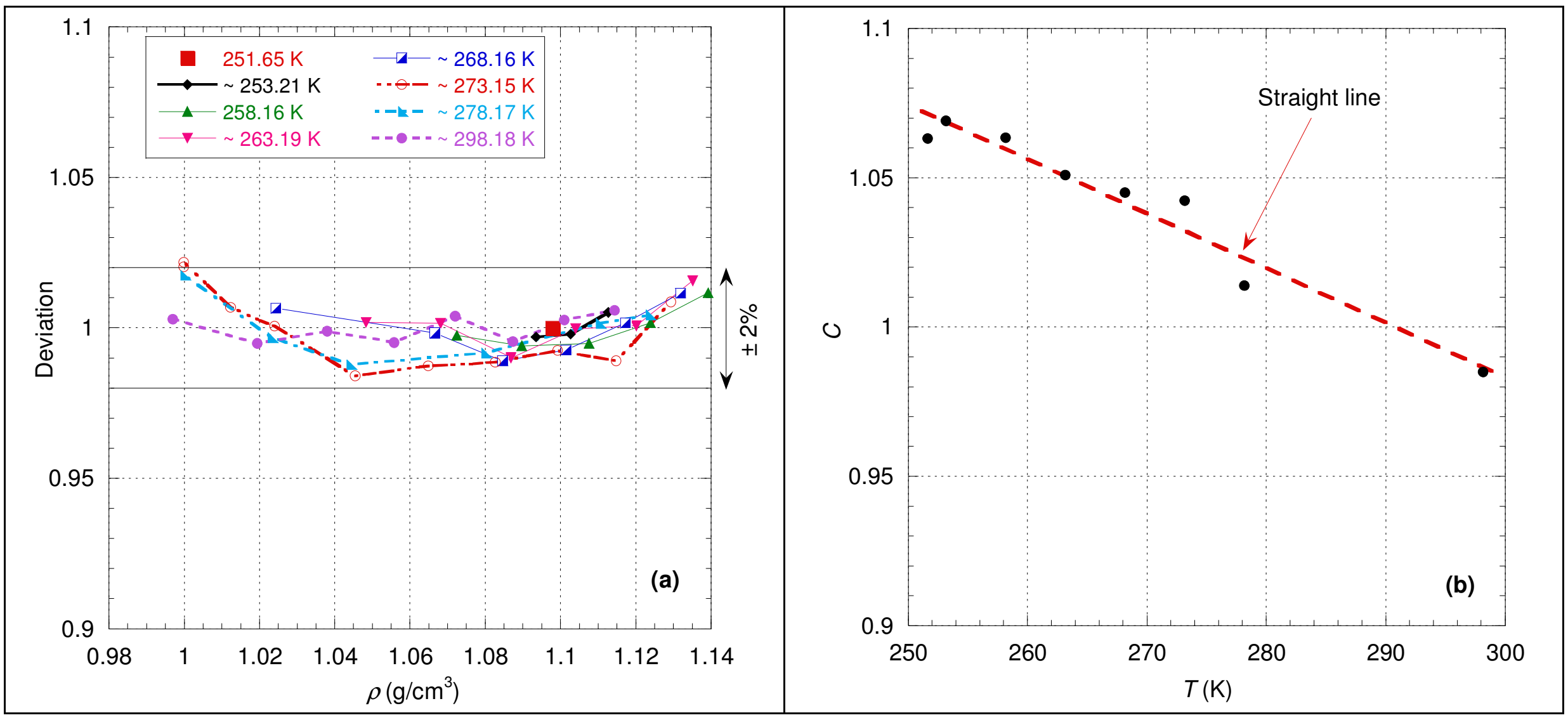

Fig. 101. (a) Ratio of Harris *et al.*'s data (Ref. 66) with the present modeling ($C_d$ = 1.824, $C_N$ = 0.3565) as a function of the density for each isotherm. The lines are eye guides. (b) Renormalization factors showing the drift as a function of the temperature of the calibration constant.

It appears that the values of $C_N$ coefficients are quite different from each other between the experiments of Prielmeier *et al.* (Ref. 65), Angell *et al.* (Ref. 80) and Harris *et al.* (Ref. 66) which shows the density dependence differs in all three cases and these differences were resumed by Harris *et al.* in the following way:

> "Satisfactory agreement was obtained with the results of Angell *et al.* (1976), though their data show more scatter (Figure 3). Good agreement was obtained with the pulsed gradient emulsion results of Prielmeier *et al.* (1987, 1988) (Figure 4), except above 200 MPa at -10 °C, where their data lie about 7-9% below ours, twice the sum of the estimated experimental errors."

We analyze now the data under pressure in normal liquid water, starting with Krynicki *et al.*'s data (Ref. 73) who realized measurements of the self-diffusion coefficient for temperatures between 275.2 and 498.2 K and at pressures up to 175 MPa. We have already mentioned that Krynicki *et al.* estimate that the uncertainty about their data is better than ±5%. As already noted, two kinds of data are present in Krynicki *et al.*'s paper: raw data in the form of points on their Fig. 1 and smoothed data in their Table 1. The previously processed SVP data are also included in the dataset analyzed here. We will therefore carry out as before for Prielmeier *et al.* (Ref. 65) a simultaneous analysis of the two types of data that reveal the influence of the smoothing function. Fig. 102 shows the comparison between the present modeling and the empirical relation Eq. (13) developed by Krynicki *et al.* to represent their raw data. In both cases, it appears that the uncertainty is about the same but twice as large as that estimated by Krynicki *et al.* In the case of the present modeling, the greatest deviation is obtained for the points corresponding to the highest densities, i.e. the coldest isotherms. However, we have shown in Fig. 95(a) that for very close temperatures around 0 °C, Krynicki *et al.*'s data are very different from those of Prielmeier *et al.* Since the present modeling represents very correctly Prielmeier *et al.*'s data, it is therefore not surprising to find such a deviation with Krynicki *et al.*'s data.
Given the fact that:

> "[…] the uncertainty in the field gradient calibration constant being given as ±10 %."

as it was written by Harris *et al.* (Ref. 81) about these data, Fig. 103 shows once again that by slightly renormalizing each isotherm by a constant $C$, the present modeling makes it possible to reproduce these data within the expected experimental uncertainty except for few points on the coldest isotherms. For warmer isotherms, the present modeling reproduces well the variations with density. Here the constant $C$ does not correspond to a systematic drift with temperature but rather represents noise in the calibration constant and therefore it is not useful to represent it on a figure.

It should be finally noted that the coefficients $C_d$ and $C_N$ are indeed identical here to those used in Fig. 91.

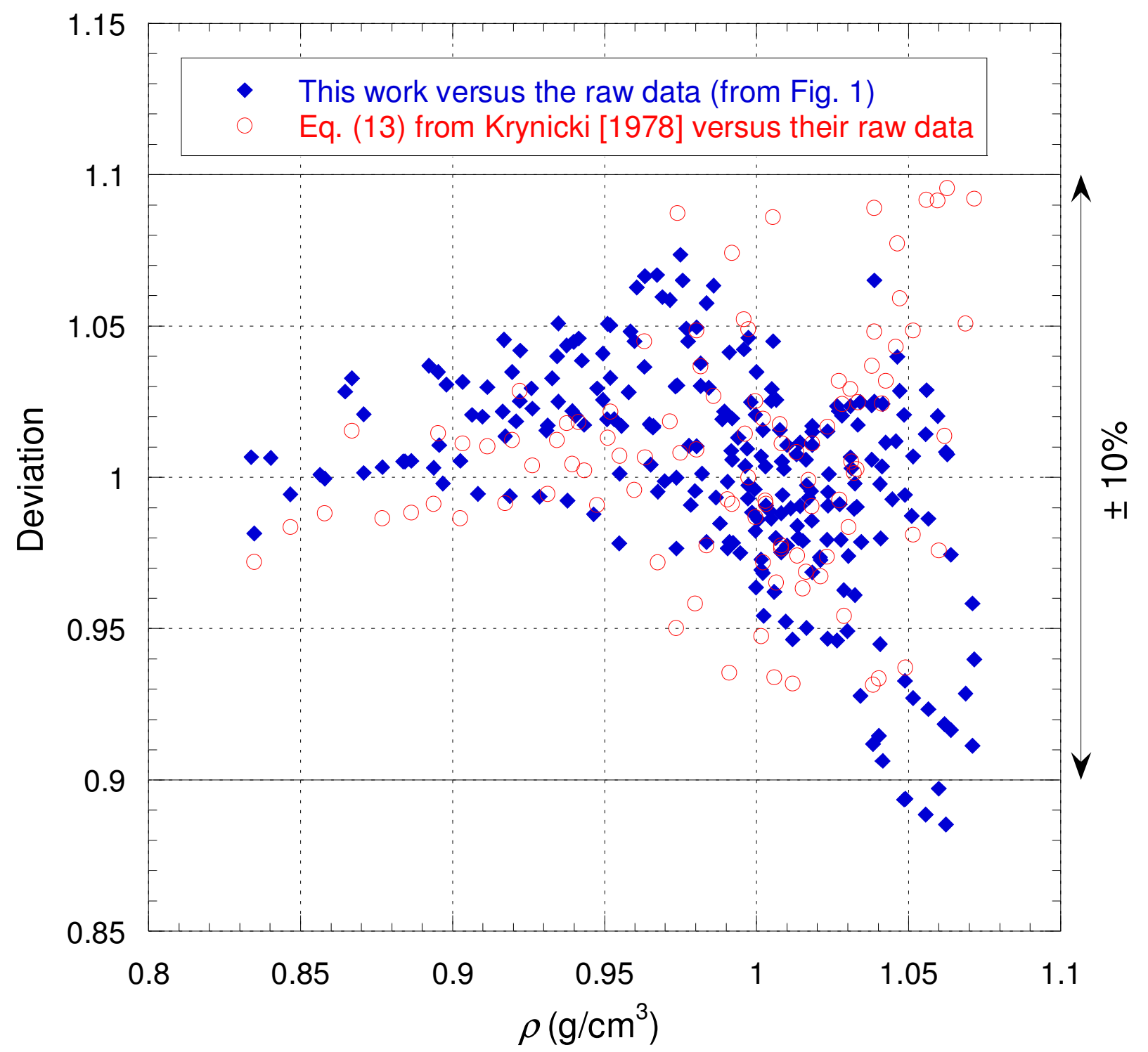

Fig. 102. Ratio of Krynicki *et al.*'s raw data (Ref. 73) with the present modeling (blue diamonds, $C_d = 1.5625$, $C_N = 0.3264$) and the empirical relation Eq. (13) developed by Krynicki *et al.* (red circles) as a function of the density for all the isotherms.

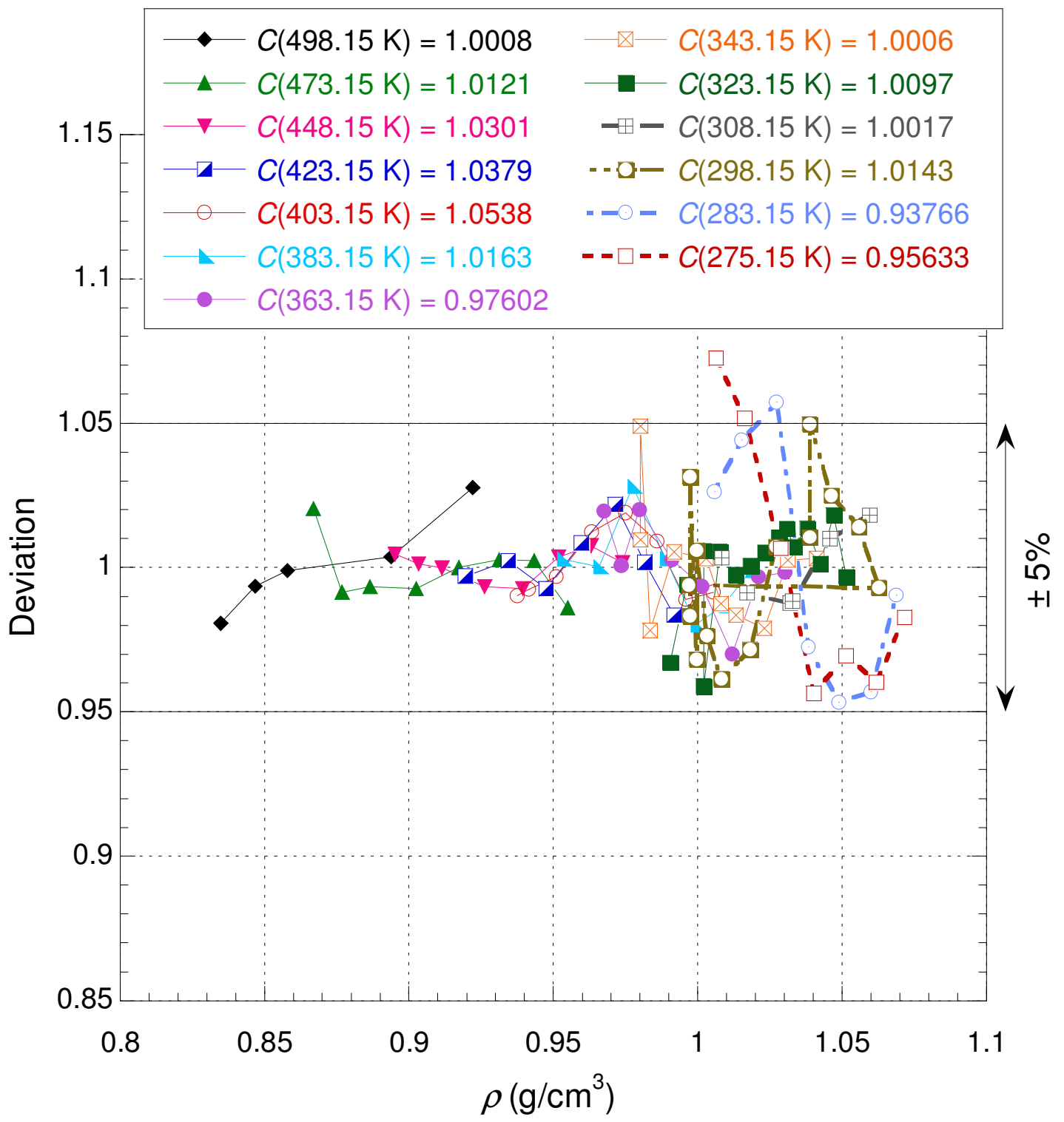

Fig. 103. Ratio of Krynicki *et al.*'s raw data (Ref. 73) with the present modeling ($C_d$ = 1.5625, $C_N$ = 0.3264) as a function of the density for each isotherm. The lines are eye guides.

Harris *et al.* (Ref. 81) also performed NMR-type measurements with a wider pressure range than Krynicki *et al.* (Ref. 73) but over a narrower temperature range. Therefore, the corresponding densities are generally more in the range of 1 to 1.1 g/cm$^3$, i.e. in the region where the deviations with Krynicki *et al.* are the highest. The authors point out that the interest of these data lies in the fact that:

> "The present results, which have a precision of ±1 %, are more accurate than earlier n.m.r. measurements."

In the experimental section the authors also wrote that:

> "[…] the agreement with tracer data obtained with diaphragm cells is ±1 %, which is within the limits of precision of the n.m.r. results. The overall accuracy of the latter is estimated to be ±2 %."

Fig. 104(a) shows the deviation obtained for each of the isotherms. On the one hand, it can be observed that variations with density are correctly reproduced and on the other hand that there is a progressive shift with the temperature of the isotherms. This shift is entirely due to the calibration of the device as we will see. The authors tell us that:

> "In this work a ceramic coil former was used and the coil was calibrated using atmospheric pressure self diffusion coefficients for benzene and water."

What is not said here is that the data for water used at atmospheric pressure were those determined by Mills (Ref. 68) by extrapolation at zero concentration of HTO self-diffusion coefficient data (see Fig. 86). Indeed, Fig. 104(b) shows that if the points at atmospheric pressure are treated with the parameters determined for the Mills' experiment, then the deviation becomes almost horizontal and compatible with the experimental uncertainty. In other words, since Harris *et al.*'s experiment is different from that of Mills, it is necessary to renormalize each isotherm with a constant that counterbalances the calibration from Mills' data. Fig. 104(c) shows that by renormalizing each isotherm by a constant, the uncertainty obtained is then compatible with that expected. It is interesting to note here that the coefficients $C_d$ and $C_N$ used are identical to those used to describe Harris *et al.*'s data (Ref. 66) in the supercooled phase (see Fig. 100).

It can be seen that for isotherms that are almost identical with those of Krynicki *et al.* (see Fig. 103), the variation with density is much better reproduced here with an uncertainty at least twice as low. This supports the idea that the low temperature data from Krynicki *et al.* are very imprecise. This fact was already noticed by Harris *et al.* (Ref. 66) in the following form:

> "The lowest isotherms of Krynicki *et al.* (1980) are also shown in Figure 2, but there are substantial differences between these data and those of Prielmeier *et al.* and ourselves, the results peaking at pressures lower than are found here and declining more rapidly with increasing pressure".

Finally, we observe that the value of $C_N$ is different for the dataset of Krynicki *et al.* and Harris *et al.* (Ref. 81), which indicates that the density dependence differs between these two experiments.

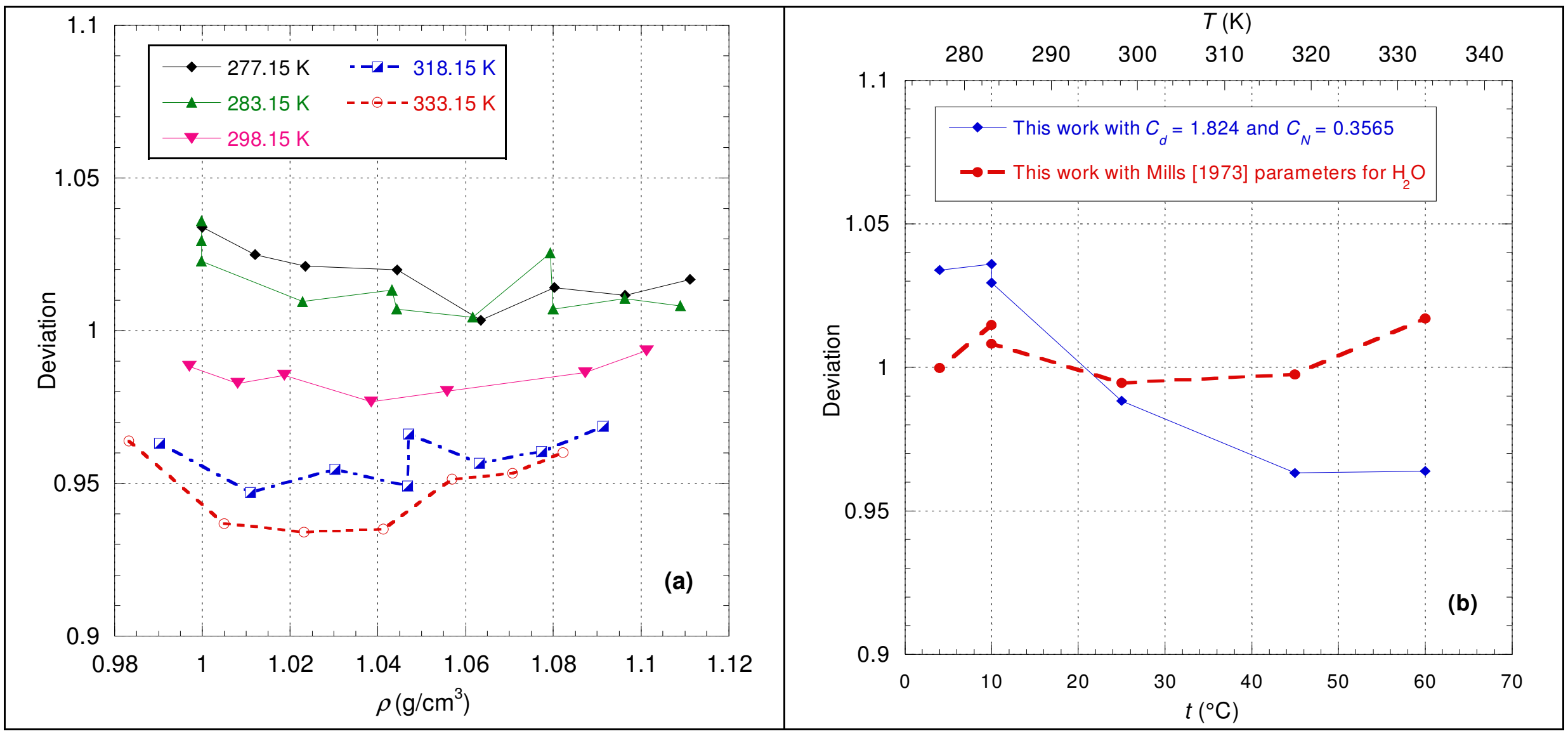

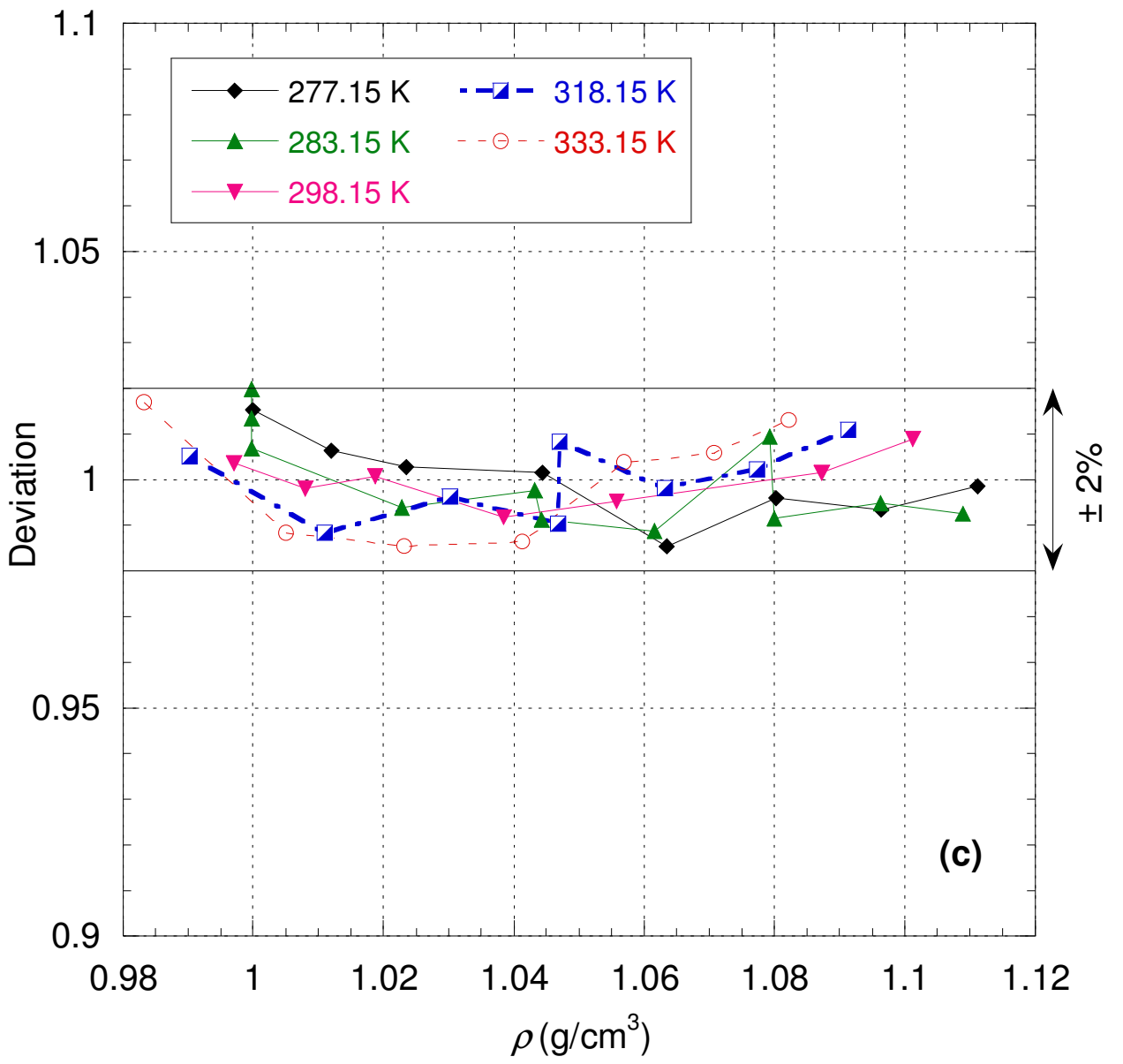

Fig. 104. (a) Ratio of Harris *et al.*'s data (Ref. 81) with the present modeling ($C_d$ = 1.824, $C_N$ = 0.3565) as a function of the density for each isotherm; (b) Ratio of Harris *et al.*'s data (Ref. 81) with the present modeling for two sets of parameters $C_d$ and $C_N$ along the atmospheric isobar; (c) Ratio of Harris *et al.*'s renormalized data (Ref. 81) with the present modeling ($C_d$ = 1.824, $C_N$ = 0.3565) as a function of the density for each isotherm. The lines are eye guides.

The highest densities in previous experiments are around 1.1 g/cm$^3$. Bove *et al.* (Ref. 82) reached a density of 1.4 g/cm$^3$ by applying pressures of a few GPa but on a single isotherm at 400 K. These data have been obtained by quasielastic neutron scattering (QENS) measurements. Fig. 105 shows that the present modeling allows these data to be reproduced correctly given their error bars with the exception of the three points at the highest densities. For these 3 points, we used to calculate the densities an extrapolation of Wiryana *et al.*'s data (Ref. 20) which certainly induces an error on the calculation of $D_t$. Since $D_t$ decreases with increasing density, the quasi-elastic broadenings decrease while the energy resolution does not change, resulting in a naturally increasing uncertainty. In addition, the values of $D_t$ determined by this method depend strongly on the chosen model to describe internal motions, in particular on the assumption made here that translation and rotation of the water molecules are not coupled, which is known to be incorrect, at least in the normal liquid phase under one atmosphere (see for example Ref. 67) discussed previously. It therefore seems difficult to conclude that the deviation on these three high-density measurement points is relevant.

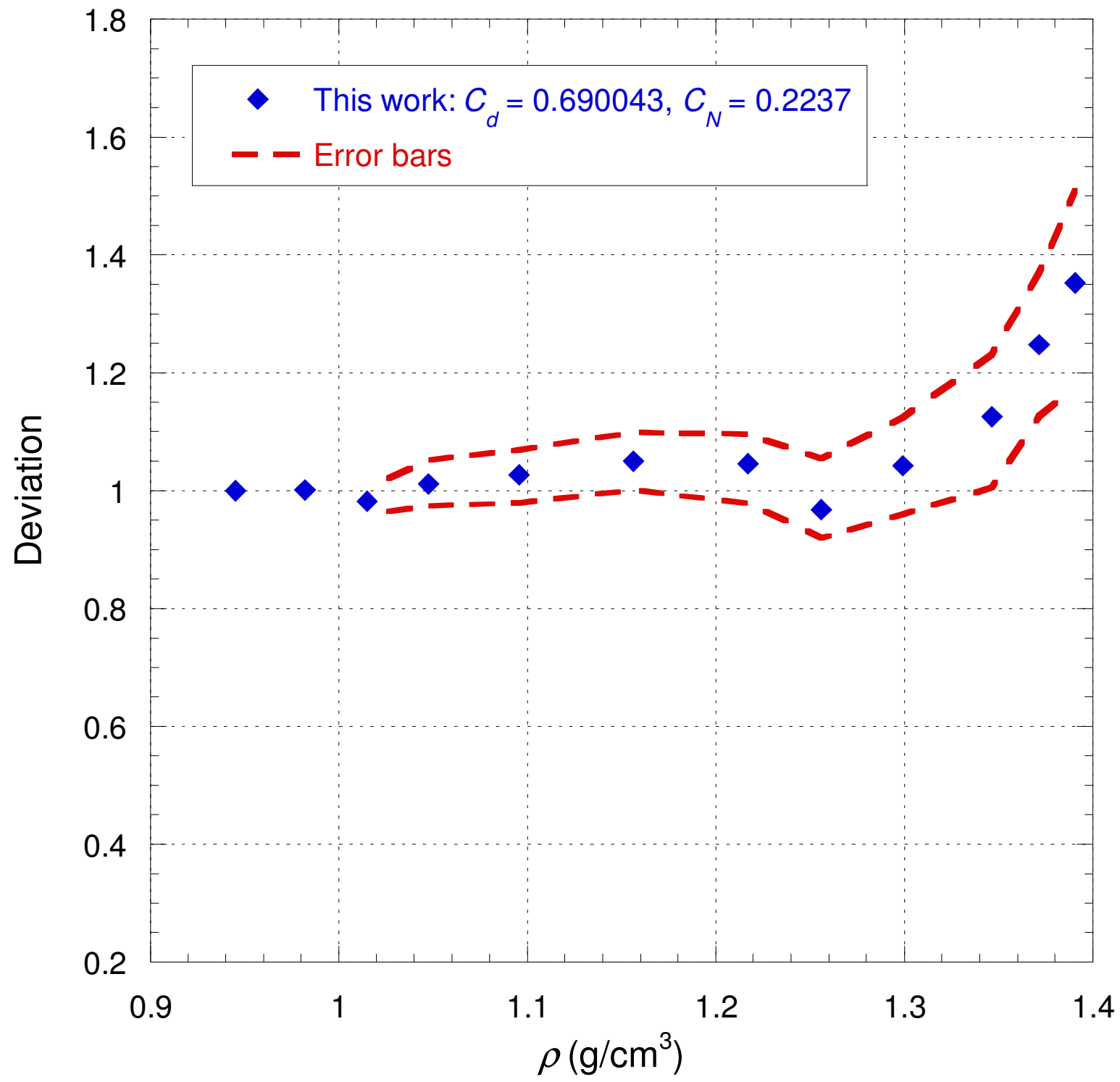

Fig. 105. Ratio of Bove *et al.*'s data (Ref. 82) with the present modeling as a function of the density on the isotherm at 400 K. The error bars for the two points at the lowest densities correspond to the size of the markers.

From their measurements, Bove *et al.* have deduced that:

> "[…] the Stokes Einstein (SE) relation, $D_t\eta = k_B T/C\pi a$, which predicts a constant product of diffusion and shear viscosity at constant temperature is violated. Here $C$ is a constant depending of the geometry of the motion, and $a$ = 1.38 Å is the hydrodynamic radius of the molecule, is violated. […]
> The failure of the SE equation observed in hot water under compression seems most likely connected to the free volume reduction and the consequent onset of the hopping phenomenon, which is clearly observed in our data by the saturation of the translational component as a function of $Q$."

In the context of the present modeling, there is no violation of the SE law but the variation of $D_t$ is just a consequence of the variation of $q_c$ compared to $q_{c0}(\rho)$ on this isotherm. Indeed, Fig. 106(a) shows that under the conditions of the Bove *et al.*'s experiment, the viscosity gas-like term can be neglected therefore the product $D_t\eta \approx D_t\eta_l$ and thus varies as $q_c$. Fig. 106(b) then shows that $q_c(\rho, 400\,\mathrm{K})$ first decreases as the density increases and then increases approximately as $q_{c0}(\rho)$. In other words, when the pressure along this isotherm is increased, the basic units first grow in size until the density corresponding to the maximum density at atmospheric pressure is reached and then decrease continuously. This second variation is in line with the idea of a more compact water structure as suggested by Bove *et al.*
We can notice that this is coherent with the previous analysis of the isotherm at 403.15 K from Krynicki *et al.*'s dataset.

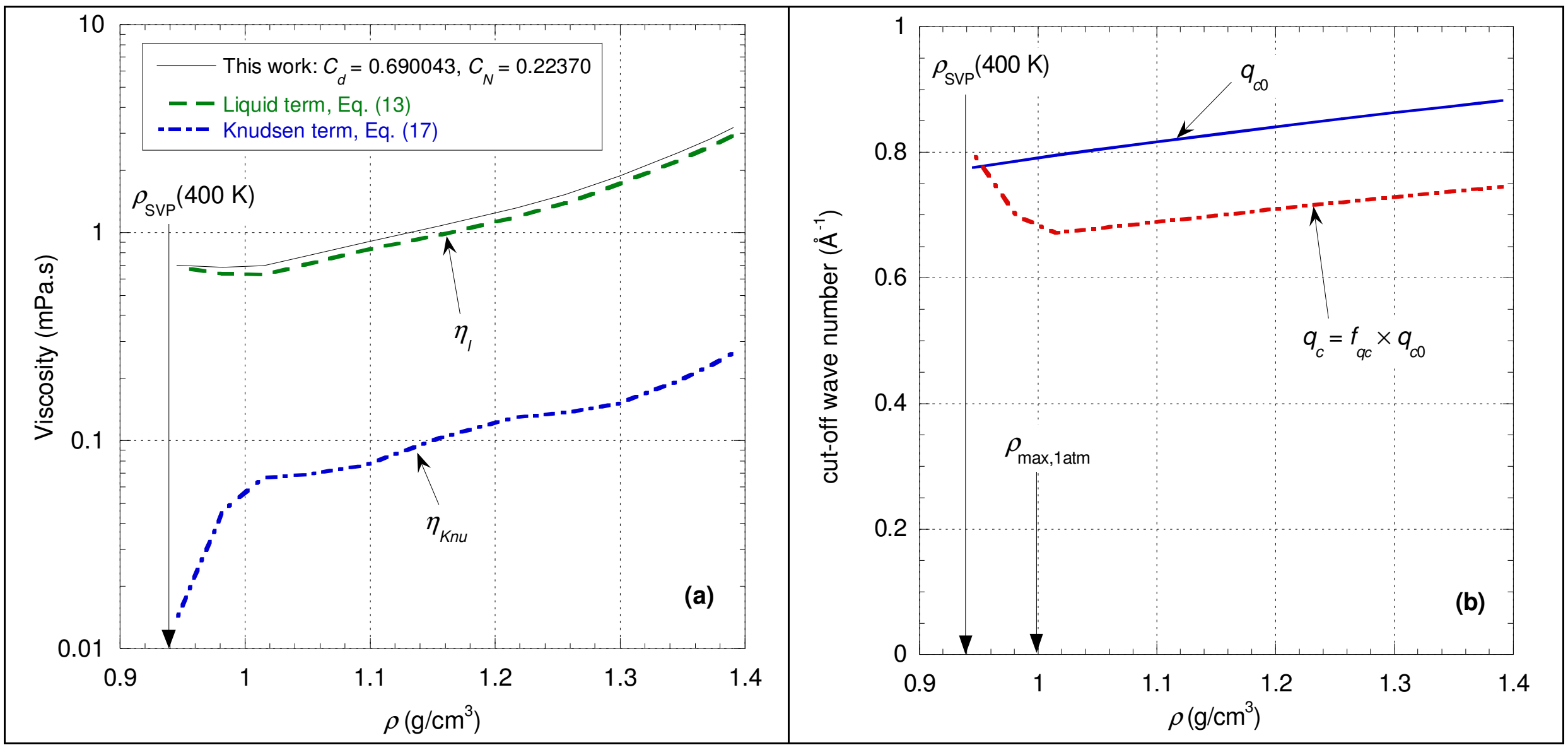

Fig. 106. Plots as function of density along the isotherm at 400 K from the liquid coexistence density $\rho_{SVP}$ with the parameters corresponding to Bove *et al.* (Ref. 82) experiment: (a) Water viscosity and the corresponding viscous terms in the present modeling; (b) Cut-off wave numbers $q_c(\rho, 400\,\mathrm{K})$ and $q_{c0}(\rho)$.

Previous experiments, in particular those of Krynicki *et al.* (Ref. 73) and Harris *et al.* (Ref. 66), were carried out to observe the "anomaly" of the self-diffusion coefficient $D_t$, i.e. the fact that along isotherms $D_t$ will reach a maximum when the pressure varies as it is shown on the isotherm example of Fig. 107. Previously we have shown that for the density range corresponding to Krynicki *et al.* (Ref. 73) and Harris *et al.* (Ref. 66) experiments, the viscosity could be reasonably approximated by the liquid part only which then presents along such an isotherm a minimum with the pressure variation (see Fig. 52(a)). In the present modeling, a minimum of the liquid term of viscosity results in a maximum of the self-diffusion coefficient if over the same pressure variation range the term $q_c$ varies very little. On the isotherm chosen as an example in Fig. 107, we have $q_c \approx q_{c0}(\rho)$, and $q_{c0}(\rho)$ increases as $\rho^{1/3}$, i.e. very little over the pressure range shown: this is indeed what we see on Fig. 107. So, when the pressure increases, the basic units systematically decrease in size but also that the coherence volume decreases so that they decrease more rapidly at low pressure than at high pressure. On the other hand, the cohesion parameter $K$ systematically increases with the increase of pressure. Therefore the "anomaly" results from a complex combination of all these variations.

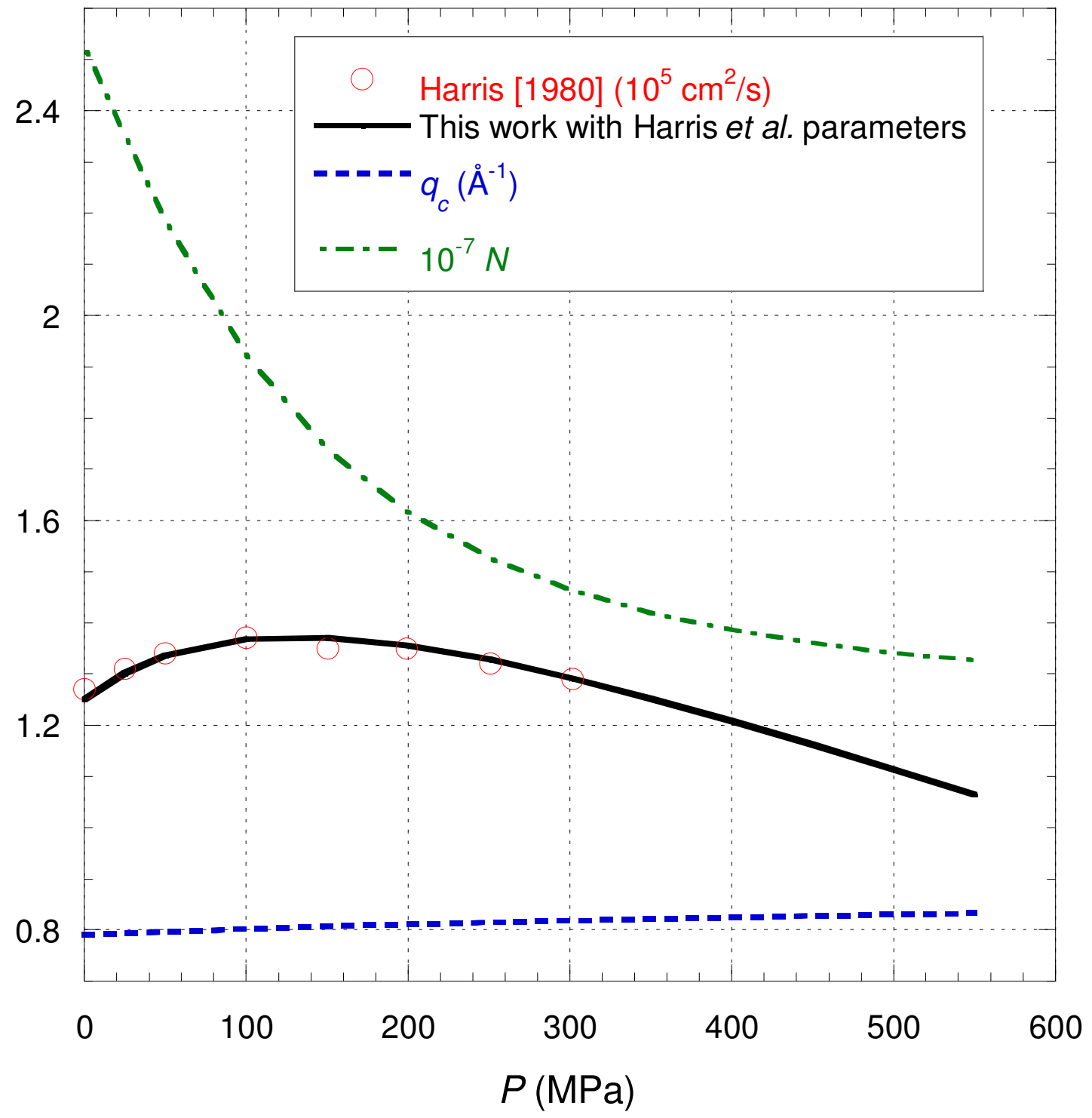

Fig. 107. Self-diffusion coefficient $D_t$, cut-off wave number $q_c$ and coherence reduced size $N$ as function of pressure along the isotherm at 277.15 K.

As with atmospheric pressure data, it exists also some pressure data obtained with tracers. The data from Woolf *et al.* (Ref. 79) concerning the diffusion of HTO in normal water between 277 and 318 K and for pressures up to about 200 MPa are given here along five isotherms. To analyze these data, the parameters $C_q$ and $C_K$ determined with Mills' data (see Fig. 86), which are specific to HTO diffusion in normal water, must be considered. Fig. 108(a) shows that when temperature increases the variation of isotherms with density is well reproduced, but it can be observed that isotherms appear as shifted as those of Harris *et al.* (see Fig. 100 and Fig. 104(a)). The origin of this shift is still due to the fact that Woolf *et al.* used also the Mills' HTO diffusion results in water to calibrate their cell at atmospheric pressure. Calibration at atmospheric pressure is perfectly seen on Fig. 108(b): indeed, we can see that by affecting all the Mills' parameters that describe HTO diffusion coefficient in water in the present modeling, the deviation of Woolf *et al.*'s atmospheric pressure data is practically flat around the unit value. So once again by introducing renormalization factors on each isotherm that counterbalance the imposed calibration (i.e. it is like horizontally straightening the blue curve of Fig. 108(b)), we observe on Fig. 108(c) that all data are finally included in the ±3% band. This deviation is consistent with the root mean square deviation estimated by Woolf *et al.* (i.e. 1.4 %). It is also consistent with that of Harris *et al.* (see Fig. 104(c)) for a very similar experiment.

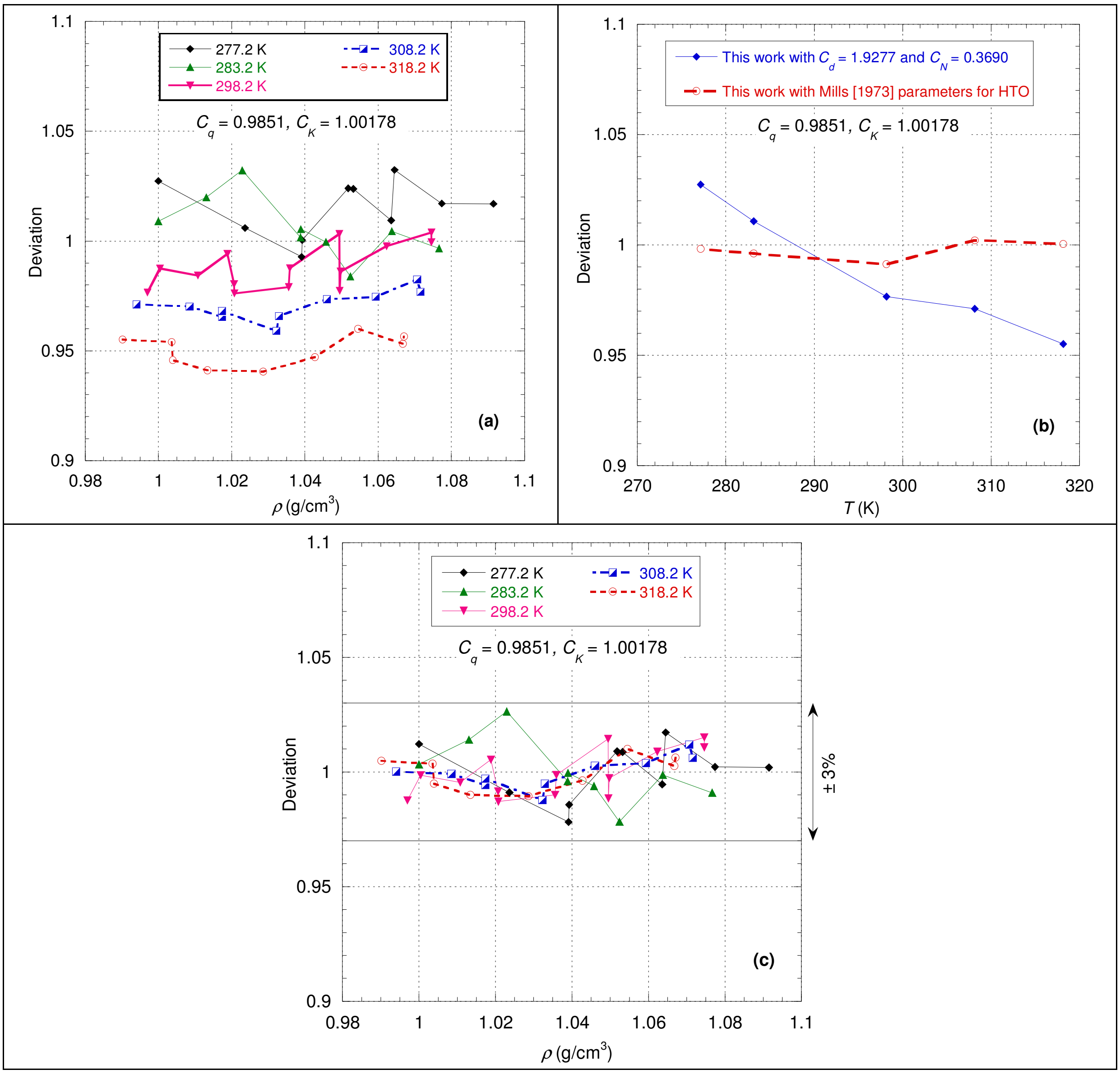

Fig. 108. (a) Ratio of Woolf *et al.*'s data (Ref. 79) with the present modeling ($C_d$ = 1.9277, $C_N$ = 0.3690) as a function of the density for each isotherm; (b) Ratio of Woolf *et al.*'s data (Ref. 79) with the present modeling for two sets of parameters $C_d$ and $C_N$ along the atmospheric isobar; (c) Ratio of Woolf *et al.*'s renormalized data (Ref. 79) with the present modeling ($C_d$ = 1.9277, $C_N$ = 0.3690) as a function of the density for each isotherm. The lines are eye guides.

In the case of viscosity, we finished section 4.1.2 by applying the present modeling in the case of salt water to show the possible extent of the data that can be analyzed with the present modeling. To show this, we analyze Harris *et al.*'s pure $H_2O^{18}$ self-diffusion coefficient data (Ref. 81). Pure $H_2O^{18}$ self-diffusion coefficient is different from that of $H_2O^{18}$ diffusion in normal water; these two coefficients should not be confused. To determine the density corresponding to the pressure and temperature of the measurement Harris *et al.* proceeded as follows:

> "The density of oxygen-18 water was obtained from the compressions of ordinary water appropriately scaled using the molar volume equations of Kell for $H_2{}^{18}O$ and $H_2O$ and the molecular weight ratio $1.1110_2$."

Considering therefore as Harris *et al*. that the molar volume is the same for normal water (i.e. oxygen-16) and for $H_2O^{18}$, we have therefore kept the molar mass and densities of normal water for the corresponding data analysis, knowing that only the molar volume is relevant in the present modeling. Harris *et al*. estimated that the overall accuracy of their experimental data is ±2%. Using the same $C_d$ and $C_N$ parameters as for the Harris *et al*.'s data in normal water (see Fig. 104(a)), we can see on Fig. 109(a) that the variations with density of each isotherm are well reproduced but once again the isotherms appear shifted with the temperature increase. This shift is also explained by the calibration imposed with the Mills' data (see Fig. 86) as shown in Fig. 109(b): we can see that by taking the $C_d$ and $C_N$ parameters of the Mills' experiment and taking into account $C_q$ and $C_K$ parameters deduced for $H_2O^{18}$, the deviation becomes practically horizontal and compatible with the estimated experimental uncertainty. So once again by introducing renormalization factors on each isotherm that counterbalance the imposed calibration, we observe on Fig. 109(c) that all the data are finally included in the ±2% band corresponding to the estimated experimental uncertainty.

It appears that the values of $C_q$ and $C_K$ are different from those for HTO diffusion in normal water but also from those for $H_2O^{18}$ diffusion in normal water. The fact that $C_K$ is slightly greater than 1 indicates that the cohesion of $H_2O^{18}$ is slightly greater than the cohesion of $H_2O^{18}$ molecules in normal water, which in the latter case behave essentially like $H_2O^{16}$ water molecules. The fact that $C_q$ is slightly smaller than 1 indicates that the basic unit in $H_2O^{18}$ is slightly larger than that formed by $H_2O^{18}$ molecules in normal water which seems to go in the "right" direction.

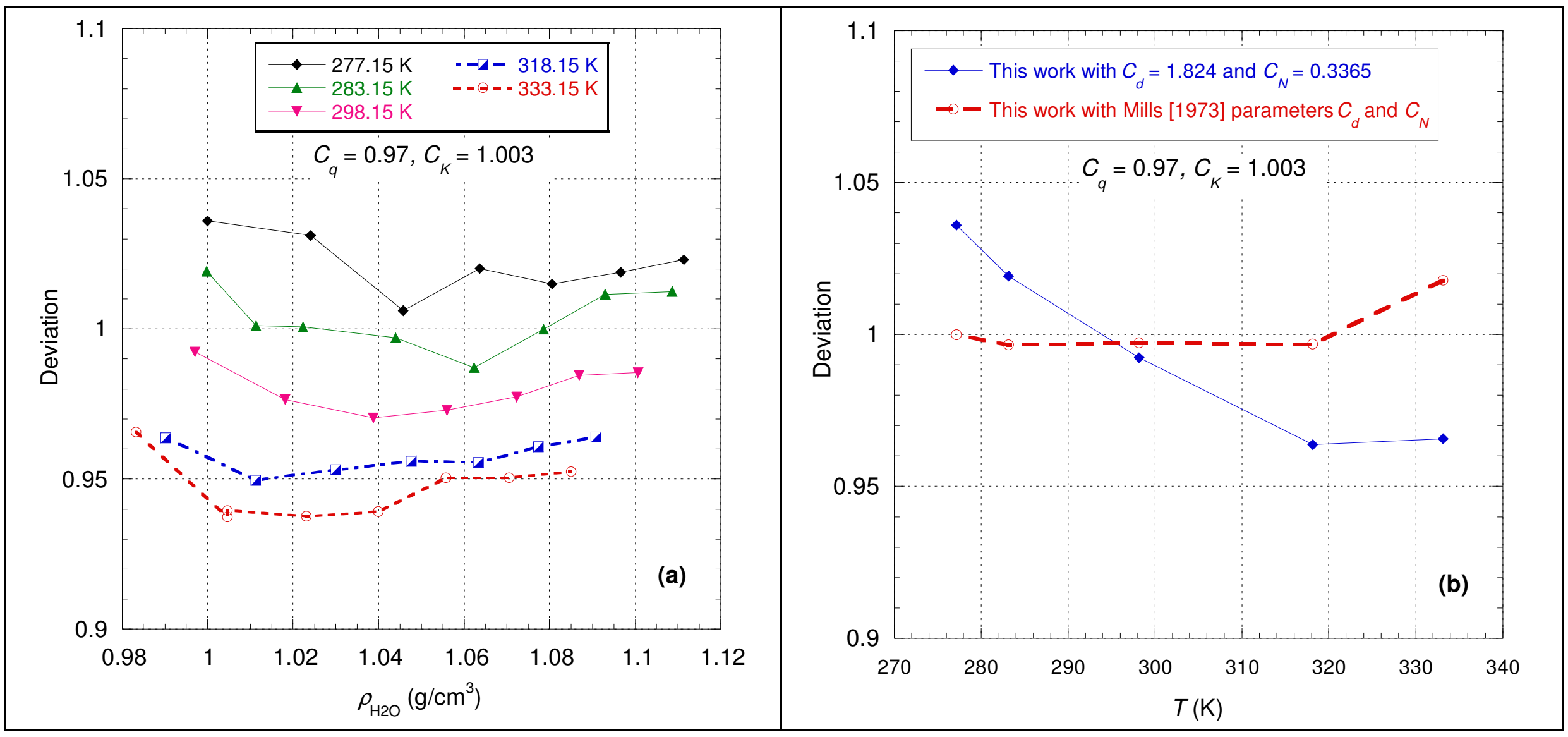

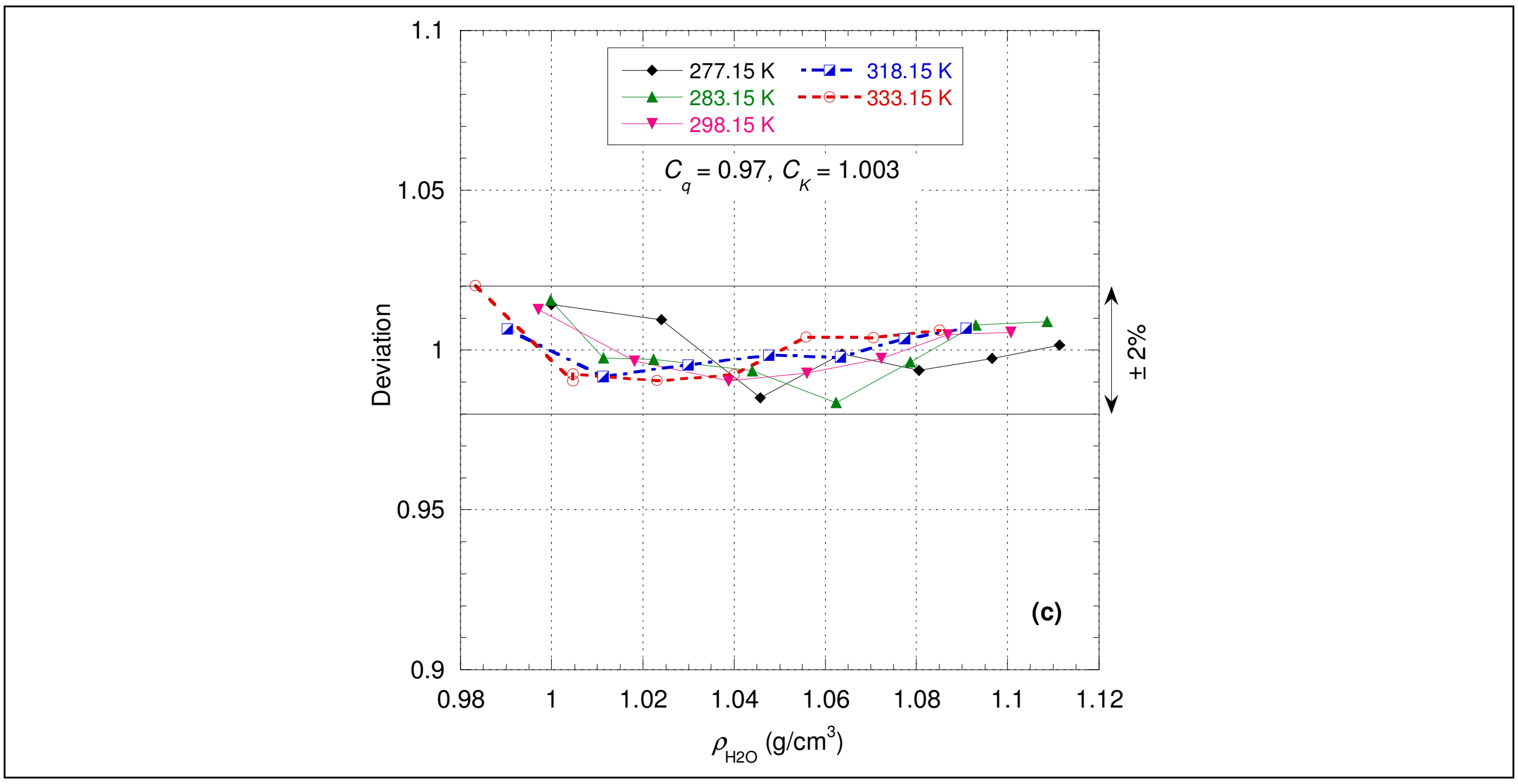

Fig. 109. (a) Ratio of Harris *et al.*'s $H_2O^{18}$ data (Ref. 81) with the present modeling ($C_d$ = 1.824, $C_N$ = 0.3565) as a function of the density for each isotherm; (b) Ratio of Harris *et al.*'s $H_2O^{18}$ data (Ref. 81) with the present modeling for two sets of parameters $C_d$ and $C_N$ along the atmospheric isobar; (c) Ratio of Harris *et al.*'s renormalized $H_2O^{18}$ data (Ref. 81) with the present modeling ($C_d$ = 1.824, $C_N$ = 0.3565) as a function of the density for each isotherm. The lines are eye guides.

### *4.2.4. Self-Diffusion of Gaseous Water*

In this section, we will analyze the data that are essentially in the gas-like density region defined in Fig. 76. In this phase, there are essentially two datasets that will allow us to understand the self-diffusion coefficient variations: these are the data from Yoshida *et al.* (Ref. 83) and Lamb *et al.* (Ref. 84). This may seem like a small amount, but it is very important to understand that Yoshida *et al.*'s data for enough low densities can be identified with the thermal diffusion coefficient $D_{th}$ defined with the isochoric heat capacity $C_V$, i.e.:

$$D_{th} = \frac{\lambda}{\rho\, C_V} \tag{32}$$

where $\lambda$ represents the thermal conductivity. The thermal conductivity was calculated here from the 2011 IAPWS formulation (Ref. 85) and the isochoric heat capacity from the IAPWS-95 formulation (Ref. 17). To illustrate this observation, we have shown in Fig. 110 the isotherm at 673.15 K which has the widest experimental data range of variation as a function of density. It is clear from Fig. 110 that the thermal diffusion coefficient coincides with Yoshida *et al.*'s data for the self-diffusion coefficient but not with the higher densities Lamb *et al.*'s data. It is reasonable that heat transport is identical to molecular transport when the density is sufficiently low. Thus, for enough low densities (globally less than 0.1 g/cm$^3$) the thermal diffusion coefficient calculated according to Eq. (32) can be used as a dataset to describe the self-diffusion coefficient.

The fact that $D_t$ coincides with $D_{th}$ built with $C_V$ justifies that the separation line (i.e. the Frenkel/Widom line) between the gas-like and liquid-like regions (see Fig. 76) should be determined from the peaks of $C_V$ and not from the peaks of $C_P$ (i.e. the isobaric heat capacity).

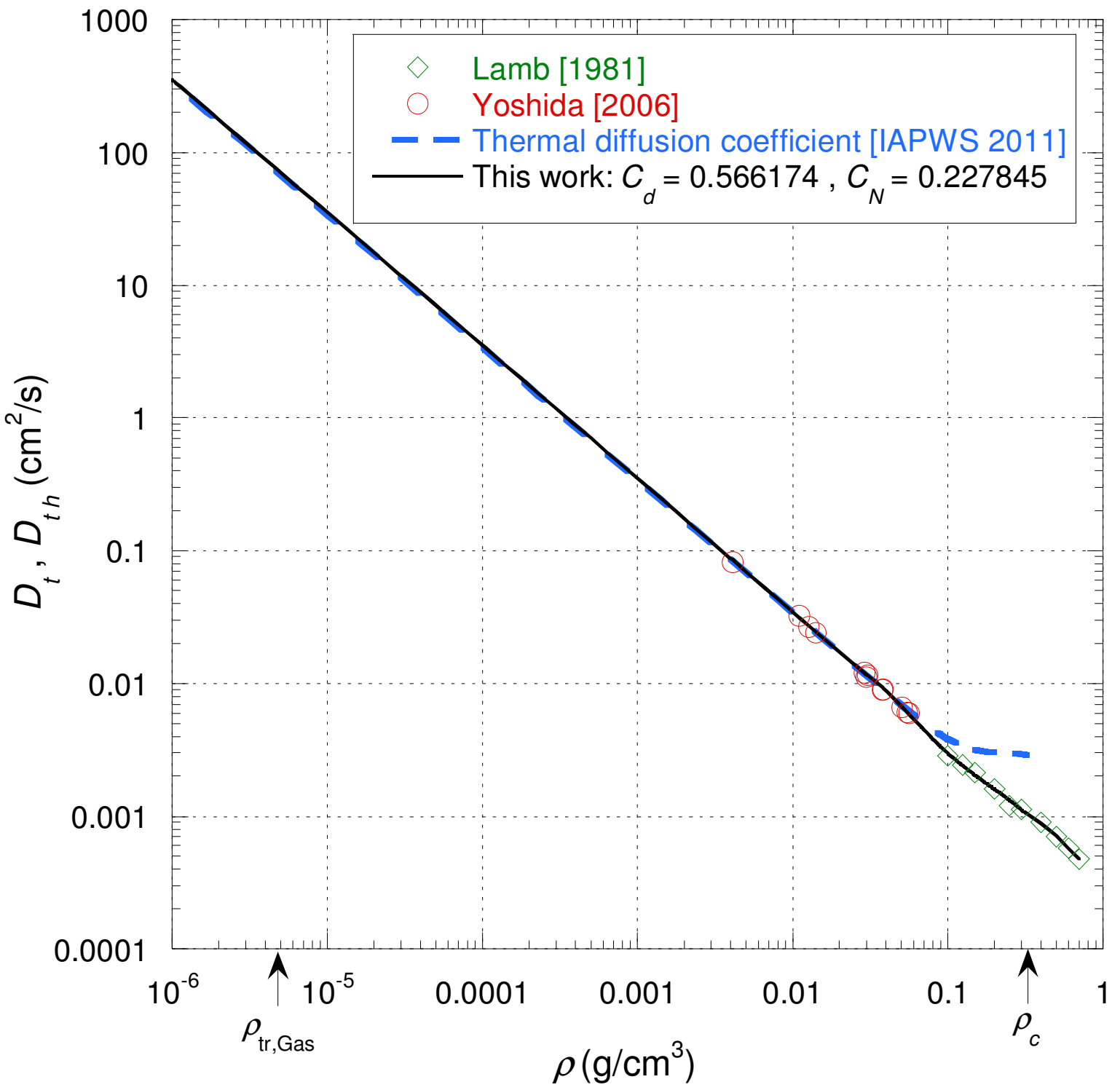

Fig. 110. Logaritmic plot of the self-diffusion coefficient $D_t$ and the thermal diffusion coefficient $D_{th}$ as function of density along the isotherm at 673.15 K.

The most recent and accurate data are those of Yoshida *et al.* (Ref. 83) obtained from a high-sensitivity multi-nuclear magnetic resonance (NMR) probe. The dataset obtained lies entirely in the gas-like density region (see Fig. 76) and cover on one isotherm the subcritical gaseous phase. Concerning the uncertainty of their data, the authors wrote:

> "The uncertainty of $D_t$ was ±1% among five runs consecutively performed for one sample. The total uncertainty of $D_t$ was estimated at about ±5% by taking into account the error due to the sample resetting."

Fig. 111 shows the deviation obtained between the present modeling and the data along two isotherms (the third isotherm will be processed further simultaneously with the Lamb *et al.*'s data). We observe that the deviation on the isotherm at 473.15 K is in agreement with what is expected while the low-density points of the isotherm at 573.15 K deviate strongly from it. This deviation would mean that $D_t$ no longer follows the law of the $D_{th}$ for these densities. However, it is difficult to admit that this law is compatible with data on isotherms at lower and higher temperatures but not in the middle. This effect is also found in Fig. 3 of the 2007 erratum paper but in inverted form: indeed, the zero-density limit at 573.15 K is in perfect agreement with the result of the molecular dynamics simulation, but the limit at 673.15 K and especially at 473.15 K is in disagreement with about 10% for the latter temperature. One possibility to explain this deviation is the determination of the density which has been reduced from 5 to 10% between the 2006 paper and the 2007 erratum paper. It may be that the estimated density is still too high and if it is reduced by a few percent more then the deviation falls into the uncertainty band.
It should be noted here that the coefficients $C_d$ and $C_N$ have identical values to the values used to describe Yoshida *et al.*'s data along SVP (see Fig. 94), which makes it possible to integrate these points into the analysis here. This is also an observation written by Yoshida *et al.* in the following form:

> "[…] it is seen that the $\rho D_t / \sqrt{T}$ values at 200 and 300 °C are in harmony with the values on the coexistence curve."

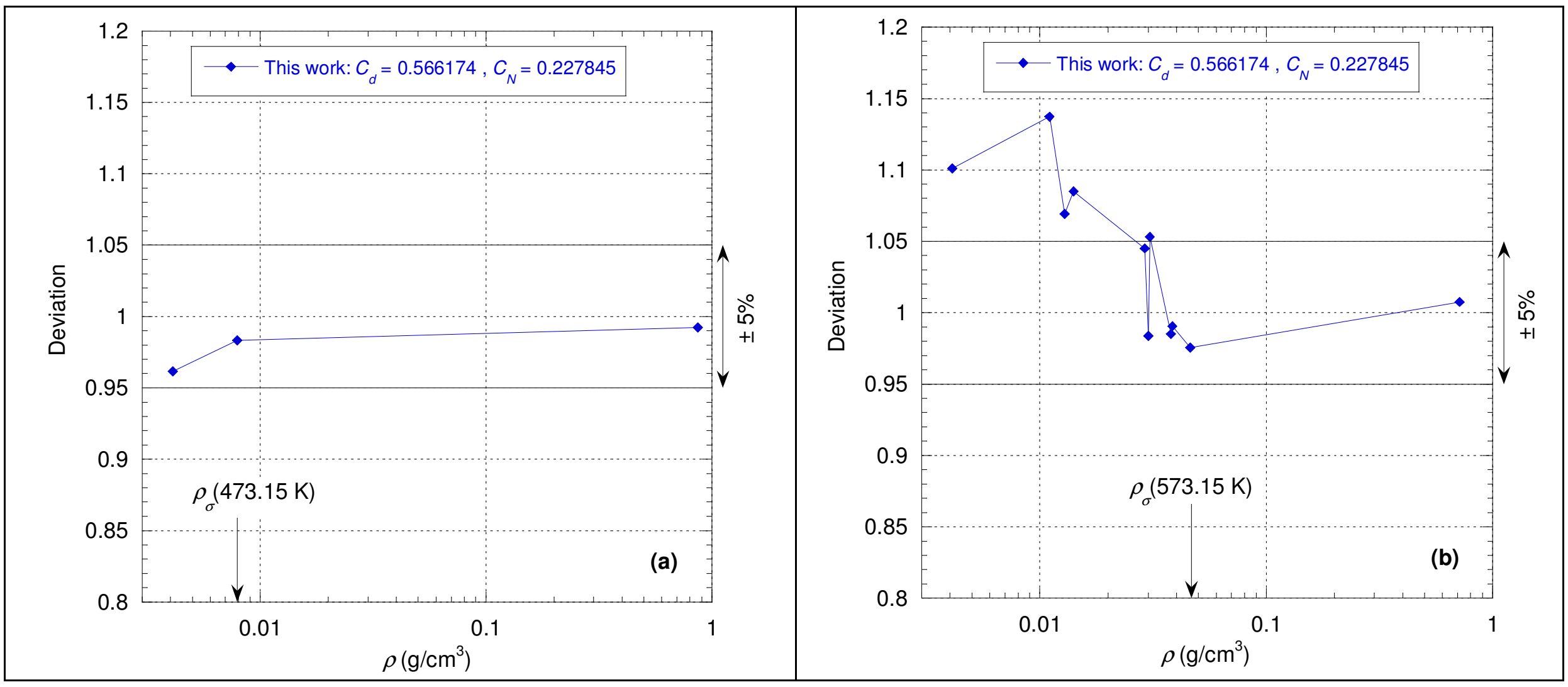

Fig. 111. Ratio of Yoshida *et al.*'s data (erratum paper of Ref. 83) with the present modeling as a function of the density for two isotherms: (a) 473.15 K; (b) 573.15 K. The point at the highest density on each of the isotherms is on SVP (i.e. liquid coexistence state) and coincides with that of Fig. 94. $\rho_\sigma$ represents the density on the vapor coexistence line for the corresponding isotherm. The lines are eye guides.

Lamb *et al.*'s data (Ref. 84) are the most important because they allow us to cross the Frenkel/Widom line (see Fig. 76). These data were also obtained from an NMR-type experiment but *a priori* are less precise than those of Yoshida *et al.* Thus, the authors wrote that:

> "The errors in the self-diffusion measurements are both random and biased. The random error is due to the inherent signal-to-noise ratio and the biased error is due to the gradient calibration. The estimated accuracy of the experimental self-diffusion constants is about ± 10%."

We begin by separately analyzing the 673.15 K isotherm that is common to the data of Lamb *et al.* and Yoshida *et al.* (Ref. 83). To this end, we have also added some isolated data from the 2005 Yoshida *et al.*'s paper (Ref. 75, Table Ia). The data on this isotherm thus extend from the subcritical region to the supercritical region. The fact that all these data can be analyzed with the same $d$ and $d_N$ (i.e. same $C_d$ and $C_N$) allows to check the relevance of the function $f_{q_c}$ which is a piecewise function with a connection on the Frenkel/Widom line: indeed, in the liquid-like density region, the function $f_{q_c}$ is fully defined and allows to reproduce, as we have seen in section 4.2.3, the water self-diffusion coefficient data under pressure. Fig. 112 then shows that the present modeling allows reproducing the different datasets on this isotherm within the estimated experimental uncertainty.

Fig. 113 shows that the present modeling allows to reproduce the other isotherms of Lamb *et al.* with the estimated experimental uncertainty except for one point on the isotherm at 873.15

K. Considering the noise on the calibration constant of the field gradient coil, it is possible to slightly renormalize the isotherms by a factor $C$ and under these conditions the deviation of the isotherms at 773.15 K and 873.15 K can be re-centered on the unit value so that finally all points are included in the ±10% band. Here the renormalization factors $C$ clearly represent noise and therefore their representation does not have a particular interest.

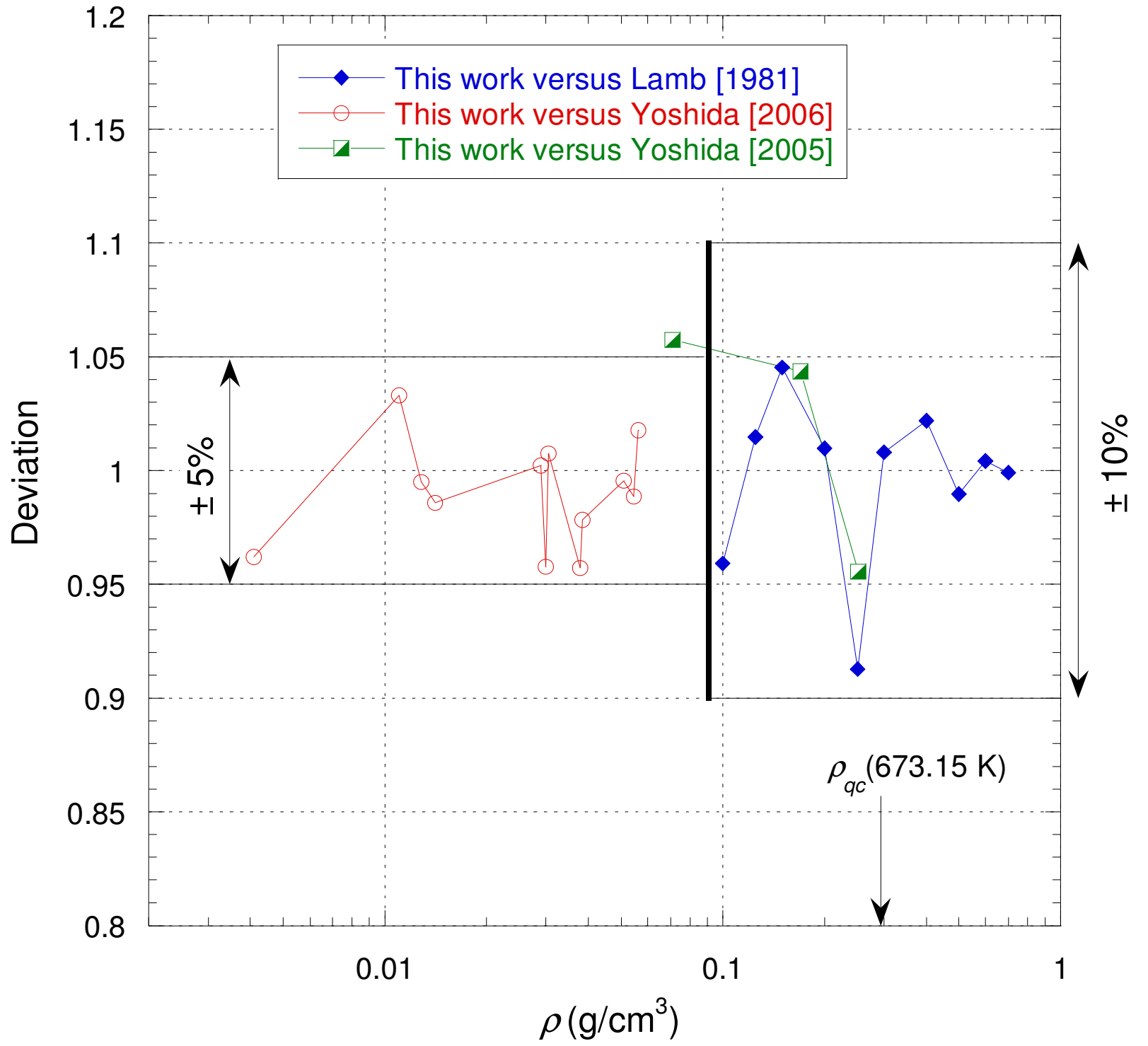

Fig. 112. Ratio of Lamb *et al.*'s (Ref. 84) and Yoshida *et al.*'s data (erratum paper of Ref. 83 and Ref. 75) with the present modeling ($C_d$ = 0.566174, $C_N$ = 0.227845) as a function of the density for the isotherm at 673.15 K. $\rho_{q_c}$ represents the density of the Frenkel/Widom line for the corresponding isotherm. The lines are eye guides.

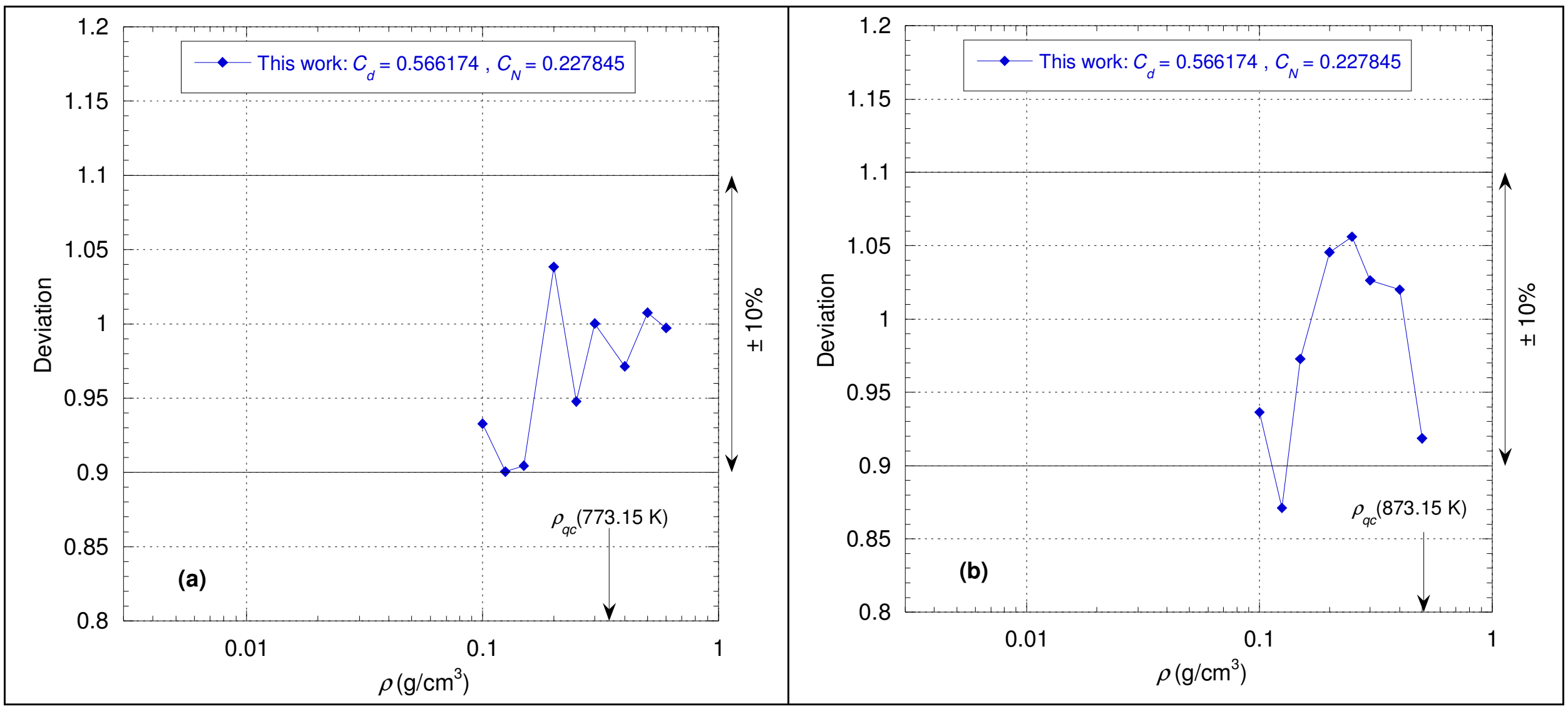

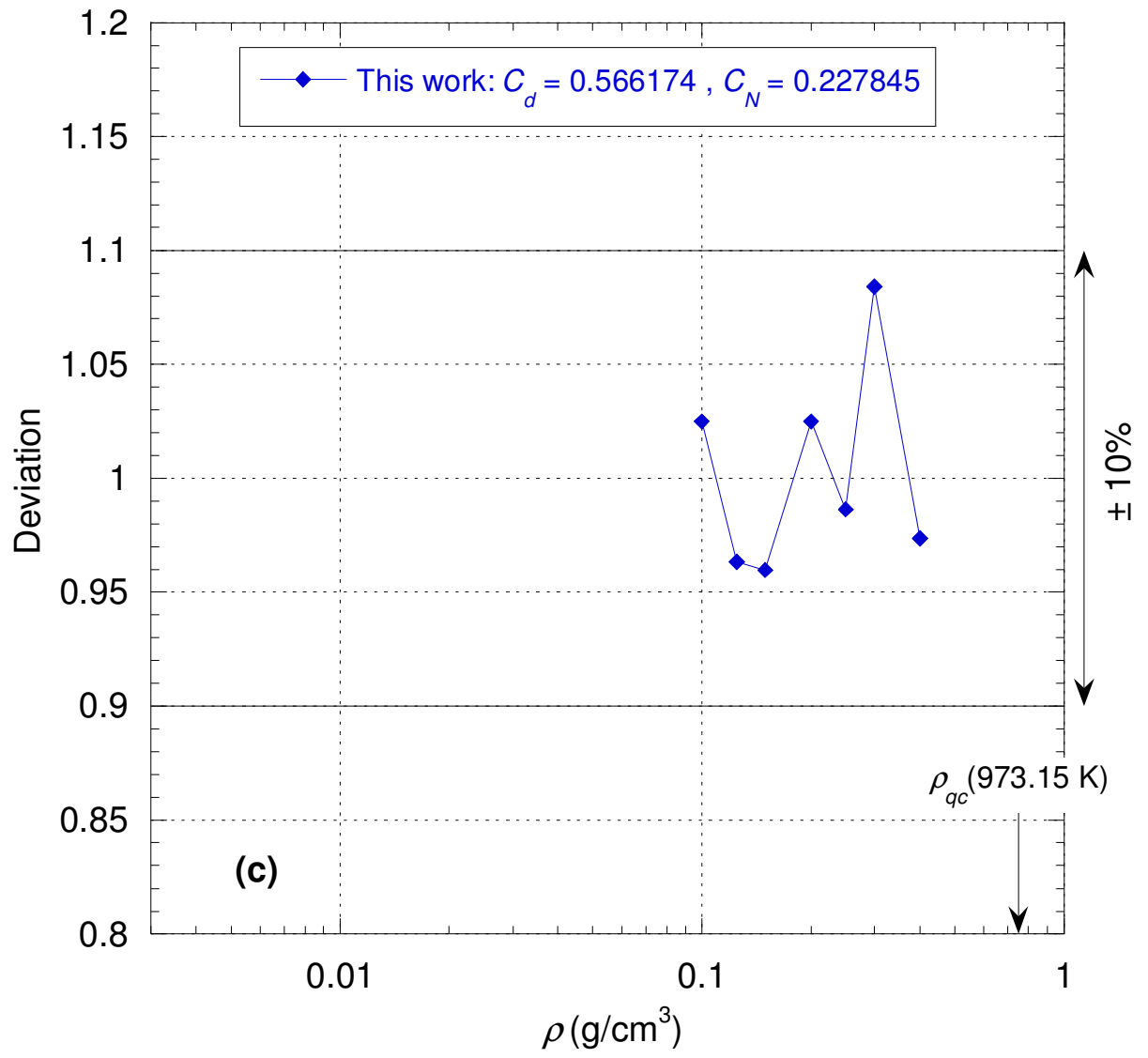

Fig. 113. Ratio of Lamb *et al.*'s data (Ref. 84) with the present modeling ($C_d$ = 0.566174, $C_N$ = 0.227845) as a function of the density for three isotherms: (a) 773.15 K; (b) 873.15 K; (c) 973.15 K. $\rho_{q_c}$ represents the density of the Frenkel/Widom line for the corresponding isotherm. The lines are eye guides.

As Lamb *et al.* (Ref. 84) wrote, one of the objectives of the previous measurements was:

> "[…] to determine the temperature and density dependence of the product $\rho D_t$ and compare the result with other gases, […]".

Firstly, Lamb *et al.*'s study has concluded that the product $\rho D_t$ is *density independent* under isothermal conditions which is conform to a prediction of the gas kinetic theory but the study of Yoshida *et al.* have still found a slight negative slope of the product $\rho D_t / \sqrt{T}$ .

Fig. 114 actually seems to show on the particular isotherm at 673.15 K that the product $\rho D_t$ tends indeed towards a constant at a sufficiently low density. However, if we limit ourselves to the range of the existing datasets, we observe that with the present modeling it is not possible to conclude when this limit property is reached which is in line with the point (4) of Yoshida *et al.*'s conclusion (corrected in the erratum). Moreover, Fig. 114 shows that the relationship $\eta = \rho D_t$ is never verified. However, at a sufficiently low density there is a proportionality coefficient between the two quantities $\eta$ and $\rho D_t$: for the example of Fig. 114, the factor between the two quantities is close to 1.5.

Now we will show that the present modeling does indeed have the limit property that $\rho D_t$ is independent of $\rho$ at the zero-density limit. According to Eq. (12), we have:

$$\rho D_t = \frac{3k_B T q_c H_N(v)}{2\pi^2 d}\sqrt{\frac{\rho}{K}} .$$

It has already been shown in section 3.1 that $\lim_{\rho\to 0} K_0^* \propto (\rho/\rho_c)^3$ and $\lim_{\rho\to 0} H_N(v) \propto \left(\frac{\rho}{\rho_c}\right)^2$ so $\lim_{\rho\to 0} \rho D_t \propto \rho q_c$. But $\lim_{\rho\to 0} q_c \propto \rho^{1/3} \lim_{\rho\to 0} f_{q_c}$ and from the expression of $f_{q_c,\mathrm{Gas}}$ in Table 4 for $\rho < \rho_{q_c}$, it follows that $\lim_{\rho\to 0} f_{q_c} \propto (\rho_c/\rho)^{4/3}$. We therefore deduce that the zero-density limit of $\rho D_t$ on an isotherm is also a constant with the present modeling.

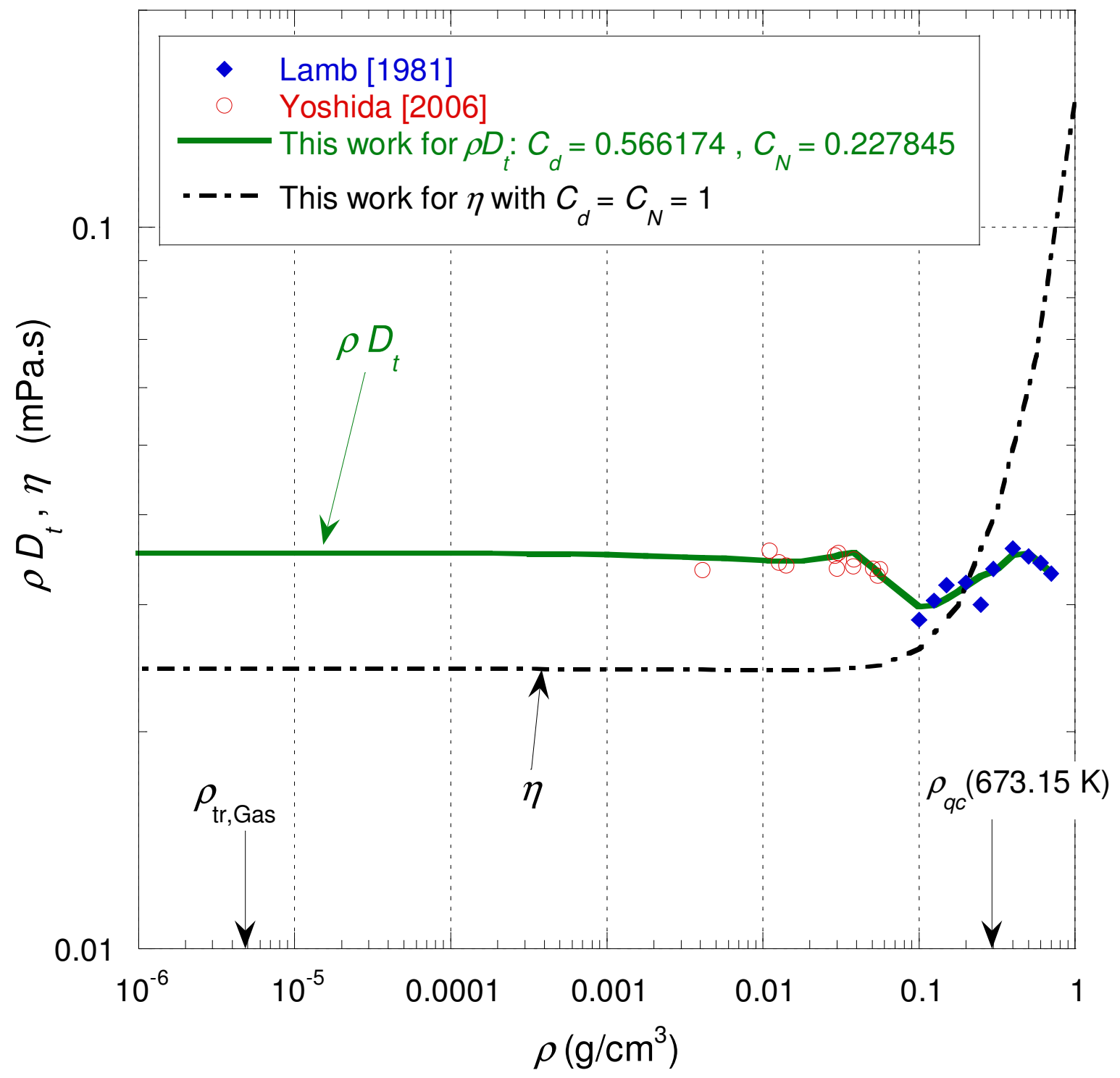

Fig. 114. Logaritmic plot of the product $\rho D_t$ and the dynamic viscosity $\eta$ as function of density along the isotherm at 673.15 K.

Secondly, we have to examine the *temperature dependence* of $\rho D_t$ at the zero-density limit. The relevance of examining this limit property has been described as follows by Yoshida *et al.* (Ref. 83):

> "Thus, the temperature dependence of the product $\rho D_t/\sqrt{T}$ in the zero-density limit is a good probe for the effect of attractive interaction on the self- diffusion."

Both Yoshida *et al.* and Lamb *et al.* have found that the temperature dependence of $\rho D_t$ can be approximated empirically by a power of the temperature such that $\rho D_t \propto T^n$. The $n$ values obtained by Yoshida *et al.* vary between 1.2 and 1.4 whereas the $n$ value for Lamb *et al.* is equal to 0.763. The explanation given by Yoshida *et al.* for this large discrepancy is as follows:

> "[…] note that [the value of $n$ = 0.763] is obtained at higher temperatures (400–700 °C) and densities (0.10–0.60 g cm$^{-3}$). The difference in the *B* [here called *n*] values between the present and previous studies can arise from the strong attractive potential effect at the low densities and temperatures. The present *B* [i.e. *n*] value is a largest among the values reported for various substances; the *B* [i.e. *n*] value is ranged from 0.75 to 1.1 for less polar substances."

As the smallest density reached by Yoshida *et al.* is not yet in the region where $\rho D_t$ is density independent, we have chosen to represent the product $\rho D_t$ on Fig. 115 along the gas triple point isochor (with the units used by Yoshida *et al.*): we observe that over the temperature range where Yoshida *et al.*'s data exist, the variation of the present modeling can be well approximated by a power law but this power law is no longer valid for lower or higher temperatures, in particular for the temperature range corresponding to Lamb *et al.*'s dataset. The value obtained for the exponent $n$ is here intermediate between that obtained by Lamb *et al.* and that obtained by Yoshida *et al.* and is more in line with the expected values. The value of the exponent $n$ depends strongly on the temperature range considered and with Fig. 115 we understand that the value of $n$ deduced from the extrapolation of Lamb *et al.*'s dataset could be very different from the values of $n$ obtained by Yoshida *et al.* and this is not related to any particular physical effect.

Finally, $\rho D_t$'s power law with temperature is only a possible approximation given the range of experimental data but does not necessarily represent an asymptotic law. In particular the present modeling is at the zero-density limit a complex function of $T$ through the function $f_{q_c}$.

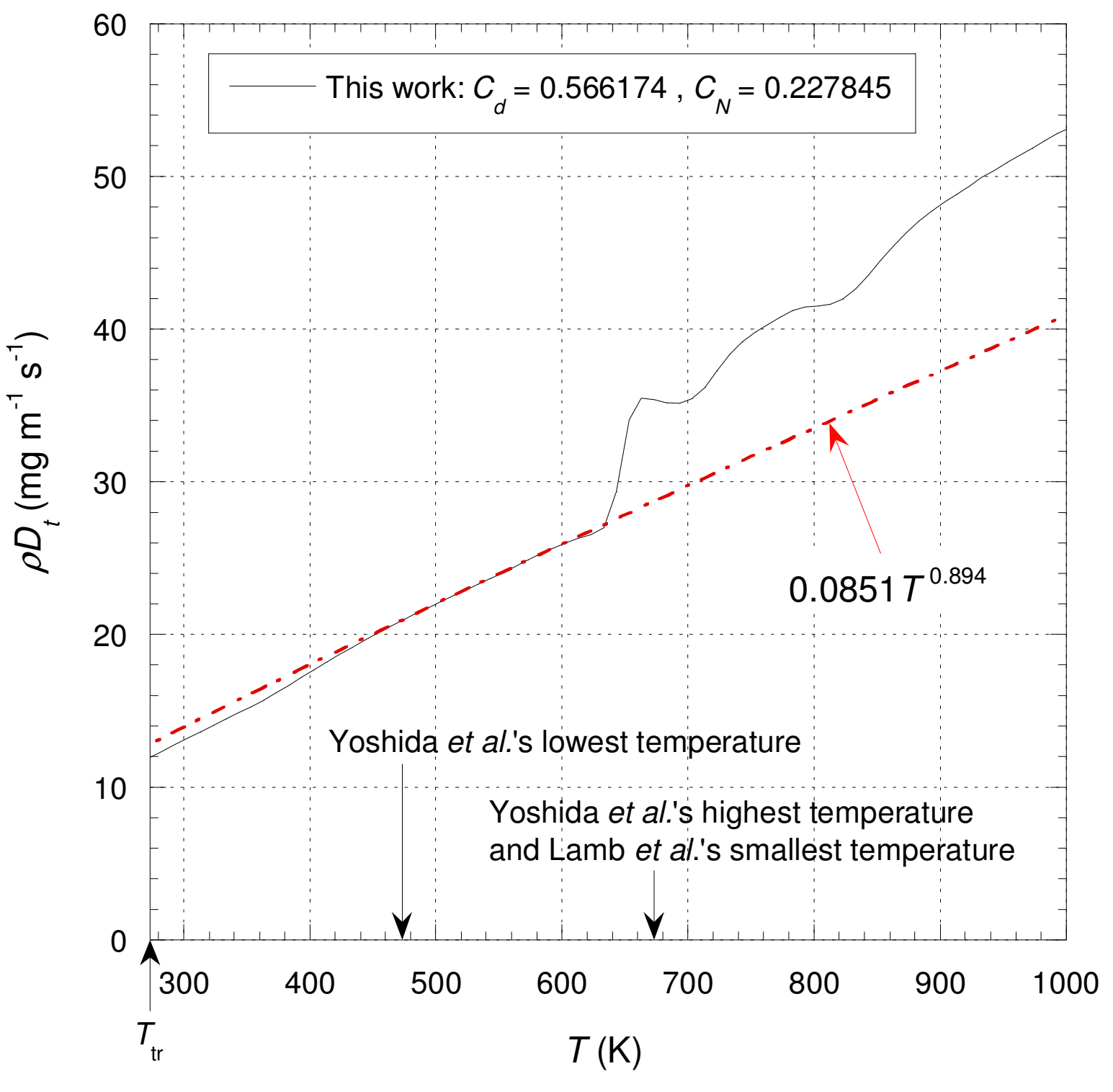

Fig. 115. Plot of the product $\rho D_t$ as function of tempeature along the gas triple point isochor (i.e. $\rho_{tr,\,Gas}$). The dot-dashed red curve corresponds to a power law such that $\rho D_t \propto T^{0.894}$. The oscillations above $T_c$ simply reflect those of the Frenkel/Widom line (see Fig. 76).

To conclude this section, it is interesting to note that the zero-density limit of $D_t$ no longer depends on experimental set-up characteristics but only on the intrinsic properties of the fluid which is consistent and similar to Eq. (21) for the dilute-gas limit of the viscosity liquid term. This highlights the great consistency between the analysis of viscosity and that of the self-diffusion coefficient. It should be noted, however, that while the dilute-gas limit of the viscosity liquid term is a constant, this is not the case for the self-diffusion coefficient, which is inversely proportional to density and remains a function of temperature.

## 5 Conclusion

In conclusion, in this paper we have presented a microscopic model that simultaneously describes the dynamic viscosity and the self-diffusion coefficient of fluids. This model is shown to emerge from the introduction of fractional calculus in a traditional model of condensed matter physics based on an elastic energy functional (see Appendix A), which allow, among other things, to integrate phase transitions directly into the modeling.

Conceptually, the formalism of this model developed to analyze experimental data consists in describing a system at thermal equilibrium perturbed by the measurement. Consequently, the transport parameters depend both on the equilibrium properties of the system and also on some "external" parameters specific to the experiment. In other words, dynamic viscosity and self-diffusion coefficient are not intrinsic properties of the medium, contrary to what is generally assumed in "usual" molecular theories. Moreover, the model does not lead to a Stokes-Einstein law in a classical form but only to an analogy between the self-diffusion coefficient and the "liquid" component of the viscosity.

It is worth noting that some relations such as Eq. (13) and Eq. (16), which seem to be defined here in a phenomenological way, are related to a more general model that will be demonstrated and justified in a forthcoming paper.

More specifically, this model has been applied to a very large collection of published water data in all its fluid phases. In particular, the data in the supercooled phase were included in the analysis, unlike the IAPWS08 formulation, which uses some of these data but does not comment on them. Through an in-depth analysis of the abondant literature data, it has been shown that the discrepancies between them are only apparent and can be explained if both geometric and viscometer calibration experimental conditions are correctly taken into account. In particular, wall effects (strong increase of the correlation length locally) are important in the supercooled phase due to the metastability of this phase. Thus, for all the datasets analyzed, it has been shown that the data can be reproduced by this model in a consistent way within their experimental uncertainty, unlike the IAPWS08 formulation, which arbitrarily discards some datasets. In addition, some features of the model appear to be physically more satisfactory than the IAPWS08 formulation, such as high temperature extrapolation of isochors with densities greater than the critical one.

It has been shown that data must be processed differently depending on whether there are raw or smoothed data. Indeed, the choice of the smoothing functions, possibly in some cases based on authors' implicit *a priori*, more or less biases the variations of the raw data; the smoothed data may then appear incompatible with the estimated experimental uncertainty.

The water "anomalies" could be related to the precise behavior of the transition temperature $T_t$ as well as the quantity of gas released in a viscosity experiment. The formalism of this model allows to understand that "anomalies" observed on the dynamic viscosity can be linked or not on "anomalies" of the self-diffusion coefficient. Moreover, the formalism of the model, which consists in describing the fluid medium as a sheared elastic medium, makes it possible to understand processes of the "memory-effect" type that can be attributed to defects that exist in all solids which can be more or less annealed according to the mechanical and/or thermal treament. But all of this clearly requires a thorough experimental study to be confirmed.

We have also shown that the usual Stokes-Einstein's law does not apply with a constant hydrodynamic radius of the water molecule, but under certain conditions this is a possible reasonable approximation. Moreover, in the gas phase, we have found that there is no equality between $\eta$ and $\rho D_t$ but at best at a sufficiently low density there is a proportionality coefficient between these two quantities. In addition, in this gaseous phase it appears that the self-diffusion coefficient $D_t$ is equal to the thermal diffusion coefficient $D_{th}$ built with the isochoric heat capacity $C_V$ as soon as the density is sufficiently small. This equality between $D_t$ and $D_{th}$ is clearly highlighted here for the first time.

Finally, we can note that we have developed new equations of state to describe the water density along the atmospheric isobar and more globally for the supercooled phase, which has validity over a much broader temperature range than similar equations in the literature (see Appendix D).

One of the objectives of a modeling is to be able to describe consistently a maximum of experimental data within their accuracy limits and with a minimum number of parameters. This objective has been achieved here better than with any other microscopic or macroscopic model.

The fact that this new theoretical approach has been shown to enable accurate analysis of a very large dataset concerning the transport properties of water opens perspectives on its ability to reproduce the transport properties of all common liquids that are often considered "simpler" than water.

## 6 ACKNOWLEDGEMENTS

We thank Prof. Dr.-Ing. Petra Först (Technische Universität München, Germany) and Dr. Masaru Nakahara (Kyoto University, Japan) for providing their experimental data.

We thank our colleague Dr. L. Noirez (Laboratoire Léon Brillouin, France) for many helpful suggestions and discusssions.

This work benefited from the support of the project ZEROUATE under Grant ANR-19-CE24-0013 operated by the French National Research Agency (ANR).

## 7 APPENDIX A

The fundamental assumption is that the excess energy due to thermal fluctuations can be described by an elastic energy functional $F$ given by:

$$F = \frac{1}{2}\int K\left[\vec{\nabla}^{\alpha}\vec{u}\right]^2 dV \qquad \text{(A.1)}$$

where $\vec{\nabla}^{\alpha}$ stands for the fractional gradient operator (that is the derivative of order $\alpha$ with respect to $\vec{r}$ ), and where the coefficient $K$ is an elastic constant (dimensionality: energy per unit volume). Since the order of derivation $\alpha$ is *a priori* allowed to take any real value, this expression generalizes the simplest form with $\alpha = 1$ used in usual condensed matter physics (Ref. 86). Since the function $\vec{u}(\vec{r})$ is expanded in Fourier series on the lattice, Eq. (A.1) involves only fractional derivatives of $\exp(i\vec{q}\cdot\vec{r})$. With $\vec{q}\cdot\vec{r} = |q| r_{\vec{q}}$ where $r_{\vec{q}}$ is the component of $\vec{r}$ on $\vec{q}$, it is easy to show that the generalization of the gradient operator can be written as:

$$\vec{\nabla}^{\alpha}\left[\exp(i\vec{q}\cdot\vec{r})\right] = q_c \frac{d^{\alpha}\left[\exp(iqr_{\vec{q}})\right]}{\left[d(q_c r_q)\right]^{\alpha}} \frac{\vec{q}}{|q|} \qquad \text{(A.2)}$$

Note that this definition of the fractional gradient of a function of a three dimensional variable $\vec{r}$ makes sense only if this function can be expanded in a three dimensional Fourier series, since for each term $\exp(i\vec{q}\cdot\vec{r})$, the relevant variable for derivation is different, being the component of $\vec{r}$ on the corresponding wave-vector $\vec{q}$.
For complex exponential functions, the fractional derivative of order $\alpha$ is defined as follows: if $\alpha$ is a real number, $p$ a complex number, the operator $\frac{d^\alpha}{dx^\alpha}$ acting on exponential functions $\exp(px)$ is given by one of the two following expressions (Ref. 87):

$$\text{If } \alpha < 0, \text{ then} \quad \frac{d^\alpha[\exp(px)]}{dx^\alpha} = \int_{-\infty}^{x} \exp(ps)\frac{(x-s)^{-\alpha-1}}{\Gamma(-\alpha)}ds \tag{A.3a}$$

$$\text{If } \alpha \geq 0, \text{ then} \quad \frac{d^\alpha[\exp(px)]}{dx^\alpha} = \frac{d^n}{dx^n}\int_{-\infty}^{x} \exp(ps)\frac{(x-s)^{n-\alpha-1}}{\Gamma(n-\alpha)}ds \tag{A.3b}$$

where for $\alpha \geq 0$, $n$ is the integer number such that $n-1 \leq \alpha < n$ and $\frac{d^n}{dx^n}$ is the usual derivative of order $n$. $\Gamma(x)$ represents the gamma function. Note that (A.3b) can be used for all $\alpha$ if one assumes that for $\alpha < 0$, $n = 0$ and that $\frac{d^0}{dx^0}$ is the identity operator. This definition shows that the fractional derivative of a function of $x$ is a usual (*i.e.* integer) derivative of this function convoluted by a fractional power kernel. For exponential functions, it is easily shown that if $p$ is a purely imaginary number (as it is the case in Eq. (A.2) with $p = iq$), then $\frac{d^\alpha[\exp(px)]}{dx^\alpha} = p^\alpha \exp(px)$, for $-\infty < \alpha < \infty$.
Inserting Eq. (1) in Eq. (A.1), using Eq. (A.2) with $\alpha = 1 + \nu/2$ and the orthogonality property of imaginary exponential functions:

$$\int \exp[i(\vec{q}-\vec{q}')\cdot\vec{r}]dV = V\delta_{\vec{q},\vec{q}'} \tag{A.4}$$

where $\delta_{\vec{q},\vec{q}'}$ is the Kronecker symbol, one obtains the following expression for $F$:

$$F = \frac{1}{2}\sum_{q=\frac{q_c}{N}}^{q_c} Kq^2\left(\frac{q}{q_c}\right)^\nu \left|u_q^2\right| V \tag{A.5}$$

where $\left|u_q^2\right| = \left|u_{xq}^2\right| = \left|u_{yq}^2\right|$ corresponding to an isotropic property. It is important to remark that if $\alpha = 1$ (*i.e.* $\nu = 0$), Eq. (A.5) becomes identical to the traditional elastic energy expression (Ref. 85). So, it appears that the introduction of fractional derivatives of order $\alpha = 1 + \nu/2$ in Eq. (A.1) is mathematically equivalent to the replacement of $K$ by $K\left(\frac{q}{q_c}\right)^\nu$ in the elastic energy functional expression in Ref. 6.

# 8 APPENDIX B

In this Appendix we will illustrate the scaling given by Eq. (7) for different liquid media by putting it in perspective with latent heat of vaporization data (i.e. by focusing on experimental data in accordance with the systematic approach of this paper). The latent heat of vaporization is written $L_v$ in this Appendix.

Given the importance of water in this paper, we begin by observing the latent heat of vaporization of this medium. Fig. 116 shows both experimental measurements and the IAPWS-95 formulation between the triple point and the critical point. Of course, different ways of extrapolating a curve from the data is always possible, but considering that the value of $K_0$ is not very different from the value of $L_v$ at the triple point, it can then be assumed that $K_0$ represents a "reasonable" extrapolation of the latent heat of vaporization.

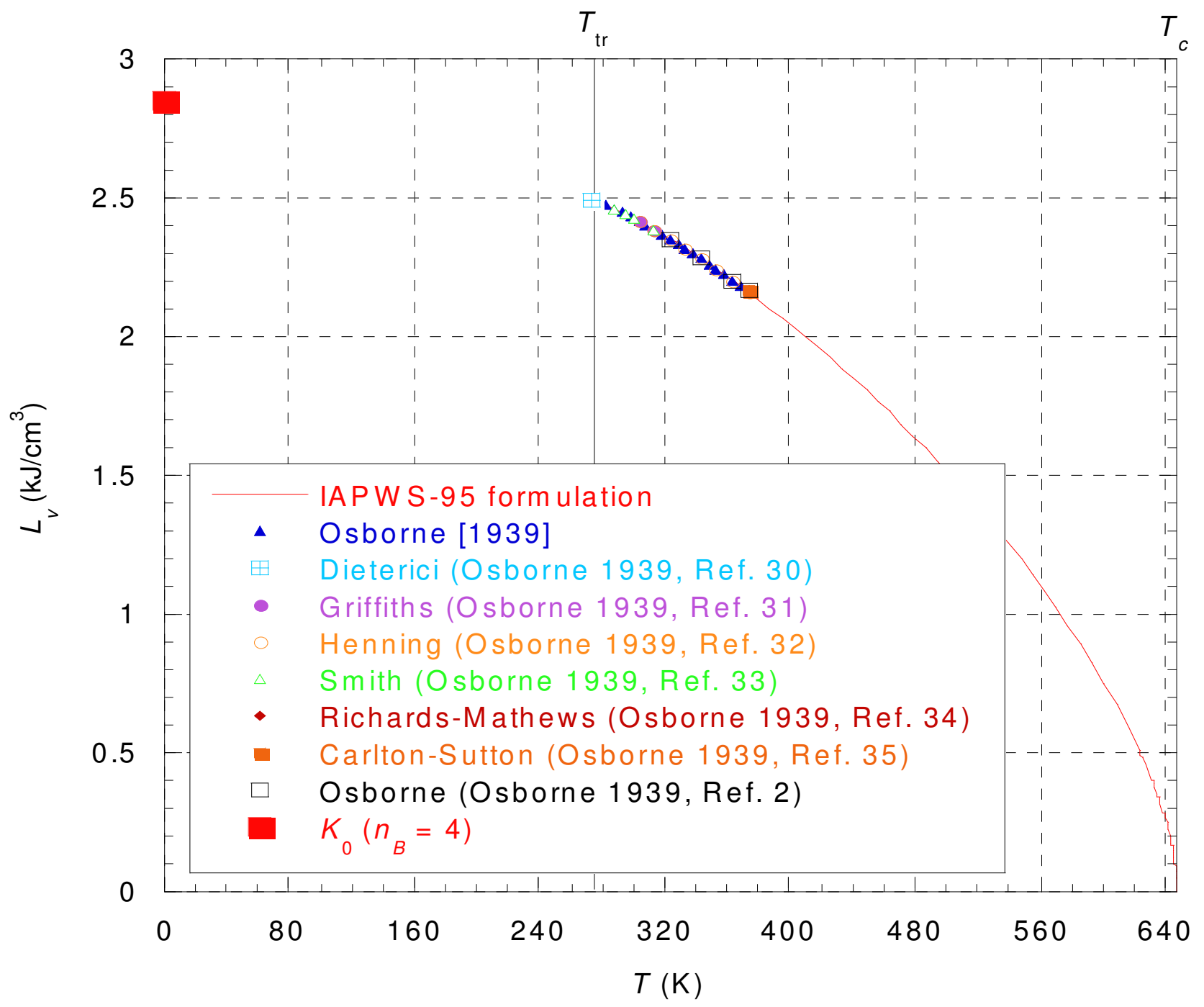

Fig. 116. Latent heat of vaporization of water versus the temperature. The red curve represents the IAPWS-95 formulation (Ref. 17) and all the data points are extracted from Osborne *et al.* (Ref. 88) therefore the reference numbers in the legend are all related to Ref. 88.

To understand how the extrapolation of latent heat of vaporization data should be considered, it is interesting to study the case of helium 4. According to Ref. 16, it is possible to calculate the latent heat between 0.8 K and the critical temperature. Using NIST data, one can model the different parameters and extrapolate the latent heat of vaporization curve using the Clapeyron equation. Fig. 117 then shows the evolution of latent heat down to 0.4 K. It appears that in the superfluid phase, $L_v$ decreases when the temperature decreases and then increases strongly. This variation comes from a subtle combination between the variation of $(dP/dT)_\sigma$ and the gaz volume $V_{\sigma,\mathrm{gas}}$, both on the saturated vapor pressure curve: indeed $(dP/dT)_\sigma$ decreases while $V_{\sigma,\mathrm{gas}}$ increases when the temperature tends towards zero. We can thus deduce that the value of $K_0$ for helium 4 is not aberrant given the function $L_v(T)$.

If we admit that the case of helium 4 is the archetype of the behavior of liquids if their solidification could be prevented, then we deduce that $K_0$ must have a value higher than the value of the latent heat of vaporization corresponding to the triple point, exactly as observed for water. On the other hand, any representation of the extrapolation of the latent heat is hazardous because of the product $V_{\sigma,\,\mathrm{gas}}\left(dP/dT\right)_\sigma$ in which each of the terms must be extrapolated without experimental background.

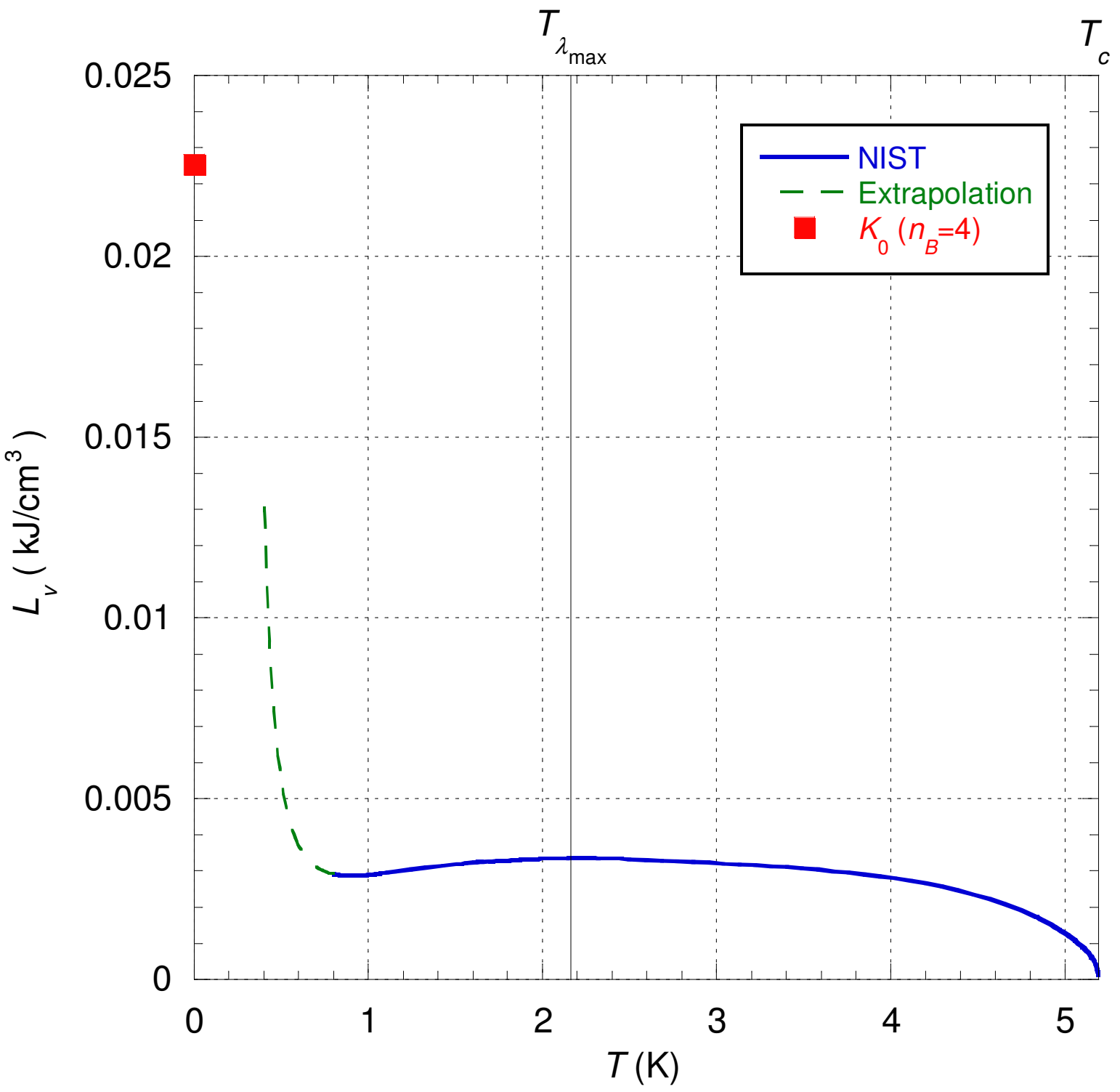

Fig. 117. Latent heat of vaporization of helium 4 versus the temperature. The blue curve represents the NIST data calculated from Ref. 16 and the dashed green curve the extrapolation down to 0.4 K of NIST data using the Clapeyron equation. $T_{\lambda_{\max}}$ represents the temperature of the lambda line which coincides with the saturated vapor pressure curve, i.e. $T_{\lambda_{\max}} = 2.1768\,\mathrm{K}$ .

Chen *et al.* (Ref. 89) measured the latent heat of vaporization of the following rare gases: argon, krypton and xenon. Therefore, a relevant extrapolation can be obtained for these media. Fig. 118 shows that for all these media the value of $K_0$ is about twice the value of the latent heat of vaporization at the triple point while there is a factor of 10 in the case of helium 4.

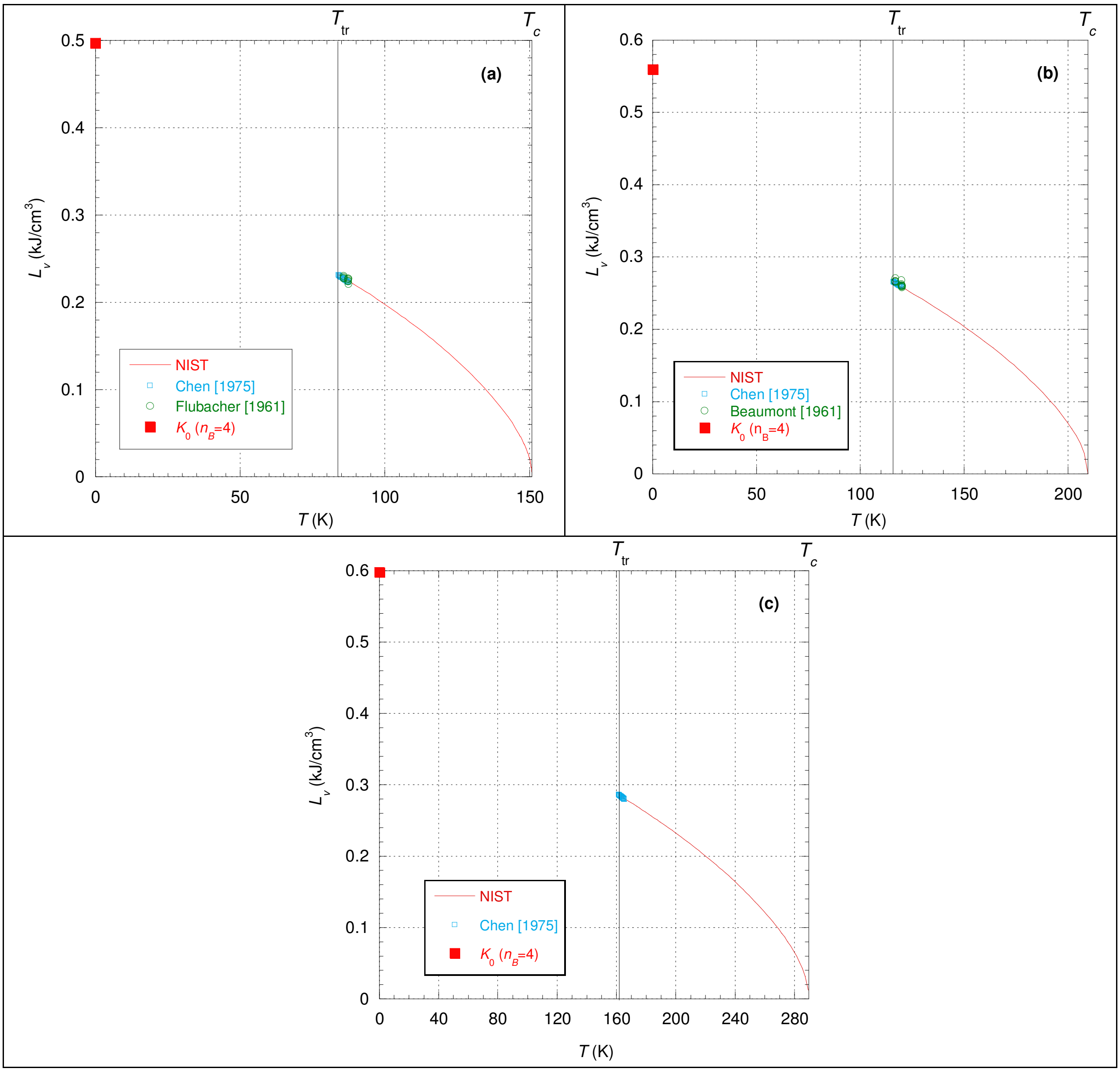

Fig. 118. Latent heat of vaporization of some rare gases versus the temperature: (a) Argon; the data are from Ref. 89 and Ref. 90. (b) Krypton; the data are from Ref. 89 and Ref. 91. (b) Xenon; the data are from Ref. 89. All the NIST red curves are calculated from Ref. 16.

To end this section, we will illustrate the case of some liquids whose molecules are more complex than simple rare gases. The figure shows for the 4 media presented that the value of $K_0$ is now less than a factor of 2 times the value of the latent heat of vaporization at the triple point. We notice that Octane is a bit different in that the latent heat necessarily implies $n_B = 2$ for this medium and moreover the function $L_v(T)$ shows an inflection point around 320 K. However, the data are not precise enough to confirm this behavior of $L_v(T)$.

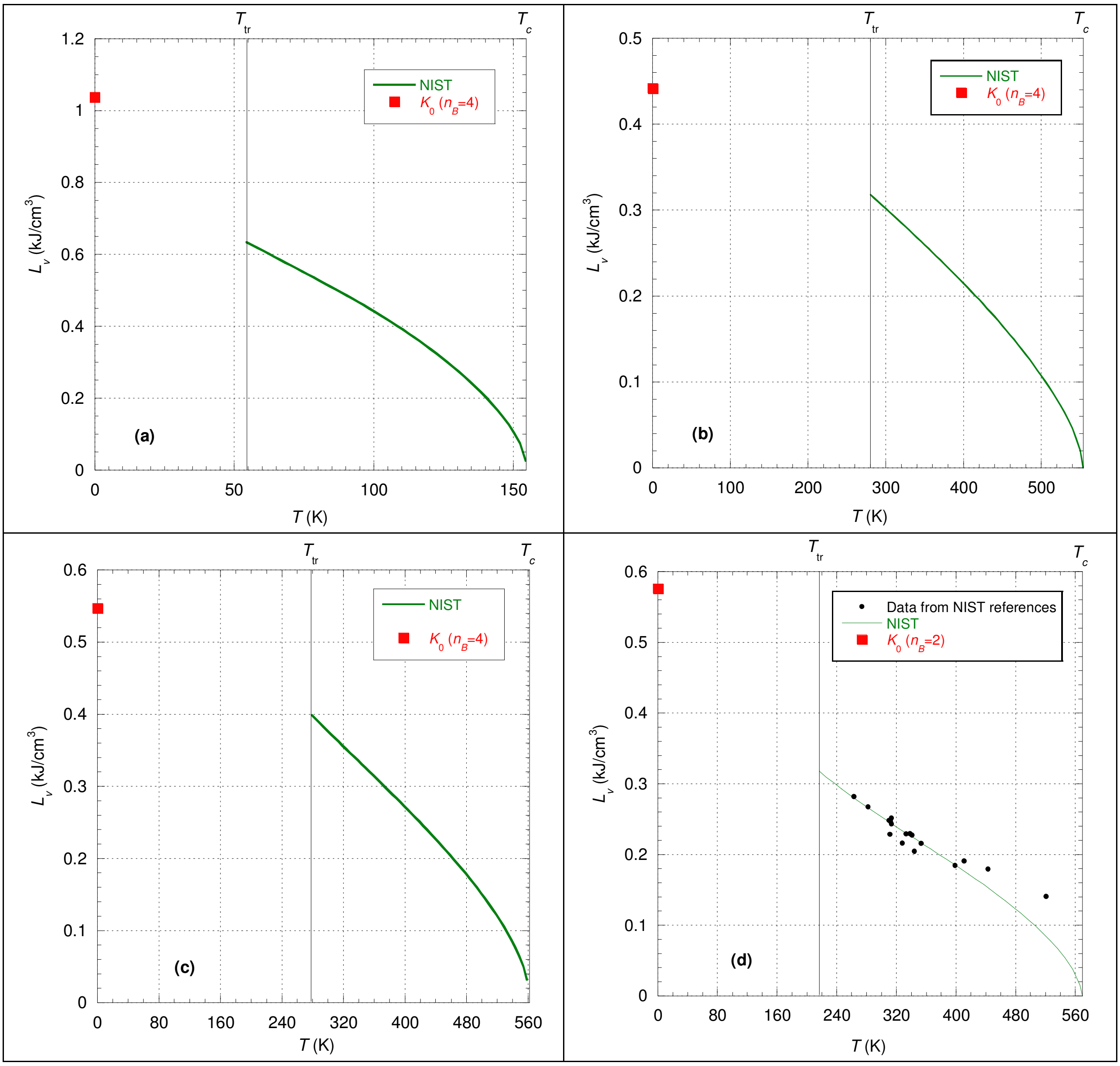

Fig. 119. Latent heat of vaporization versus the temperature for different liquids: (a) Oxygen; (b) Cyclohexane; (c) Benzene; (d) Octane. All the NIST green curves are calculated from Ref. 16.

The various examples presented in this section show that the values of $K_0$ are much systematically greater than the value of the latent heat of vaporization at the triple point and this condition makes it possible to fix the value of $n_B$ when the crystallography of the medium is poorly known. In any case the values of $K_0$ are not aberrant with respect to the latent heat of vaporization.

## 9 APPENDIX C

The main point of this Appendix is to show that the correlation length defined by Eq. (10) can be identified with experimentally "observed" lengths. As an example, Xie *et al.* (Ref. 88) measured the structure factor $S(q)$ of water using synchrotron-based small-angle X-ray scattering in the supercooled phase. These authors deduced a density fluctuations correlation length from a fit of their $S(q)$ measurements using their Eq. (2) which consists of a decomposition of the structure factor $S(q)$ into a sum of two components called "normal" and "abnormal". The additive decomposition in the frame of the model presented is inadequate and therefore the $S(q)$ data need to be re-analyzed. Thus, we have chosen to represent the $S(q)$ dataset by a unique term which is a simple "stretched lorentzian" function such as:

$$S(q) = C\left(\frac{\rho}{\rho_{\mathrm{tr,Liq}}}\right)^{1/3} \frac{\xi}{1+(\xi\, q)^{3/2}} \quad \text{(C.1)},$$

where $C = 0.04$ represents a device constant. Fig. 120(a) shows that this relation with a single free parameter $\xi$ not only allows the data of $S(q)$ to be correctly reproduced, but is particularly better than the Xie *et al.*'s fitting funtion on the isotherm at 0 °C.

We now compare the results deduced from the above analysis of $S(q)$ and calculation from Eq. (10) with those in Xie *et al.*'s Fig. 2. Taking into account the parameters of the model for water and those for the particular experiment of Xie *et al.* (i.e. $d_N$ must be rescaled by a factor $C_N = 0.01$), and using Eq. (10) it gives the black curve in Fig. 120(b): we can see in this figure that the results of the analysis of $S(q)$ and the calculation of Eq. (10) are in agreement. However, we note that the value of the red square at -34 °C is slightly too low, but this is explained by the fact that on this isotherm there are two points of $S(q)$ which are very outside their error bar, so the uncertainty relative to this point is greater than the size of the square. Now Fig. 120(b) shows that the values deduced by Xie *et al.* are practically two times larger than our values. Their correlation length seems to decrease very slightly on cooling contrary to the variation given by the present modeling: an increase of the correlation length when the temperature decreases seems physically more understandable than the opposite. One possible explanation for these differences may be found in the model used by Xie *et al.* as the authors themselves point out:

> "If no normal component is assumed in the fit [of $S(q)$], the correlation lengths are much smaller—from 2.0 to 2.8 Å, in good agreement with Dings, Michielsen, and van der Elsken [Rev. A 45, 5731, 1992]."

Finally, we can also note that both the results of Xie *et al.* and ours exclude any divergence for this correlation length near 228 K (i.e. -45 °C) where many measurable quantities seem to diverge when considering conventional analysis although our modeling predict that $\xi$ continue to increase on cooling.

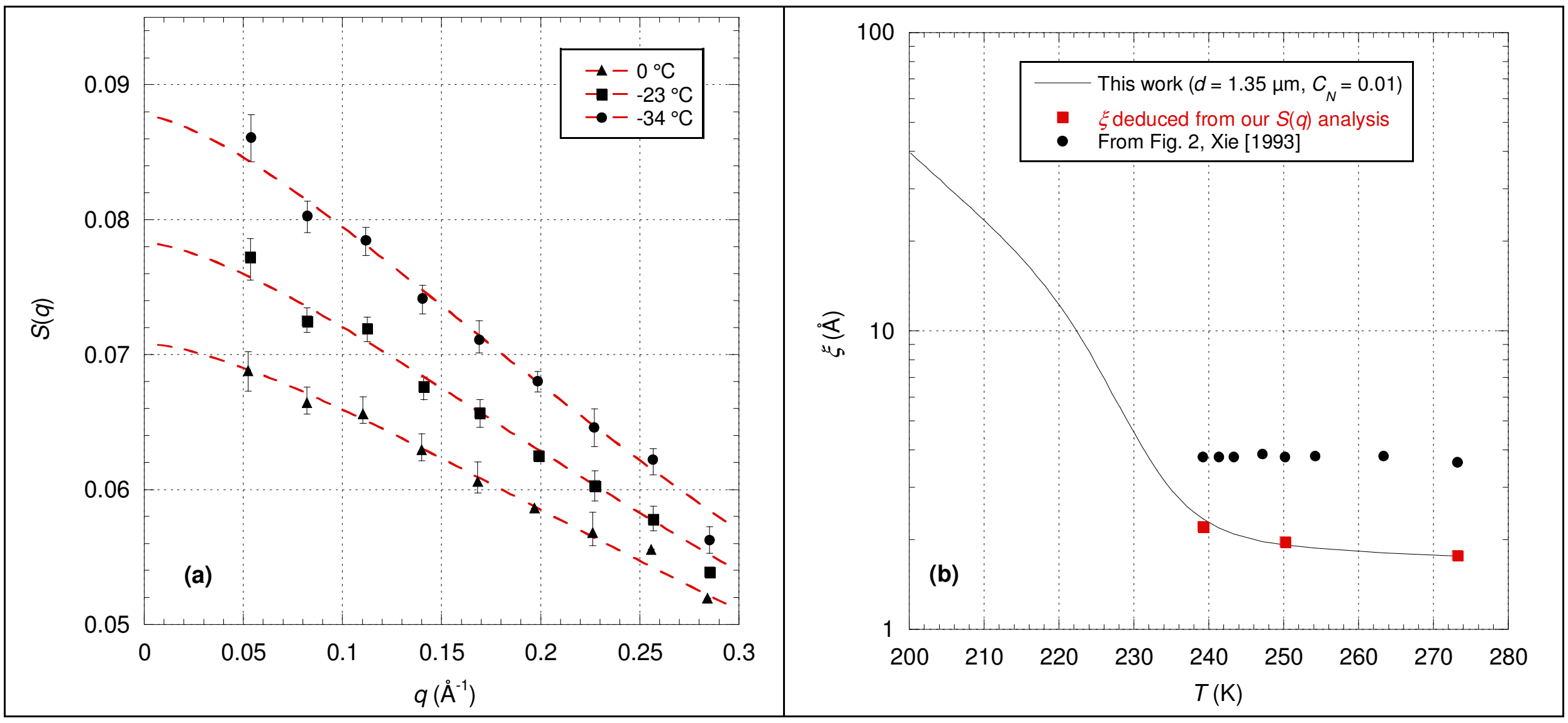

Fig. 120. (a) Plot of the structure factors of water measured by Xie *et al.* (Ref. 88) (black markers) along three isotherms in the supercooled phase and their representations (red dashed curves) with Eq. (C.1); (b) Comparison of the correlation lengths deduced by Xie *et al.* (black points) and calculated from Eq. (10) of the present modeling (black curve). The red squares represent the values obtained from $S(q)$ data analysis using Eq. (C.1). $d$ is a characteristic length of the system introduced in section 2.2.

# 10 APPENDIX D

In this appendix we will present a new state equation in the form of a Tait-Tammann equation which allows us to describe continuously the supercooled states and those of the normal liquid phase. In Ref. 93, it has been shown that such an approach makes it possible to represent, with a precision equivalent to the IAPWS-95 formulation (Ref. 17), the $P$-$V$-$T$ data of water in the normal liquid phase for pressures lower than 100 MPa. For higher pressures, it has been shown that the modified Tait equation should be used. But given the greater compressibility of the supercooled phase compared to the normal liquid phase for the same pressures, Sotani *et al.* (Ref. 94) and Asada *et al.* (Ref. 95) showed that it is possible to represent, using a Tait-Tammann equation, the data in the supercooled phase up to 380 MPa for temperatures above 253 K. Here we will extend the temperature range up to 200 K and the pressure range up to 400 MPa.

Following the notations of Ref. 93, we will write the Tait-Tammann equation such as:

$$V(T,P)=V_0(T)-\tilde{J}(T)\ln\left(\frac{P+\tilde{\Pi}(T)}{P_0+\tilde{\Pi}(T)}\right) \qquad \text{(D.1)},$$

where $V(T,P)$ represents the specific volume, $\tilde{J}(T)$ and $\tilde{\Pi}(T)$ are the two Tait-Tammann parameters. In order to describe the supercooled phase, we consider here that $P_0 = 1\,\text{atm}$.and $V_0(T)=V_{1\text{atm}}(T)$ where $V_{1\text{atm}}(T)$ represents the specific volume along the atmospheric isobar. The determination of the state equation $V_{1\text{atm}}(T)$ is the subject of section 10.1 and the state equations of $\tilde{J}(T)$ and $\tilde{\Pi}(T)$ are given in section 10.2.

## 10.1. Representation of densities along the atmospheric isobar

In this section, we present a new equation of state used in this paper to describe the states from the supercooled phase to the saturated vapor pressure point along the atmospheric isobar. The new density formula, with density in g/cm$^3$, is:

$$\begin{aligned} \rho_{1\mathrm{atm}}(T \le T_\sigma)^{-1} = {} & 1.2236 - \frac{0.1125}{1+(T/395.939)^{12.835}} + 0.00946\exp\left(-\left|\frac{T-359.613}{49.408}\right|^{2.3079}\right)\left(1-\exp\left(-\left(\frac{T}{291.624}\right)^{40.969}\right)\right) \\ & - 0.00946\exp\left(-\left(\frac{T-242.553}{15.265}\right)^{2}\right) + \frac{0.08335\exp\left(-\left|\frac{T-242.931}{12.327}\right|^{1.9211}\right)}{1+(T/273.788)^{-11.274}} - 6.2409\times10^{-5}\exp\left(-\left|\frac{T-341.45}{9.324}\right|^{3.32}\right) \\ & - \frac{0.000353\exp\left(-\left(\frac{T}{302.99}\right)^{2.96}\left|\frac{T-302.99}{7.786}\right|^{2.698}\right)}{1+(T/302.99)^{35}} + 2.8363\times10^{-5}\exp\left(-\left|\frac{T-320.45}{5.3936}\right|^{3.35}\right) \end{aligned} \qquad \text{(D.2)}$$

where $T$ represents the temperature in Kelvin and $T_\sigma$ represents the temperature on SVP at the atmospheric pressure. From the IAPWS-95 formulation (Ref. 17), $T_\sigma \cong 373.124\ \mathrm{K}$. Eq. (D.2) can be now compared with some experimental datasets and other models in the literature.

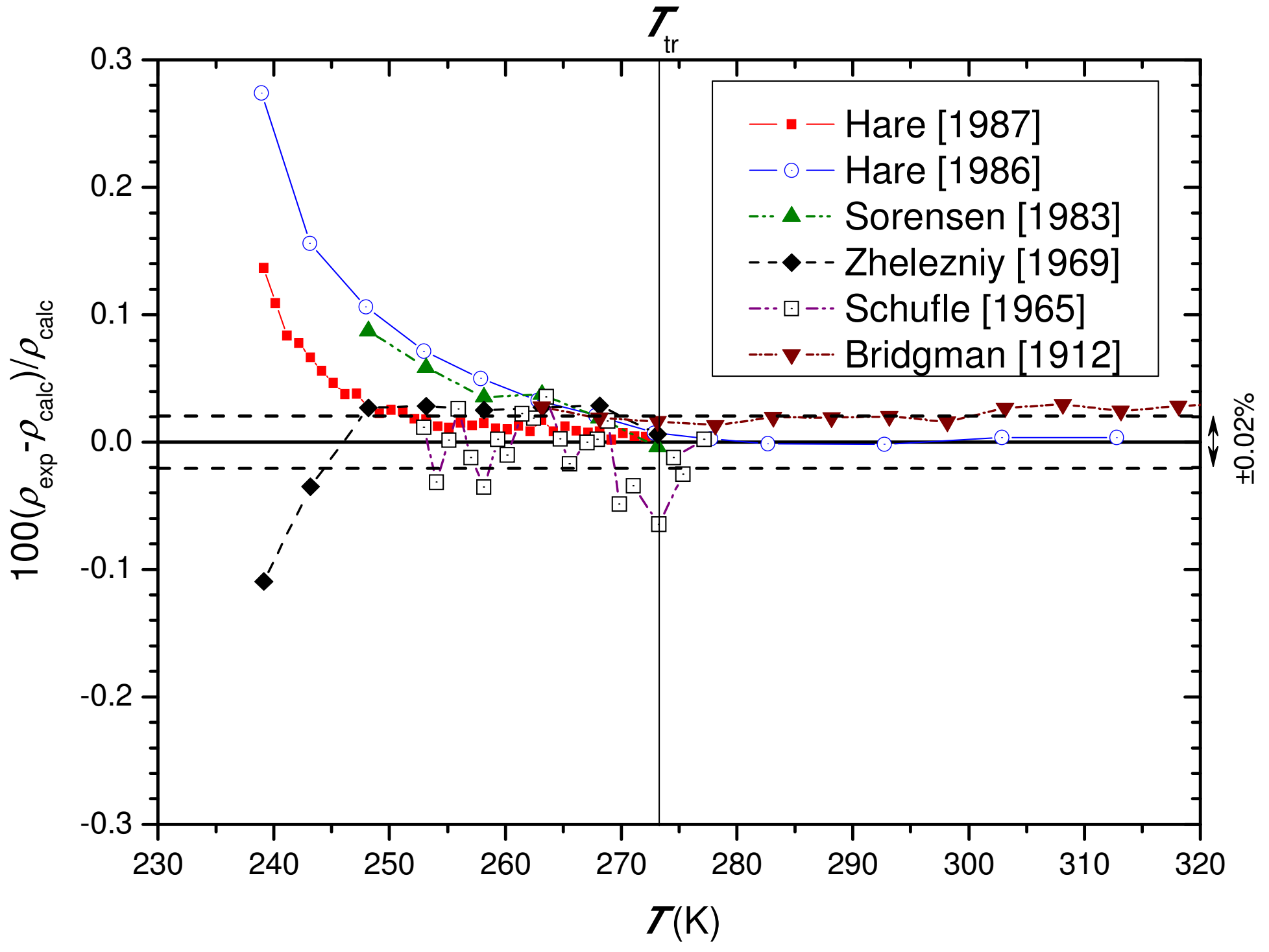

Fig. 121. Percentage deviations of experimental density data (Ref. 18 and 93 to 100) at atmospheric pressure from values calculated with Eq. (D.2). The lines are eye guides.

Fig. 121 shows the deviation of Eq. (D.2) from some experimental data considered to be very accurate. We have seen that Bridgman's data (Ref. 18) have been used in this paper to calculate densities in the supercooled phase. Although they are not very extended in temperature in the supercooled phase, it is interesting to observe that they seem shifted by

+0.02% but this shift appears to be compatible with the uncertainties of other experimental datasets. Indeed Sorensen (Ref. 93) wrote about his data:

> "From our reproducibility with temperature cycling mentioned above, we estimate the overall accuracy of all our density measurements to be $3\times10^{-4}$ g/cm$^3$."

In other words, considering the density values, this represents an uncertainty of about ±0.03%. With the exception of Sorensen's two coldest points, these data points are therefore compatible with our formula. The greatest deviation is obtained with Hare *et al.*'s data (Ref. 97) measured in 1986. These authors wrote:

> "Deviations greater than one part in $10^4$ appear for $H_2O$ for $T \leq$ - 20 °C, and for $D_2O$ for $T \leq$ - 5 °C. These deviations become as large as 30 parts in $10^4$ at the lowest temperatures."

In other words, considering the density values, this represents an uncertainty of about ±0.3%. Also, these authors found deviations of about ±0.2% with some other experimental datasets (see their Fig. 1). These authors have also concluded that:

> "[…] our densities may be as much as several parts in $10^4$ too large at -34.2 °C […]"

Therefore, the deviations from Eq. (D.2) with 1986 Hare *et al.*'s data in the supercooled phase are in line with their comments. From their conclusion, Hare *et al.* performed a new experiment in 1987 (Ref. 98) with much larger inner diameter capillary tubes. We have presented on Fig. 121 the smoothed data of Hare *et al.*: the deviation of their polynomial fit compared to their data according to their Fig. 2 is ±0.02%. We observe that our formula is compatible with the smoothed data above 250 K within the estimated experimental uncertainty.

It is interesting to note that Hare *et al.*'s data measured in 1987 are very different from those measured in 1986. The difference is mainly attributed to capillary effects that could strongly disturb the measurement in small diameter capillaries. This argument has indeed been used on a hypothetical basis to exclude the data of Zheleznyi (Ref. 99) and Schufle (Table I from Ref. 100) as Hare *et al.* wrote:

> "These authors showed that 4 and 10 μm diameter capillaries used by Schufle and Venugopolan had considerable surface effect problems. They also argued that the data of Zheleznyi, who used 2 μm i.d. capillaries, are probably unreliable."

But finally, it appears on Fig. 121 that Zheleznyi's data above 250 K are between the 1986 and 1987 datasets and that the 1987 data are almost identical to those of Schufle. In other words, the argument put forward to exclude these data is unfounded. In view of this observation, we have chosen to switch between the 1987 Hare *et al.*'s data and those of Zheleznyi. This variation is also imposed by the expansivity variations of Ter Minassian *et al.* (Ref. 101).

Indeed, Ter Minassian *et al.* (Ref. 101) have determined the expansivity along 14 isotherms up to 600 MPa by using different setups. All the 14 isotherms are above the triple point temperature and with these data the authors have built an equation of state such that:

> "The fitting of our experimental points with an equation like Eq. (8) allowed us to determine $\alpha$ [i.e. the expansivity named here $\beta$], $\kappa$ [i.e. the isothermal compressibility], and $\Delta C_P$ in the range of temperature and pressure investigated. The extrapolation into the supercooled region of water has permitted us to obtain a fair agreement with the compressibility data of other authors (Fig. 5)."

Fig. 122 shows the comparison between Eq. (D.2) and Holten *et al.*'s state equation (Ref. 102) with Ter Minassian *et al.*'s state equation (i.e. their Eq. (8)): we observe that the global deviation over the entire temperature range is in the order of $2\times10^{-5}$ $K^{-1}$ for Eq. (D.2) while Holten *et al.*'s state equation diverges both at low and at high temperature by deviating widely from the experimental uncertainties.

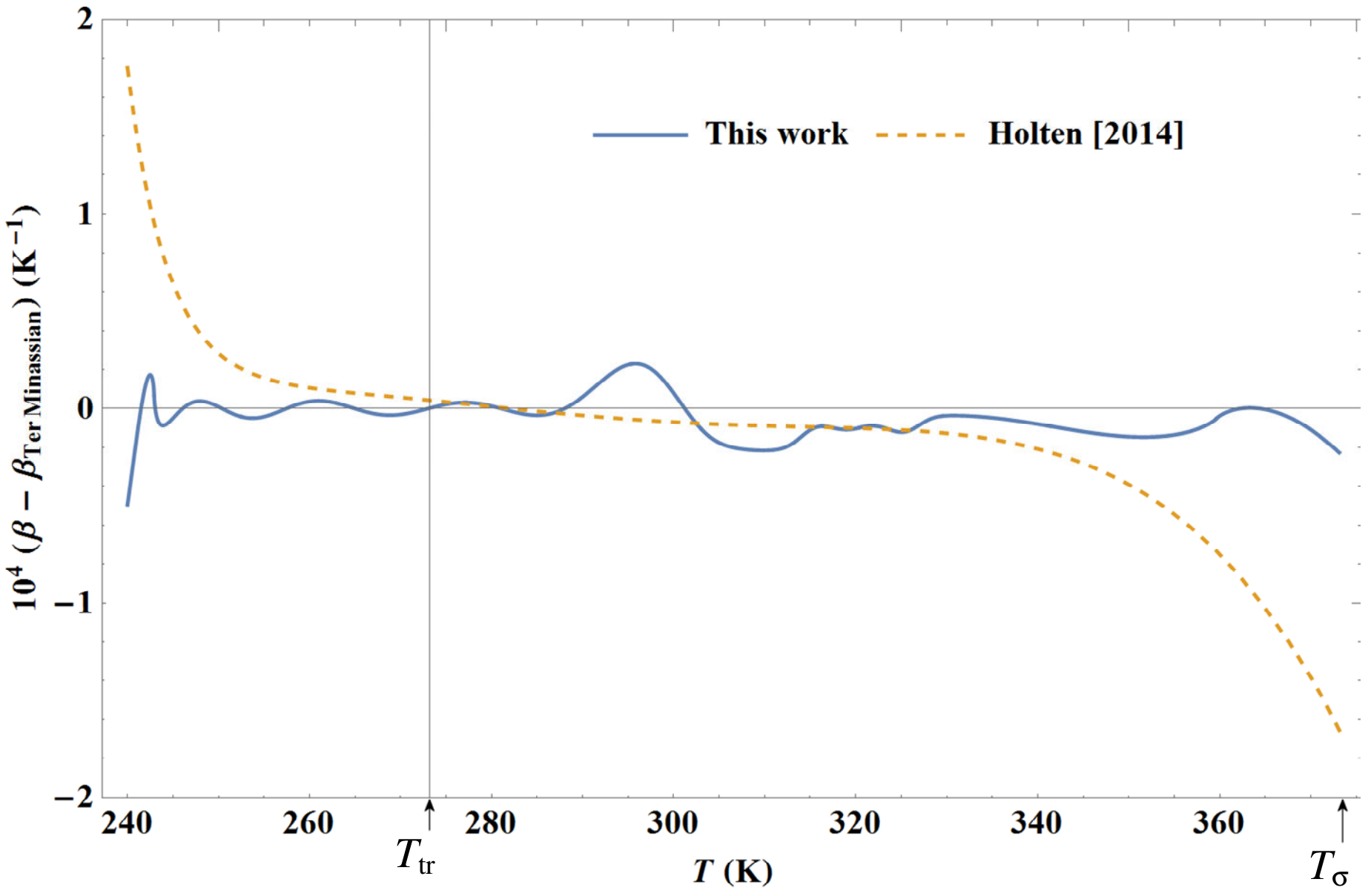

Fig. 122. Expansivity difference at atmospheric pressure where values from Ter Minassian *et al.*'s equation of state (Ref. 101) have been subtracted from Eq. (D.2) (blue curve) and from Holten *et al.*'s equation of state (dashed curve, Ref. 102). $T_\sigma$ represents the saturated vapor temperature value corresponding to this isobar.

Finally, we can compare Eq. (D.2) with different empirical equations of state of the literature. Most of the equations of state in the literature are only valid above 240 K and do not cover the entire temperature range up to $T_\sigma$(1 atm.). Apart from Eq. (D.2) which can be extrapolated to 0 K, one finds the equation of state developed by Kell (Ref. 103) which can be extrapolated down to 200 K. In the same way as Ter Minassian *et al.* (Ref. 101), Kell's state equation was developed by considering density data only above 0 °C and its extrapolation for lower temperatures is described as follows by the author:

> "The entries for the region below 0° C., metastable relative to ice, are based on extrapolations of the equation for the density and the equation for the compressibility outside the range where they were fitted."

Fig. 123 shows that Eq. (D.2) is quite comparable with Kell's state equation: the deviation in the supercooled phase is less than 0.02%, i.e. is lower than the uncertainty of the experimental data discussed previously. In the normal liquid phase, Kell's state equation is almost identical to the IAPWS-95 formulation (Ref. 17) and with these two formulations Eq. (D.2) has a deviation of about 0.0007%. This deviation can be considered too high if we accept the uncertainty of 0.0001% estimated by the IAPWS-95 formulation between $T_{tr}$ and 359 K. Over a narrower temperature range from $T_{tr}$ to 325 K, the state equation developed by Holten *et al.* (Ref. 102) is better in that it can be identified with the IAPWS-95 formulation but it can certainly not be extrapolated beyond 325 K as mentioned above. In the supercooled phase, the 1995 IAPWS and Holten *et al.* state equations follow the 1987 Hare *et al.*'s data (Ref. 98), hence the deviation observed in Fig. 123.

Sotani *et al.* (Ref. 94) measured the specific volumes of water along 11 isotherms at temperatures from 253 to 298 K and pressures up to 200 MPa. These authors wrote about their data that:

> "The uncertainty in the specific volumes obtained was estimated to be less than ±0.01%. The specific volume data obtained were correlated with an empirical Tait [i.e. Tait-Tammann] equation as a function of temperature and pressure. […]
> The deviation between our specific volume data and those of Hare and Sorensen tends to increase with decreasing temperature. The deviation of our data from those of Hare and Sorensen at 253 K is about 0.025%."

It can be deduced from this that at 253 K the data of Sotani *et al.* are outside the uncertainty estimated by Hare *et al.* (Ref. 98). But from Sotani *et al.*'s Fig. 5 it appears that the smallest temperature reached at atmospheric pressure is only 263 K. So, their uncertainty at 253 K is deduced from the extrapolation of their Eq. (3) which represents their data at atmospheric pressure. In Fig. 123 we have represented the deviation of Eq. (D.2) with the Tait-Tammann equation developed by Sotani *et al.* So, if we limit ourselves to temperatures above 263 K, then the deviation between Eq. (D.2) and Sotani *et al.*'s Eq. (3) is compatible with the estimated uncertainty of the experimental data.

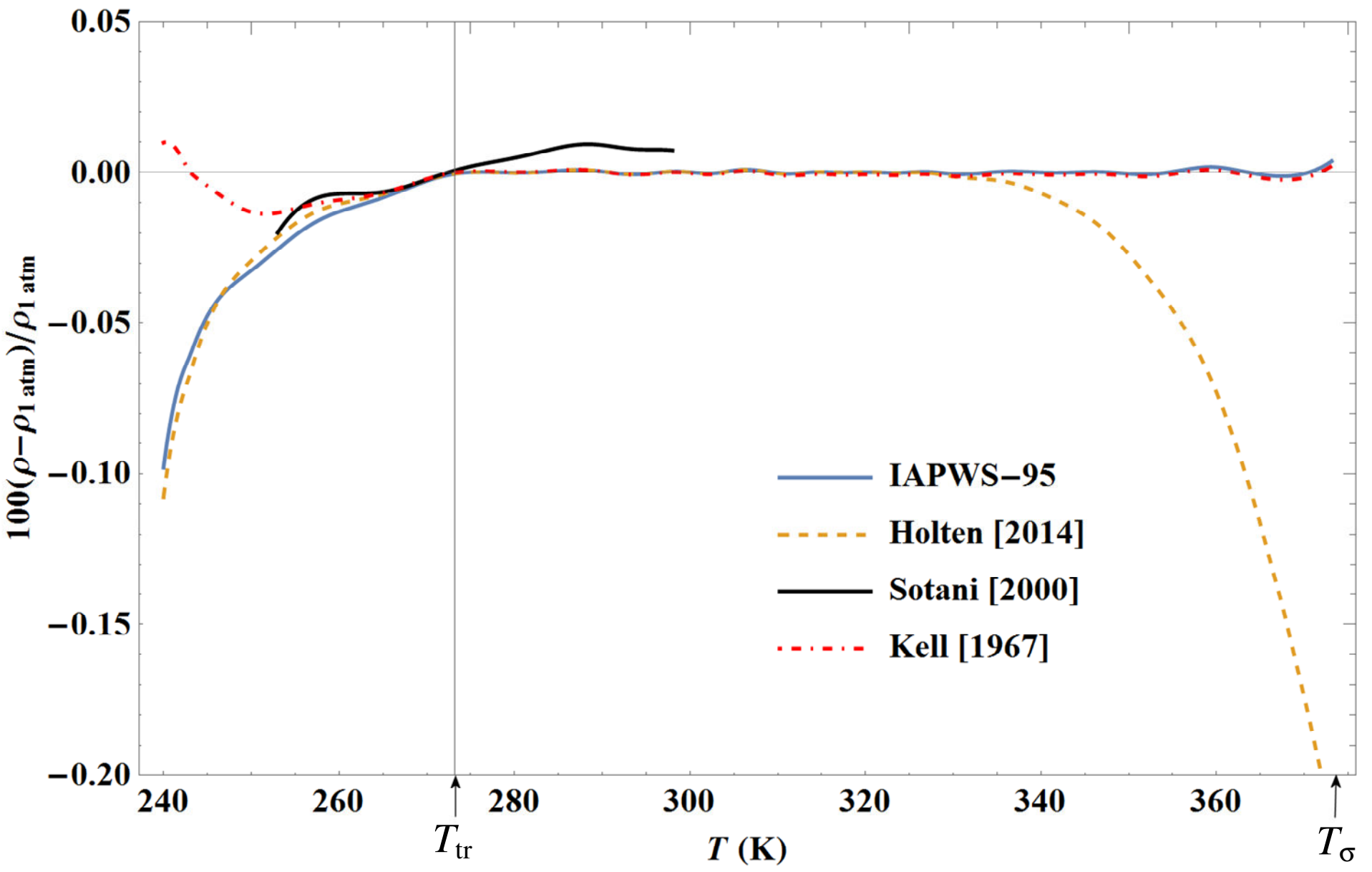

Fig. 123. Percentage deviations at atmospheric pressure of the IAPWS-95 formulation (blue curve, Ref. 17), Holten *et al.*'s state equation (dashed curve, Ref. 102), Sotani *et al.*'s Eq. (3) (black curve, Ref. 94) and Kell's state equation (dot-dashed curve, Ref. 103) from values calculated with Eq. (D.2). $T_\sigma$ represents the saturated vapor temperature value corresponding to this isobar.

To conclude this section, we have developed here a new equation of state that describes the liquid water density at atmospheric pressure. Even if this equation may seem to have a somewhat low accuracy in the normal liquid phase, it has the advantage of being able to calculate all states from 0 K to $T_{\sigma}$(1 atm.), unlike the literature equations which are generally very limited for their temperature range. This equation represents a large number of experimental data in accordance with their uncertainties especially in the supercooled phase. The extrapolation of the equation below 240 K is performed according to the prediction of Holten *et al.*'s Fig. 3 (Ref. 104) which is particularly important for analyzing some viscosity and self-diffusion coefficient data.

## 10.2. The Tait-Tammann equation of state

In order to realize a continuous description from the supercooled phase to the normal liquid phase, we have to determine a unique expression of the Tait-Tammann parameters. The expression of these parameters having been determined in Ref. 93 (Chapter 3), we will consider a deviation from the extrapolation of these expressions for the supercooled phase such that:

$$\begin{cases} \tilde{\Pi}(T) = \tilde{\Pi}_{\substack{\text{normal} \\ \text{liquid} \\ \text{phase}}}(T) + \Delta\tilde{\Pi}(T) \\ \tilde{J}(T) = \tilde{J}_{\substack{\text{normal} \\ \text{liquid} \\ \text{phase}}}(T) + \Delta\tilde{J}(T) \end{cases} \tag{D.3}$$

where $\tilde{\Pi}_{\substack{\text{normal} \\ \text{liquid} \\ \text{phase}}}(T)$ and $\tilde{J}_{\substack{\text{normal} \\ \text{liquid} \\ \text{phase}}}(T)$ are Eq. 3.12a and Eq. 3.12b of Ref. 93, respectively.

Only the expressions of $\Delta\tilde{\Pi}(T)$ and $\Delta\tilde{J}(T)$ are written in this Appendix because they do not appear in Ref. 93. Given the high compressibility of the supercooled phase, this makes it necessary to determine $\Delta\tilde{\Pi}(T)$ and $\Delta\tilde{J}(T)$ with very high numerical precision, which is difficult to achieve without a sufficient number of terms. The expressions of the deviation terms are:

$$\Delta\tilde{\Pi}(T)=\left(1.04931\times10^{7}-\frac{1.04931\times10^{7}}{1+(T/2.1887)^{-2.219}}\right)\left(1-\exp\left(-\left(\frac{230.059}{T}\right)^{30}\right)\right)\left(1-\exp\left(-\left(\frac{T_{\mathrm{tr}}}{T}\right)^{70}\right)\right)$$
$$-491.015\exp\left(-\left|\frac{T-227.678}{14.889}\right|^{1.632}\right)\left(1-\exp\left(-\left(\frac{236.185}{T}\right)^{30}\right)\right)-63.1931\exp\left(-\left|\frac{T-247.226}{12.878}\right|^{5.926}\right)$$
$$-42.5062\exp\left(-\left|\frac{T-247.709}{7.822}\right|^{2.499}\right)-\frac{158.617\exp\left(-\left|\frac{T-155}{44.7736}\right|^{3.085}\right)}{1+(T/155)^{30}}+1.14522T\exp\left(-\left|\frac{T-112.492}{43.7102}\right|^{5.084}\right)$$
$$+\frac{1.1477\exp\left(-\left|\frac{T-231.27}{0.895}\right|^{3.5}\right)}{1+(T/231.27)^{145}}-\frac{2.1476\exp\left(-\left|\frac{T-234.6}{1.805}\right|^{2.824}\right)}{1+(T/234.6)^{145}}-1.002\exp\left(-\left(\frac{T}{227.38}\right)^{38.132}\left|\frac{T-227.38}{2.437}\right|^{6.066}\right)$$
$$+1.7367\exp\left(-\left|\frac{T-227.65}{0.742}\right|^{1.553}\right)+\frac{0.98415\exp\left(-\left|\frac{T-218.37}{5.329}\right|^{9.812}\right)}{1+(T/218.37)^{-33.796}}-\frac{1.1839\exp\left(-\left|\frac{T-219.05}{2.735}\right|^{2.058}\right)}{1+(T/219.05)^{-33.796}} \quad \text{(D.4a)}$$

$$\Delta\tilde{J}(T)=\left(0.726583-\frac{0.781413}{1+(T/106.719)^{-3.62}}\right)\exp\left(-\left(\frac{T}{233.486}\right)^{10.432}\right)$$
$$+\left\{\frac{0.010835\exp\left(-\left(\frac{T}{624.799}\right)\left|\frac{T-624.799}{390.871}\right|^{118.41}\right)}{1+(T/253.861)^{89.963}}-\frac{0.081106\exp\left(-\left(\frac{T}{241.683}\right)^{9.337}\left|\frac{T-241.683}{19.1606}\right|^{2.916}\right)}{1+(T/212.237)^{8.237}}\right\}$$
$$\times\left(1-\exp\left(-\left(\frac{T_{\mathrm{tr}}}{T}\right)^{130.483}\right)\right)+5.1857\times10^{-5}\exp\left(-\left|\frac{T-231.01}{1.2977}\right|^{3.053}\right)-1.44\times10^{-4}\exp\left(-\left|\frac{T-235.62}{2.342}\right|^{2.952}\right) \quad \text{(D.4b)}$$
$$-1.9937\times10^{-4}\exp\left(-\left(\frac{T}{226.38}\right)^{11.5}\left|\frac{T-226.38}{2.51}\right|^{3.655}\right)+\frac{3.5705\times10^{-4}\exp\left(-\left|\frac{T-226.37}{2.4131}\right|^{3.299}\right)}{1+(T/226.37)^{9.584}}$$
$$+4.7888\times10^{-5}\exp\left(-\left(\frac{T}{219.63}\right)^{4.92}\left|\frac{T-219.63}{2.33}\right|^{2.6}\right)$$

where $\Delta\tilde{\Pi}(T)$ is in bar and $\Delta\tilde{J}(T)$ is in cm$^3$/g.

Fig. 124 shows that the deviation terms cancel each other out well in the normal liquid phase and therefore there is a connection between the representations of the normal liquid phase when we take as reference $V_0=V_{1\mathrm{atm}}$ or $V_0=V_\sigma$, $V_\sigma$ corresponding to the specific volume of the liquid on the coexistence curve, because it appears that $V_\sigma$ is numerically almost identical to $V_{1\mathrm{atm}}$ in the temperature range from $T_{\mathrm{tr}}$ to 300 K. Fig. 124 also shows that the extrapolation of $\tilde{\Pi}_{\substack{\text{normal}\\ \text{liquid}\\ \text{phase}}}(T)$ from the expression of Ref. 93 is such that $\lim_{T\to0}\tilde{\Pi}_{\substack{\text{normal}\\ \text{liquid}\\ \text{phase}}}(T)=-P_c$ which is not physically correct at low temperatures. Finally, we observe the presence of a minimum of the Tait-Tammann parameters around 223 K which is generally considered as the glass transition temperature of water. This change in the parameters variations is the consequence of the function $V_{1\text{atm}}(T)$ which below 223 K does not vary practically any more according to Holten *et al.*'s Fig. 3 (Ref. 104).

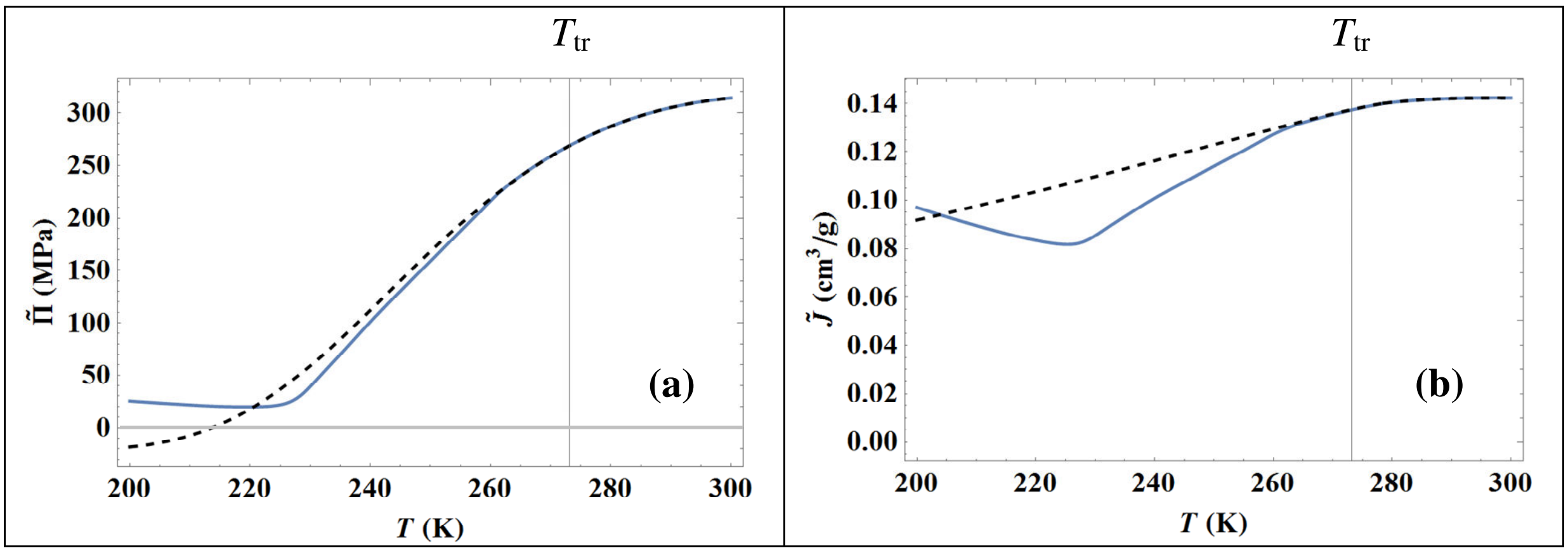

Fig. 124. Temperature variation of the Tait-Tammann parameters from extrapolation of the expressions in Ref. 93 (black dashed curve) and from Eq. (D.3) with Eq. (D.4) (blue curve): (a) representation of the function $\tilde{\Pi}(T)$; (b) representation of the function $\tilde{J}(T)$.

In order to compare the Tait-Tammann parameters with those of the litterature, it is necessary to identify the expressions of $V(T,P)$ of Sotani *et al.* (Ref. 94) and Asada *et al.* (Ref. 95) with Eq. (D.1). We deduce that $V_0 = V_{1\text{atm}}$ and $\tilde{\Pi}$ is identified with $B$ while $\tilde{J}$ is identified with $V_0 \times C$. Taking into account the fact that the expression of $V_0$ by Asada *et al.* (Ref. 95) contains errors on the numerical coefficients, we are going to compare the evolutions of the parameters $B(T)$ with $\tilde{\Pi}(T)$ and $C(T)$ with $\tilde{J}(T)/V_{1\text{atm}}(T)$. Fig. 125 shows that Tait-Tammann's parameters of the present modeling are quite close to those of Sotani *et al.* between 240 K and 300 K. The $\tilde{\Pi}(T)$ parameter is however closer to the $B$ parameter of Asada *et al.* between 240 K and 230 K. The parameters of Sotani *et al.* as well as those of Asada *et al.* were determined from measurements above 253 K. Fig. 125 also shows that if the relations of Sotani *et al.* and Asada *et al.* are extrapolated to lower temperatures up to 200 K then parameters $B$ and $C$ become physically inadmissible, in particular the fact that $C$ and $B$ become null or negative at low temperatures.

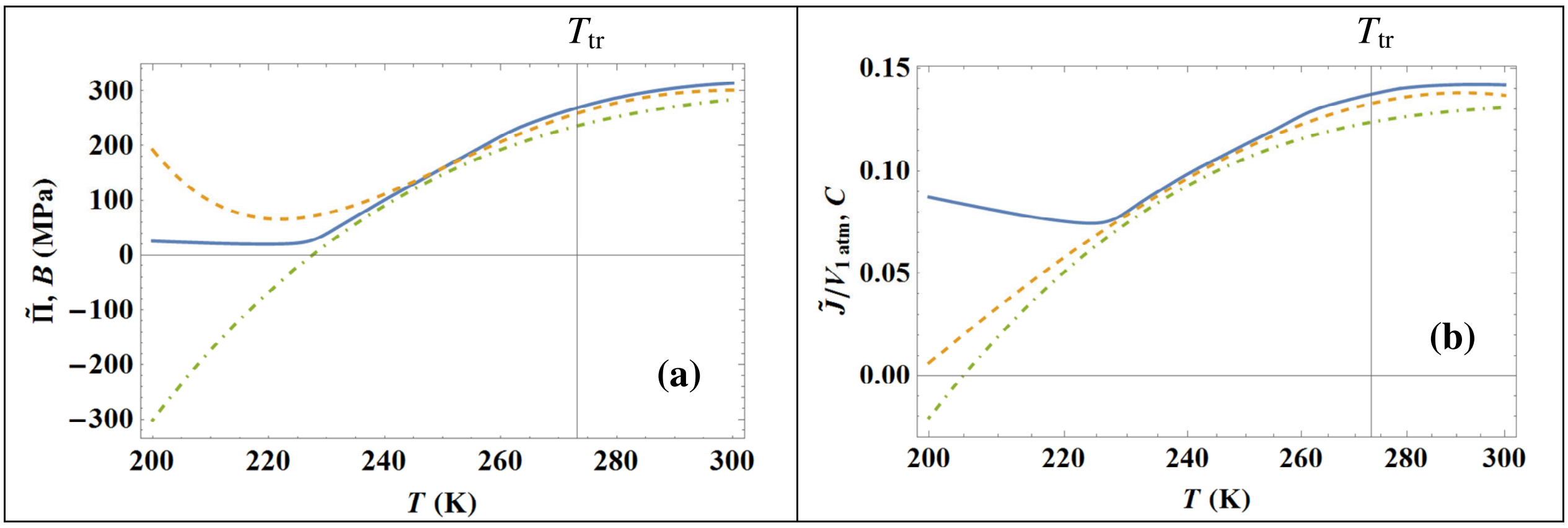

Fig. 125. Temperature variation of the Tait-Tammann parameters from Eq. (D.3) with Eq. (D.4) (blue curve), from Sotani *et al.* (Ref. 94, orange dashed curve) and from Asada *et al.* ((Ref. 95, green dot-dashed curve): (a) representation of the functions $B(T)$ and $\tilde{\Pi}(T)$; (b) representation of the functions $C(T)$ and $\tilde{J}(T)/V_{1\text{atm}}(T)$.

Sotani *et al.* (Ref. 94) wrote:

> "This equation [i.e. their Tait-Tammann equation] can reproduce the experimental data within the experimental error."

We quoted in section 10.1 that the experimental uncertainty admitted by Sotani *et al.* is of the order of ±0.01%. Fig. 126 shows that the difference between the two Tait-Tammann representations is greater than the experimental uncertainty proposed by Sotani *et al.* However, it is likely that Sotani *et al.* were overly optimistic about the value of the uncertainty. Indeed, Asada *et al.* (Ref. 95) who used the same experimental set-up as Sotani *et al.* and in the same pressure and temperature range wrote that:

> "The final uncertainties in the specific volume measurement are estimated to be less than ±0.1%."

This uncertainty value seems more compatible to characterize the data set. It is then deduced that Eq. (D.1) allows a correct representation of Sotani *et al.*'s data.

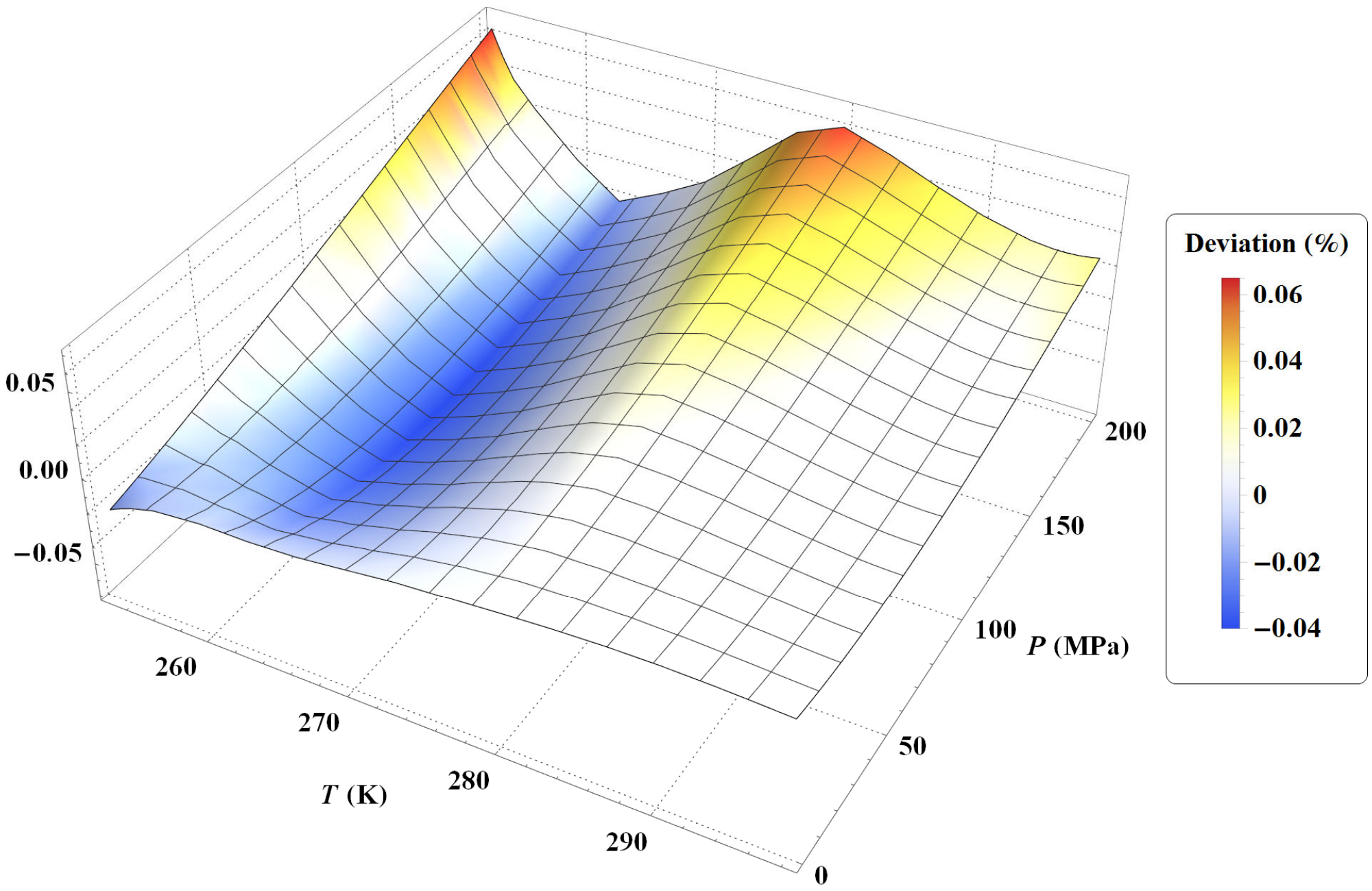

Fig. 126. Percentage deviation of the Tait-Tammann equation of state from Sotani *et al.* (Ref. 94) with Eq. (D.1), in the temperature range from 253 K to 300 K. The vertical axis represents $100\left(V_{\text{Sotani}} - V_{\text{Eq.(D.1)}}\right)/V_{\text{Eq.(D.1)}}$ .

The largest experimental data set (i.e. covering the widest range of temperature and pressure) is that of Mishima (Ref. 19). It is therefore important to be able to reproduce this data set correctly. Given what we have shown above, it is not possible to analyze such data set by using the relations of Sotani *et al.* and Asada *et al.* since Mishima's data covers a temperature range from 200 K to 270 K.

Mishima considers that the uncertainty of its experimental data is of the order of 2% maximum. Fig. 127 shows that one can reproduce, from Eq. (D.1), Eq. (D.3) and Eq. (D.4), Mishima's data with an overall deviation of ±0.4% with the exception of a few points at low pressure for some isotherms. The Tait-Tammann equation of state proposed here is the most accurate representation of these data.

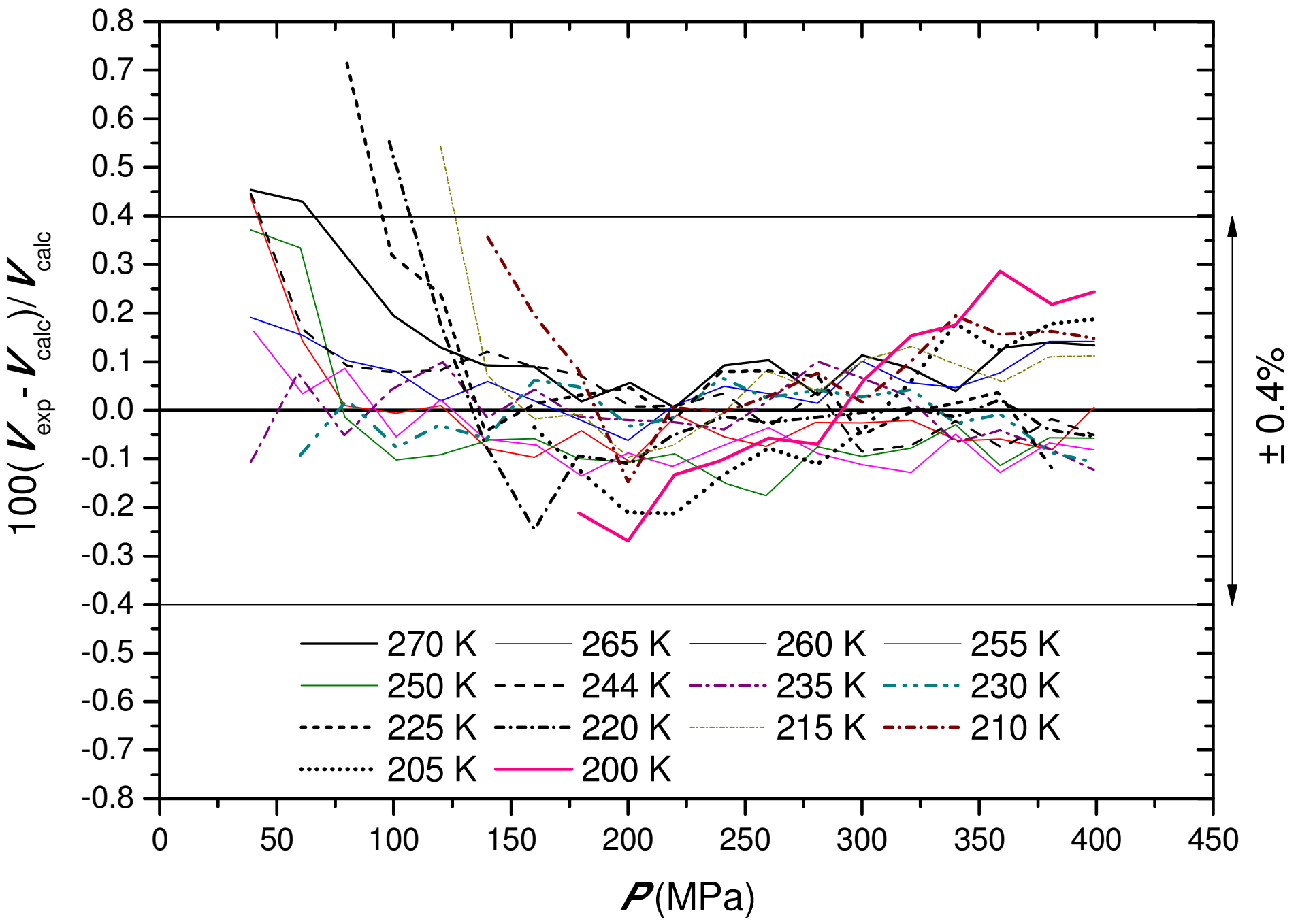

Fig. 127. Percentage deviations of Mishima's experimental specific volume data (Ref. 19) from values calculated with Eq. (D.1) for all experimental quasi-isotherms. The lines are eye guides.

We also mentioned in section 4 that we used Bridgman data (Ref. 18) to calculate densities in the supercooled phase. These densities extend over a small temperature range but go much higher in pressure. It is therefore interesting to compare the Tait-Tammann equation with these data. It can be seen in Fig. 128 that the deviation remains compatible with the uncertainty of Mishima's data. However, there is a systematic deviation in the same trend which shows that a Tait-Tammann equation of state is not a physically acceptable model for a too wide pressure range as described in Ref. 93 for the water normal liquid phase. It is therefore not recommended to use such an approach for pressures well above 400 MPa.

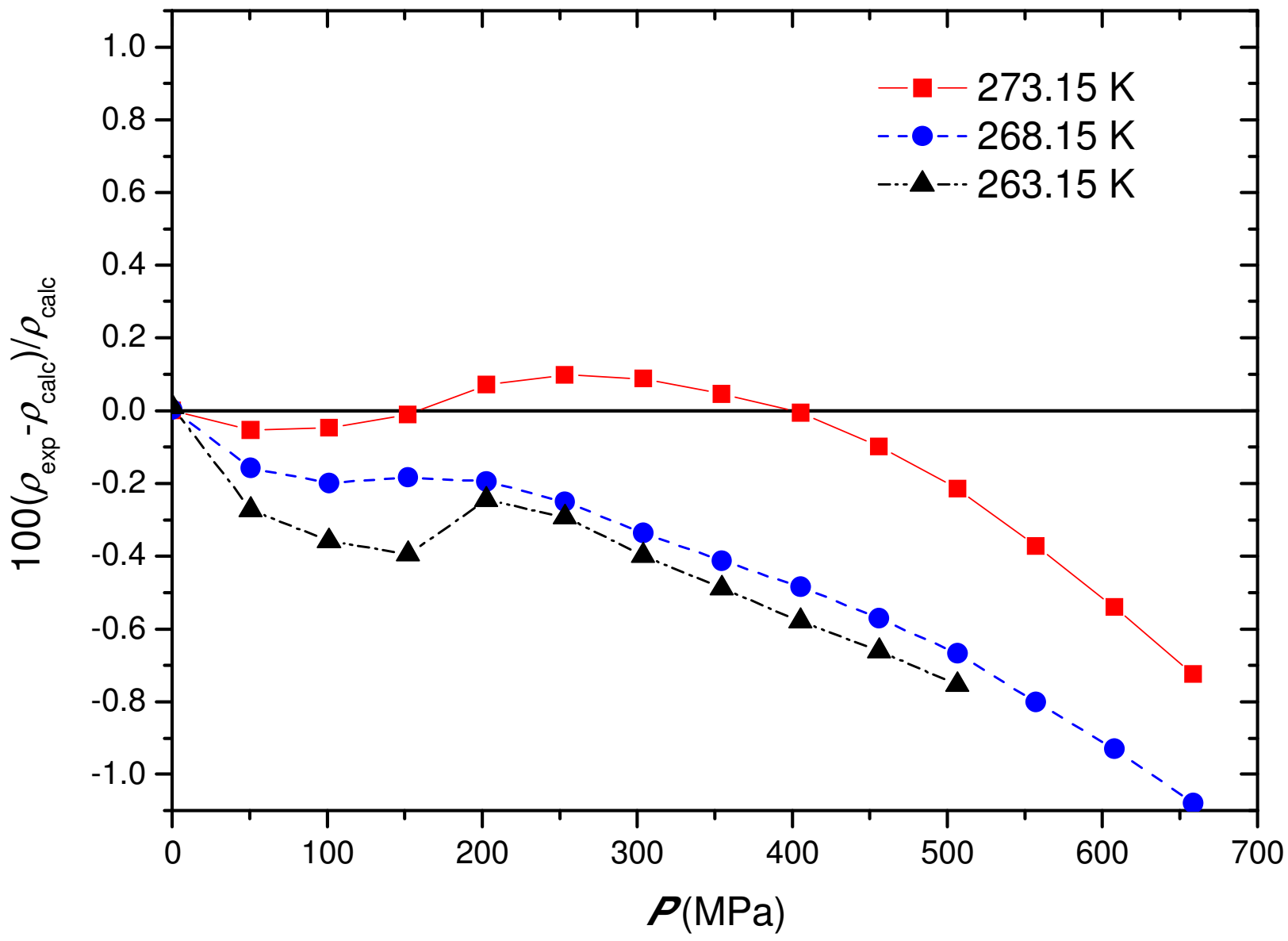

Fig. 128. Percentage deviations of Bridgman's experimental density data (Ref. 18) from values calculated with Eq. (D.1) for 3 isotherms. The lines are eye guides.

By staying within sufficiently low-pressure values, one can also analyze the compressibility data of Kanno *et al*. (Ref. 105). Fig. 129 shows a progressive shift in absolute values between the data and the calculation with the Tait-Tammann equation as the pressure increases but the variations seem identical given the error bars. This shift can be perfectly explained by the drift of the calibration constant of the instrument which was determined at 25°C and 1 atm. The data have also been corrected to agree with the data of Fine *et al*. (Ref. 106) and Grindley *et al*. (Ref. 107) along the 25°C isotherm while Mishima indicates that:

> "Since the uncertainty in the matrix weight caused an error in *V* (estimated to be less than ~2%), all of the above-mentioned averages *V* are slightly shifted by a fixed value within its experimental uncertainty in order to be in accord with the known volume of water [i.e. from Kell *et al*. (Ref. 108)]"

Or Fine *et al*. worked on an earlier reference by Kell *et al*. (Ref. 109) from 1965 and they wrote about the model they developed (i.e. for the temperature range from 0°C to 100°C):

> "Our specific volume results are in reasonable agreement with the work of Kell and Whalley at low pressures; our results at high pressures (1000 bar) disagree as much as 169 ppm [i.e. 0.0169%]."

There is therefore no incompatibility between Mishima's data and those of Kanno *et al*. since the discrepancy comes from different calibration choices that define the absolute values. It is therefore reasonable to assume that the Kanno *et al*. data should be renormalized with those of Mishima.

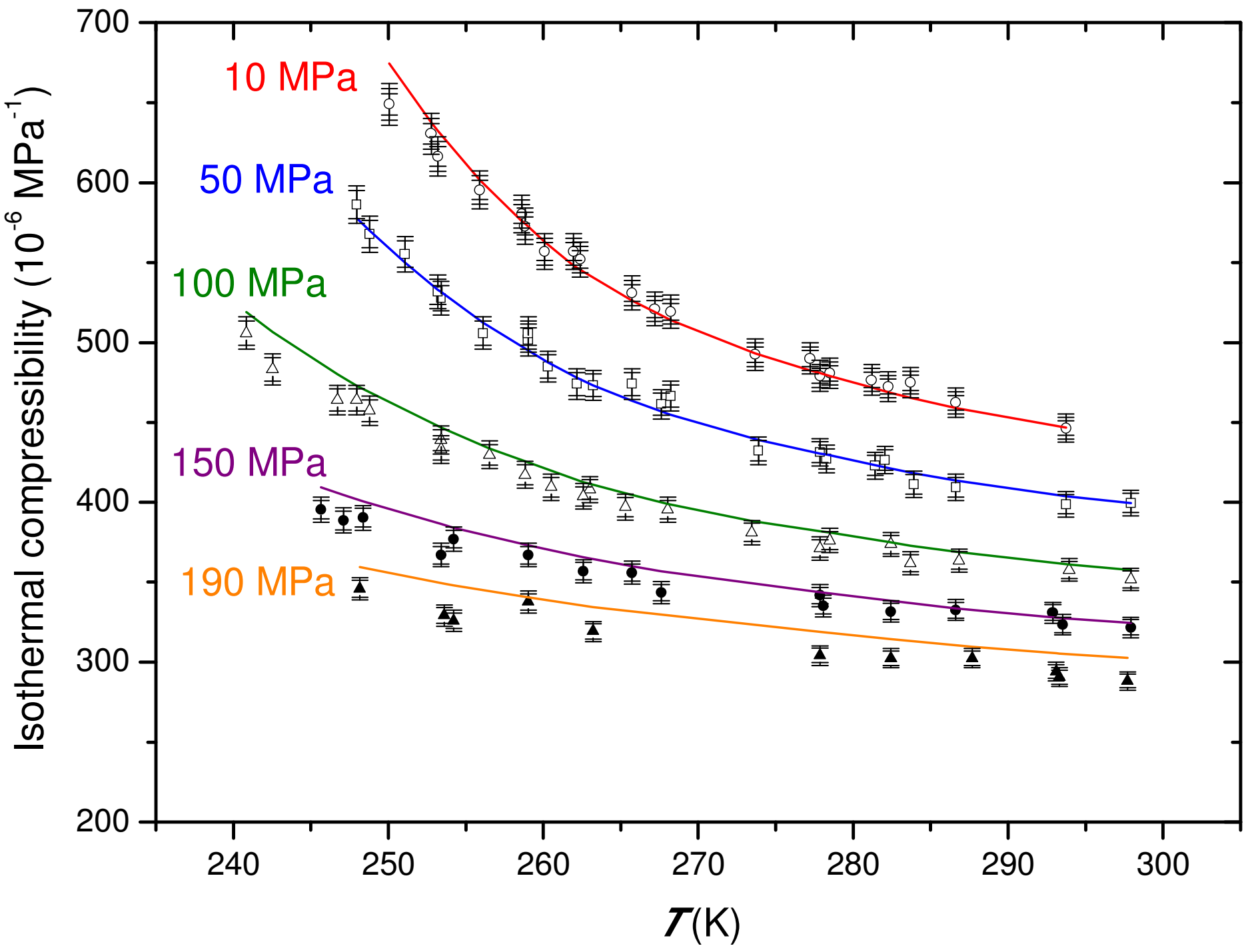

Fig. 129. Plot of the isothermal compressibility factor of water measured by Kanno *et al.* (Ref. 105) (black markers) along five isobars and their representations (colored curves) with Eq. (D.1).

In the literature, there are various models for approaching *P-V-T* data in the supercooled phase. We have already mentioned Ter Minassian *et al*.'s approach (Ref. 101), which from isobaric expansion coefficient data in the normal liquid phase extrapolates its expression in the supercooled phase. As the most recent approach is that of Holten *et al.* (Ref. 102), we will end this section with a comparison between the model of Holten *et al.* and the Tait-Tammann equation of state Eq. (D.1) in the supercooled phase for the pressure range corresponding to Mishima's data. Fig. 130 shows that the deviation between the two models remains consistent with the uncertainty of the Mishima's data. However, there is a systematic deviation in the same trend when the pressure increases whatever the temperature, which indicates a systematic shift of the Holten *et al.*'s model with respect to Mishima's data since the error with Eq (D.1) is centered.
The calculation of the density in the model of Holten *et al.* is delicate because of the determination of the parameter $x_e$ by iteration (i.e. Eq. (18) of Ref. 102). The advantage of the Eq (D.1) is that the density can be easily calculated directly and the calculation can be easily extrapolated to any temperature below 200 K.

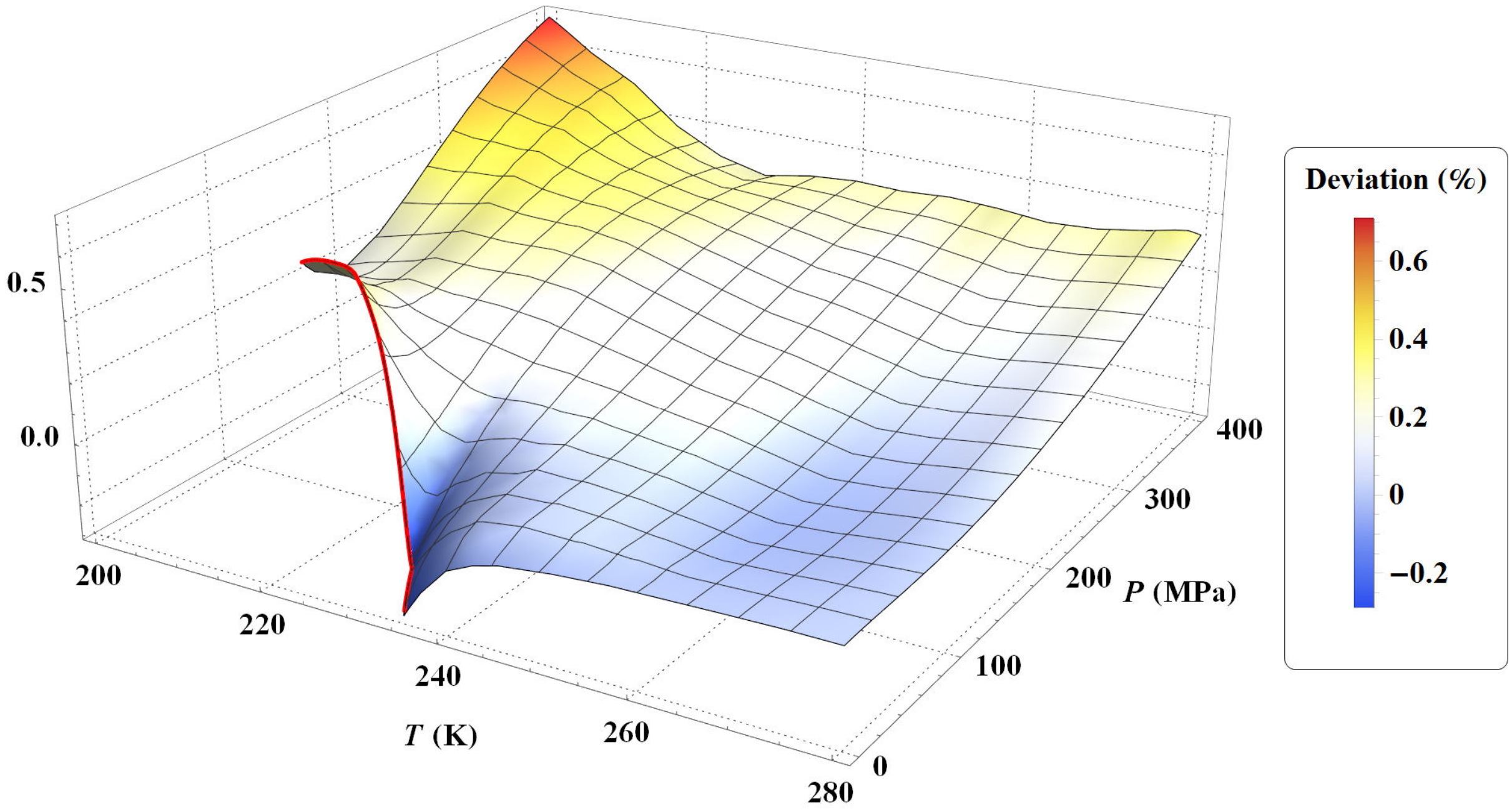

Fig. 130. Percentage deviation of the equation of state from Holten *et al.* (Ref. 102) with Eq. (D.1), in the temperature range from 200 K to 280 K. The vertical axis represents $100\left(V_{\text{Holten}} - V_{\text{Eq.(D.1)}}\right)/V_{\text{Eq.(D.1)}}$ . The red curve represents the lower part of the homogeneous nucleation curve corresponding to Eq. (A1) of Ref. 102.

In conclusion, it appears that the Tait-Tammann equation of state proposed here is therefore suitable for calculating *P*-*V*-*T* data in the supercooled phase down to the lowest temperatures for viscosity and diffusion coefficient data of water. In addition, this equation of state provides a perfect match with the values of the IAPWS-95 formulation (Ref. 17) in the normal liquid phase.

## 11 APPENDIX E

This Appendix contains the density formula corresponding to the Frenkel/Widom line (determined from the isochoric heat capacity peaks of the IAPWS-95 formulation, Ref. 17):

$$\rho_{\text{Supercrit}}(T \ge T_c) = \left\{ 0.23652 + 1.5303\left(\frac{T}{T_c} - 1\right)^{1.616} + 0.4535\exp\left(-\left(\frac{T - T_c}{6.401}\right)^{1.0077}\right) + 0.043463\exp\left(-\left|\frac{T - 670.523}{10.885}\right|^{1.944}\right) + 0.0192\exp\left(-\left|\frac{T - 705.5}{33.81}\right|^{1.52}\right)\right\}\left[1 + \left(\frac{648}{T}\right)^{100}\right]^{-1} \quad \text{(E.1)},$$

where the density is in g/cm$^3$ and the temperature is in Kelvin.

Eq. (E.1) allows, using the IAPWS-95 formulation, to calculate the corresponding pressure states. Fig. 131 shows that in a *P*-*T* diagram the Frenkel/Widom line is in the continuation of the saturated vapour pressure curve as expected.

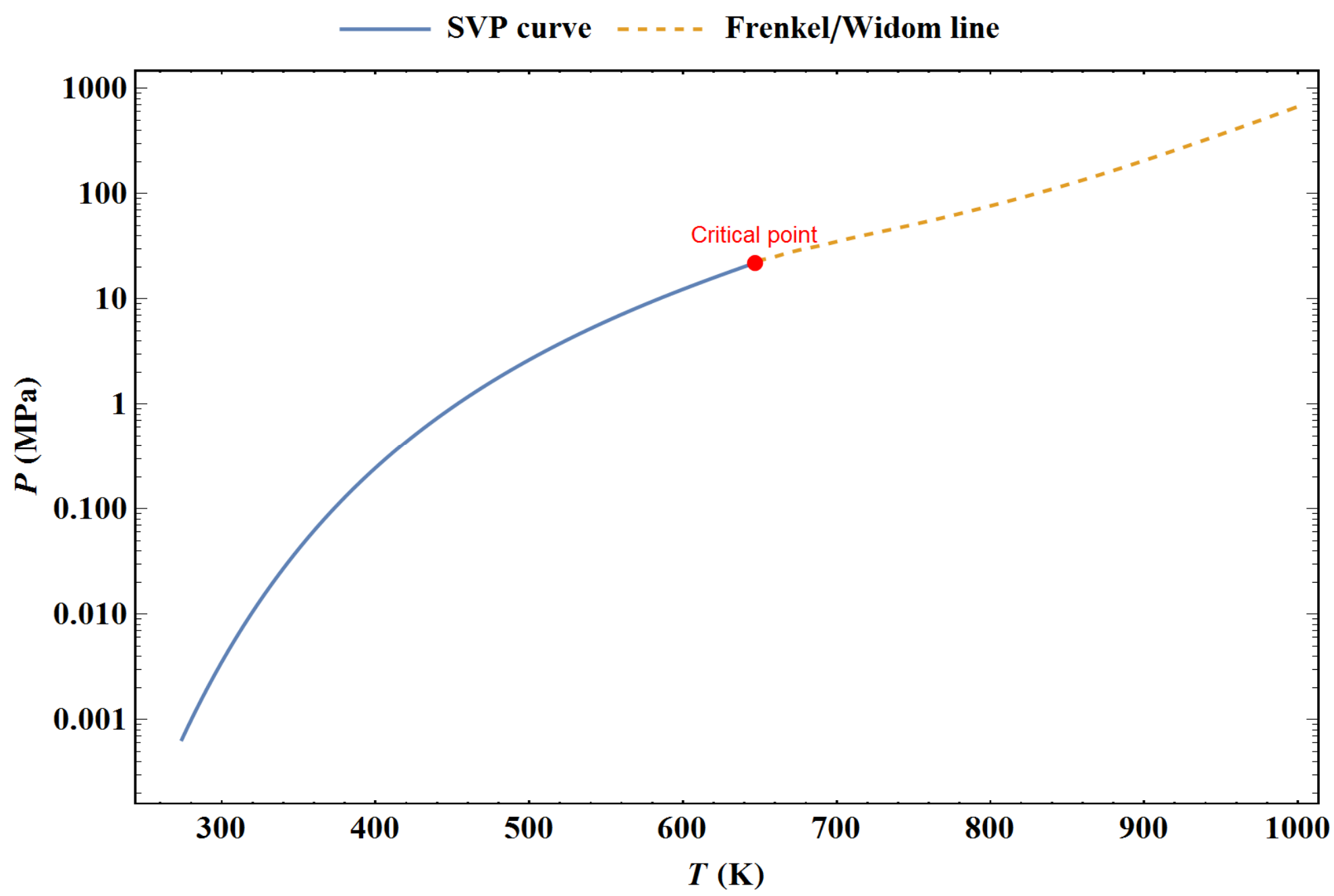

Fig. 131. *P*-*T* diagram of water calculated from the IAPWS-95 formulation: the blue line represents the Saturated Vapour Pressure curve (SVP) and the dashed line represents the Frenkel/Widom line determined by using Eq. (E.1).